\documentclass[reprint,footinbib,amsmath,amssymb,aps,prx,floatfix,twocolumn,superscriptaddress]{revtex4-2}

\usepackage{physics}
\usepackage{printlen} 
\usepackage{amsthm}
\usepackage{mathtools}

\usepackage{graphicx}
\graphicspath{{images/}}
\usepackage{wrapfig}

\usepackage{dcolumn}
\usepackage{bm}
\usepackage[hidelinks]{hyperref}
\usepackage{tikz}
\usepackage{dsfont}
\usetikzlibrary{quantikz}

\usepackage{diagbox}

\usepackage{lipsum} 

\usepackage[labelformat=simple]{subcaption}

\usepackage{ragged2e}
\DeclareCaptionJustification{justified}{\justifying}
\usepackage{printlen}

\usepackage{etoolbox}

\usepackage[innercaption]{sidecap}
\sidecaptionvpos{figure}{c}

\usepackage{siunitx}
\AtBeginDocument{\RenewCommandCopy\qty\SI}
\makeatletter
\renewcommand{\p@subsubsection}{}
\makeatother

\newcommand{\n}[1]{\mathrm{#1}}
\newcommand{\cl}[1]{\mathcal{#1}}
\newcommand{\bb}[1]{\mathbb{#1}}
\newcommand{\h}[1]{\hat{#1}}

\newcommand{\s}{\hat{\sigma}}

\newcommand{\be}{\begin{equation}} 
\newcommand{\ee}{\end{equation}}

\newcommand{\dif}{\scalebox{0.7}{\ensuremath{\Delta}}}

\newcommand{\SB}{\mathtt{SB}}

\newcommand*\pFqskip{8mu}
\catcode`,\active
\newcommand*\pFq{\begingroup
        \catcode`\,\active
        \def ,{\mskip\pFqskip\relax}%
        \dopFq
}
\catcode`\,12
\def\dopFq#1#2#3#4#5{%
        {}_{#1}F_{#2}\biggl[\genfrac..{0pt}{}{#3}{#4};#5\biggr]%
        \endgroup
}

\newcommand{\PRLsep}{\medskip\noindent\makebox[\linewidth]{\resizebox{0.3333\linewidth}{1pt}{$\bullet$}}\medskip}

\newcommand{\cz}{\text{c}_\zeta}
\newcommand{\sz}{\text{s}_\zeta}

\begin{document}
\preprint{APS/123-QED}

\title{Stabilization of cat-state manifolds using nonlinear reservoir engineering}

\author{Ivan Rojkov}
  \email{irojkov@phys.ethz.ch}
  \thanks{These authors contributed equally.}
\author{Matteo Simoni}
  \email{masimoni@phys.ethz.ch}
  \thanks{These authors contributed equally.}
\author{Elias Zapusek}
\affiliation{
    Institute for Quantum Electronics,
    ETH Z\"{u}rich, Otto-Stern-Weg 1, 8093 Z\"{u}rich, Switzerland}
\affiliation{Quantum Center, ETH Zürich, 8093 Zürich, Switzerland}
\author{Florentin Reiter}
\affiliation{
    Institute for Quantum Electronics,
    ETH Z\"{u}rich, Otto-Stern-Weg 1, 8093 Z\"{u}rich, Switzerland}
\affiliation{Quantum Center, ETH Zürich, 8093 Zürich, Switzerland}
\affiliation{Fraunhofer Institute for Applied Solid State Physics IAF, 
             Tullastr. 72, 79108 Freiburg, Germany}
\author{Jonathan Home}
  \email{jhome@phys.ethz.ch}
\affiliation{
    Institute for Quantum Electronics,
    ETH Z\"{u}rich, Otto-Stern-Weg 1, 8093 Z\"{u}rich, Switzerland}
\affiliation{Quantum Center, ETH Zürich, 8093 Zürich, Switzerland}

\date{\today}


\begin{abstract}

We introduce a novel reservoir engineering approach for stabilizing multi-component Schr{\"o}dinger's cat manifolds. The fundamental principle of the method lies in the destructive interference at crossings of gain and loss Hamiltonian terms in the coupling of an oscillator to a zero-temperature auxiliary system, which are nonlinear with respect to the oscillator's energy. The nature of these gain and loss terms is found to determine the rotational symmetry, energy distributions, and degeneracy of the resulting stabilized manifolds. Considering these systems as bosonic error-correction codes, we analyze their properties with respect to a variety of errors, including both autonomous and passive error correction, where we find that our formalism gives straightforward insights into the nature of the correction. We give example implementations using the anharmonic laser-ion coupling of a trapped ion outside the Lamb-Dicke regime as well as nonlinear superconducting circuits. Beyond the dissipative stabilization of standard cat manifolds and novel rotation symmetric codes, we demonstrate that our formalism allows for the stabilization of bosonic codes linked to cat states through unitary transformations, such as quadrature-squeezed cats. Our work establishes a design approach for creating and utilizing codes using nonlinearity, providing access to novel quantum states and processes across a range of physical systems.

\end{abstract}

\maketitle

\section{Introduction}

Dissipation is ubiquitous in nature. In the context of control of quantum systems, it is an essential resource for pumping into quantum states or manifolds, however often constitutes a hindrance to quantum coherence. The last decade has seen the development of increasingly intricate methods for engineering dissipation to directly produce quantum states (including entangled states) \cite{plenio_cavity_1999,diehl_quantum_2008,vacanti_cooling_2009,krauter_entanglement_2011,cho_optical_2011,lin_dissipative_2013,shankar_autonomously_2013,navarrete_inducing_2014,reiter_scalable_2016,cole_resource_efficient_2022,malinowski_generation_2022}, as well as to achieve protected manifolds of states required for quantum error correction \cite{paz_continuous_1998,barnes_automatic_2000,sarovar_continuous_2005,pastawski_quantum_2011,kapit_hardware_2016,reiter_dissipative_2017,lihm_implementation_2018, de_neeve_error_2022}. These methods involve engineering the coupling to a bath system which removes entropy, and are known as \textit{reservoir engineering}~\cite{poyatos_quantum_1996}. Such dissipative techniques have also been applied to the implementation of quantum simulation \cite{barreiro_open_2011,raghunandan_initialization_2020} and have been shown to be universal for quantum computing \cite{verstraete_quantum_2009}.

Bosonic harmonic oscillator systems were among the first considered for reservoir engineering~\cite{poyatos_quantum_1996}, which formed the basis for experiments demonstrating stabilization of a wide range of states, including displaced coherent~\cite{kienzler_quantum_2015}, squeezed~\cite{cirac_dark_1993,kronwald_arbitrarily_2013,kienzler_quantum_2015,lo_spinmotion_2015,wollman_quantum_2015,kienzler_quantum_2017}, entangled~\cite{behrle_phonon_2023}, and many-body states~\cite{ma_dissipatively_2019}. These works have been extended to the rich field of bosonic error correction codes, in which the infinite-dimensional Hilbert space of a quantum harmonic oscillator is used to encode quantum information redundantly with a low hardware overhead. The two primary classes of bosonic codes are translation-symmetric codes and rotation-symmetric codes~\cite{grimso_rotational_2020}, exemplified by Gottesman-Kitaev-Preskill (GKP)~\cite{gottesman_encoding_2001} and cat codes~\cite{cochrane_macroscopically_1999,mirrahimi_dynamically_2014}, respectively. In this context, reservoir engineering techniques have been used to stabilize code manifolds and to perform error correction in numerous experiments using superconducting circuits~\cite{leghtas_confining_2015,touzard_coherent_2018,grimm_stabilization_2020,campagne-ibarcq_quantum_2020,lescanne_exponential_2020,gertler_protecting_2021,kwon_autonomous_2022,sivak_real-time_2023,ni_beating_2023,berdou_one_2023,reglade_quantum_2024} and trapped ions~\cite{fluhmann_encoding_2019,de_neeve_error_2022}.

One prominent bosonic code is based on Schrödinger's cat states, defined as superpositions of two displaced coherent states of a quantum harmonic oscillator. The stabilization of such states using engineered two-photon dissipation ~\cite{hach_iii_generation_1994,gilles_generation_1994,poyatos_quantum_1996,de_matos_filho_even_1996} has since been the subject of numerous studies~\cite{sarlette_stabilization_2011,mundhada_generating_2017,mamaev_dissipative_2018,putterman_stabilizing_2022}. These methods effectively stabilize two-dimensional manifolds of coherent states~\cite{mirrahimi_dynamically_2014,albert_symmetries_2014}, giving rise to the aforementioned cat code~\cite{cochrane_macroscopically_1999,mirrahimi_dynamically_2014,albert_pair_cat_2019}. The engineered dissipation offers inherent protection against dephasing errors of the harmonic oscillator~\cite{mirrahimi_dynamically_2014}, and enables universal quantum computation in parallel with the stabilization~\cite{mirrahimi_dynamically_2014,goto2016universal,albert_holonomic_2016,touzard_coherent_2018,puri_bias_preserving_2020,grimm_stabilization_2020,gautier_designing_2023}. Furthermore, studies have theoretically shown that squeezed Schrödinger's cat states could be stabilized using reservoir engineering~\cite{schlegel_quantum_2022,hillmann_quantum_2023}, providing additional protection against bosonic loss errors~\cite{xu_autonomous_2023}. 

Protocols for continuous reservoir engineering of bosonic modes have primarily been derived analytically, using an approach along the following lines. A Hamiltonian 
\be \label{eq:hamiltonian}
    \h{H} = \Omega(\h{K} \h{A}^\dagger + {\rm h.c.})
\ee 
couples a designed operator $\h{K}$ on the target oscillator to excitation of a second auxiliary system for which the raising operator is $\h{A}^\dagger$. This second system is damped, either naturally, or through optical pumping, with a resulting dissipation operator $\propto \sqrt{\gamma} \h{A}$. {Assuming that this damping happens at a shorter timescale than the Hamiltonian evolution,} adiabatic elimination allows the dynamics of the target oscillator to be described by a dissipative Lindblad equation acting on the oscillator alone $\dot{\h{\rho}} = \h{L}\h{\rho} \h{L}^\dagger - (\h{L}^\dagger \h{L}\h{\rho}  + \h{\rho} \h{L}^\dagger\h{L} )/2$ with $\h{L} \propto \sqrt{\Omega^2/\gamma}\,\h{K}$ \cite{poyatos_quantum_1996}. Steady state solutions $\ket{D}$ are dark states of the dynamics satisfying ${\h{K}\ket{D}=0}$, which depending on the form of $\h{K}$ may be unique or belong to a manifold. Choices for $\h{K}$ include $\h{a}$, $\h{a}-\alpha$, $\mu\h{a}^\dagger+\nu\h{a}$ or $\h{a}^2-\alpha^2$ which stabilize the ground, displaced coherent, squeezed (if $\mu^2-\nu^2=1$ and $|\nu|>|\mu|$) and Schrödinger cat states, respectively \cite{cirac_dark_1993,poyatos_quantum_1996,kronwald_arbitrarily_2013,wollman_quantum_2015,kienzler_quantum_2015,kienzler_quantum_2017,lo_spinmotion_2015,mirrahimi_dynamically_2014}. The simple nature of these operators makes predicting the steady-states straightforward. Such $\h{K}$ operators are also consistent with the typical operating regimes for quantum mechanical control, which result from low-order expansions of native Hamiltonians near the quantum ground state. As an example, trapped ions are primarily controlled in the Lamb-Dicke regime, which considers low-order expansion in $\eta$ of the interaction $e^{i\eta (\h{a} + \h{a}^\dagger)}\h{\sigma}_+$ terms produced by the light-matter coupling~\cite{wineland_experimental_1998}. For superconducting microwave resonators, similar low-order expansions (e.g. transmon) can be engineered by introducing Josephson junctions to a primarily linear resonator~\cite{blais_circuit_2021}. {However, these standard operating regimes have strong limitations. The $\h{K}$ operators above rely on low-order boson processes, yet higher-order processes enable codes with superior error correction capabilities, protecting against a broader range of errors~\cite{leghtas_hardware_2013,mirrahimi_dynamically_2014,grimso_rotational_2020,gertler_protecting_2021}. For instance, $d$-legged cat~\cite{leghtas_hardware_2013,mirrahimi_dynamically_2014} are stabilized via ${\h{K}=\h{a}^d-\alpha^d}$, requiring a $d$th-order boson loss process. Moreover, operations on bosonic codes often rely on higher-order processes~\cite{albert_holonomic_2016,gautier_designing_2023} as well. In the standard low-excitation operating regimes, such processes are inaccessible due to their coupling strength decreasing exponentially with $d$ (see Appendix~\ref{app:nonlinear_important}). By relaxing the low-excitation approximation, the strength of higher order processes is no longer negligible, which enables broader operational capabilities. While experimentally accessible~\cite{mcdonnell_long-lived_2007,stutter_sideband_2018,hrmo_sideband_2019,jarlaud_coherence_2020,hofheinz_bright_2011,rolland_antibunched_2019,peugeot_generating_2021,menard_emission_2022,cohen_degeneracy-preserving_2017,smith_superconducting_2020,smith_magnifying_2022,smith_spectral_2025}, these strong coupling regimes introduce operators of the form $f(\h{n})\h{a}^d$ where $f$ is a non-trivial function of the number operator $\h{n}$. Such terms do not fit within the analytically well-behaved $\h{K}$ operators studied so far and remain challenging to analyze with existing methods. This highlights the need for more general and intuitive theoretical frameworks to describe strong coupling regimes. Such a framework could systematically elucidate the fundamental mechanisms behind reservoir engineering, enable novel experimentally accessible bosonic codes, and address the open question of how many boson processes are required for autonomous correction of both dephasing and loss errors.
}

In this Paper, we introduce a general scheme for identifying processes which stabilize multi-component bosonic cat manifolds using a method that we refer to as \emph{Nonlinear Reservoir Engineering} (NLRE). The method involves identifying crossings in the strengths of terms inducing boson gain and loss in dissipators formed as a coherent sum of these two. Engineering the nature of these crossings allows the production of various dark-state distributions, which feature manifolds of superposed displaced (number) squeezed states with discrete rotational symmetry. We analyze the convergence into the manifolds and relate the error-correction properties to the types of gain and loss terms employed. We propose experimental implementations based both on the nonlinear interaction induced by a laser between the internal and motional states of an ion, as well as on recently developed nonlinear superconducting elements in a circuit QED system. We then show how these schemes can be extended to more complex stabilization operators, allowing for stabilization of more general bosonic manifolds protected from a larger range of errors. Our work represents a design technique which departs significantly from the established analytic treatment, opening up the control and  stabilization of states in regimes where nonlinearities are strong. In this sense, our approach provides a new avenue for the efficient generation and protection of a range of bosonic codes by identifying and engineering relevant nonlinear processes.

\begin{figure*}
    \includegraphics[]{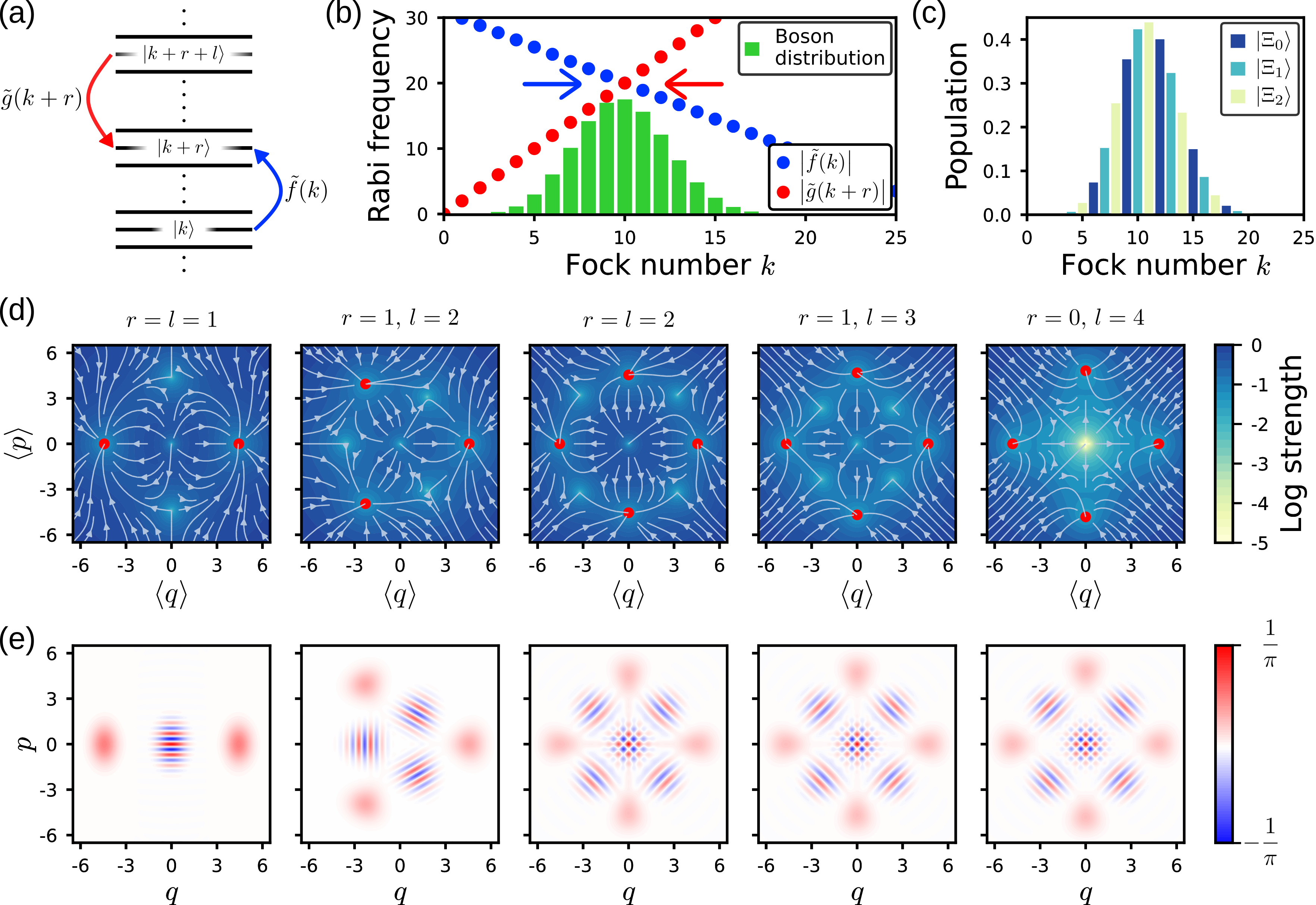}
    {\phantomsubcaption\label{fig1:competition}}
    {\phantomsubcaption\label{fig1:rabi_frequency}}
    {\phantomsubcaption\label{fig1:different_steady_states}}
    {\phantomsubcaption\label{fig1:classical_dynamics}}
    {\phantomsubcaption\label{fig1:wigner_functions}}
    \caption{Toy model of the nonlinear reservoir engineering method.
    (a) The interaction in Eq.~\eqref{eq:hamilton_modif} is engineered such that the bosonic mode is simultaneously subject to competing nonlinear boson raising $\tilde{f}$ and lowering $\tilde{g}$ processes of orders $r$ and $l$, respectively. 
    (b) The boson distribution is stabilized around the crossing of the strengths (termed Rabi frequency) of $\tilde{f}$ and $\tilde{g}$. 
    (c) The dissipative dynamics stabilizes a manifold of dimension ${d=r+l}$ around the crossing point (here, $r=1$ and $l=2$) with basis states $\ket{\Xi_\mu}$ that have Fock state population occupied every $d$ states.
    (d) Classical phase-space trajectories. The stable critical points of the classical dynamics are highlighted by red dots, suggesting the existence of $d$ attractors exhibiting rotational symmetry relative to the origin.
    (e) Wigner quasiprobability of steady states for the processes $\tilde{f}$ and $\tilde{g}$ shown in (c) for various orders. The dimension $d$ sets the number of nonlinear coherent states composing the stabilized state.}
    \label{fig1}
\end{figure*}

\section{Summary of main results and structure of the paper} \label{sec:summary}

In this paper, we introduce the paradigm of Nonlinear Reservoir Engineering (NLRE). We consider a quantum harmonic oscillator coupled to a damped auxiliary system as described in Eq.~\eqref{eq:hamiltonian} with the bosonic operator $\h{K}$ given by
\be
\label{eq:hamilton_modif}
\begin{split}
    \h{K} &:= \h{a}^{\dagger\,r} f(\h{n}) - g(\h{n}) \h{a}^{l} \\
    &\equiv \sum_{k} \tilde{f}(k)\dyad{k+r}{k} - \tilde{g}(k)\dyad{k}{k+l}\,.
\end{split}
\ee
This operator is comprised of one boson raising and one boson lowering processes of orders $r$ and $l$ whose action is illustrated in Fig.~\ref{fig1:competition}. The nonlinear aspect of the paradigm consists in the fact that $\h{K}$ is a nonlinear combination of the ladder operators $\h{a}^{\dagger\,r}$ and~$\h{a}^{l}$. The scalar functions ${\tilde{f}(k):=\bra{k+r}\h{a}^{\dagger\,r} f(\h{n}) \ket{k}}$ and ${\tilde{g}(k):=\bra{k} g(\h{n}) \h{a}^{l} \ket{k+l}}$ characterize the strength of the processes and are referred to as \emph{Rabi frequencies}~{\footnote{{The term originates from the description of coherent driving in closed systems without dissipation, and we retain it here for historical consistency.}}}. Our approach relies in engineering these functions to induce a destructive interference between the boson raising and lowering processes which results in the stabilization of a manifold. We do this by noting that in regions for which the gain ${G\equiv|\tilde{f}(k)|/|\tilde{g}(k + r)| > 1}$, the bosonic system is driven towards higher energy states {leading to a dynamically unstable system}, while for ${G < 1}$, the system is driven towards lower energies. Thus, by engineering the functions $\tilde{g}(k+r)$ and $\tilde{f}(k)$ such that $G$ switches from $>1$ to $<1$ at a point ${k = k^*}$, the population of the system will accumulate around the crossing as depicted in Fig.~\ref{fig1:rabi_frequency}. We denote the value of functions at this point $|\tilde{f}(k^*)| = |\tilde{g}(k^*+r)| \equiv h^*$. Since both raising and lowering processes are part of the same jump operator, they will destructively interfere. As a consequence, the dissipator allows for a dark subspace of pure states localized near $k^*$ (see Fig.~\ref{fig1:different_steady_states}) that we refer to as a manifold.

Figure~\ref{fig1} illustrates three key characteristics of these stabilized states: 
\begin{enumerate} \setlength{\itemsep}{1pt}
    \item the average boson number, determined by the position of the crossing, 
    \item the variance of the boson number distribution, dictated by the rate at which the strengths of the two processes change relative to each other near the crossing point, 
    \item the $d = r + l$ dimensional manifold spanned by states with a $d$ fold rotational symmetry in phase space. 
\end{enumerate}
This shows that the method can produce a rich variety of multi-component (number-squeezed) cat manifolds. This is complemented by semi-classical phase space attractor diagrams given in Fig.~\ref{fig1:classical_dynamics}, which are obtained from a classical model formed using the Husimi Q representation of the operators (further details are given in Appendix~\ref{app:classical_phase_space}). This classical picture indicates the presence of multiple critical points (i.e. points with vanishing first derivatives) of various natures (stable, saddle, and unstable). Empirically, we observe that the number of stable points corresponds to $d=r+l$ which suggests the existence of a set of $d$ dark states which are rotation symmetric relative to the origin. The classical picture does not, however, provide information about the nature of the dark states. Examining these in more detail, we show that the dark states are pure and that they constitute a stabilized subspace of the model.

Such stabilized subspaces can serve as rotation-symmetric error correction codes~\cite{grimso_rotational_2020}. Indeed, the well-known cat codes~\cite{cochrane_macroscopically_1999,mirrahimi_dynamically_2014,albert_pair_cat_2019} are a special case, with the operator ${\h{K}=\h{a}^2-\alpha^2}$ corresponding in our framework to $(r,l) = (0, 2)$ and $f(\h{n}) = \n{const} = g(\h{n})$~\cite{poyatos_quantum_1996,leghtas_confining_2015}. {For stabilization of general cat codes of dimension $d$ one usually requires a boson lowering process of order $d$, i.e, ${\h{K}=\h{a}^d-\alpha^d}$. Our observations show however that this is not a unique method for obtaining a $d$-dimensional cat manifold. Choices such as ${(r,l) = (0, 2), (1,1), (2, 0)}$ will all result in a cat qubit (i.e. $d=2$) for suitable $\tilde{g}$ and $\tilde{f}$ dependencies. Similarly, schemes such as $(1,3), (2,2)$ and $(3, 1)$ will stabilize a four-legged cat code, while $(5, 3)$ will result in a manifold of dimension $d=8$. This previously unreported feature has a key practical implication: stabilizing $d$-dimensional cat manifolds no longer requires engineering boson processes of order $d$. This relaxes the experimental requirements for stabilizing $d>2$ codes, which has been challenging in the standard low-excitation operating regime of quantum systems (see Appendix~\ref{app:nonlinear_important}).}

The remainder of the paper investigates these observations in more detail. First, in Sec.~\ref{sec:simplest_nlre_setting}, we analyze the foundations giving theoretical underpinning to the observations above. By linearizing the nonlinear functions $\tilde{f}$ and $\tilde{g}$ around their crossing point $k^*$, we offer an approximate analytical expansion of the stabilized state distributions that closely matches their actual boson distributions. This approximation enables us to treat these states analytically and derive various properties, including the mean and variance of boson number, along with the linewidth of the engineered dissipative process, often referred to as the effective confinement rate~\cite{mirrahimi_dynamically_2014,chamberland_building_2022,gautier_combined_2022}. We connect our work to standard cat-state stabilization as well as nonlinear coherent states by showing that these are special limits of the resulting distributions  
\cite{de_matos_filho_nonlinear_1996, dodonov_nonclassical_2002, manko_trapped_2000}.

In Sec.~\ref{sec:error_correction}, we explore the error correction capabilities of these nonlinearly stabilized cat states. Leveraging the rotational symmetry inherent to these manifolds, we establish the conditions for correcting various bosonic errors. Specifically, we compare cases where the parameters ${d,\,h^*,\,k^*}$ are held constant with different $r$ and $l$, which stabilize the same manifolds but have different error correction capabilities depending on the commutation relation between $\hat{K}$ and the rotational symmetry operator $e^{2\pi i \hat{n}/d}$. We find that cases with $(r,l)=(0,d)$ or $(d,0)$ possess both the ability to autonomously correct against dephasing errors, thus generalizing the established results of the standard cat codes. Furthermore, reducing the variance of the boson distribution of the steady states enhances the error protection thanks to a greater confinement rate. {This in turn allows to increase the noise bias of cat codes by several orders of magnitude.} Additionally, we find that models with $r=l=1$ have a specific error correction property of being able to protect against errors in one of the oscillator's quadratures, $\hat{q} \propto \hat{a}^\dagger + \hat{a}$ or $\hat{p} \propto i(\hat{a}^\dagger - \hat{a})$. When the cat states lie along $\hat{q}$ in phase space, the stabilization corrects incoherent momentum errors, and vice versa. We show that this feature is central to the robustness of the quadrature-squeezed cat states against boson loss and gain errors~\cite{xu_autonomous_2023}.

In Secs.~\ref{sec:trapped_ions} and~\ref{sec:cqed_setting}, we illustrate the applicability of our methods by proposing implementations in both trapped ions and superconducting circuits.  In the trapped-ion setting, we find that for the standard ion-light interaction the functions $\tilde{f}$ and $\tilde{g}$ are well approximated by the Bessel functions produced using resonant driving of motional sideband transitions \cite{wineland_experimental_1998, kienzler_quantum_2015}, and show how their intersection points and derivative at these points can be adjusted by manipulating experimental parameters. NLRE thus opens the avenue for realizing a wider-range of bosonic error-correction codes in trapped ion systems and emphasizes the richness of trapped ion physics beyond the Lamb-Dicke regime~\cite{mcdonnell_long-lived_2007,stutter_sideband_2018,hrmo_sideband_2019,jarlaud_coherence_2020}. In the superconducting circuit QED architecture, we reinterpret previous cat state stabilization experiments~\cite{leghtas_confining_2015,touzard_coherent_2018} through the lens of NLRE. Furthermore, we show that combining recently realized methods involving a cavity capacitively coupled to an asymmetrically threaded superconducting quantum interference device (ATS)~\cite{lescanne_exponential_2020, berdou_one_2023, reglade_quantum_2024}, together with a voltage bias~\cite{hofheinz_bright_2011,rolland_antibunched_2019,peugeot_generating_2021,menard_emission_2022}, provides suitable nonlinear control to realize the toy model in Eq.~\eqref{eq:hamilton_modif}. The nonlinearities of the circuit are then similar to those in the trapped-ion setting and can be exploited to stabilize a variety of dark state manifolds in the cavity. Both in trapped ions and superconducting circuits, we note that these methods include proposals for realizing rotation-symmetric bosonic codes not demonstrated to date on any platform.

While the form of Eq.~\eqref{eq:hamilton_modif} contains only one raising and one lowering term, it is also possible to consider scenarios in which multiple terms of each are engineered. In Sec.~\ref{sec:bogoliubov_transf} we provide a first example of this by transforming the ladder operators isomorphically, enabling the identification of more general stabilized manifolds. As an example, we employ linearly (Bogoliubov) transformed $\h{a}$ and $\h{a}^\dagger$~\cite{kienzler_quantum_2015,lo_spinmotion_2015,kienzler_quantum_2017} and illustrate the use of the NLRE method to stabilize quadrature-squeezed cat state manifolds. Additionally, we show that the recently proposed stabilization method, which autonomously corrects these states against boson loss~\cite{xu_autonomous_2023}, corresponds to an isomorphic transformation of the $r=l=1$ scheme. In this case, the noise is transformed into momentum errors, explaining its correctability under stabilization. {{This suggest that} a minimum of four boson processes, two first-order and two third-order, is required to autonomously protect a code against both dephasing and loss errors.} The method offers a novel systematic approach for both studying the experimental requirements for realizing such codes and conducting theoretical investigations of new bosonic codes with enhanced error correction capabilities. 

Finally, Sec.~\ref{sec:conclusion} gives a review of the results and possibilities for further extensions, highlighting open questions before concluding.

\section{Steady states and manifolds} \label{sec:simplest_nlre_setting}

In this section, we investigate in more detail the steady states of the Lindblad master equation created by the dissipator in Eq.~\eqref{eq:hamilton_modif}. We refer to this as the NLRE toy model. The steady states are the zero-eigenvalue eigenstates of the $\h{K}$ operator. We establish the recurrence relation for the superpositions of Fock states occurring in the dark states, which determines their rotation symmetry and number state distributions. By focusing on crossings in which $\tilde{f}$ and $\tilde{g}$ vary linearly in the vicinity of the crossing, we find analytical boson distributions and relate these to simple examples previously found in the literature.

\subsection{Steady states distributions and symmetry} \label{subsec:number_of_steady_states}
The dissipative process ${\cl{L}(\cdot):=\h{L}\cdot\h{L}^\dagger - 1/2 \{\h{L}^\dagger\h{L},\cdot\}}$ admits a \textit{pure} stationary state ${\ket{\Xi}=\sum_k \xi_k \ket{k}}$ if and only if ${\h{L}\ket{\Xi}=0}$~\cite{kraus_preparation_2008}. Using the operator in Eq.~\eqref{eq:hamilton_modif}, this condition leads to a recurrence relation
\begin{align}
    &\xi_{k+r+l}  \quad\,=\,\, \displaystyle\frac{\tilde{f}(k)}{\tilde{g}(k +r)} \, \xi_{k}
    &&\text{for} \quad k \geq r \,, \label{eq:recusive_relation} \\[3pt]
    &\xi_{k+l}\,\,\tilde{g}(k) =\,\, 0 &&\text{for} \quad k < r \,.
    \label{eq:recusive_relation_leakage}
\end{align}
Eq.~\eqref{eq:recusive_relation} reflects the destructive interference between $\tilde{f}$ and $\tilde{g}$ depicted in Fig.~\ref{fig1:competition}. The absolute value of the ratio sets the relation between the boson population in states $\ket{k+r+l}$ and $\ket{k}$, whereas its argument gives the condition on their phase relationship. Assuming that $\arg(\tilde{f}(k))$ and $\arg(\tilde{g}(k))$ are constant for all $k$~\footnote{We keep this assumption for the remaining of the paper.}, we find that ${\arg(\xi_{k}) = \frac{k}{r+l}\left(\!\arg(\tilde{f}) - \arg(\tilde{g})\!\right)}$. The recurrence equation for $k \geq r$ needs $d=r+l$ initial values to define a sequence $\{\xi_k\}$. The normalization condition for the state $\ket{\Xi}$ requires this sequence to be convergent or in other words the recurrence equation to be stable. From its characteristic polynomial, we determine the constraints on the functions $\tilde{f}$ and $\tilde{g}$,
\be \label{eq:convergence_criterion}
    \lim_{k\rightarrow\infty} \abs{\frac{\tilde{f}(k)}{\tilde{g}(k+r)}} < 1 \, .
\ee

\begin{figure}
    \includegraphics[]{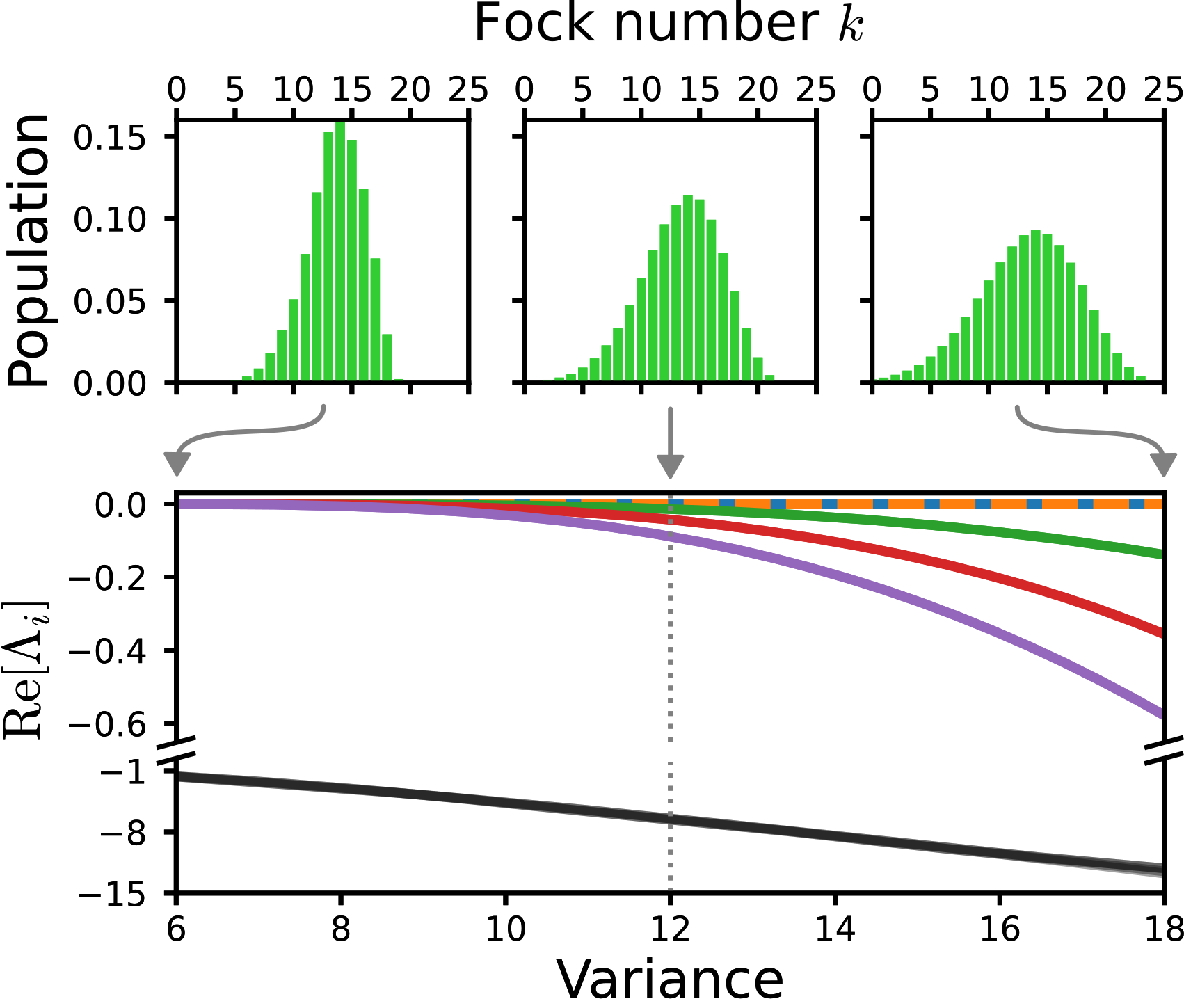}
    \caption{Change from five to two steady states as the state's variance increases. The lower panel illustrates the real part of the nine lowest eigenvalues $\Lambda_i$ of the effective Liouvillian $\cl{L}$ for the case ${(r,l)=(3,2)}$ with $\tilde{f}$ and $\tilde{g}$ as in Fig.~\ref{fig1:rabi_frequency}. The system exhibit two dark states associated to $\Lambda_{1/2}=0$ and three metastable states associated to $\Lambda_{3/4/5}\approx0$. Leakage from the latter into the former starts when the steady state predicted solely by Eq.~\eqref{eq:recusive_relation} has some non negligible population in the first five Fock states. {The lower broken axis illustrates the real parts of the eigenvalues $\Lambda_{6/\ldots/9}$ corresponding to the first excited eigenstates of the Liouvillian.}}
    \label{fig3_new}
\end{figure}

\begin{figure*}
    \includegraphics[]{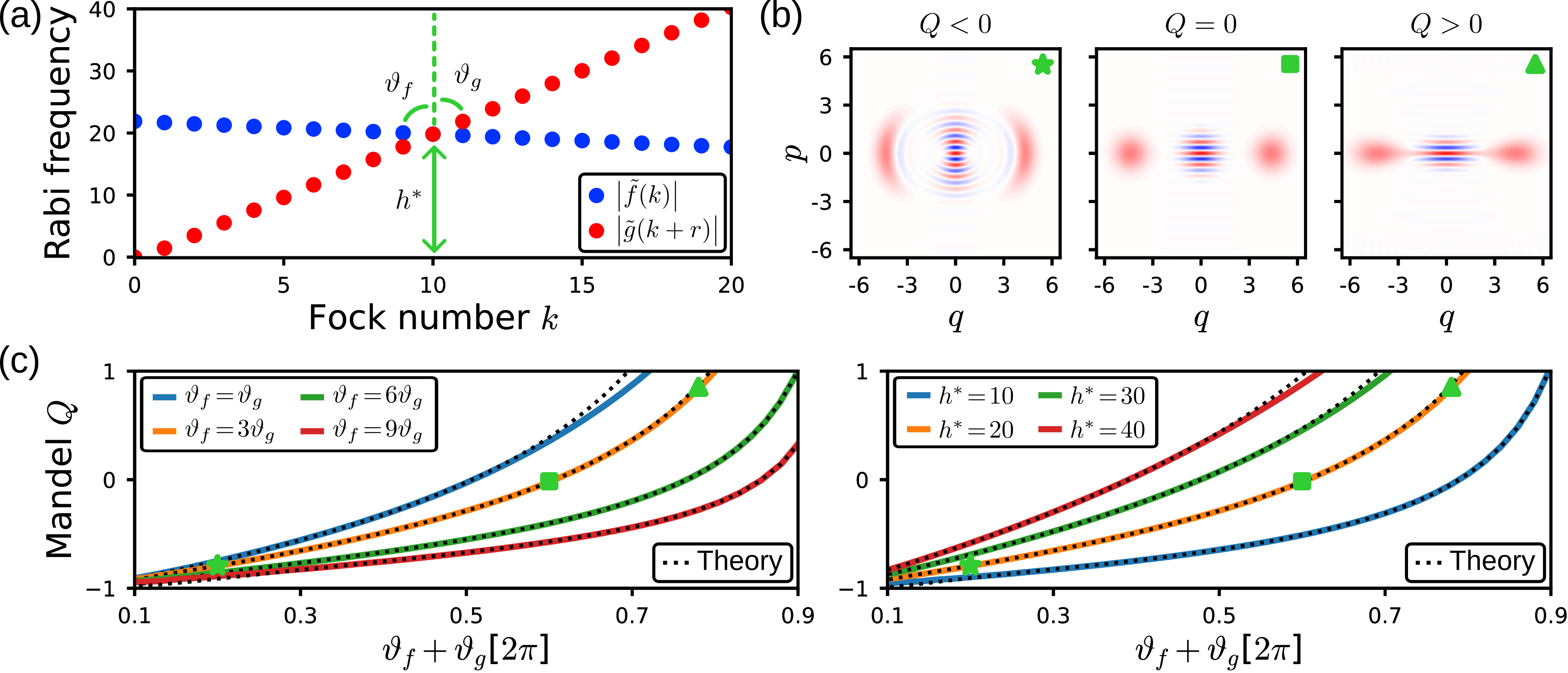}
    {\phantomsubcaption\label{fig3:rabi_frequency}}
    {\phantomsubcaption\label{fig3:wigner_functions}}
    {\phantomsubcaption\label{fig3:vary_angle_height}}
    \caption{Shape of the stabilized states. (a) Illustration of the parameters affecting the form of the steady states, specifically the angle between the functions $\tilde{f}$ and $\tilde{g}$, and the value $h^*$ of the Rabi frequency at the intersection point $k^*$. (b) Classification of the shapes into three primary categories based on the Mandel $Q$ parameter: number-squeezed states for $Q<0$, coherent shaped states for $Q=1$, and phase-squeezed shaped states for $Q>0$. (c) The shape can be modified by varying the angle $\vartheta_f+\vartheta_g$ between the functions. An increase in the ratio of $\vartheta_f$ to $\vartheta_g$ impacts the skewness of the stabilized boson distribution and decreases $Q$. The shape of the steady state is also governed by $h^*$, with larger values resulting in an increase in $Q$. Theoretical curves were obtained using variance and mean boson number given in Eq.~\eqref{eq:mean_and_variance_CMB}. In all situations $k^*=10$ and $(r,l)=(0,2)$, on the left $h^*=20$ while on the right $\vartheta_f=3\vartheta_g$.}
    \label{fig3}
\end{figure*}

Provided adequate $\tilde{f}$ and $\tilde{g}$ and ignoring for now the condition in Eq.~\eqref{eq:recusive_relation_leakage}, $\cl{L}$ seems to admit a set of $d$ pure dark states ${\ket{\Xi_\mu}=\sum_{m\in\bb{N}}\,\, \xi_{\mu+d\,m} \,\,\ket{\mu+d\,m}}$ with ${\mu\in\{0,\ldots,d-1\}}$ (e.g. Fig.~\ref{fig1:different_steady_states} for $d=3$). These states possess a  $\bb{Z}_d$ rotational symmetry as they are invariant (up to a global phase) under the transformation ${\h{P}=\exp\left(i\,2\pi\!/\!d \, \h{n}\right)}$. 

We now examine the constraint imposed by Eq.~\eqref{eq:recusive_relation_leakage}, which requires that either the dark state components $\xi_\mu$ for ${l \leq \mu < d}$ or the values of the $\tilde{g}$ function for the first $l$ Fock states vanish identically. This constraint emerges from the absence of a raising process for these Fock states. Thus, if this condition is not met, the destructive interference between the two processes cannot occur and the states $\ket{\Xi_\mu}$ for ${l \leq \mu < d}$ are no longer true dark states of the Liouvillian. The eigenvalues of $\cl{L}$ associated to these states will have a non-zero real part meaning that initializing the system in one of them would result in an exponential decay of the population into the true dark states~\cite{minganti_spectral_2018}. The rate of this undesired process is proportional to $\abs{\xi_{\mu}}$ and to the value of $\tilde{g}$ near the Fock ground state~\footnote{We can in fact identify two processes: when $l>0$, the population in $\ket{\Xi_\mu}$ with ${l \leq \mu < d}$ will leak into the other dark states; when $l=0$, an incoherent dephasing will happen.}. Consequently, it can be mitigated by engineering {the Hamiltonian in Eq.~\eqref{eq:hamiltonian}} such that ${\tilde{g}(k) = 0}$ $\forall\,k<r$, or by stabilizing states with sufficiently small population in Fock states ${\{\ket{\mu}\,|\,l\leq\mu< d\}}$. { Linearly decreasing the population of these Fock states (or, equivalently, changing the mean and/or variance of the boson distribution) results in an exponential increase of the lifetime of $\ket{\Xi_\mu}$ with ${l \leq \mu < d}$. We therefore refer to these approximate dark states as \emph{metastable}.} In summary, the system possesses $l$ true dark states and $r$ {metastable} states. In the special case of ${r=0}$, all the $d$ steady states are true dark states. This phenomenon is evident when tracking the $d$ lowest magnitude eigenvalues $\Lambda_i$ of the effective Liouvillian $\cl{L}$ while varying the steady-state variance. Fig.~\ref{fig3_new} illustrates this for the specific case of $(r,l)=(3,2)$. We observe that $\n{Re}\left[\Lambda_i\right]$ for $i>2$ deviates from zero as soon as the steady state obtained using Eq.~\eqref{eq:recusive_relation} exhibits non-negligible population in Fock states $\ket{k}$ with $k\in\{2,3,4\}$. Similar results are obtained if we keep the variance constant but change instead the position of the crossing point $k^*$. These observations suggest the occurrence of a potential dissipative phase transition~\cite{minganti_spectral_2018} when varying the parameters of some Rabi frequency $\tilde{g}$. Furthermore, one can utilize this feature to devise dissipative schemes with unique cat-like steady states which could be valuable as state preparation protocols. A particular instance of such a scheme was proposed in Ref.~\cite{mamaev_dissipative_2018}. {This phenomenon also explains why dissipative stabilization schemes for squeezed vacuum states~\cite{cirac_dark_1993,kienzler_quantum_2015} have unique dark states. In Appendix~\ref{app:squeeze_vacuum}, we show this by linking these schemes to the NLRE framework. Note that similar multi-timescale effects have been predicted in the presence of dissipative leakage channels~\cite{labay-mora_quantum_2023,labay-mora_quantum_2024,labay-mora_theoretical_2025}.}

Importantly, due to the effective Liouvillian{, obtained after adiabatic elimination of the auxiliary system acting as the bath,} being purely dissipative (i.e. no Hamiltonian part), we conclude that the set of dark states constitutes an exact or {metastable} subspace of dimension~$d$~\cite{kraus_preparation_2008}.

\subsection{Number-squeezed cat states} \label{subsec:number_squeezed_states}
We established above that the sum of orders $r$ and $l$ sets the dimension of the stabilized manifold. We are now interested in the shape of the states in both Fock and phase spaces.

In the presence of a crossing point $k^*$ where the ratio $\lvert\tilde{f}(k)/\tilde{g}(k+r)\rvert$ transitions from being larger than unity to lower than 1 with increasing $k$, every state $\ket{\Xi_\mu}$ peaks around it. The shape of the distribution is governed by the behavior of the two functions near $k^*$. When both $\tilde{f}$ and $\tilde{g}$ exhibit monotonic behavior, the shape is determined by the ratio of the slopes of the two functions at the intersection point as well as by their value at $k^*$. Modifying the slopes while keeping ${h^*:=\tilde{f}(k^*)=\tilde{g}(k^*+r)}$ constant enables the stabilization of bosonic states displaying Poissonian (like standard cat states), sub-Poissonian or super-Poissonian boson distributions. A similar effect can be achieved by varying $h^*$ and keeping the slopes constant. We illustrate this property in Fig.~\ref{fig3} where we choose $\tilde{f}$ and $\tilde{g}$ to be linearly decreasing and increasing functions, respectively.

In order to quantify the departure of boson distributions from Poissonian statistics we use the Mandel $Q$ parameter~\cite{mandel1979sub}, defined as $Q=\langle(\Delta \h{n})^2\rangle/\langle \h{n} \rangle - 1$ with $\langle(\Delta \h{n})^2\rangle$ and $\langle \h{n} \rangle$ the variance and expectation value of the number operator, respectively. The minimum value of $Q$ is $-1$, corresponding to single Fock states. States with ${Q=0}$ are similar to coherent states with potential non-zero higher order moments. Fig.~\ref{fig3:wigner_functions} presents Wigner quasiprobability distributions of three quantum states stabilized using NLRE method with different Mandel $Q$ values: $Q<0$, $Q=0$, $Q>0$.  We illustrate in Fig.~\ref{fig3:vary_angle_height} the dependence of this parameter on the characteristics of $\tilde{f}$ and $\tilde{g}$, i.e. their slopes (alternatively, their angles with a vertical line) and the height $h^*$, while keeping the crossing point at $k^*=10$. The total angle between $\tilde{f}$ and $\tilde{g}$ sets the variance $\langle(\Delta \h{n})^2\rangle$ of the stabilized states, while the ratio of one slope compared to the other sets the skewness of the boson distribution which also impacts $Q$. Increasing the height $h^*$ results in an increase of the variance and thus larger Mandel $Q$ parameters. This demonstrates that leveraging a system's nonlinearities can lead to the stabilization of a variety of bosonic states. The ability to characterize these states using only a few parameters adds a simplicity in comprehending and representing these systems.

\subsection{Crossings linear in $\tilde{f}$ and $\tilde{g}$} \label{subsec:linear_tilde_f_g}

When the Rabi frequencies are arbitrary continuously differentiable functions, we can derive a first-order approximation of the actual steady-state boson distribution by considering the first-order expansion of $\tilde{f}$ and $\tilde{g}$ around the crossing point $k^*$. This approximation holds when the functions exhibit sufficient linearity across the variance of the stabilized states. We therefore proceed to analytically investigate the case of linearly decreasing and increasing functions $\tilde{f}$ and $\tilde{g}$ (see Fig.~\ref{fig3:rabi_frequency}). {Notably, the original functions $f$ and $g$ are no longer essential and can be reconstructed if needed (see Appendix~\ref{app:f_g_linear}).} In this scenario, the recurrence relation in Eq.~\eqref{eq:recusive_relation} can be reformulated as
\be \label{eq:recusive_relation_linear}
 \xi_{k+d} = \frac{h^* - s_f ( k - k^*)}{h^* + s_g ( k - k^*)} \, \xi_{k}\,,
\ee
where ${s_f,\,s_g>0}$ denote the slopes of $\tilde{f}$ and $\tilde{g}$, and ${h^*=\tilde{f}(k^*)=\tilde{g}(k^*+r)}$~\footnote{When $\tilde{f}$ and $\tilde{g}$ are nonlinear, the slopes $s_f$ and $s_g$ are defined as $\frac{\partial}{\partial k}\,\tilde{f} (k^*)$ and $\frac{\partial}{\partial k}\,\tilde{g}(k^*+r)$, respectively.}. Solving this equation, we find that the boson distribution of the steady states follows a linearly transformed Conway--Maxwell--Binomial (CMB)~\cite{shmueli_useful_2005,borges_compoisson_2014,kadane_sums_2016,daly_conway-maxwell-poisson_2016} distribution ${\abs{\xi_{k}}^2 = \bb{P}[X=x(k)|m,\theta,2]}$ with
\be \label{eq:CMB_distribution}
 \begin{split}
     \bb{P}[X=x|m,\theta,2] &\propto \frac{\theta^x}{(x!(m-x)!)^2} \\[2pt]
     &\propto  \binom{m}{x}^{\!2} \,p^x\, (1-p)^{m-x}
 \end{split}
\ee
and parameters defined as
\be \label{eq:state_distribution_params}
 \begin{split}
     &x(k) = \frac{1}{d} \left( k - k^* + \frac{h^*}{s_g} \right) -1  \,, \\[2pt]
     &m = \frac{1}{d}\left(\frac{h^*}{s_f}+\frac{h^*}{s_g}\right) - 1 \,, \\[2pt]
     &\theta = \frac{p}{1-p} = \left(\frac{s_f}{s_g}\right)^2 \,.
 \end{split}
\ee
The proportionality constants in Eq.\eqref{eq:CMB_distribution} correspond to the normalization constant of the respective definition of the CMB probability density function. A detailed derivation of the boson distribution is given in Appendix~\ref{app:cmb_derivation}. The CMB distribution has been introduced by \citet{shmueli_useful_2005} as a means of capturing random variables following an ordinary binomial distribution, but with variances that deviate from the expected one. The original distribution uses $x,m\in\bb{N}$, however, since the parameters in Eq.~\eqref{eq:state_distribution_params} are in general nonintegers, we extend the distribution's definition by replacing factorials with Gamma functions. Moments of the CMB distribution have previously been studied~\cite{borges_compoisson_2014,kadane_sums_2016,daly_conway-maxwell-poisson_2016}, allowing us to obtain analytical expressions for the expectation value and variance of $\h{n}$ in this scenario; 
\be \label{eq:mean_and_variance_CMB}
\begin{split}
    &\expval{\h{n}} = d\,\bb{E}[X] + k^* - \frac{h^*}{s_g} + d, \\ 
    &\expval{(\Delta \h{n})^2} = d^2\,\n{Var}(X),
\end{split}
\ee
where $\bb{E}[X]$ and $\n{Var}(X)$ express the mean and variance of a random variable ${X\sim \n{CMB}(m,\theta,2)}$ (see Appendix~\ref{app:cmb_derivation}). In Fig.~\ref{fig3:vary_angle_height} we show that the Mandel $Q$ parameter obtained using the analytical expressions in Eq.~\eqref{eq:mean_and_variance_CMB} agrees very well with the one obtained numerically. {The discrepancy observed at large $Q$ values originates from the increased variance of the boson number distribution, which leads to a nonzero overlap between $\ket{\Xi_\mu}$ and the Fock ground state~\footnote{This behavior mirrors that of standard bosonic cat states, whose mean and variance deviate from the expected Poissonian value $\alpha^2$ for small $\alpha$. The deviation vanishes exponentially with $\alpha^2$.}.}

An interesting regime, which will be relevant in the next subsection, is when one of the slopes approaches~0. In the case ${s_f\rightarrow0}$, the value of $m$ diverges and the CMB distribution given in Eq.~\eqref{eq:CMB_distribution} approaches the Conway--Maxwell--Poisson (CMP) distribution ~\cite{conway1961queueing,borges_compoisson_2014,kadane_sums_2016,daly_conway-maxwell-poisson_2016}
\be \label{eq:CMP_distribution}
  \bb{P}[X=x|\lambda,2] \propto \frac{\lambda^x}{(x!)^2},
\ee
with ${\lambda=m^2\,p}$ and $x$, $m$ and $p$ defined in~Eq.~\eqref{eq:state_distribution_params}. In the opposite limit, namely $s_g\rightarrow0$, the distributions are given by
\be \label{eq:new_distribution}
  \bb{P}[Y=y|\lambda,2] \propto \frac{(y!)^2}{\lambda^y},
\ee
where $\lambda=m^2\,(1-p)$ and $y=x-m-1$. This probability density function has in general a divergent behavior, though if ${\lvert\tilde{f}(k)\rvert\rightarrow0}$ after crossing $\abs{\tilde{g}(k+r)}$ then the boson population will be confined in Fock space. The asymptotic analysis of the CMB distribution as $x,m\rightarrow\infty$ has, to the best of our knowledge, not been previously derived in existing literature. The exploration of the characteristics of the new distribution defined in Eq.~\eqref{eq:new_distribution} is left for future investigations.

\begin{figure}
    {\phantomsubcaption\label{fig5:rabi_frequency}}
    {\phantomsubcaption\label{fig5:relative_entropy}}
    \includegraphics[]{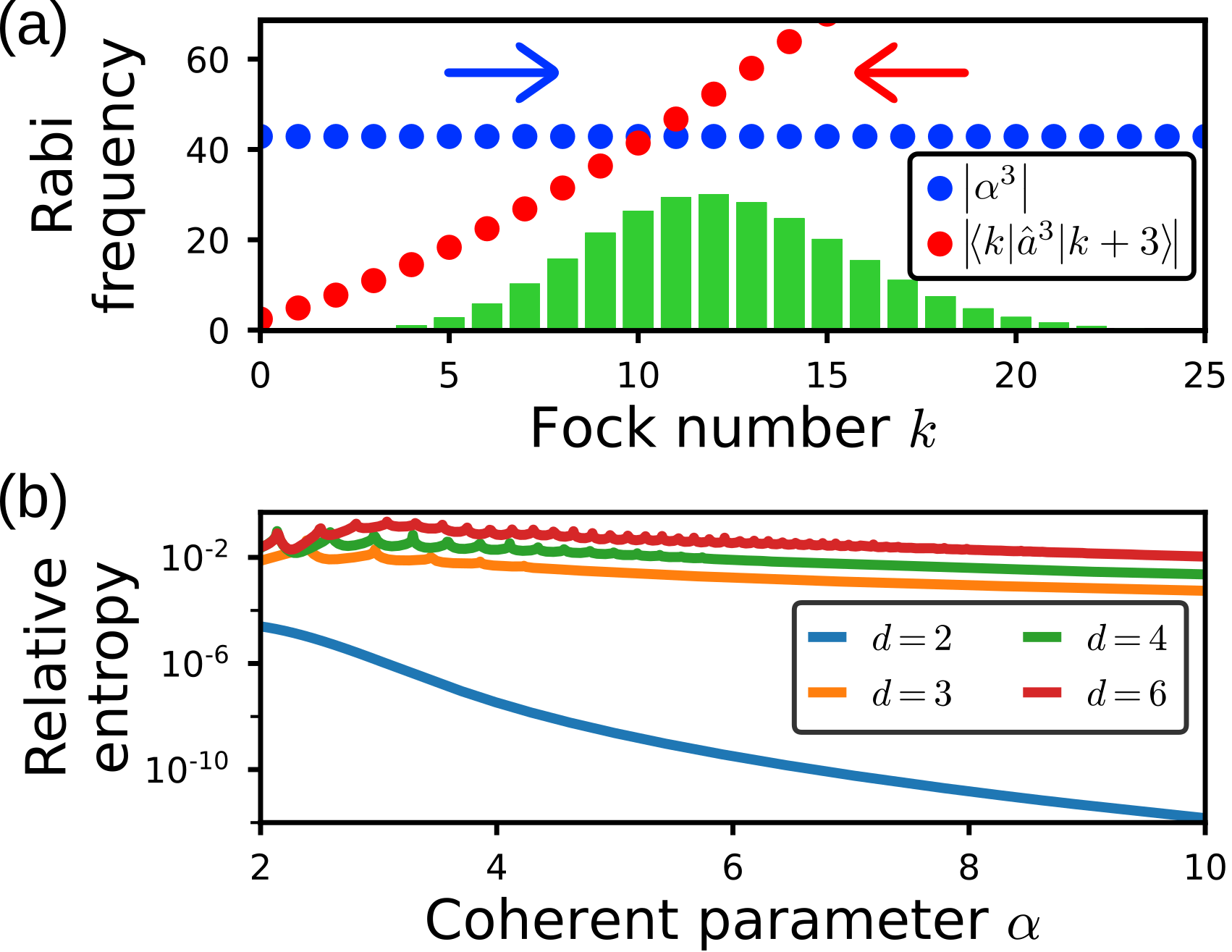}
    \caption{ (a) Standard dissipative cat-state stabilization explained using the nonlinear reservoir engineering method. Rabi frequencies and stabilized boson distribution of the jump operator ${\h{a}^3-\alpha^3}$ with $\alpha=3.5$. (b)~Relative entropy of a Poisson distribution, parameterized by $\alpha$, with respect to its approximation using a CMP distribution given in Eq.~\eqref{eq:CMP_distribution}. The Poisson distribution is stabilized using ${\h{a}^d-\alpha^d}$, where $d$ represents the dimension of the coherent-state manifold. The low relative entropy suggests that the true and approximate distributions contain nearly identical amounts of information. This demonstrates that the proposed nonlinear reservoir engineering method effectively generalizes well-established reservoir engineering cases.}
    \label{fig5}
\end{figure}

\subsection{Comparison to standard dissipative cat-state stabilization schemes} \label{subsec:comparison}

{
Our analytical description of the boson distribution of the stabilized states presented above relies on the fact that the Rabi frequencies of both boson raising and lowering processes are linear in vicinity of the crossing point $k^*$. As we noted, in practice $\tilde{f}$ and $\tilde{g}$ are not necessarily linear functions making our analytical formulas in Eqs.~\eqref{eq:CMB_distribution},~\eqref{eq:CMP_distribution} and~\eqref{eq:new_distribution} approximate. However, as we will now show, this approximation is accurate in practice. 

Let us consider,} as an illustrative example, the standard dissipative stabilization process of coherent-state manifolds, accomplished by engineering a jump operator ${\h{K}=\h{a}^d-\alpha^d}$~\cite{gilles_generation_1994,hach_iii_generation_1994,poyatos_quantum_1996,de_matos_filho_even_1996,mirrahimi_dynamically_2014,albert_symmetries_2014}. The stabilized $d$--dimensional manifold, for such a jump operator, is spanned by coherent states $\ket{\alpha\,e^{i2\pi k/d}}$ with ${k\in\{1,\ldots,d\}}$ and ${\alpha>0}$. The boson distribution associated with these states follows a Poisson probability density function ${\abs{\xi_{k}}^2 = \bb{P}[Z=k\vert\alpha^2]}$, where ${\bb{P}[Z=z\vert\lambda]:=e^{-\lambda}\lambda^{z}/z!}$. 

{Through the lens of the NRLE framework,} we can straightforwardly affirm that this system indeed stabilizes a $d$--dimensional manifold {given that $r=0$ and $l=d$}. The boson addition and removal functions correspond here to ${\tilde{f}(k)=\alpha^d}$ and $\tilde{g}(k)=\sqrt{(k+1)^{(d)}}$ with $(k)^{(j)}:=(k)(k+1)\cdots(k+j-1)$ denoting the rising factorial. These functions cross at $k^*$ satisfying the equation $\tilde{g}(k^*)-\alpha^d=0$ (see example in Fig.~\ref{fig5:rabi_frequency} for $d=3$). Solving this equation for $k^*$ and evaluating $\frac{\partial}{\partial k}\,\tilde{g}(k^*)$ allows us to access the approximate representation of stabilized states following the CMP distribution in Eq.~\eqref{eq:CMP_distribution}. {For conciseness, we omit here the expression of these two quantities.}

{To compare the true stabilized states, characterized by the Poisson distribution $\bb{P}[Z=k\vert\alpha^2]$, with their approximate counterpart,} we use the quantum relative entropy defined as $S(\h{\rho}\parallel\h{\sigma}):=\n{Tr}\h{\rho}(\log \h{\rho} - \log \h{\sigma})$. This serves as a measure of distinguishability between states $\h{\rho}$ and $\h{\sigma}$ with ${S(\h{\rho}\parallel\h{\sigma})=0}\,\Leftrightarrow\,{\h{\rho}=\h{\sigma}}$, and reduces to the relative entropy (a.k.a. Kullback-Leibler divergence) for classical (non-entangled) states. In our case, we choose $\h{\sigma}=\dyad{\alpha}$ as the reference state and evaluate the relative entropy of the approximate state for $2\leq\alpha\leq10$ and $d\in\{2,3,4,6\}$. The results, depicted in Fig.~\ref{fig5:relative_entropy}, reveal that ${S(\h{\rho}\parallel\h{\sigma})}$ remains consistently low (i.e., $<0.1$) across all dimensions. Notably, the approximation of stabilized states improves exponentially with increasing $\alpha$. For $d=2$, the approximation is most accurate thanks to the fact that $\tilde{g}(k)\equiv\sqrt{(k+1)(k+2)}$ is very close to a linear function. As mentioned previously, higher-order approximations can be derived using higher-order derivatives of $\tilde{f}$ and $\tilde{g}$. However, delving into this analysis surpasses the scope of the present paper, and we leave it for future work.

\subsection{Connection to nonlinear coherent states}

The solution to $\cl{L}$ is readily obtained when linked to the standard dissipative stabilization process of coherent state manifolds. In the case of our model, we can rewrite the effective jump operator using 
\be \label{eq:jump_rewritten}
\begin{split}
    \h{K} &= \h{a}^{\dagger\,r} 
    \left[ f(\h{n}) - \left(\h{a}^{\dagger\,r}\right)^{-1} g(\h{n}) \h{a}^{l} \right]  \\
    &\approx \h{a}^{\dagger\,r}\,\frac{g(\h{n}+r)}{(\h{n}+1)^{(r)}} \left[ \frac{f(\h{n})\,(\h{n}+1)^{(r)}}{g(\h{n}+r)} - \h{a}^{r+l} \right]\,.
\end{split}
\ee 
We are interested in the term that is enclosed in the brackets as it resembles the jump operator for coherent state stabilization mentioned above. To obtain this term we required the inverse of $\h{a}^{\dagger\,r}$. Although the operator is unbounded and singular as ${\bra{k}\h{a}^{\dagger\,r}=0\,\,\,\forall\,k<r}$, we make use of its Moore-Penrose pseudoinverse~\cite{penrose_generalized_1955}, that is given by ${\sqrt{(\h{n}+1)^{(r)}}\,(\h{a}^{\dagger\,r})^{-1}\approx\h{a}^{r}}$. To derive Eq.~\eqref{eq:jump_rewritten}, we additionally assume that ${g(\h{n}+r)}$ does not vanish in the considered Fock domain and use the relations ${\h{a}^{r} g(\h{n})=g(\h{n}+r)\h{a}^{r}}$.

From the bracket operator in Eq.~\eqref{eq:jump_rewritten} we can infer previously determined properties of the steady-state subspace such as the dimension of the manifold and the boson distribution. Indeed we can see that Eq.~\eqref{eq:recusive_relation} comes out straightforwardly once we use the raising factorial operator from $\h{a}^{r+l}$ to obtain $\tilde{f}$ and $\tilde{g}$. This operator is however not precisely equal to ${\h{a}^d-\alpha^d}$ as the coherent parameter is in our case non-constant in $\h{n}$. In fact, the bracket operator we derived corresponds to the jump operator stabilizing so-called nonlinear coherent states. These states were formally introduced in Ref.~\cite{de_matos_filho_nonlinear_1996} as the right-hand eigenstates of $f(\h{n})\h{a}$ (with $f$ being an arbitrary function) and helped to generalize a variety of nonclassical bosonic states~\cite{dodonov_nonclassical_2002}. In particular, the dark states of $\h{a}^d-f(\h{n})$ were studied in Ref.~\cite{manko_trapped_2000}. The authors provided a solution for the steady states in terms of $f(\h{n})$, however up to now the empirical understanding of the stabilization mechanism and a first-order approximation of the boson distribution were not known. 

\medskip
To conclude this section, we established that the toy model defined in Eq.~\eqref{eq:hamilton_modif} stabilizes a variety of cat-like state manifolds of dimension $d$ when the Rabi rates $\tilde{f}$ and $\tilde{g}$ intersect appropriately. We showed that the NLRE method provides an intuitive understanding of the stabilization as well as of the nature and shape of the steady states, and facilitates analytical treatment of such operators, which have proven challenging with known methods.

\section{Quantum Error Correction capabilities} \label{sec:error_correction}

After determining the static properties of the stabilization operator $\h{K}$, we now focus on the dynamics it induces, particularly in the presence of common errors of a quantum harmonic oscillator.

Open quantum systems hosting a stabilized subspace can exhibit passive or autonomous quantum error correction (QEC) against certain type of errors~\cite{paz_continuous_1998,barnes_automatic_2000,sarovar_continuous_2005,pastawski_quantum_2011,mirrahimi_dynamically_2014,kapit_hardware_2016,reiter_dissipative_2017,puri_engineering_2017,leghtas_confining_2015,lihm_implementation_2018,albert_pair_cat_2019,lieu_symmetry_2020,lebreuilly_autonomous_2021,lieu_candidate_2023}. We distinguish passive and autonomous error correction as follows. The former is defined according to Refs.~\cite{lieu_symmetry_2020,lieu_candidate_2023} as the ability of a dissipative process to correct an erroneous state with unit fidelity in the thermodynamic (i.e. ${\langle\h{n}\rangle\rightarrow\infty}$) and the infinite-time limits. {To determine if it does so, one relies on symmetry of the system's dynamics with respect to a unitary operator $\h{P}$. A Liouvillian $\cl{L}$ is said to exhibit a weak symmetry (or be weakly symmetric) if $\left[\cl{P},\cl{L}\right]=0$ where $\cl{P}(\cdot)=\h{P}(\cdot)\h{P}^\dagger$ for which it is sufficient that $[\h{P},\h{H}]=0=\{\h{P},\h{L}\}$ with $\h{H}$ and $\h{L}$ being the Hamiltonians and jump operators in play in $\cl{L}$. If instead $[\h{P},\h{H}]=0=[\h{P},\h{L}]$ then we say the system is strongly symmetric~\cite{baumgartner_analysis_2008,buca_note_2012,albert_symmetries_2014,minganti_spectral_2018,lieu_symmetry_2020}.} \citet{lieu_symmetry_2020} conjectured that codes constructed from strongly symmetry-broken models {(i.e. with more than one stable state)} demonstrate passive QEC against errors that maintain the system within the strong-symmetry phase. An exemplary illustration of this concept is the dissipatively stabilized cat code, which passively protects against both coherent and incoherent dephasing errors. {To see that let us recall that their standard dissipative stabilization is realized using a purely dissipative Liouvillian $\cl{L}$ (i.e. no Hamiltonian $\h{H}$) with a single jump operator $\h{L}\propto\h{a}^d-\alpha^d$. Since cat states are rotationally symmetric around the phase space origin, the symmetry operation that we shall consider is the parity operator $\h{P}=e^{2\pi i \hat{n}/d}$. In this case, $\cl{L}$ exhibits a strong symmetry and allows a passive quantum error correction against strongly symmetric errors (i.e. those commuting with $\h{P}$) such as dephasing errors $\h{n}$.} 

{In the NLRE context, the states stabilized using $\h{L}\propto\h{K}$ with $\h{K}$ in Eq.~\eqref{eq:hamilton_modif} exhibit a rotational symmetry with respect to the same parity operator $\h{P}$ where now ${d=r+l}$. It is straightforward to verify that} ${\h{K} \h{P} = \exp(i\,2\pi\,r/d) \h{P} \h{K}}$. Thus, our model manifests strong symmetry only when $r=0$ or $r=d$ as the jump operator commutes with the symmetry operator. In these cases, similarly to the cat code, dephasing errors result in $\h{L}$ remaining within the strongly broken phase, providing passive protection against such errors. Conversely, if ${0<r<d}$, the model is only weakly symmetric, lacking passive QEC against dephasing errors. Other typical oscillator errors such as boson loss $\h{a}$ and gain $\h{a}^\dagger$ are not strongly symmetric with respect to the parity operator $\h{P}$. Consequently, the model does not inherently possess passive QEC capabilities against these types of errors for any value of $r$. These conclusions can be understood empirically. Consider the situation $d=2$. When $r=0$ or $r=2$, at every $\h{L}$ jump, boson populations are redistributed exclusively between even or odd Fock states, that is no mixing of populations is happening between even and odd parity states. It is not the case when $r=1$ as at every jump even Fock state population will become odd one (and vice versa). This mixing is the reason why the weakly symmetric model does not passively protect against dephasing, as well as loss and gain errors. This empirical picture straightforwardly extends to situations with $d>2$.

\subsection{Autonomous quantum error correction}
In contrast, a code is said to be autonomously corrected if there exists a dissipative process that given a natural error rate $\kappa$, improves the decoherence rate of the logical information to $\cl{O}(\kappa/R)$ for some constant $R\gg1$ that depends on the engineered dissipation rate~\cite{lebreuilly_autonomous_2021}. Autonomous QEC is thus characterized by the ability of a dissipative process to improve the decoherence rate of logical information without necessitating the code to be in the thermodynamic and infinite-time limits. Instead, it assumes that error processes are consistently present in the system's dynamics. The subsystem decomposition method serves as a valuable tool for analyzing the logical decoherence rate and QEC codes in general. Formally introduced in Ref.~\cite{pantaleoni_modular_2020}, this method has proven instrumental in the description and analytical treatment of discrete and continuous variable codes~\cite{glancy_error_2006,raynal_encoding_2010,ketterer_quantum_2016,duivenvoorden_single_2017,weigand_generating_2018,matsuura_equivalence_2020,albert_robust_2020,chamberland_building_2022,putterman_stabilizing_2022,xu_autonomous_2023}. The method relies on decomposing the system's Hilbert space into a tensor product between a finite-dimensional logical subspace and a gauge subsystem of possibly infinite dimension: $\cl{H}=\bb{C}^d\otimes\cl{H}_g$. It is important to note that the choice of the gauge mode basis is not unique. For example, in the case of the cat code with $d=2$, one can opt for a basis derived from the modular decomposition of the Fock space~\cite{pantaleoni_modular_2020} or one based on displaced Fock states~\cite{chamberland_building_2022,putterman_stabilizing_2022}. In our case, we focus on the latter and generalize the method to nonlinear coherent states following the derivation from~\citet{chamberland_building_2022}. We define a parametrized set of states 
\be \label{eq:nonlin_disp_fock}
    \ket{\varphi_{k,\pm}} = \frac{1}{\cl{N}_k} \, R(\h{n}) \left(\h{D}(\alpha)\pm (-1)^{k}\h{D}(-\alpha)\right)\ket{k}
\ee
where ${\h{D}(\alpha):=\exp(\alpha\h{a}^\dagger-\alpha^*\h{a})}$ is the displacement operator, $R(\h{n})$ -- a function of the number operator and $\cl{N}_k$~-- the normalization factor. For ${R(\h{n})=\mathds{1}}$ and the Fock state $\ket{k=0}$, these states are equivalent to the even and odd cat states of size $\alpha$, ${\ket{\varphi_{0,\pm}}\equiv\ket{C_\pm(\alpha)}}$. Since the steady states of $\cl{L}$ are nonlinear coherent states, we use $R(\h{n})$ to reshape the boson distribution of the displaced Fock states from a Poissonian distribution to the stabilized one, namely 
\be \label{eq:reshaping_func}
    R(\h{n}) = \sqrt{\frac{\bb{P}[K=\h{n}]}{\bb{P}[Z=\h{n}\vert\alpha^2]}}
 \ee
where ${\bb{P}[Z=z\vert\lambda]:=e^{-\lambda}\lambda^{z}/z!}$, is the distribution of a coherent state, and ${\bb{P}[K=k]:=\abs{\xi_k}^2}$ is the stabilized boson distribution following the recurrence relation in~Eq.~\eqref{eq:recusive_relation}~\footnote{Using the number operator $\h{n}$ as an input of these probability density functions is strictly speaking an abuse of notation which should be understood as ${\bb{P}[K=\h{n}]=\sum_k\bb{P}[K=k]\dyad{k}}$. The reshaping operator given in Eq.~\eqref{eq:reshaping_func} is then a well-defined non-singular diagonal operator.}. In this situation, the states $\ket{\varphi_{0,\pm}}$ are equivalent to $\ket{\Xi_0}$ and $\ket{\Xi_1}$, respectively. Moreover, the choice of $\alpha$ is arbitrary, but as we will see later it is convenient to choose $\abs{\alpha}^2=\expval{\h{n}}$.

\begin{figure*}
    \includegraphics[]{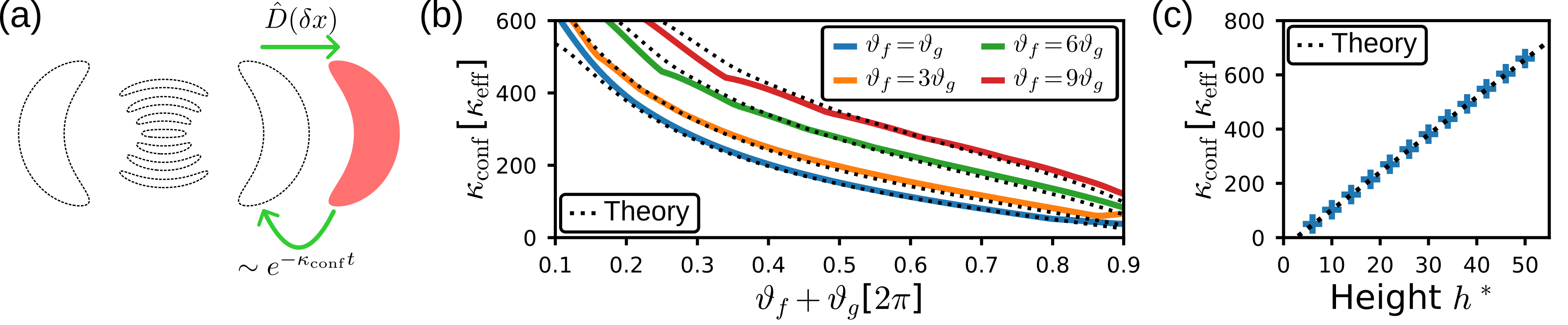}
    {\phantomsubcaption\label{fig4:displ_scheme}}
    {\phantomsubcaption\label{fig4:conf_vs_angle}}
    {\phantomsubcaption\label{fig4:conf_vs_height}}
    \caption{(a) We evaluate the effective confinement rate $\kappa_\n{conf}$ as the exponential rate at which a dark state initially displaced ($\delta x = 0.001$) returns to the stabilized manifold. (b) Effective confinement rate with respect to the angle $\vartheta_f+\vartheta_g$ between the functions $\tilde{f}$ and $\tilde{g}$. One can increase this rate by increasing the ratio $\vartheta_f/\vartheta_g$ which effectively increases the skewness of the stabilized boson distribution (the height is constant $h^*=20$). The theoretical curves have been obtained using Eq.~\eqref{eq:confine_rate} additionally corrected for the skewness. (c) The relationship between the height of the crossing point $h^*$ and $\kappa_\n{conf}$ is linear (${\vartheta_f=\vartheta_g=\pi/3}$). In both (b) and (c), we keep the crossing point at $k^*=20$.}
    \label{fig4}
\end{figure*}

The displaced Fock state method relies on the assumption that $\ket{\varphi_{k,\pm}}$ are orthonormal states. Although states with different parity are by definition orthogonal, those with the same one are not. In the case ${R(\h{n})=\mathds{1}}$, the orthonormality condition holds in the limit of large cat qubits (i.e. ${\abs{\alpha}^2\gg1}$) with a normalization factor ${\cl{N}_k=\sqrt{2}}$~\cite{chamberland_building_2022}. For nonlinear shifted Fock states $\ket{\varphi_{k,\pm}}$, the normalization constant will be different for every $k$ and depend both of on the mean and variance of the stabilized boson distribution. To orthogonalize the states, one can apply the Gram-Schmidt orthonormalization procedure over a sufficiently large Fock domain. However, when focusing solely on the dynamics between the ground and first excited states, we can restrict ourselves to normalized states only. This so-called ``cold'' gauge mode assumption is justified when the stabilization dominates the error processes and is often considered in the standard shifted Fock state method \cite{pantaleoni_modular_2020,chamberland_building_2022,xu_autonomous_2023}.

The action of the annihilation operator on $\ket{\varphi_{k,\pm}}$ is described by 
\be \label{eq:destroy_disp_fock}
  \h{a}\ket{\varphi_{k,\pm}} = \frac{R(\h{n}+1)}{R(\h{n})}\!\left(\!\sqrt{k}\,\,\frac{\cl{N}_{k-1}}{\cl{N}_{k}}\ket{\varphi_{k-1,\mp}}+\alpha\ket{\varphi_{k,\mp}}\right)\,.
\ee
The expression within the parentheses is compactly expressed as ${\h{Z}_L\otimes\!\big[ F_{1}(\h{A}^\dagger\h{A})\,\h{A}+\alpha\big]\ket{\varphi_{k,\pm}}}$ where $\h{Z}_L$ corresponds to the logical Pauli Z operator, $\h{A}^\dagger$ and $\h{A}$ are the creation and annihilation operators for the gauge mode, and the function ${F_{l}(k) \equiv \cl{N}_k/\cl{N}_{k-l}}$. This function is a crucial aspect of the nonlinear shifted Fock state method, indicating that $\h{a}$ translates into a nonlinear gauge mode annihilation operator. 

To express the first term of Eq.~\eqref{eq:destroy_disp_fock} as a function of logical and gauge mode operators we proceed similarly with the number operator. We simply quote the result in the subsystem picture
\be \label{eq:number_disp_fock}
    \h{n} \equiv \h{I}_L\otimes\!\Big[\h{A}^\dagger\h{A}+\abs{\alpha}^2+
    \alpha^* F_{1}(\h{A}^\dagger\h{A})\,\h{A} +
    \alpha\h{A}^\dagger\,F_{1}(\h{A}^\dagger\h{A})\Big]\,.
\ee
Consequently, any functions of the number operator, which can be written as a Taylor series $\sum_j c_j \h{n}^j$, act as an identity on the logical subsystem. In other words, functions such as $R(\h{n}+1)/R(\h{n})$ in Eq.~\eqref{eq:destroy_disp_fock} leave the logical information invariant and can be rewritten as $\h{I}_L\otimes R(\h{O}+1)/R(\h{O})$ with $\h{O}$ being the gauge operator from Eq.~\eqref{eq:number_disp_fock}. With this, it becomes evident that
\be \label{eq:a2_and_at_disp_fock}
\raisetag{2ex}
\begin{aligned}
    &\h{a}^2 \equiv \h{I}_L\otimes\!\frac{R(\h{O}+2)}{R(\h{O})}\!\Big[F_{2}(\h{A}^\dagger\h{A})\,\h{A}^2 + 2 \alpha F_{1}(\h{A}^\dagger\h{A})\,\h{A} +\alpha^2\Big]\,, \\
    &\h{a}^\dagger \equiv \h{Z}_L\otimes\!\frac{R(\h{O}-1)}{R(\h{O})}\!\Big[\h{A}^\dagger\,F_{1}(\h{A}^\dagger\h{A}) +\alpha^*\Big]\,,
\end{aligned}
\ee
implying that each single-boson loss or gain flips the parity of the state and potentially excites the gauge mode. In contrast, coherent or incoherent dephasing errors or two-boson loss (or gain, not explicitly presented here) preserve the logical information. Combining these insights with Eq.~\eqref{eq:jump_rewritten}, the term in the parenthesis removes excitation from the gauge mode, directing the system towards $\ket{\varphi_{0,\pm}}$ without changing its initial parity, whereas the term $\h{a}^{\dagger\,r}$ would influence the logical information stored in the system when $r$ is an odd integer. In other words, in a noiseless scenario, a system starting in the stabilized manifold will remain in this manifold for any $(r,l)\in\{(0,2),(1,1),(2,0)\}$. In the presence of dephasing errors, only the $(0,2)$ and $(2,0)$ schemes have the ability to bring back the state to the steady-state manifold without changing the parity, while the $(1,1)$ scheme would eliminate the parity information (i.e. it can only stabilize a classical bit). Single-boson loss or gain errors are detrimental for all three configurations. This demonstrates that the model in Eq.~\eqref{eq:hamilton_modif} including just one raising and one lowering term is incapable of correcting the broader set of bosonic errors using any type of nonlinear functions $f$ and $g$. Additional resources would be required for this purpose, as we will explore in Sec.~\ref{sec:bogoliubov_transf}.

\subsection{Effective confinement rate}
The nonlinear shifted Fock state method devised above allows us to evaluate quantitatively the dynamics of the system. Similar to the standard dissipative cat-qubit stabilization scheme, the evolution generated by the Liouvillian $\cl{L}$ (provided an adequate choice of $f$ and $g$) exponentially converges to the dark state manifold~\cite{azouit_well_posedness_2016}. The rate of convergence $\kappa_\n{conf}$ is referred to as the effective confinement rate~\cite{mirrahimi_dynamically_2014,chamberland_building_2022,gautier_combined_2022}. We are in particular interested in the convergence of states in a close neighborhood of this manifold. To begin with, we consider the scenario in which the noise is negligible compared to the engineered dissipation. To find the lowest order approximation of $\kappa_\n{conf}$, we use the nonlinear shifted Fock state method presented above with the setting ${r=0}$. Disregarding gauge mode excitations higher than two, we find that in the subsystem picture the operator in Eq.~\eqref{eq:jump_rewritten} can be well approximated by
\be \label{eq:Leff_disp_fock}
    \h{K} \approx 2 \, \frac{\tilde{f}(\alpha^2)}{\alpha}\,\frac{\cl{N}_0}{\cl{N}_{1}} \, \h{I}_L\otimes \h{A}\,.
\ee
This expression depends  on the normalization constants of $\ket{\varphi_{0,\pm}}$ and $\ket{\varphi_{1,\pm}}$. By the definition of nonlinear shifted Fock states $\cl{N}_0=\sqrt{2}$ and $\cl{N}_1^2 = 2 \alpha^{-2} \mel{\Xi_{0,1}}{(\h{n}-\abs{\alpha}^2)^2}{\Xi_{0,1}}$ where $0/1$ depends on the parity of $\ket{\varphi_{1,\pm}}$. It is thus convenient to fix the free parameter $\alpha$ to be the square root of the mean of the stabilized distribution, $\abs{\alpha}^2=\expval{\h{n}}$. Then using Eq.~\eqref{eq:Leff_disp_fock}, we have that $\cl{L}=\kappa_\n{conf}\h{I}_L\otimes\cl{D}[\h{A}]$ with the confinement rate
\be \label{eq:confine_rate}
    \kappa_\n{conf} \approx 4\,\kappa_\n{eff}\,\frac{\tilde{f}(\expval{\h{n}})^2}{\expval{(\Delta \h{n})^2}}\,.
\ee
where $\kappa_\n{eff}=\Omega^2/\gamma$ is the rate of the effective jump operator ${\h{L}=\sqrt{\kappa_\n{eff}}\h{K}}$ obtained after adiabatic elimination of the reservoir (see Eq.~\eqref{eq:hamiltonian}). Note that this lowest-order approximation is agnostic to the specific shape of the functions $\tilde{f}$ and $\tilde{g}$ as long as those satisfy the conditions for the stabilization of a cat-state manifold. This relation can be interpreted as following: since $\expval{\h{n}}$ is in the vicinity of the crossing point $k^*$, we can approximate ${\tilde{f}(\expval{\h{n}}) \simeq h^*}$, which implies that the Rabi frequency at the crossing point sets the depth of the potential well that hosts the stabilized state. The variance on the other hand quantifies the steepness of the well. Consequently, one can arbitrarily increase $\kappa_\n{conf}$ by either increasing the height of the crossing point or decreasing the variance of the stabilized state -- effectively altering the relative angle between $\tilde{f}$ and $\tilde{g}$.

We confirm the validity of this formula through numerical verification. Specifically, we examine the stability of the state $\ket{\psi}\propto\ket{\Xi_0}+\ket{\Xi_1}$, displaced infinitesimally along the $\h{q}$ quadrature, and observe the rate at which it exponentially returns into the stabilized manifold (see Fig.~\ref{fig4:displ_scheme}). As previously, we use non-negative linearly increasing and decreasing Rabi frequencies $\tilde{f}$ and $\tilde{g}$ with slopes $s_f$ and $s_g$, respectively, crossing at a point $k^*$ and height $h^*$. Varying the slopes (or alternatively the angles $\vartheta_f$ and $\vartheta_g$ as depicted in Fig.~\ref{fig3:rabi_frequency}) while keeping other parameters constant changes the variance of the stabilized states independent of their mean value. Qualitatively, we observe in Fig.~\ref{fig4:conf_vs_angle} that for increasing variance the confinement rate decreases, a behavior that aligns with the formula in Eq.~\eqref{eq:confine_rate}. Quantitatively, we see that the formula represents very well the confinement rate in the situation when ${s_f=s_g}$. When one of the slopes is larger than the other, the stabilized distribution is skewed. Empirically, we find that the confinement rate is given by $\kappa_\n{conf}\sqrt{1-\n{Skew}(\h{n})}$, where
\be
    \n{Skew}(\h{n}) = \frac{ \expval{ \big(\h{n} - \expval{\h{n}}\!\big)^3 } }{ \expval{(\Delta\h{n})^2}^{3/2} } \ .
\ee
The skewness $\n{Skew}(\h{n})$ is greater than 0 for ${s_f < s_g}$, equal to 0 for ${s_f = s_g}$, and less than 0 for ${s_f > s_g}$. In Fig.~\ref{fig4:conf_vs_angle}, the theoretical curves for ${s_f<s_g}$ are obtained using the $\sqrt{1-\n{Skew}(\h{n})}$ correction. It is important to note that although $\kappa_\n{conf}$ is enhanced for skewed distributions, this correction is valid only when ${s_f<s_g}$. In the opposite situation (i.e. when the skewness is negative), a different correction term is required. Alternatively, we can enhance the confinement rate using $h^*$. Varying only the height of the crossing point not only changes the variance of the stabilized distribution but also sets the first-order approximation of the stabilization rate (see Fig.~\ref{fig4:conf_vs_height}). Note that for the previously studied dissipative 2--component cat manifolds this parameter corresponds directly to the $\alpha^2$ parameter, with the effective confinement rate given by $\kappa_\n{conf}\approx4\kappa_\n{eff}\alpha^2$~\cite{mirrahimi_dynamically_2014,chamberland_building_2022}.

\subsection{Correction capability of $(0,2)$ and $(2,0)$ models}
\begin{figure}
    \includegraphics[]{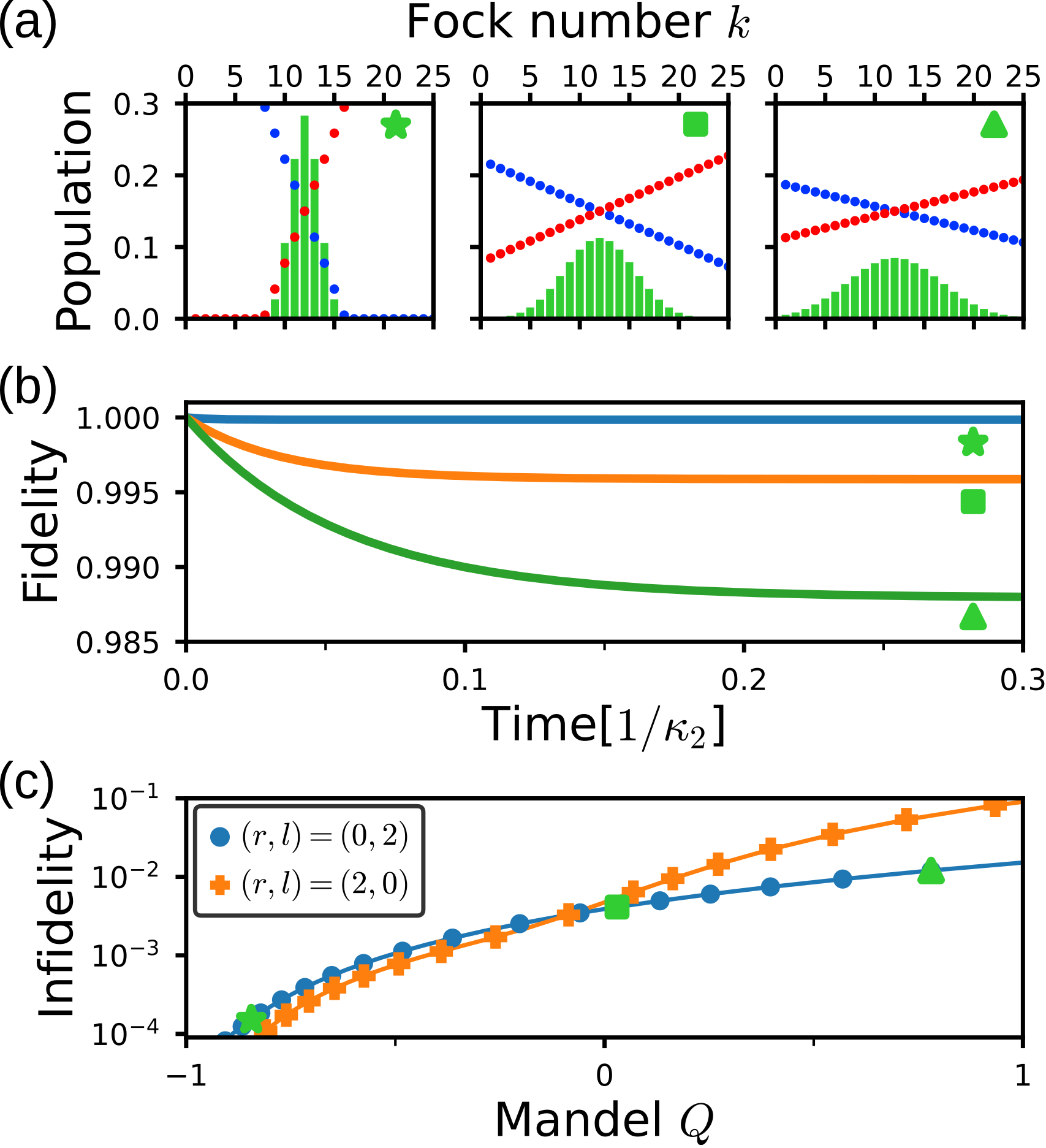}
    {\phantomsubcaption\label{fig6:states_diff_Q}}
    {\phantomsubcaption\label{fig6:fidel_vs_time}}
    {\phantomsubcaption\label{fig6:infidel_vs_mandel}}
    \caption{Reducing the Mandel $Q$ parameter improves the correction against dephasing. (b) The evolution of the fidelity of a dark state of $\cl{L}$ with itself when subject of the stabilization and incoherent dephasing processes for $(0,2)$ schemes ($\kappa_\phi=10^{-2}\kappa_\n{eff}$). This evolution is shown for different slopes ${s_f=s_g}$ of the linearly decreasing and increasing processes $\tilde{f}$ and $\tilde{g}$ crossing at a fixed Fock state ${k^*=11}$ and height ${h^*=10}$ as depicted in (a). Changing these slopes changes the Mandel $Q$ parameter (see Fig.~\ref{fig3}). We observe an enhanced stabilization when $Q\rightarrow-1$ which is explained by a lower variance of the stabilized distribution and thus a higher confinement rate $\kappa_\n{conf}$ (see Fig.~\ref{fig4}). (c) Steady state infidelity as a function of the Mandel $Q$ parameter for both $(0,2)$ and $(2,0)$ schemes.}
    \label{fig6}
\end{figure}

Based on the analysis above, we conclude that the confinement rate can be enhanced by number-squeezing the stabilized boson distribution in Fock space offering a faster recovery from dephasing errors without increasing $h^*$. We illustrate this enhancement in Fig.~\ref{fig6}. We simulate the dynamics of a system prepared in the stabilized state $\ket{\Xi_0}$ which evolves under the Liouvillian $\cl{L} + \cl{D}[\sqrt{\kappa_\phi}\h{n}]$ with $\kappa_\phi=10^{-2}\kappa_\n{eff}$ being the rate of dephasing of the harmonic oscillator. We perform this evolution for models with a fixed crossing point $k^*$ and height $h^*$ but different slopes $s_f=s_g$. This effectively allows us to vary the variance of the boson distribution which we quantify using the Mandel $Q$ parameter (see Fig.~\ref{fig3:vary_angle_height}). In Fig.~\ref{fig6:fidel_vs_time}, we observe over time a slight decay of system's fidelity with the initial state. The level to which the fidelity converges depends on the degree of squeezing in Fock space of the stabilized state; the lower the Mandel $Q$ parameter, the lower the final infidelity of the state under dephasing noise (see Fig.~\ref{fig6:infidel_vs_mandel}). Similar results are obtained if we consider stabilized states with skewed boson distribution (i.e. when ${s_f \neq s_g}$).

The results presented above concern schemes with parameters ${r=0}$ and ${l=2}$, i.e. nonlinear counterparts of the standard dissipative stabilization of two-component cat qubits. Let us now discuss the case of $(r,l)=(2,0)$. These schemes can also stabilize and correct for dephasing errors. Indeed, similarly to the stabilizing operator for $(0,2)$ given in Eq.~\eqref{eq:Leff_disp_fock}, $\h{K}$ acts as the identity in the logical subspace and as a destruction operator in the gauge space. This means that neither dephasing (see Eq.~\eqref{eq:number_disp_fock}), nor the stabilization scramble the logical information. This correction capability of schemes based on a two-boson gain term (rather than a two-boson loss) was not noted in the literature before and is made possible only thanks to the nonlinear nature of the stabilizing jump operators (i.e. appropriate crossing of $\tilde{f}$ and $\tilde{g}$). Identically to the $(0,2)$ scheme, the effective confinement rate for the $(2,0)$ scheme can be enhanced by reducing the variance of the stabilized boson distribution. We show this in Fig.~\ref{fig6:infidel_vs_mandel}. In both situations, reducing the Mandel $Q$ parameter decreases state infidelity under dephasing noise. However, when $Q>0$ the fidelity for the case $(2,0)$ starts to decrease with $Q$. This comes from the fact that the stabilized distribution becomes sufficiently wide that the Fock states near the ground state begin to be populated, leading to a non-negligible mixing effect as discussed in Sec.~\ref{subsec:number_of_steady_states}.
Finally, we note that the same conclusions apply to $(0,d)$ and $(d,0)$ schemes due to their strong symmetry under discrete rotation, as discussed at the beginning of Sec.~\ref{sec:error_correction}.

{
In practice, a system is subject to various errors, and while standard stabilized cat codes cannot correct boson loss or gain, their key advantage lies in their asymmetric response to noise, which strongly depends on the logical states. Consider a two-component cat qubit with parameter $\alpha$ and codewords $\ket{\pm\alpha}$, which we can see as the $\pm1$ eigenstates of the logical operator $\h{Z}_L$. Since $\ket{\pm\alpha}$ are coherent states, they are also eigenstates of the annihilation operator $\h{a}$ with eigenvalues $\pm\alpha$, which offers them a natural protection against these errors. Moreover, these logical states are separated in phase space such that incoherent processes flipping $\ket{\alpha}$ and $\ket{-\alpha}$ are exponentially suppressed with $\alpha$~\cite{mirrahimi_dynamically_2014}. On the other hand, the logical states $\ket{C_{\pm}}\propto\ket{\alpha}\pm\ket{-\alpha}$, which are $\pm1$ eigenstates of $\h{X}_L$, are robust against dephasing but highly susceptible to boson loss and gain. As discussed earlier, these errors, in the subsystem decomposition picture, increase gauge excitations and induce a logical phase flip (see Eq.~\eqref{eq:destroy_disp_fock}). This makes $\ket{C_{\pm}}$ particularly vulnerable to these noise processes. Therefore, when initializing a system in these states and evolving it under dissipative stabilization and noise, one observes a strong difference between bit-flip and phase-flip error rates. This asymmetry, known as the noise bias, is a fundamental feature of cat codes. Exploiting this bias can significantly reduce the experimental overhead in integrating cat codes into larger error-correction schemes, as only one type of error needs to be actively corrected~\cite{guillaud_repetition_2019,chamberland_building_2022,ruiz_ldpc-cat_2025}.
}

\begin{figure}[b!]
    \includegraphics[]{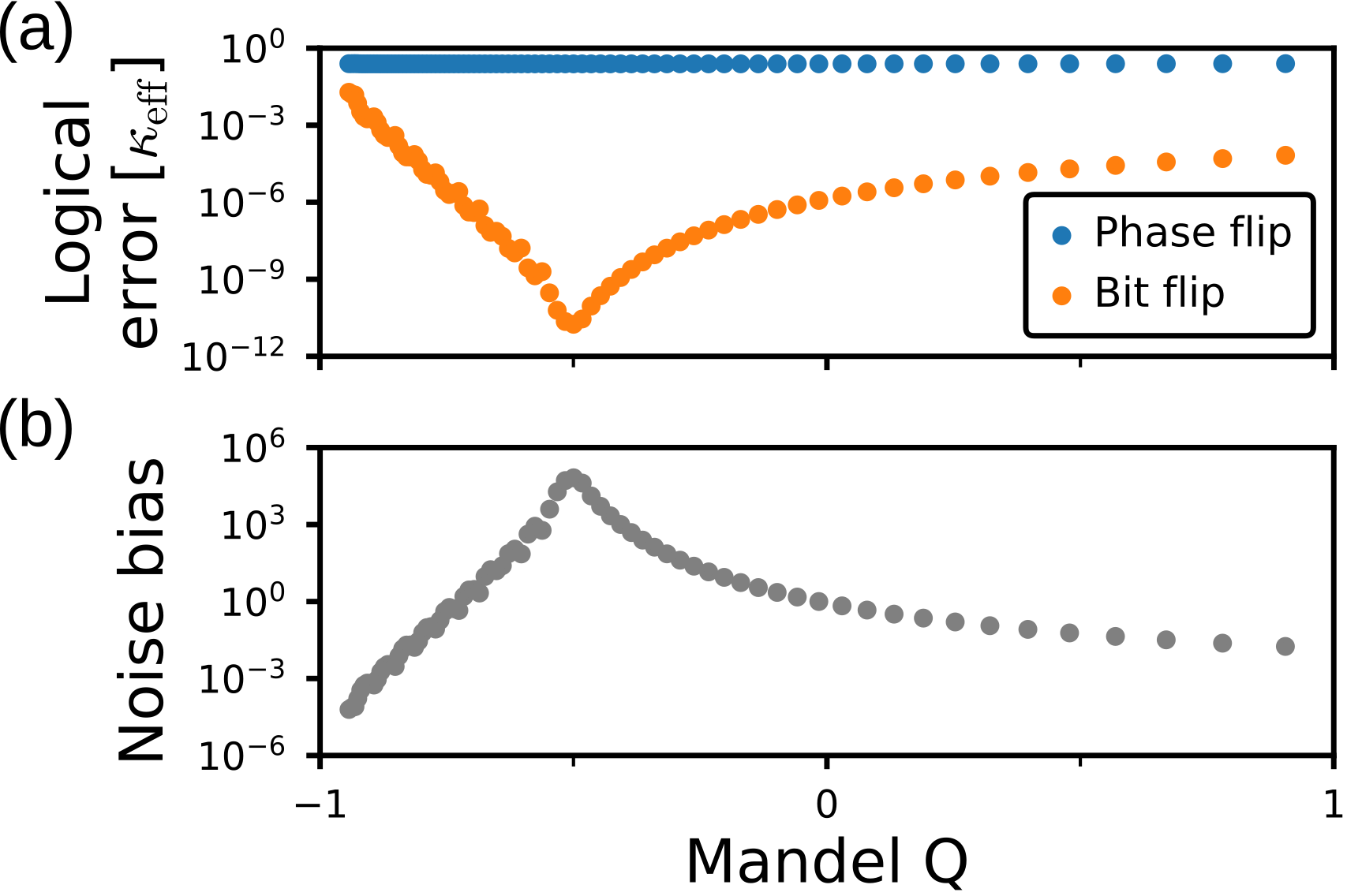}
    {\phantomsubcaption\label{fig_nb:error_rates}}
    {\phantomsubcaption\label{fig_nb:noise_bias}}
    \caption{{Logical error rates and noise bias. (a) Logical bit- and phase-flip rates for $(0,2)$ schemes with $k^*=11$ and $h^*=10$ and different Mandel $Q$ parameters obtained by varying the slopes $s_f=s_g$. We simulate the dynamics of the system in the presence of noise $\kappa_\phi\cl{D}[\h{n}]+\kappa_1(1+n_\n{th})\cl{D}[\h{a}]+\kappa_1 n_\n{th}\cl{D}[\h{a}^\dagger]$ where $\kappa_\phi=\kappa_1=10^{-2}\kappa_\n{eff}$ and $n_\n{th}=0.01$. The bit- and phase-flip rates are calculated using initial states $\ket{\Xi_0}+\ket{\Xi_1}$ and $\ket{\Xi_0}$, respectively, and extracting the exponential decay rate of their fidelity over time. (b) Noise bias as a function of the Mandel $Q$ parameter for $(0,2)$ schemes normalized by its value at $Q=0$. The bias is defined as the ratio of phase-flip to bit-flip rates.}}
    \label{fig_nb}
\end{figure}

{
One way to increase the noise bias is to enlarge the cat qubit. As the mean boson number $\expval{\h{n}}=\abs{\alpha}^2$ increases, the bit-flip rate decreases exponentially, while the phase-flip rate grows only linearly~\cite{mirrahimi_dynamically_2014}. A similar behavior is obtained in our toy model by shifting the crossing point $k^*$ to higher Fock numbers. However, engineering dissipators for $\abs{\alpha}\gtrsim4$~\cite{reglade_quantum_2024} is challenging with established methods due to intrinsic nonlinearities in the interactions used to implement the target Hamiltonian in Eq.~\eqref{eq:hamiltonian} (see Appendix~\ref{app:nonlinear_important}). Here, we show that nonlinear coherent states can be used to enhance the noise bias without increasing the size of the cat code. We consider NLRE schemes $(0,2)$ with the same parameter values as in Fig.~\ref{fig6}, along with the noise Liouvillian $\kappa_\phi\cl{D}[\h{n}]+\kappa_1(1+n_\n{th})\cl{D}[\h{a}]+\kappa_1 n_\n{th}\cl{D}[\h{a}^\dagger]$ where we assume equal dephasing and boson loss rates, $\kappa_\phi=\kappa_1=10^{-2}\kappa_\n{eff}$, and a thermal occupation number $n_\n{th}=0.01$. For each slope value $s_f=s_g$, we perform two simulations with initial states $\ket{\Xi_0}+\ket{\Xi_1}$ (nonlinear coherent state) and $\ket{\Xi_0}$ (nonlinear cat state), respectively. We then extract the logical bit-flip and phase-flip error rates by fitting the exponential decay of fidelity over time. The results, shown in Fig.~\ref{fig_nb:error_rates}, reveal a clear separation between the two error rates, confirming the presence of noise bias. Additionally, we observe a several-orders-of-magnitude decrease in the bit-flip rate as the Mandel $Q$ parameter decreases, compared to coherent-like states ($Q=0$). This reduction directly stems from the increase in the confinement rate $\kappa_\n{conf}$ with decreasing slope values (see Fig.~\ref{fig4:conf_vs_angle}). The bit-flip rate reaches a minimum at $Q\approx-0.5$ before rising again. This behavior arises because, for very small $Q$, the stabilized distribution becomes too squeezed, making small rotations sufficient to flip a right-coherent state into a left-coherent state and vice versa. In Fig.~\ref{fig_nb:noise_bias}, we plot the noise bias, defined as the ratio of phase-flip to bit-flip rates, as a function of the Mandel $Q$ parameter, normalized by its value at $Q=0$. We see that the bias can be increased by six orders of magnitude by decreasing the Mandel $Q$ parameter, without changing the crossing point $k^*$. This behavior is observed in other $(r,l)$ schemes and can be further enhanced by increasing the height of the crossing point~$h^*$.  
}

\subsection{Correction capability of $(1,1)$ models} \label{subsec:corr_11}

As seen previously, although the scheme $(1,1)$ does not have the ability to correct for dephasing errors, it can still stabilize the bosonic system into one of the two nonlinear coherent states or their mixture. This difference between a stabilization and correction power can again be straightforwardly concluded from the form of the operator $\h{K}$ when written in the nonlinear shifted Fock state formalism
\be \label{eq:Leff_disp_fock_11}
    \h{K} \approx 2 \, 
    \frac{\sqrt{\tilde{g}(\expval{\h{n}}-r)\tilde{f}(\expval{\h{n}})}}{\expval{(\Delta \h{n})^2}} \, 
    \h{Z}_L \otimes \h{A}\,.
\ee
We see that in this case the operator $\h{K}$ is a combination of the logical Pauli $Z$ operator and the gauge mode annihilation operator. This implies that when $(r,l)=(1,1)$, every application of the stabilization jump operator $\h{K}$ does not only remove excitation from the gauge mode but also scrambles the logical information. 

Interestingly, however, this does not imply that $(1,1)$ schemes are incapable of correcting any type of errors. To show this, we must consider error processes composed of the operator $\h{p}\propto\h{a}-\h{a}^\dagger$ that we will refer to as momentum errors. The Liouvillian $\cl{D}\!\left[\sqrt{\kappa_\n{diff}}\, \h{p}\right]$ represents a diffusive process of the bosonic state in position with a diffusion constant $\kappa_\mathrm{diff}$~\cite{gottesman_encoding_2001}. Phenomenologically, this process spreads the state distribution equally to lower and higher Fock states. Given this, and the fact that the phase relationship between the boson raising and lowering processes is identical in both $\h{p}$ and $\h{K}$ as presented in Eq.~\eqref{eq:hamilton_modif}, we expect the interference between $\tilde{f}$ and $\tilde{g}$ to remain unchanged by the noise. This allows the stabilization to refocus the boson population around the crossing point without disturbing the logical information.

\begin{figure}[b!]
    \includegraphics[]{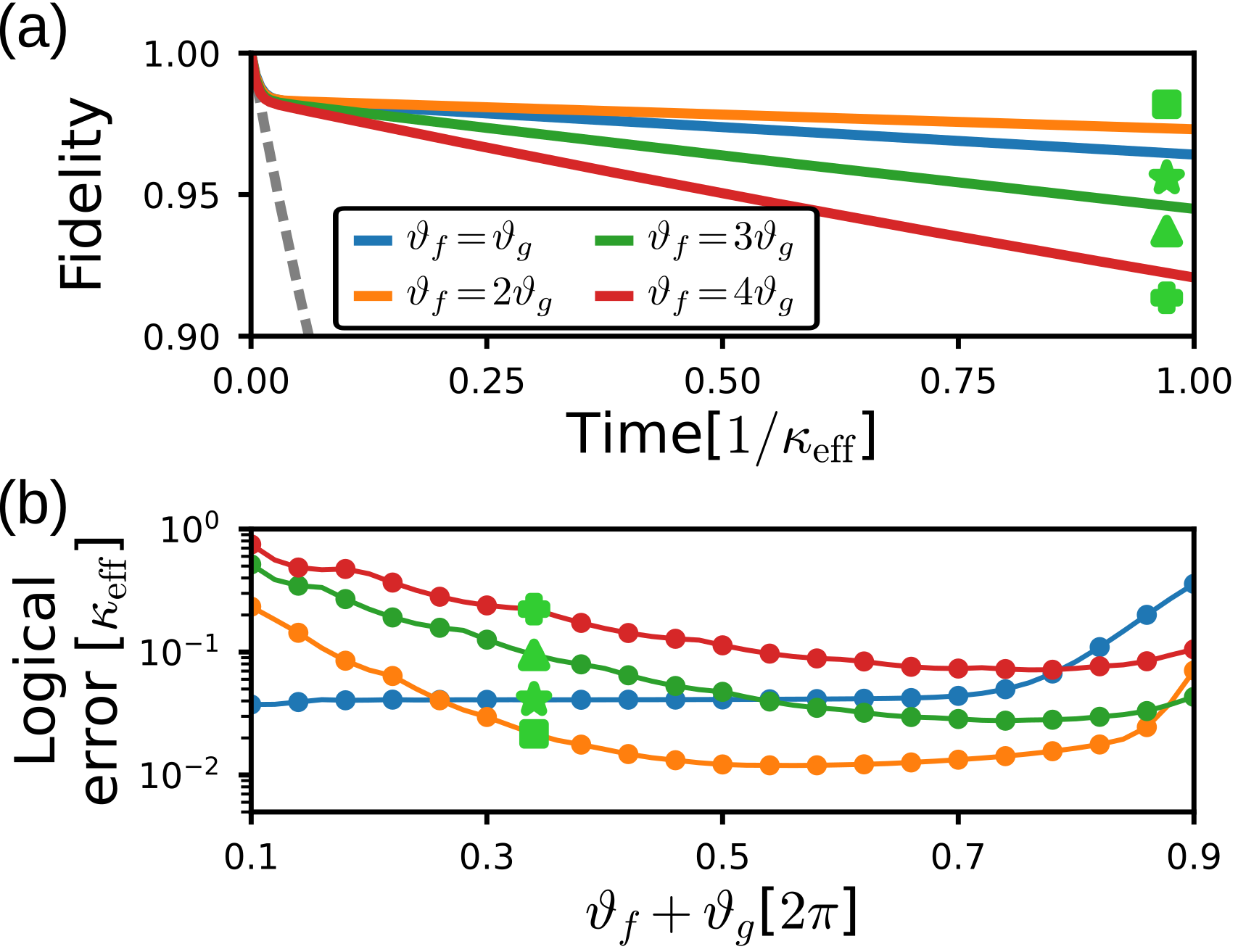}
    {\phantomsubcaption\label{fig_corr_11:fidel_vs_time}}
    {\phantomsubcaption\label{fig_corr_11:logical_rate}}
    \caption{Correction capability of $(1,1)$ models. (a) The fidelity of the initial state $\ket{\Xi_0}$ with itself over time under simultaneous stabilization and momentum errors modeled using jump operators ${\sqrt{\kappa_\mathrm{diff}}\,\h{p}}$ with $\kappa_\n{diff} = 0.5 \kappa_\n{eff}$. The stabilization is performed using linear $\tilde{f}$ and $\tilde{g}$ intersecting at $k^*=10$, $h^*=10$ and $\vartheta_f+\vartheta_g=2\pi/3$ (see Fig.~\ref{fig3}). The evolution is also shown for the $(0,2)$ model (dashed line). (b) The logical error rate for $(1,1)$ schemes as a function of the relative angle between $\tilde{f}$ and $\tilde{g}$ at the crossing point. For symmetric boson distributions, i.e. $\vartheta_f=\vartheta_g$, the rate stays constant. For asymmetric distributions, the rate can be improved by increasing their variance. However, when the relative angle is too large, the rate increases due to the mixing effect discussed in Sec.~\ref{subsec:number_of_steady_states}.}
    \label{fig_corr_11}
\end{figure}

The correction mechanism can also be understood more rigorously through the lens of the nonlinear shifted Fock state method. Combining Eqs.~\eqref{eq:destroy_disp_fock} and~\eqref{eq:a2_and_at_disp_fock}, we see that $\h{a} - \h{a}^\dagger$ does not include a purely logical $\h{Z}_L$ operator, meaning it will necessarily be accompanied by a change in the gauge mode excitation. This excitation, along with the logical error, is then removed using the correction operator given in Eq.~\eqref{eq:Leff_disp_fock_11}.

We demonstrate this in Fig.~\ref{fig_corr_11:fidel_vs_time}, which illustrates the fidelity of the initial state $\ket{\Xi_0}$ with itself over time under simultaneous stabilization and error processes ($\kappa_\mathrm{diff} = 0.5 \kappa_\mathrm{eff}$). We also illustrate the evolution when the $(0,2)$ model is used. The dynamics are drastically different, with the $(1,1)$ scheme showing robustness against momentum errors. In Fig.~\ref{fig_corr_11:logical_rate}, we plot the logical decay rate for both models (obtained through an exponential fit of the fidelity time evolution) as a function of the error rate. This highlights the robustness of the $(1,1)$ scheme against $\h{p}$ errors, even at an error rate of $0.5 \kappa_\mathrm{eff}$. Qualitatively, under such strong noise, a cat state stabilized using $(0,2)$ or $(2,0)$ would be completely mixed after a time $1/\kappa_\mathrm{eff}$, while with the $(1,1)$ scheme the fidelity remains consistently above $95\%$.

In terms of the parameters of the NLRE method, we find that for symmetric boson distributions, i.e. $s_f=s_g$, the logical decay rate remains unaffected by the variance of the states. Conversely, in other cases, increasing the variance enhances protection against $\h{p}$ errors (see Fig.~\ref{fig_corr_11:logical_rate}). Finally, increasing the height $h^*$ results in slower decoherence of the logical information, corroborating our earlier findings on the effective confinement rate.

\medskip
In summary, we established that the NLRE toy model from Eq.~\eqref{eq:hamilton_modif} exhibits different error correction properties depending on the values of $r$ and $l$. We determined these properties through symmetry arguments, subsystem decomposition picture or the modeling of the effective potential well. Interestingly, we find that $(2,0)$ schemes allow for autonomous correction of dephasing errors similarly to the well known property of the standard cat-code stabilization schemes. While $(1,1)$ schemes do not protect against dephasing errors, we demonstrate that they allow for autonomous correction of quadrature errors. By exploiting the nonlinear nature of the stabilization operator, performance of these rotation symmetric codes can be enhanced. Our analysis can be readily generalized to design $(r,l)$ schemes able to correct against more complex error operators. {Specifically, a general NLRE model would protect the cat-like codes against bosonic operators of the form $\h{E} \propto \h{a}^{\dagger\,\star} \, F_{\!\star}(\h{n}) + F_{\!\diamond}(\h{n}) \,\h{a}^{\diamond}$, i.e. similar to the operator $\h{K}$ from Eq.~\eqref{eq:hamilton_modif}. The conditions that the integers $\star$ and $\diamond$ and functions $F_{\!\star/\!\diamond}$ must satisfy are determined by the symmetry of the dynamics and extend beyond cat code stabilization and the scope of the current work. We leave their exploration for future studies.}

\section{Trapped ion setting} \label{sec:trapped_ions}

In the following section, we present how NLRE can be implemented in trapped ion settings and, as an example, propose a feasible physical realization of cat-like code stabilization schemes using a single trapped ion.

We consider a single trapped ion of mass $m$ and electronic states $\ket{g}$ and $\ket{e}$ constituting a two-level system. The bosonic degree of freedom corresponds to the quantized motion of the ion along one of the trap axes, which we call $z$. In the resolved sideband regime, the spin and the bosonic mode can be coupled by driving a laser field with polarization satisfying the selection rule for the ${\ket{g}\leftrightarrow\ket{e}}$ transition and detuned from the resonant transition frequency by $k_\SB\,\omega_{m}$, with $k_\SB\in\mathbb{Z}$ and $\omega_{m}$ the motional frequency.
By adopting the rotating wave approximation, the light-matter Hamiltonian in the interaction picture reads 
\be \label{eq:trapped_ions_hamilton}
    \h{H}(k_\SB,\eta) \!= \Omega\, e^{i\eta(\h{a}^{\dagger}e^{-i\omega_m t} + \h{a} e^{i\omega_m t})} e^{i k_\SB \omega_m t} \h{\sigma}_+ + \n{H.c.}
\ee
where $\eta$ is the so-called Lamb-Dicke parameter which quantifies the strength of the motional coupling. For ${k_\SB\geq0}$ (${k_\SB<0}$), the exponential term within $\h{H}(k_\SB,\eta)$ can be reformulated, using the rotating wave approximation a second time, as $\h{a}^{\dagger\,\abs{k_\SB}} f(\h{n},k_\SB,\eta)$ \big($f(\h{n},k_\SB,\eta) \h{a}^{\abs{k_\SB}}$\big) with
\be \label{eq:f_function_ions}
    f(\h{n},k_\SB,\eta) = e^{-\frac{1}{2}\eta^2} (i\eta)^\abs{k_\SB} \sum_{l=0}^{\infty} 
    \frac{(-1)^l \eta^{2 l}}{l! (l+\abs{k_\SB})!} \h{a}^{\dagger l} \h{a}^l \,.
\ee
This implies that when ${k_\SB>0}$ (${k_\SB<0}$) the process consists in the addition of ${r = k_\SB}$ (removal of ${l=\abs{k_\SB}}$) bosons. {Importantly, Eqs.\eqref{eq:trapped_ions_hamilton} and\eqref{eq:f_function_ions} are derived not through Floquet engineering and high-frequency expansion (or other approximate methods) but by directly reformulating the full light-matter interaction, incorporating accurate rotating wave approximations (see Appendix~\ref{app:nonlinear_important}).} We obtain the desired Hamiltonian in Eq.~\eqref{eq:hamiltonian} by driving a bichromatic laser field with detunings $r\omega_m$ and $-l\omega_m$ from the ${\ket{g}\leftrightarrow\ket{e}}$ transition frequency, where we identify ${f(\h{n},r,\eta_r)\equiv f(\h{n})}$ and ${f(\h{n},l,\eta_l)\equiv g(\h{n})}$. The relaxation of the two-level system is engineered {through optical pumping, i.e. the} coupling $\ket{e}$ to an additional short lived electronic state {$\ket{f}$. Assuming that $\ket{f}$ decay quickly to the spin ground state $\ket{g}$, we can adiabatic eliminate the former which} leads to the desired jump operator ${\h{L}=\sqrt{\gamma}\h{\sigma}_{-}}$ with a rate $\gamma$ {that can be adjusted by tuning the coupling strength between $\ket{e}$ and $\ket{f}$}.

\begin{figure*}
    \includegraphics[]{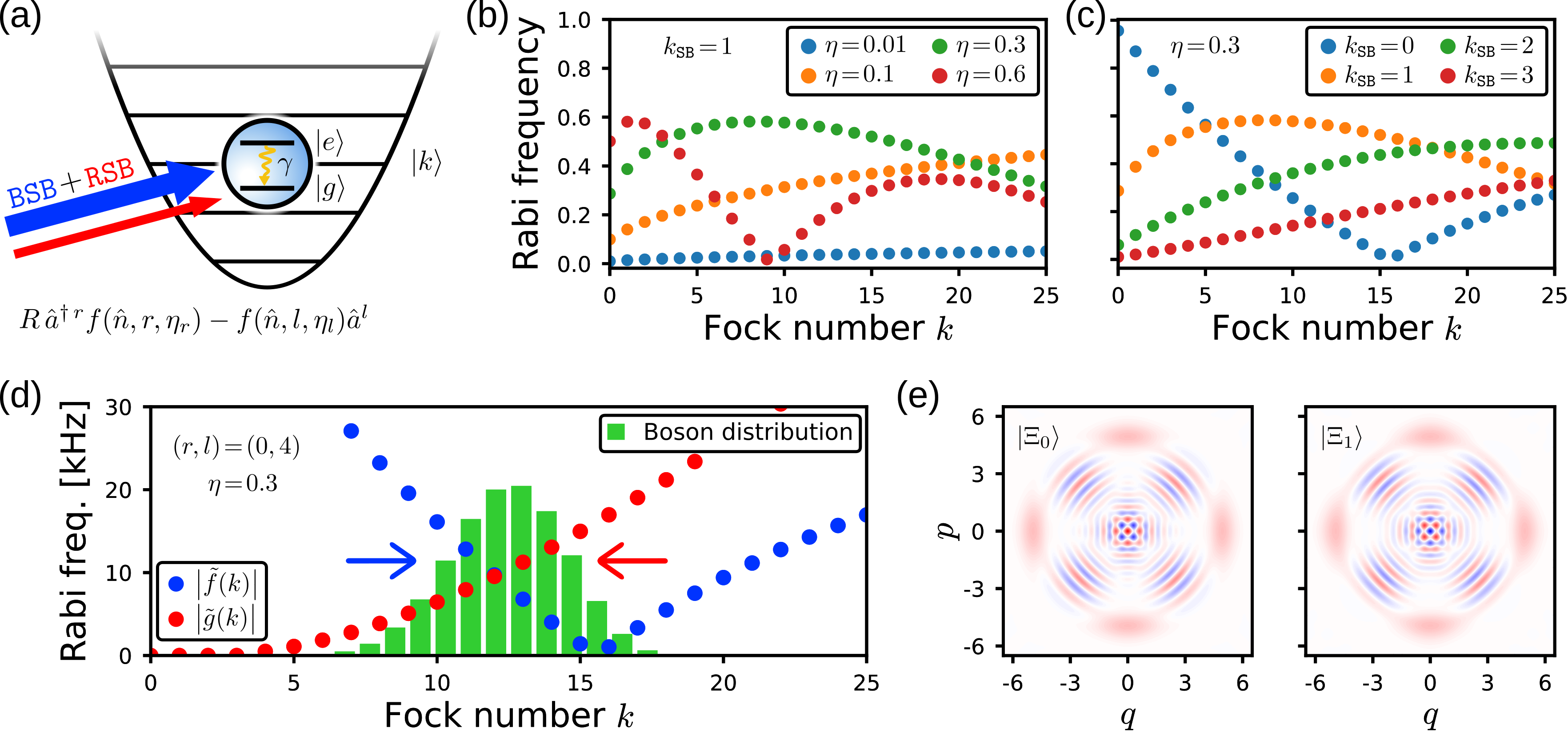}
    {\phantomsubcaption\label{fig_trap_ions:scheme}}
    {\phantomsubcaption\label{fig_trap_ions:sideband_vs_LD}}
    {\phantomsubcaption\label{fig_trap_ions:sideband_vs_order}}
    {\phantomsubcaption\label{fig_trap_ions:rabi_frequency}}
    {\phantomsubcaption\label{fig_trap_ions:wigner_functions}}
    \caption{NLRE method for trapped ion settings. (a) Schematic representation of the trapped ion setup with states $\ket{e}$ and $\ket{g}$ being two electronic states the ion and the bosonic part corresponding to its motion. The toy model of NLRE is engineered using a bichromatic laser field with the nonlinear functions $f$ given in Eq.~\eqref{eq:f_function_ions}. (b-c) Rabi frequencies $\tilde{f}$ and $\tilde{g}$ (which take the same functional form of Eq.~\eqref{eq:Bessel_approx_rabi_freq}) for various Lamb-Dicke parameters $\eta$ and sideband orders~$k_\SB$. (d) Example of the $(r,l)=(0,4)$ scheme using a laser-ion interaction with $\eta=0.3$ and the ratio between laser field strengths $R=\Omega_r/\Omega_l$.}
    \label{fig_trap_ions}
\end{figure*}

Using Eq.~\eqref{eq:f_function_ions} we find the analytical expressions for $\tilde{f}$ and $\tilde{g}$. These are commonly referred to as the Rabi frequencies and are written using factorials and generalized Laguerre polynomials~\cite{wineland_experimental_1998}. However, we find that the analytical expression for these functions can be well approximated using a unique Bessel function of first kind (see Appendix~\ref{app:bessel_approx} for details and numerical comparison)
\be \label{eq:Bessel_approx_rabi_freq}
    \tilde{f}(k) \approx J_{r}\!\left( 2\eta_r\sqrt{k+\frac{r+1}{2}} \right) \,.
\ee
Since $\tilde{f}$ and $\tilde{g}$ take the same functional form, $\tilde{g}$ is obtained by substituting $r$ with $l$ and $\eta_r$ with $\eta_l$ in Eq.~\eqref{eq:Bessel_approx_rabi_freq}. Examples of the Rabi frequencies for a trapped ion system are shown in Figs.~\ref{fig_trap_ions:sideband_vs_LD} and~\ref{fig_trap_ions:sideband_vs_order} for various $\eta$ and $k_\SB$. Hence, in this implementation the crossing points $k^*$ correspond to the intersections of two Bessel functions. Finding an analytical expression of such intersections remains, to the best of our knowledge, an unresolved question in the field of orthogonal polynomials. In practice, the simple form of the expression in Eq.~\eqref{eq:Bessel_approx_rabi_freq} allows for a more straightforward numerical determination of both the crossing points and the slopes of the Rabi frequencies at these points than the exact analytical formula based on Laguerre polynomials.

The position and slopes at the crossing points can be engineered using several parameters. First, the strengths $\Omega_r$ and $\Omega_l$  of the boson addition and removal processes can be controlled by the intensity of the laser fields. Without loss of generality, we can set $\Omega_l$ to be the global strength $\Omega$ from Eq.~\eqref{eq:hamiltonian}, then $\tilde{f}$ gets a factor $R=\Omega_r/\Omega_l$ which corresponds to the relative intensity of the electric fields. Second, the Lamb-Dicke parameter reads $\eta=k_{eg}z_0\cos{\theta}$ with $k_{eg}$ being the amplitude of the wave vector of the field driving the transition ${\ket{g}\leftrightarrow\ket{e}}$, $z_0=\sqrt{\hbar/\!\left(2m\omega_{m}\right)}$ the zero-point motion of the ion and $\theta$ the angle between the ion's motion and the laser beam. While $k_{eg}$ and $z_0$ affect the Lamb-Dicke parameter of both raising and lowering processes equally, the angle $\theta$ allows us to tune $\eta_r$ and $\eta_l$ independently. In total, four parameters can be used to engineer the toy model of the NLRE method. We show an example of such engineering in Fig.~\ref{fig_trap_ions:rabi_frequency} for $(r,l)=(0,4)$, $R=0.2$ and $\eta_r=\eta_l=0.3$ together with the Wigner quasiprobability of the stabilized cat-like states $\ket{\Xi_0}$ and $\ket{\Xi_1}$ in Fig.~\ref{fig_trap_ions:wigner_functions}. 

From Fig.~\ref{fig_trap_ions:sideband_vs_LD}, we note that to observe the nonlinear aspect of the light-matter interaction in Eq.~\eqref{eq:f_function_ions} one requires a sufficiently large $\eta$, which means operating the trapped ion setup outside the Lamb-Dicke regime. The Lamb-Dicke regime is defined by restricting the ion's motional excursion to be much smaller than the laser wavelength. A necessary condition for this is ${\eta\ll1}$~\cite{wineland_experimental_1998}, in which case the light-matter interaction is characterized by ${f(\hat{n})=g(\hat{n})=\n{const}}$ for Fock states near the ground state. However, remaining in the Lamb-Dicke regime also implies that processes involving orders $\abs{k_\SB}>1$ are challenging to realize, due to the unfavorable scaling of the Rabi frequency's strength with $\eta^{\abs{k_\SB}}$. This leads to slow interactions that can exceed both the oscillator and spin coherence times. In order to implement the NLRE schemes we go beyond the Lamb-Dicke regime and consider the nonlinear light-matter interaction in Eq.~\eqref{eq:f_function_ions} as a feature rather than a limitation. This approach enables the utilization of processes of arbitrary $k_\SB$ orders to stabilize motional subspaces through a adequate engineering of $\tilde{f}$ and $\tilde{g}$. Although the concept of trapped ion operations beyond the Lamb-Dicke regime is not novel, with several theoretical proposals~\cite{vogel_nonlinear_1995,morigi_ground_state_1997,morigi_laser_1999,wallentowitz_high_order_1999,carvalho_decoherence_2001,cheng_nonlinear_2018,joshi_population_2019,puebla_quantum_2019} and a handful of experimental works~\cite{mcdonnell_long-lived_2007,stutter_sideband_2018,hrmo_sideband_2019,jarlaud_coherence_2020} (primarily focused on cooling and state preparation), the NLRE method provides a straightforward and well defined methodology to explore such operations in the context of manifold stabilization and also quantum error correction. 

Let us now consider a specific example of a single beryllium-9 ion (${}^{9}\mathrm{Be}^+$) confined in a Penning micro-trap~\cite{jain_penning_2024}, with a motional frequency $\omega_m=2\pi\times\qty{2.5}{\MHz}$. We denote the two electronic states from the $S_{1/2}$ manifold as $\ket{g}$ and $\ket{e}$, which we optically couple using two Raman beams through a virtual level \qty{313}{\nm} away from $S_{1/2}$. The engineered relaxation is achieved via a fast-decaying state in the $P_{3/2}$ manifold with a effective rate set to ${1/\gamma=\qty{7}{\mu s}}$. By aligning counter-propagating Raman beams with a certain angle to the motion of the ion, we can vary the Lamb-Dicke parameter in the range $\eta\in[0,0.6]$. We fix the strength of the boson raising and lowering processes to be $\Omega_l=2\pi\times\qty{50}{\kHz}$ and $\Omega_r=R\,\Omega_l$ with $R=0.2$. {For these parameter values, the Rabi frequency near the crossing point will be on the order of $\qty{10}{\kHz}$, which is lower than $\gamma$, ensuring that the spin degree of freedom can be adiabatically eliminated. Consequently, the stabilization rate $\kappa_\n{eff}$ will also be on the order of $\qty{10}{\kHz}$, which, as we show below, exceeds the rates of the decoherence processes present in the system.} Among the natural disturbances, we consider four types of errors: motional dephasing, motional heating/cooling, spin dephasing, and photon recoil effect. The first three are modeled as Lindblad jump operators ${\h{L}_\phi=\sqrt{\kappa_\phi}\h{n}}$, ${\h{L}_{h}=\sqrt{\kappa_h}\h{a}^\dagger}$, ${\h{L}_{c}\approx\sqrt{\kappa_h}\h{a}}$, and ${\h{L}_\n{s}=\sqrt{\kappa_\n{s}}\h{\sigma}_{z}}$, with measured rates $1/\kappa_\phi=\qty{66}{ms}$, $1/\kappa_h=\qty{10}{\s}$, and ${1/\kappa_\n{s}=\qty{1.12}{ms}}$. The photon recoil effect results from the spontaneous emission of the $P_{3/2}$ level. Specifically, at every jump ${\h{L}=\sqrt{\gamma}\h{\sigma}_{-}}$, a photon is emitted in a random direction according to the dipolar angular distribution~\cite{stenholm_semiclassical_1986,cirac_laser_1992}, imparting a momentum kick to the ion. We find that this recoil effect and the engineered jump operator can be efficiently modeled together as a single superoperator (see Appendix~\ref{app:recoil} for more details).

\begin{figure*}
    \includegraphics[]{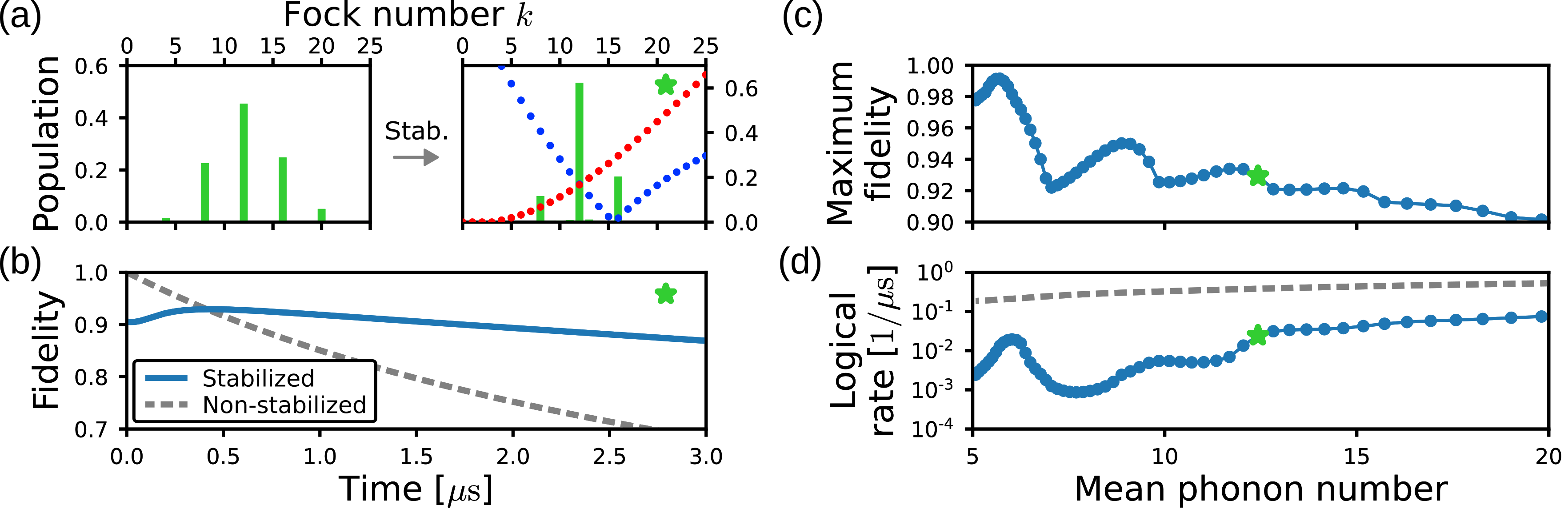}
    {\phantomsubcaption\label{fig8:recoil_cat}}
    {\phantomsubcaption\label{fig8:fidelity_time}}
    {\phantomsubcaption\label{fig8:max_fidelity}}
    {\phantomsubcaption\label{fig8:logical rate}}
    \caption{Stabilization starting from a four-component cat state. (a) Initial and final boson distributions after approximately $\qty{0.5}{\mu\s}$ of stabilization for a single trapped beryllium ion in a Penning micro-trap. The system is stabilized using the $(r,l)=(0,4)$ scheme with a Lamb-Dicke parameter $\eta=0.3$ and a ratio $R=\Omega_r/\Omega_l = 0.2$ (see Fig.~\ref{fig_trap_ions}). The system experiences motional dephasing, heating/cooling, spin dephasing, and photon recoil due to spontaneous emission. The initial motional state is a four-component cat state with the same parity and mean phonon number as the target state $\ket{\Xi_0}$. (b) System's fidelity with the target state as a function of time. The dashed curve represents the fidelity of the cat state with itself when only noise is present. (c) Maximum fidelity of the stabilized state as a function of the mean phonon number, with $\eta\in[0.22,0.5]$. (d) Logical decay rate of the stabilized state as a function of the mean phonon number. The dashed line represents the decay rate of the non-stabilized cat state.}
    \label{fig8}
\end{figure*}

In order to demonstrate the efficacy of NLRE method for the trapped ion setting, we simulate manifold stabilization using the $(r,l)=(0,4)$ scheme and account for natural disturbances of the system. We do this for different Lamb-Dicke parameters $\eta\in[0.22,0.5]$ (here $\eta_r=\eta_l=\eta$) which affects both the mean and variance of the desired steady states while keeping the height $h^*$ of the crossing point constant. We set the electronic state to be $\ket{g}$ and initialize the oscillator in a four-component cat state with the same modular-parity and mean phonon number as the desired target state $\ket{\Xi_0}$ (see Fig.~\ref{fig8:recoil_cat}). This initialization can be performed through the use of standard techniques~\cite{lo_spinmotion_2015,kienzler_observation_2016,fluhmann_sequential_2018}. In the absence of disturbances, we expect the system's fidelity with the target state to exponentially approach unity over time. However, in the presence of disturbances, the fidelity will reach a finite maximum before exponentially decaying due to error accumulation. This evolution is depicted in Fig.~\ref{fig8:fidelity_time} for $\eta=0.3$. Fig.~\ref{fig8:max_fidelity} illustrates the maximal fidelity as a function of the mean phonon number $\expval{\h{n}}$. We observe that the fidelity decreases with $\expval{\h{n}}$, i.e. when the system starts in large cat states. This is because the initial (Poissonian) boson distribution is much broader than the stabilized one, necessitating more occurrences of the effective jump operator $\hat{K}$ to reach the dark state manifold. Consequently, this leads to more photon recoils and a greater mixing of population between the logical states. {Once the system is in the dark manifold, the fidelity will remain constant until the system is perturbed by an external source embodied by the phonon loss, gain or dephasing. Such perturbations will trigger once again the execution of stabilization jump operator $\h{K}$, thus more photon recoils and a slow exponential decay of the fidelity.} Nonetheless, our results demonstrate that even with this basic state initialization, cat-like states with large $\expval{\hat{n}}$ can be stabilized above $90\%$ fidelity. The oscillations in Fig.~\ref{fig8:max_fidelity} are explained by the fact that some initial states are closer to the desired one and thus lead to less photon recoil. For all the cases, we extract the logical error rate from an exponential fit of the time evolution. As shown in Fig.~\ref{fig8:logical rate}, the rate for the stabilized evolution is consistently lower compared to that of the non-stabilized cat state. {Moreover, we observe that the rate decreases with $\eta$ which is due to the fact that motional dephasing becomes more important for large cat states.} To further improve the fidelity, one can consider the $(r,l)=(3,1)$ scheme which, as discussed in Sec.~\ref{subsec:number_of_steady_states}, can stabilize the system into $\ket{\Xi_0}$, the only true dark state of the system, and then switch to the $(0,4)$ scheme. Another possibility is to adiabatically change the strength ratio $R$ from $0$ to the desired value during the stabilization process.

\medskip
With this section, we demonstrated that the light-matter interaction natively present in trapped ion systems provides nonlinearities suitable for the NLRE method. Our non-perturbative treatment of this interaction (using Bessel functions) offers a new approach to the study of trapped ions outside the Lamb-Dicke regime.

\begin{figure*}
    \includegraphics[]{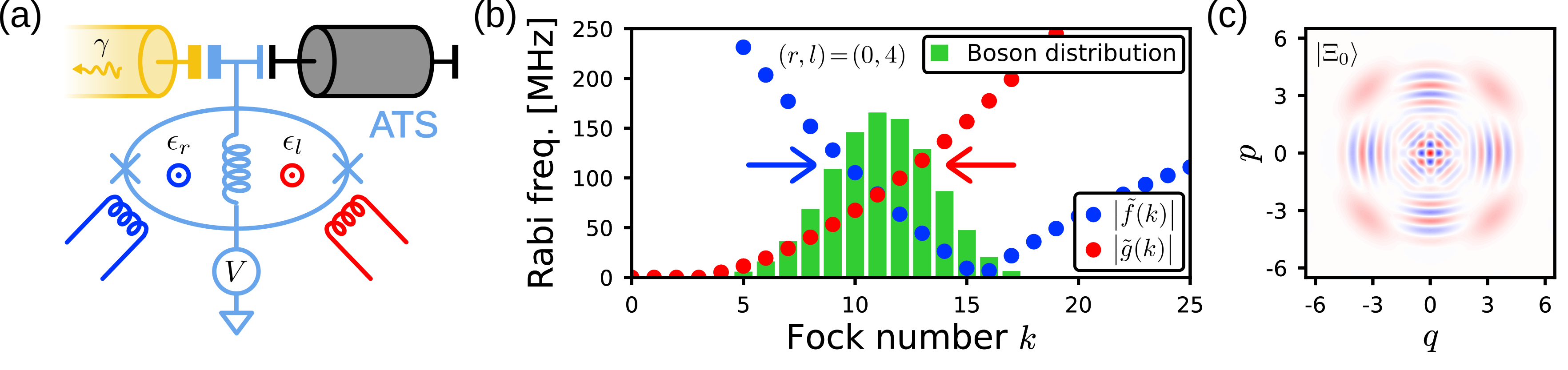}
    {\phantomsubcaption\label{fig_ATS:circuit}}
    {\phantomsubcaption\label{fig_ATS:rabi_frequency}}
    {\phantomsubcaption\label{fig_ATS:wigner_functions}}
    \caption{NLRE method for superconducting circuit QED settings. (a) Schematic representation of cQED with a high-quality $LC$ oscillator (black) storing the cat-like state manifold and a nonlinear circuit (blue) coupled to a dissipative environment represented here by a lossy $LC$ oscillator (yellow). The nonlinear elements is a DC-biased asymmetrically threaded superconducting quantum interference device (ATS). The toy model of NLRE is engineered using a two-tone flux drive. (b) Example of the $(r,l)=(0,4)$ scheme using a voltage-biased ATS circuit with parameters specified in the main text. (c) Wigner quasiprobability of the stabilized cat-like states $\ket{\Xi_{0}}$ for the scheme in (b).}
    \label{fig_ATS}
\end{figure*}

\section{Superconducting circuit QED setting} \label{sec:cqed_setting}

We now demonstrate how the NLRE method can facilitate a more intuitive understanding and design of dissipative stabilization schemes for rotation-symmetric codes within the framework of superconducting circuit QED systems. As shown in Sec.~\ref{subsec:comparison}, the standard dissipative stabilization of coherent state manifolds is a special case of our toy model. Consequently, pioneering works~\cite{leghtas_confining_2015,touzard_coherent_2018} and recent advances~\cite{lescanne_exponential_2020,berdou_one_2023,reglade_quantum_2024,cohen_degeneracy-preserving_2017,smith_superconducting_2020,smith_magnifying_2022,smith_spectral_2025} in cat codes can be straightforwardly reinterpreted through the prism of NLRE. Interested readers can find more details in Appendix~\ref{app:nlre_cqed}. Here, we will present a novel and versatile stabilization scheme for nonlinear cat-like manifolds using DC-biased asymmetrically threaded superconducting quantum interference devices (ATS)~\cite{lescanne_exponential_2020}.

We consider a generic dissipative system consisting of a high-quality factor ($Q$) quantum $LC$ oscillator capacitively coupled to a superconducting nonlinear element, which in turn is coupled to a dissipative environment comprising a low-$Q$ oscillator or a transmission line. A schematic of the circuit is depicted in Fig.~\ref{fig_ATS:circuit}. The high-$Q$ resonator, described by the ladder operator $\h{a}$ and frequency $\omega_a$, serves as a storage medium for the stabilized manifolds, while the environment is modeled as a bosonic reservoir of frequency $\omega_c$ characterized by the ladder operator $\h{c}$. We assume that the nonlinear element remains in its ground state throughout the dynamics and that the system-bath coupling is weak relative to the inverse characteristic correlation time of the bath. Under these conditions, the dissipative dynamics of the system can be accurately modeled using the Lindblad master equation with a single jump operator ${\h{L}=\sqrt{\gamma}\,\h{c}}$.

As a nonlinear element we consider an ATS device which corresponds to a superconducting quantum interference device (SQUID) shunted in its center by a large inductance. This device, demonstrated in Ref.~\cite{lescanne_exponential_2020}, offers additional tunability compared to a single Josephson junction or a SQUID thanks to the use of two external fluxes. It is often assumed that the Josephson energies $E_J$ of the junctions composing the ATS are equal (in practice the asymmetry amounts to $<1\%$ relative to $E_J$)~\cite{lescanne_exponential_2020,berdou_one_2023,reglade_quantum_2024}. Finally, we bias the circuit with an additional DC voltage $V$ which induces a voltage drop across the nonlinear element, influencing the number of Cooper pairs flowing through the Josephson junctions. Assuming $V$ to be much smaller than the superconducting gap voltage of the junctions composing the ATS, the effective Hamiltonian in the frame rotating with the two oscillators reads~\footnote{We consider $\h{a}$ and $\h{c}$ to be already the dressed modes of the circuit.}
\be \label{eq:hamil_dc_biased_ATS}
\begin{split}
    \h{H} = -
     2 E_J \cos\!\left(\varphi_\Sigma\right) \cos\!\left( \h{\varphi}(t) + \varphi_\Delta + \omega_V t \right)
\end{split}
\ee
where $\varphi_\Sigma$ and $\varphi_\Delta$ denote the sum and difference of the external fluxes, ${\omega_V\equiv2eV/\hbar}$ represents the frequency associated with voltage bias~\cite{armour_universal_2013,gramich_coulomb-blockade_2013,trif_photon_2015}, and $\h{\varphi}(t) = \varphi_a e^{-i\omega_a t} \h{a} + \varphi_c e^{-i\omega_c t} \h{c} + \n{H.c.}$ is the phase operator across the ATS device with $\varphi_a$ and $\varphi_c$ being the dimensionless zero-point phase fluctuations of each mode. 

For context, such ATS devices allowed the demonstration of the dissipatively stabilized cat codes to exponentially suppress dephasing errors with increasing mean photon number~\cite{lescanne_exponential_2020,berdou_one_2023,reglade_quantum_2024}. On the other hand, single DC-biased Josephson junctions coupled to an $LC$ oscillator have been investigated in the context of quantum heat engines~\cite{hofer_quantum_2016} and the preparation and emission of Fock states~\cite{souquet_fock-state_2016}. In these studies, the authors exploit the multi-photon capability of voltage-biased junctions by setting $\omega_V$ to a multiple of the oscillator's frequency, a capability that has been demonstrated in experiments~\cite{hofheinz_bright_2011,rolland_antibunched_2019,peugeot_generating_2021,menard_emission_2022}. The application of both ATS devices and DC-biased circuits in quantum information processing is an active area of research.

In our approach, we drive the external fluxes using a modulated radio frequency signal such that $\varphi_\Delta=\pi/2$ and $\varphi_\Sigma=\pi/2+\epsilon(t)$. Assuming a two-tone modulation $\epsilon(t)=\epsilon_r\cos(\omega_rt)+\epsilon_l\cos(\omega_lt)$ with amplitude $\abs{\epsilon(t)}\ll1$, we use the small-angle approximation and simplify the Hamiltonian in Eq.~\eqref{eq:hamil_dc_biased_ATS} as
\be \label{eq:hamil_ATS_rot}
\begin{split}
    \h{H} \approx &- E_J \epsilon_r \left( e^{-i\omega_r t} + e^{i\omega_r t}\right) 
    \sin\!\left( \h{\varphi}(t) + \omega_V t \right) - \\
    &- E_J \epsilon_l \left( e^{-i\omega_l t} + e^{i\omega_l t}\right) 
    \sin\!\left( \h{\varphi}(t) + \omega_V t \right)\,.
\end{split}
\ee
Writing the trigonometric functions in their exponential form allows us to identify four resonant terms with frequencies $\omega_V \pm \omega_{r/l}$. We set both the bias voltage and flux drive frequencies to be linear combinations of $\omega_{a}$ and $\omega_{c}$.  Without loss of generality, we choose $\omega_V + \omega_r = r\omega_a - \omega_c$ and $\omega_V + \omega_l = l\omega_a - \omega_c$ with $r, l \in \mathbb{Z}$. To ensure that only these two processes are resonant, we propose using $\omega_V - \omega_{r/l} = k_a\omega_a/2 + k_c\omega_c/2$ with $k_{a/c} \in \mathbb{Z}$ arbitrarily chosen. This choice causes all undesired processes to be off-resonant and fast-rotating, allowing them to be neglected using the rotating wave approximation. In the case of $r>0$ and $l<0$, the Hamiltonian can be rewritten in terms of the same nonlinear functions $f(\h{n},k_\SB,\eta)$ used in the trapped ion setting (see Eq.~\eqref{eq:f_function_ions}). Assuming the reservoir's temperature to be 0, we obtain
\be \label{eq:hamil_cQED_NLRE}
\begin{split}
    \h{H} \approx - 
    \left(\Omega_r\,\h{a}^{\dagger\,\abs{r}} f(\h{n}) + \Omega_l\,g(\h{n}) \, \h{a}^{\abs{l}}\right) 
    \h{c}^{\dagger} + \n{H.c.}
\end{split}
\ee
with ${\Omega_{r/l} = i E_J\,e^{-\varphi_c^2/2}\,\varphi_c\,\epsilon_{r/l}}$, ${f(\h{n})=f(\h{n},r,\varphi_a)}$ and ${g(\h{n})=f(\h{n},l,\varphi_a)}$. This expression shows that the Hamiltonian of the NLRE toy model is straightforwardly realized in such systems. Instead of the usual expansion of $f(\h{n})$ and $g(\h{n})$ in terms of the number operator $\h{n}$ and the phase fluctuations $\varphi_a$, we tune the system's parameters $\varphi_{a/c}$, $E_J$ and $\epsilon_{r/l}$ to engineer a desired crossing point between the scalar functions $\tilde{f}$ and $\tilde{g}$. The zero-point phase fluctuations typically range from $0.1$ to $1.0$ and are fixed together with $E_J$ by the experimental device parameters (capacitances and inductances). The strengths of the microwave drives $\epsilon_{r/l}$ are thus the free parameters that can allow an in situ engineering of the crossing points. The range of their tunability is limited by the condition ${\abs{\epsilon(t)}\ll1}$. Overall, the combined voltage bias and two-tone drive approach enables the realization of Hamiltonians akin to those derived from the light-matter interaction and thus provide a novel way to engineer bosonic interactions in superconducting circuit QED systems. A detailed derivation and analysis of the Hamiltonian for this setting is provided in Appendix~\ref{app:dc_bias_cqed}.

Let us now consider a specific example of an ATS with Josephson energy ${E_J=2\pi\times\qty{45}{\GHz}}$ (see Eq.~\eqref{eq:hamil_dc_biased_ATS}) and $LC$ oscillators with typical frequencies ${\omega_a=2\pi\times\qty{7.9}{\GHz}}$ and ${\omega_c=2\pi\times\qty{5.5}{\GHz}}$. We assume that both modes have a relatively high impedance, resulting in zero-point phase fluctuations ${\varphi_a=0.3}$ and ${\varphi_c=0.8}$ (arbitrarily chosen). Our goal is to realize a stabilization of a four-dimensional cat-like state manifold using a scheme with $(r,l)=(0,4)$, which requires the voltage bias and flux drives to satisfy $\omega_V+\omega_{r}=-\omega_c$ and ${\omega_V+\omega_{l}=4\omega_a-\omega_c}$. We fix the voltage bias such that ${\omega_V=3.25\omega_a-3.25\omega_c=2\pi\times\qty{7.8}{\GHz}}$, corresponding to a voltage ${V=\qty{16.13}{\uV}}$ (well below the typical junction's superconducting gap of ${>\qty{0.1}{\mV}}$). With this choice of parameters, we find that $\omega_r=-3.25\omega_a+2.25\omega_c=-2\pi\times\qty{13.3}{\GHz}$ and $\omega_l=0.75\omega_a+2.25\omega_c=2\pi\times\qty{18.3}{\GHz}$. In the rotating frame, after applying the rotating wave approximation, we are left with the desired resonant processes $\h{a}^4\h{c}^{\dagger}$ and $\h{a}^{\dagger\,0}\h{c}^{\dagger}$, while all undesired ones are off-resonant. The remaining parameters to select are the drive amplitudes $\epsilon_{r,l}$ ensuring that the Rabi frequencies $\tilde{f}$ and $\tilde{g}$ intersect appropriately. Fig.~\ref{fig_ATS:rabi_frequency} illustrates the Rabi frequencies for the particular choice of $\epsilon_{r}=0.005$ and $\epsilon_{l}=0.04$. These functions are expressed as $\Omega_r \tilde{f}(k)$ and $\Omega_l \tilde{g}(k)$, where $\Omega_{r/l} = 0.5 E_J \varphi_c \, e^{-\varphi_c^2/2} \, \epsilon_{r/l}$. For the given values, $\Omega_r$ and $\Omega_l$ correspond to $2\pi\times\qty{65}{\MHz}$ and $2\pi\times\qty{522}{\MHz}$, respectively. The functions $\tilde{f}$ and $\tilde{g}$, similar to those in the trapped ion case, can be approximated by the Bessel function of Eq.~\eqref{eq:Bessel_approx_rabi_freq}. The crossing point of the Rabi frequencies is located at $k^{*}=11$ with a height of $h^*=\qty{83.8}{\MHz}$. With an engineered dissipation $\gamma=2\pi\times\qty{15}{\MHz}$, this configuration stabilizes four-legged cat-like states in approximately ${\sim\qty{25}{\us}}$ after initializing the system in the first Fock states. The steady states of the system are shown in Fig.~\ref{fig_ATS:wigner_functions}, showing the Wigner functions of the dark states $\ket{\Xi_\mu}$ for $\mu=0,1$ achieved after initializing the system in the Fock states $\ket{\mu}$.

It is important to note that the examples discussed above are specific instances, and that the NLRE method can be applied to a wide range of superconducting circuit QED systems, including those with more complex nonlinear elements and drive mechanisms. Notably, a recent theoretical proposal~\cite{nathan_self-correcting_2025} suggested the use of a controllable switch to activate the coupling between the storage quantum $LC$ oscillator and the nonlinear dissipative circuit to stabilize a GKP code space. This controllable switch approach could be combined with the NLRE method to engineer dissipative stabilization of a cat-like state manifold. Therefore, the NLRE method provides a versatile tool for designing and understanding the stabilization of rotation symmetric state manifolds in superconducting circuit QED systems.

\begin{figure*}
    \includegraphics[]{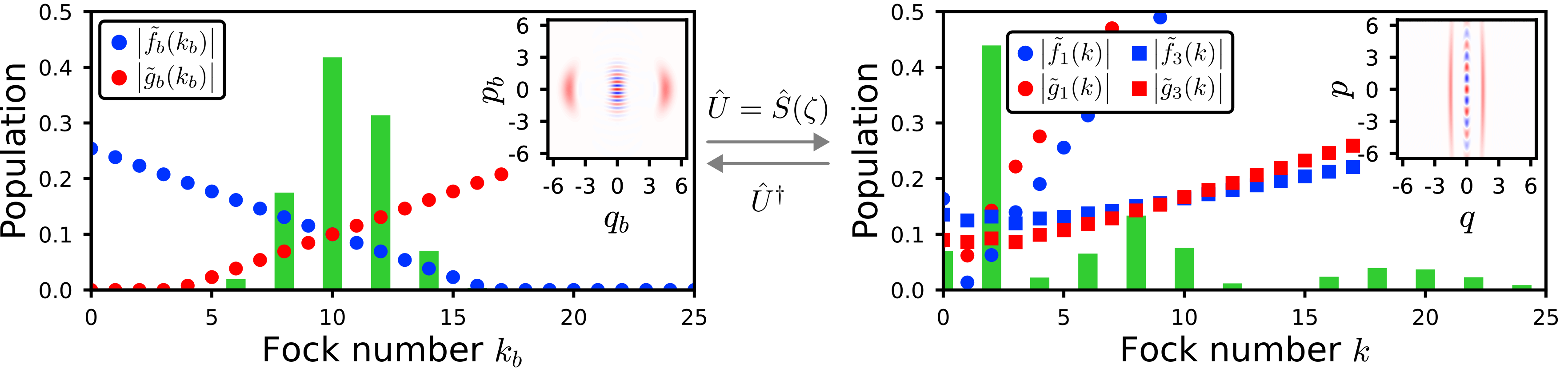}
    \caption{Generalization of the NLRE method through unitary transformations. The figure shows the Rabi frequencies, the boson distribution of a steady state, and its Wigner function in both the transformed frame (left) and the standard ladder operator framework (right), with the transformation being a squeezing along the $\h{q}$ quadrature. The stabilization scheme in the transformed frame corresponds to $(r,l)=(1,1)$, offering protection against momentum errors. In the standard ladder picture, these errors are translated to boson loss and dephasing processes (see Eq.~\eqref{eq:bogol_noise}).}
    \label{fig_bogoliubov}
\end{figure*}

\section{General nonlinear reservoir engineering} \label{sec:bogoliubov_transf}

Sections~\ref{sec:summary}\,--\!~\ref{sec:cqed_setting} demonstrated how the NLRE method can be used to stabilize a variety of cat-like codes, both theoretically and experimentally. We identified which errors these codes can autonomously correct and how to leverage nonlinearities to enhance the logical decoherence rate. However, up to this point we considered $\h{K}$ to be single-mode bosonic operators characterized by a combination of one boson raising and one lowering processes. In this section, we extend the NLRE method to operators $\h{K}$ comprising more processes and show that this allows for the stabilization of subspaces beyond cat codes, such as quadrature-squeezed cat states. Our approach relies on the fact that an operator's eigenvalues remain invariant under an isomorphic transformation of the canonical commutation relation algebra (i.e. ${[\h{a},\h{a}^\dagger]=1}$). Therefore, stabilized manifolds can be unitarily mapped between each other. The dimension of the transformed manifold remains invariant, but the dark states are in general different from the cat-like ones. Appropriately choosing the transformation allows us to map the dominant noise processes of the system onto errors, e.g. dephasing or momentum, that in the transformed basis are correctable using an instance of the toy model in Eq.~\eqref{eq:hamilton_modif}. In the original basis, this NLRE model corresponds to a transformed operator $\h{K}$ capable of correcting the noise processes of interest.

As an example, consider the single-mode operator ${\h{K}_b = \h{b}^{\dagger\,r} f(\h{n}_b) - g(\h{n}_b) \h{b}^{l}}$ with ${\h{n}_b = \h{b}^{\dagger}\h{b}}$ and ${\h{b}=\h{U}\h{a}\h{U}^\dagger}$ obtained using a linear unitary transformation $\h{U}$ of the canonical ladder operators (also known as Bogoliubov transformation). In general, such a transformation $\h{U}$ is decomposed as $\h{R}(\theta)\h{S}(\zeta)$ where ${\h{R}(\theta)=\exp(i\theta\h{a}^\dagger\h{a})}$ with ${\theta\in[0,2\pi]}$ and  ${\h{S}(\zeta)=\exp(\zeta/2(\h{a}^{2}-\h{a}^{\dagger\,2}))}$ with ${\zeta\in\bb{C}}$ correspond to a rotation and squeezing of phase space, respectively~\cite{weedbrook_gaussian_2012}. Since the orientation of the cat states can be adjusted using the relative phase of the boson raising and lowering processes (see Sec.~\ref{sec:simplest_nlre_setting}), we focus exclusively on squeezing transformations. The transformed ladder operators are then explicitly given by ${\h{b}=u\h{a}+v\h{a}^\dagger}$ with $\abs{u}^2-\abs{v}^2=1$ and similarly for $\h{b}^\dagger$. Thus, in the standard ladder picture, the effective jump operator will in general have the following form
\be \label{eq:hamilton_bogoliubov_a_at}
    \h{K} = \sum_{j=0} \h{a}^{\dagger\,j} f_j(\h{n}) + g_j(\h{n}) \h{a}^{j}\,.
\ee
where all $f_j$ and $g_j$ depend eventually on the functions $f$ and $g$, the transformation $\h{U}$, as well as the orders $r$ and~$l$ (see Appendix~\ref{app:fun_in_general_NLRE}). 

The dimensions of stabilized subspaces remain invariant under the considered transformations. Consequently, analyzing the problem using the standard or modified ladder operators framework yields equivalent results~\cite{poyatos_quantum_1996,kienzler_quantum_2015,lo_spinmotion_2015,kienzler_quantum_2017}. In the Fock basis ${\{\ket{k}_b = \h{U}\ket{k} \mid k \in \mathbb{N}\}}$, the transformed boson raising and lowering processes compete, leading to a cat-like steady state manifold when the functions $\tilde{f}_b$ and $\tilde{g}_b$ intersect adequately. Therefore, the same conclusions regarding the dimension of the manifold, the form of the dark states $\ket{\Xi_\mu}_b$, and their error-correction capability as discussed in previous sections hold true. In the standard ladder operator framework, the steady states are succinctly defined by ${\ket{\Xi_\mu}_b =  \h{U} \ket{\Xi_\mu}}$ with $\ket{\Xi_\mu}$ derived in Sec.~\ref{sec:simplest_nlre_setting}.

Although the conclusions drawn in the transformed picture remain consistent with those in the standard ladder framework, their interpretations differ. The transformation $\h{U}$ not only modifies the stabilized states but also alters the coupling between the system and the environment, thereby changing the form of the noise processes. For instance, if $\h{U} = \h{S}(\zeta)$ with $\zeta\in\bb{R}$ (i.e. a squeezing of the $\h{q}$ quadrature) the most common oscillator errors that are boson loss $\h{a}$ and dephasing $\h{n}$ become
\begin{equation} \label{eq:bogol_noise}
\begin{split}
    &\h{a} \mapsto i\,e^{\zeta}\,\h{p}_b + e^{-\zeta}\,\h{q}_b  \,, \\
    &\h{n} \mapsto e^{2\zeta}\,\h{p}_b^2 + e^{-2\zeta}\,\h{q}_b^2 - 1 \,,
\end{split}
\end{equation}
where $\h{q}_b$ and $\h{p}_b$ are the quadrature operators of mode $\h{b}$. This shows that for $\zeta\gg1$, the unitary $\h{U}$ transforms the leading noise processes primarily into momentum errors. Thus, selecting the $(r,l) = (1,1)$ scheme in the transformed picture will stabilize a cat-like state manifold that autonomously protects against such errors (see Sec.~\ref{subsec:corr_11}). In the standard ladder picture, the steady states will resemble squeezed Schrödinger cat states. This sheds light on recent works on the subject. Refs.~\cite{schlegel_quantum_2022,hillmann_quantum_2023} proposed and analyzed the stabilization of such states using a jump operator $\h{K}=\h{S}(\zeta)(\h{a}^2-\alpha^2)\h{S}(\zeta)^\dagger$, but without achieving autonomous correction of boson loss/gain errors. This can be understood in our framework, as this stabilization corresponds to a $(0,2)$ scheme in the transformed picture which only protects against dephasing errors. In contrast, Ref.~\cite{xu_autonomous_2023} has demonstrated autonomous correction of boson loss/gain errors using a modified jump operator $\h{K}=(c_1\h{a}^\dagger+c_2\h{a})\h{S}(\zeta)(\h{a}^2-\alpha^2)\h{S}(\zeta)^\dagger$. This model is equivalent to a $(1,1)$ scheme (see Appendix~\ref{app:fun_in_general_NLRE}) in the transformed picture and thus aligns with our conclusions. The NLRE method provides additional insights into this scheme. First, the values of $c_{1}$ and $c_2$ must be selected to ensure an appropriate intersection of the functions $\tilde{f}_b$ and $\tilde{g}_b$. Second, Eq.~\eqref{eq:bogol_noise} shows that while increasing the squeezing parameter improves the approximation of the leading error by $\h{p}_b$, it also increases the overall rate of the noise processes. Therefore, there exists an optimal $\zeta$ (dependent on the natural error rates) for which the $(1,1)$ scheme is most efficient. This optimum has also been highlighted in Ref.~\cite{xu_autonomous_2023}.

Given this understanding, we can see how the generalization of the NLRE method can help to engineer stabilization processes with specific correction capabilities. By first identifying an isomorphism that transforms the leading errors of the system into either dephasing or momentum errors, one can then choose the appropriate stabilization scheme in the transformed picture to correct these errors. Finally, using the chosen $\h{U}$ and $\h{K}_b$, it becomes possible to determine the experimental requirements necessary to realize the designed stabilized manifold. For instance, to stabilize a two-dimensional squeezed cat manifold, the operator $\h{K}$ must include at least two pairs of bosonic raising and lowering processes. To implement in the transformed basis the schemes $(0,2)$ and $(2,0)$, the orders of these processes in the standard ladder picture must correspond to $j \in \{0, 2\}$. To stabilize the $(1,1)$ model, $j$ should be equal to $\{1, 3\}$. The parity of $r$ and $l$ matches the orders $j$ due to the parity-preserving nature of the squeezing operation. Consequently, it becomes possible to reverse-engineer the system parameters for which the functions $\tilde{f}_b$ and $\tilde{g}_b$ intersect appropriately (see Appendix~\ref{app:fun_in_general_NLRE}). {This suggests that four boson processes, two first-order and two third-order, may constitute the minimal requirement for a bosonic code to autonomously correct both dephasing and loss errors. We leave the proof of this sufficient condition for future work.}

The generalized NLRE method is universal as we can always find a unitary operator mapping the steady states $\ket{\Xi_\mu}$ onto a subset of another orthonormal basis defining a new code. In general, this unitary mapping does not transform $\h{a}$ and $\h{a}^\dagger$ linearly. For example, in dissipative stabilization of GKP states~\cite{royer_stabilization_2020,de_neeve_error_2022}, stabilization operators take the form $\h{K}_{q} = \alpha\h{q}_\n{mod}+i\beta\h{p}$ and $\h{K}_{p} = \alpha\h{p}_\n{mod}+i\beta\h{q}$, with constants $\alpha$ and $\beta$, and modular quadrature operators $\h{q}_\n{mod}$ and $\h{p}_\n{mod}$. The operators $\h{K}_{q}$ and $\h{K}_{p}$ can be brought to the form of Eq.~\eqref{eq:hamilton_bogoliubov_a_at}, showing that mapping cat states to GKP codewords employs a nonlinear transformation of canonical ladder operators. This method extends naturally to multi-mode systems using multi-mode Bogoliubov transformations.

\medskip
This generalization underscores the NLRE method's versatility, applicable for studying and developing a variety of bosonic stabilization models. It can be helpful for the realization of such models in experiments but also for theoretical investigations of novel bosonic codes with enhanced robustness against oscillator errors. Lastly, it prompts a significant question: given a certain rate of boson dephasing, loss, heating, or quadrature errors in a system, what constitutes the optimal bosonic encoding and dissipative stabilization procedure? This question exceeds the current paper's scope and is left for future exploration.

\section{Conclusions and outlook} \label{sec:conclusion}

The nonlinear reservoir engineering approach we described in this paper introduces an innovative method for engineering quantum states with broad applicability. It relies on a novel paradigm where the stabilization process is not only analytically defined by the existence of a zero eigenstate of an operator but also corresponds to a destructive interference between excitation raising and lowering processes. 
Our construction provides a new intuitive framework for designing these stabilization schemes, by identifying crossing points in the rates of the aforementioned processes, and analyzing their error-correcting properties. We demonstrated that this encompasses existing schemes previously covered in the literature, while also extending to new possibilities. Our framework enables the design of quantum codes robust against arbitrary noise processes and sheds light on a connection between autonomous error correcting codes.

Although this paper focuses on reservoir engineering, we also note the intriguing connection between dissipative and Hamiltonian (also known as Kerr) stabilization of cat states~\cite{puri_engineering_2017,grimm_stabilization_2020}. This invites the exploration of whether similar considerations of crossing points in Kerr schemes can produce error-robust states. It is evident that this holds true for simple examples, with the stabilized states located at the point in phase space where the derivative of the Hamiltonian vanishes, in analogy to the crossings of processes composing the operator $\h{K}$ in the dissipative case. This insight may lead to new Hamiltonian-based approaches for stabilizing various bosonic codes.

By departing from the standard Lamb-Dicke regime commonly used in trapped ions and atoms, our approach offers a method to stabilize systems into specific manifolds of states. For these states, the higher-order sideband operations available provide a variety of Hamiltonians with intriguing possibilities. For instance, when measuring the parity of bosonic states, it is advantageous to have a standard dispersive interaction Hamiltonian~\cite{rosenblum_fault_2018,grimm_stabilization_2020}, for which the dependence of the Rabi frequency on the spin transition is proportional to the occupancy of the motional state. For trapped ions in the Lamb-Dicke regime, such an interaction is typically weak. However, outside this regime, it is feasible to select a Hamiltonian with an appropriate dependence over the limited range of occupied Fock states, thus enabling such measurements without relying on a native dispersive Hamiltonian.

DC-biased circuits are often avoided due to the challenges of working with highly non-Gaussian bosonic states. In this paper, we provide a technique to realize multi-photon dissipators that stabilize a cavity into a manifold of states with well-defined structure (i.e. rotational symmetry) and specific behavior under common errors, and whose non-Gaussianity can be well captured using a finite number of system parameters. On top of this, we showed that DC-biased circuits can realize a variety of multi-photon Hamiltonians without spurious terms (e.g., AC Stark shift, self-Kerr, and cross-Kerr) caused by undesired always-on resonances and present in previous implementations. This novel approach to engineering bosonic interactions makes a direct parallel between superconducting circuit QED and atomic systems.

Although we illustrated the nonlinear reservoir engineering method using trapped ions and superconducting circuits, the nonlinearities discussed in this work are prevalent in a variety of quantum system. Bulk acoustic resonators coupled to Josephson junctions are examples where quantum harmonic oscillators inherit nonlinearities through their interaction with a nonlinear element~\cite{bild_schrodinger_2023,marti_quantum_2024}. Cavity QED, spin ensembles and optomechanical systems are thus other areas where the NLRE method could be applied. Our study may advance  the understanding of non-Gaussian states across a broad range of platforms. These states are essential for quantum information processing and for describing various many-body phenomena, making them an active field of research~\cite{shi_variational_2018,walschaers_non-gaussian_2021}.

While we have established here the means to generate states using nonlinear reservoir engineering, devising schemes for controlling these states would be facilitated by analytical descriptions of Hamiltonians acting on them. Their desired action on codewords can be understood from the logical and gauge subsystem decomposition picture used in Sec.~\ref{sec:error_correction}.
The ideal operation on our cat-like manifolds are exactly the same as those required by standard cat codes, namely the Zeno-type operations~\cite{mirrahimi_dynamically_2014,touzard_coherent_2018,gautier_combined_2022,gautier_designing_2023} and holonomic gates~\cite{albert_holonomic_2016}. Nonetheless, better methods would be highly advantageous to implement these gates in practice as they require specific high-order nonlinear interactions. It would be interesting to ask whether there exist logical operators which can be defined as an expansion around the stable points of dissipative stabilized states and could in practice simplify their engineering. These questions remain for future research.

{
\textit{Note:} During the review process, we became aware of several new preprints that align with aspects of our work. These include studies employing a similar representation of nonlinearities as in Eq.~\eqref{eq:hamilton_modif} for implementing gates on bosonic error-correcting codes~\cite{wetherbee_mathematical_2025}, exploring the full nonlinear regime of quantum systems for Floquet engineering of bosonic codes~\cite{guo_engineering_2025}, and using DC-biased superconducting circuits for cat qubit stabilization~\cite{aissaoui_cat_2024,danner_quantum_2025}. Additionally, an experimental team recently demonstrated the stabilization of a cat-state manifold in a circuit QED setup using the NLRE scheme $(r,l)=(0,2)$ proposed in our work, showing improved error-correction bias when the Fock distribution is squeezed~\cite{rousseau_enhancing_2025}. Finally, during the review process, we experimentally demonstrated stabilization of rotation-symmetric manifolds in a trapped ion setup using NLRE~\cite{simoni_non_linear_2025}.
}

\section{Acknowledgements} \label{sec:acknowledgements}

We would like to express their gratitude to S.~Lieu and V.~V.~Albert for their valuable insights on passive error correction. Additionally, we thank L.~Jiang, G.~Zheng and G.~Lee for the fruitful discussion about the necessary operations for qudit cat codes, as well as A.~Grimm and F.~Adinolfi for their valuable insights regarding circuit QED systems. We thank T.~Nadolny and T.~Kehrer for the valuable exchanges on quantum optics and synchronization, and their continuous interest in the project. Finally, we also acknowledge A.~Clerk, A.~Imamoglu, and S.~Girvin for helpful and insightful discussions. 

We acknowledge funding from the Swiss National Science Foundation (Ambizione Grant No. PZ00P2$\_$186040) and the ETH Research Grant ETH-49 20-2.

I.R. and M.S. contributed equally to this work. M.S. discovered the stabilization effect. I.R. conducted the theoretical analysis, with input from M.S. and E.Z., under guidance from J.H. and F.R. All authors discussed the theoretical conclusions and experimental implications. I.R. wrote the manuscript, with input from all authors.

\PRLsep
\appendix

{
\section{Inaccessibility of high-order processes in the low-excitation operating regimes} \label{app:nonlinear_important}

In this appendix we give a quantitative explanation of why high-order boson processes are hard to engineer using current operating regimes of quantum systems and how nonlinear interactions of the form considered in this work resolve this. It is well established that boson processes of the form $\h{a}^d$ are necessary for advanced bosonic quantum information processing tasks such as quantum error correction and operation of logical qubits. The four-legged cat codes which offers protection against both dephasing and boson-loss errors is a typical example~\cite{leghtas_hardware_2013,ofek_extending_2016}. Its standard dissipative stabilization requires a jump operator $\h{K}=\h{a}^4-\alpha^4$. However, this process is hard to engineer in practice as demonstrated by several experiments~\cite{mundhada_generating_2017,vanselow_dissipating_2025}. Why is this the case?

Let us illustrate it on the example of a trapped ion system. We start with the standard Hamiltonian of a trapped ion system in the interaction picture given in Eq.~\eqref{eq:trapped_ions_hamilton}. We restate it below and refer the reader to Ref.~\cite{wineland_experimental_1998} for its detailed derivation
\[
    \h{H}(k_\SB,\eta) \!= \Omega\, e^{i\eta(\h{a}^{\dagger}e^{-i\omega_m t} + \h{a} e^{i\omega_m t})} e^{i k_\SB \omega_m t} \h{\sigma}_+ + \n{H.c.}
\]
with $k_\SB\in\bb{Z}$. The element that interests us in this expression is the exponential operator. For simplification, let us first consider its time independent version
\be \label{eq:exp_expansion}
\begin{split}
  &e^{i\eta(\h{a}^{\dagger} + \h{a})} = e^{-\frac{1}{2}\eta^2} e^{i\eta\h{a}^\dagger} e^{i\eta\h{a}}
  = e^{-\frac{1}{2}\eta^2}\!\sum_{n,m=0}^{\infty}\!\!\frac{(i\eta)^{n+m}}{n!m!}\,\h{a}^{\dagger n}\h{a}^m  \\[3pt]
  &\equiv e^{-\frac{1}{2}\eta^2} \h{a}^{\dagger k_\SB}
  \left[\sum_{n,m=0}^{\infty}\frac{(i\eta)^{n+k_\SB+m}}{(n+k_\SB)!(m)!}\,\h{a}^{\dagger n}\h{a}^m \right]\\[3pt]
  &\equiv e^{-\frac{1}{2}\eta^2}
  \left[\sum_{n,m=0}^{\infty}\frac{(i\eta)^{n+m+k_\SB}}{(n)!(m+k_\SB)!}\,\h{a}^{\dagger n}\h{a}^m \right] \h{a}^{k_\SB} \,.
\end{split}
\ee 
Here, the first equality is exact and follows from the Baker-Campbell-Hausdorff formula. The second equality results from the Fourier expansion of the exponential terms while preserving normal ordering. This expression is exactly equivalent to the following two formulas, where we isolate $k_\SB$ powers of $\h{a}^\dagger$ and $\h{a}$ from the Fourier expansions and redefine the summation indices. We can now move back to the oscillator's interaction picture where $\h{a}^{\dagger}\mapsto\h{a}^{\dagger}\, e^{-i \omega_m t}$ and $\h{a}\mapsto\h{a}\,e^{i\omega_m t}$. The terms in the summations within the square brackets rotate rapidly except those indexed by $n=m$. Thus, the square brackets can be approximated using the rotating wave approximation, leading to the function $f(\h{n},k_\SB,\eta)$ given in Eq.~\eqref{eq:f_function_ions} in main text.

\begin{figure*}
    {\phantomsubcaption\label{figRabi:linear}}
    {\phantomsubcaption\label{figRabi:nonlinear}}
    \includegraphics[width=\textwidth]{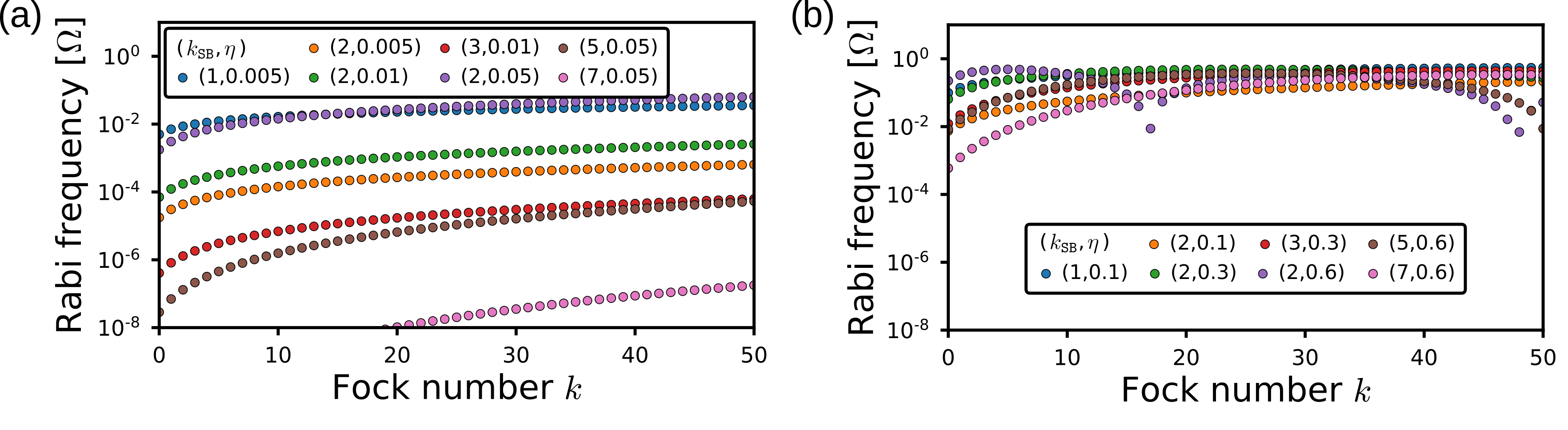}
    \caption{Comparison of the Rabi frequencies for the (a) linearized and (b) nonlinear Hamiltonians in Eqs.~\eqref{eq:light_matter_ions_linear} and~\eqref{eq:trapped_ions_hamilton}, respectively, for different $k_\SB$ and $\eta$. Curves in (a) follow Eq.~\eqref{eq:rabi_freq_lin} while those in (b) follow Eq.~\eqref{eq:rabi_freq_nonlin}.}
\end{figure*}

Due to the nontrivial form of this operator, the main drive of the trapped ion field has been to linearize the interaction Hamiltonian, i.e. to consider the leading order terms in $\eta$ in the expansion of \eqref{eq:exp_expansion}
\be \label{eq:light_matter_ions_linear}
  \h{H}(k_\SB,\eta) \!
  \approx\,\Omega\,e^{-\frac{1}{2}\eta^2} \frac{1}{k_\SB!} 
  (i\eta)^{k_\SB} \h{a}^{\dagger k_\SB} \h{\sigma}_+ 
  + \n{H.c.}
\ee
where we assume $k_\SB\geq0$ (for $k_\SB<0$ use $\h{a}^{k_\SB}$). Operating a trapped ion setup under such an approximation is known as operating in the Lamb-Dicke regime. It can be attained by choosing ${\eta\expval{\h{a}^\dagger + \h{a}}\ll1}$. This linearization is a key assumption in the derivation of the Jaynes-Cummings models and led to the demonstration of key advancements in trapped ions physics. However, the Hamiltonian in Eq.~\eqref{eq:light_matter_ions_linear} does not allow to drive boson processes with $k_\SB>1$. Indeed, the strength of these processes decreases as a power law in $\eta$. Given the perturbative expansion assumption, this parameter must remain small. Consequently, achieving a measurable effect requires a coupling strength $\Omega\propto\eta^{-k_\SB}$. Practically speaking, this means that the EM field power must scale exponentially with the bosonic process order $k_\SB$, posing a significant limitation. 
  
To illustrate this point we evaluate the Rabi frequency $f(k,k_\SB,\eta)=|\bra{n+k_\SB,\uparrow}\h{H}(k_\SB,\eta)\ket{n,\downarrow}|$. For the Hamiltonian in Eq.~\eqref{eq:light_matter_ions_linear} it reads 
\be \label{eq:rabi_freq_lin}
    f_\n{approx}(k,k_\SB,\eta) = \Omega \, e^{-\frac{1}{2}\eta^2} \frac{1}{k_\SB!} \, \eta^{k_\SB} \, \sqrt{\frac{(k+k_\SB)!}{k!}}
\ee
where the square root term arises from the matrix elements of the operator $\h{a}^{\dagger k_\SB}$. We represent this quantity as a function of $k$ and $k_\SB$ in Fig.~\ref{figRabi:linear} for different $\eta\ll1$. We observe that the Rabi frequency for $k_\SB>1$ is often several orders of magnitude smaller than for $k_\SB=1$ showing that driving high-order boson processes is experimentally challenging in the standard operational regime. The strength of high-order bosonic processes can be compensated by longer interaction times, but this increases sensitivity to noise and decoherence creating a trade-off between driving time and
coherence lifetime.

Operating outside the perturbative expansion assumption, i.e. $\eta\sim1$, allows to overcome this. Including all terms from the expansion in Eq.~\eqref{eq:exp_expansion}, the Rabi frequency becomes
\be \label{eq:rabi_freq_nonlin}
\begin{split}
    f_\n{full}(k,k_\SB,\eta) = \Omega \, \sqrt{\frac{(k+k_\SB)!}{k!}} \,\abs{f(k,k_\SB,\eta)} \\[3pt]
    \approx \Omega\,J_{k_\SB}\left(2\eta\sqrt{n+\frac{k_\SB+1}{2}}\right)
\end{split}
\ee
where $f(n,k_\SB,\eta)$ is the operator in Eq.~\eqref{eq:f_function_ions} evaluated for the state $\ket{k}$, while $J_{k_\SB}$ is the Bessel function of the first kind as in Eq.~\eqref{eq:Bessel_approx_rabi_freq} (see its derivation in Appendix~\ref{app:bessel_approx}). Fig.~\ref{figRabi:nonlinear} shows the expression in Eq.~\eqref{eq:rabi_freq_nonlin} for different $\eta$ and $k_\SB$. Compared to the linear case, this Rabi frequency is larger by several orders of magnitude and almost of the same order of magnitude for all $k_\SB$, enabling bosonic processes with $k_\SB>1$ at reasonable coupling strengths $\Omega$. A drawback however is that now the Rabi frequency depends non trivially on $k$ which requires novel intuitive theoretical methods to treat such operators. This is exactly the problem that the NLRE method solves in the context of reservoir engineering.

This difference between the approximate low-excitation (named after the condition ${\eta\expval{\h{a}^\dagger + \h{a}}\ll1}$) and the full nonlinear regimes of operation arises in other experimental platforms too. The only difference would be in the shape of the bosonic potential. For example, in circuit QED setups the interaction often takes form of a trigonometric function of $\h{a} + \h{a}^\dagger$. However, these potential function can often be represented, omitting terms that are fast rotating in the interaction picture,
\be 
    F(\h{a} + \h{a}^\dagger) = \sum_{i=-\infty}^{\infty} F_i(\h{n})\,\h{a}^i
\ee
where negative powers of $\h{a}$ correspond to powers of $\h{a}^\dagger$. In the low-excitation regime, the leading term in the expansion corresponds to $F_i(\h{n})\approx c_i \h{I}$ with $c_i$ being a constant that becomes exponentially low with $i$. This results in the same problem as in case of trapped ions illustrated above. It is therefore essential to operate outside such regimes to engineer high-order bosonic processes. 

}

\section{Classical phase space diagrams} \label{app:classical_phase_space}

To obtain the semi-classical phase space attractor diagram shown in Fig.~\ref{fig1:classical_dynamics}, we first derive the equations of motion for the canonical quadratures, $\hat{q} \propto \hat{a}^\dagger + \hat{a}$ and $\hat{p} \propto i(\hat{a}^\dagger - \hat{a})$. These are given by $\dot{\hat{q}} = \mathcal{D}[\hat{L}_\text{eff}^\dagger](\hat{q})$ and a similar expression for $\dot{\hat{p}}$. We then define the corresponding classical model using the Husimi Q representation of these operators. Specifically, we use $\dot{Q}(\alpha) = \bra{\alpha} \dot{\hat{q}} \ket{\alpha} / \pi$ and $\dot{P}(\alpha) = \bra{\alpha} \dot{\hat{p}} \ket{\alpha} / \pi$, where $\ket{\alpha}$ with $\alpha \in \mathbb{C}$ is a coherent state of the bosonic mode. It is straightforward to show that this approach is equivalent to a semi-classical (mean-field) approximation of quantum dynamics. Fig.~\ref{fig1:classical_dynamics} illustrates these classical equations of motion through their associated vector field in phase space, where we use $\langle q \rangle = \text{Re}(\alpha)$ and $\langle p \rangle = \text{Im}(\alpha)$. This dynamics is depicted for the same nonlinearities as in Fig.~\ref{fig1:wigner_functions}. The classical picture reveals multiple critical points (i.e., points with vanishing derivatives) of various natures: stable, saddle, and unstable. Empirically, we deduce that the number of stable points corresponds to $d = r + l$, suggesting the existence of a set of $d$ dark states that are rotationally symmetric relative to the origin. However, the classical picture does not provide information about the quantum aspect of these dark states.

{ 
\section{Squeezed vacuum stabilization in the NLRE framework} \label{app:squeeze_vacuum}

The NLRE framework gives an intuitive explanation of the squeezed vacuum state stabilization scheme proposed in Ref.~\cite{cirac_dark_1993} and first demonstrated in Ref.~\cite{kienzler_quantum_2015}. The scheme stabilizes a squeezed vacuum state $\h{S}(\zeta)\ket{0}$, where $\h{S}(\zeta) = \exp(\zeta/2(\h{a}^{2}-\h{a}^{\dagger\,2}))$ is the squeezing operator and $\zeta=r e^{i\varphi}$ is the squeezing parameter. The stabilization operator reads $\h{K}=\sinh(r)\h{a}^\dagger+e^{i\varphi}\cosh(r)\h{a}$. Casting it into the NLRE framework in Eq.~\eqref{eq:hamilton_modif}, it is straightforward to see that $f$ and $g$ are simply identity operators. Thus, their tilde versions reads simply $\abs{\tilde{f}(k)} = \sinh(r)\sqrt{k+1}$ and $\abs{\tilde{g}(k)} = \cosh(r)\sqrt{k+1}$ where the square roots come from the matrix elements of the ladder operators. These functions cross at $k=-1$, meaning that, according to the NLRE method, the system must be stabilized around this non-physical point. However, the hard boundary of Fock space and $\abs{\tilde{f}}<\abs{\tilde{g}}$ force the distribution to accumulate around the origin ($k=0$). In addition, NLRE sets the dimension of the stabilized manifold to $2$ with one true dark state (with even parity) and one {metastable} dark state (with odd parity). However, as explained in Sec.~\ref{subsec:number_of_steady_states}, the fact that the distribution is located at the origin the {metastable} dark state is in fact a bad dark one that decays quickly in the true dark state, that corresponds to $\h{S}(\zeta)\ket{0}$. For interested readers, this boundary effect also appears in other systems, such as spin ensembles, and is responsible for the difference between dissipatively stabilizing squeezed states of ensembles with an even or odd number of spins (stabilizing a pure or mixed state, respectively)~\cite{groszkowski_reservoir_2022}.

}

{ 
\section{Exact form of $f$ and $g$ for linear $\tilde{f}$ and $\tilde{g}$} \label{app:f_g_linear}

Up to Sec.~\ref{sec:trapped_ions} of the main text, we assumed that the Rabi frequencies $\tilde{f}$ and $\tilde{g}$ are linear functions of the Fock number $k$. However, this does not imply that the original functions $f$ and $g$ are linear in $\h{n}$. Their explicit forms can be determined by reverse engineering the definitions of $\tilde{f}$ and $\tilde{g}$, i.e. ${\tilde{f}(k):=\bra{k+r}\h{a}^{\dagger\,r} f(\h{n}) \ket{k}}$ and ${\tilde{g}(k):=\bra{k} g(\h{n}) \h{a}^{l} \ket{k+l}}$. This yields
\be \label{eq:linear_f_g}
  f(\h{n}) = \tilde{f}(\h{n}) \sqrt{\frac{\h{n}!}{(\h{n}+r)!}}
\ee
with an analogous expression for $g(\h{n})$, replacing $r$ with $l$. One can see that these take highly non-physical forms that are challenging to implement in practice. Nevertheless, as discussed in Sec.~\ref{sec:simplest_nlre_setting}, the stabilization of a cat-like state manifold does not rely on $f(\h{n})$ and $g(\h{n})$ being of this exact form. Instead, stabilization occurs as long as the resulting Rabi frequencies intersect appropriately allowing the usage of Bessel-like functions in the examples of Secs.~\ref{sec:trapped_ions} and~\ref{sec:cqed_setting}. For our analytical results to remain valid, particularly the formulas of boson distributions Eqs.~\eqref{eq:CMB_distribution} and~\eqref{eq:CMP_distribution}, the functions $\tilde{f}$ and $\tilde{g}$ must be sufficiently linear around the crossing point. Any higher-order nonlinearities would then introduce correction terms affecting the shape of the distributions.

In most figures (except in Secs.~\ref{sec:trapped_ions} and~\ref{sec:cqed_setting}), we used Rabi frequencies $\tilde{f}$ and $\tilde{g}$ that are exactly linear at the crossing point and exponentially converging towards $0$ when they approach the y-axis from the right and left, respectively. This choice is motivated to avoid the introduction of other crossing points that could arise due to the reflection of the absolute value of the functions from the y-axis. 
}

\section{Boson distribution for linear $\tilde{f}$ and $\tilde{g}$} \label{app:cmb_derivation}

In this appendix, we derive the probability density function describing the boson distribution of the stabilized states in the case of linearly decreasing and increasing Rabi frequencies $\tilde{f}$ and $\tilde{g}$ (see Sec.~\ref{subsec:linear_tilde_f_g}). Let us start the derivation from the recurrence relation given in Eq.~\eqref{eq:recusive_relation}. The general solution of this equation reads
\be \label{eq:recursive_rel_general_sol}
    \xi_{k} = \xi_{k\,\n{mod}\,d}\,\,\prod_{j=1}^{\lfloor k/d \rfloor}
    \!\!\left(\frac{\tilde{f}(k-j\,d)}{\tilde{g}(k-j\,d+r)}\right)
\ee
where $\xi_{k\,\n{mod}\,d}$ corresponds to an initial condition which also encompasses the normalization constant of the steady state $\ket{\Xi}$. Disregarding for now this term and the floor function $\lfloor \cdot \rfloor$, we find that for linear functions $\tilde{f}$ and $\tilde{g}$ as defined in Eq.~\eqref{eq:recusive_relation_linear} the product in Eq.~\eqref{eq:recursive_rel_general_sol} can be compactly expressed as
\be 
    \xi_{k} \propto \left(-\frac{s_f}{s_g}\right)^{\!\frac{k}{d}-1} 
    \frac{\left(1 - \frac{1}{d}(k^* + \frac{h^*}{s_f})\right)^{\left(\frac{k}{d}-1\right)}}
    {\left(1 - \frac{1}{d}(k^* - \frac{h^*}{s_g})\right)^{\left(\frac{k}{d}-1\right)}}
\ee
Here, $(a)^{(j)}:=(a)(a+1)\cdots(a+j-1)$ denotes the rising factorial which can be in turn represented as a ratio of factorials, or alternatively a ratio of Gamma functions when the argument is noninteger. We simplify this expression by ignoring all the constant terms (captured by the normalization factor) and employing properties of the rising factorial 
\be 
    \xi_{k} \propto \left(-\frac{s_f}{s_g}\right)^{\!\frac{k}{d}} 
    \left(\frac{1}{d}\left(k - k^* + \frac{h^*}{s_g}\right)\right)^{\!-\frac{1}{d}\left(\frac{h^*}{s_f}+\frac{h^*}{s_g}\right)}
\ee
The last step of the derivation consists in introducing (removing) the right constants in order to rewrite the rising factorial as a binomial coefficient (one over the product of two factorials)
\be 
    \xi_{k} \propto \left(-\frac{s_f}{s_g}\right)^{\!\frac{k}{d}} 
    \left(
    \begin{matrix}
        \frac{1}{d}\left(\frac{h^*}{s_f}+\frac{h^*}{s_g}\right) - 1 \\[5pt]
        \frac{1}{d}\left(k - k^* + \frac{h^*}{s_g}\right) - 1
    \end{matrix}
    \right)\,.
\ee
We see appearing here the parameters defined in Eq.~\eqref{eq:state_distribution_params}. Squaring the absolute value of this expression yields the probability density function of the Conway–Maxwell–Binomial (CMB) distribution, presented in Eq.~\eqref{eq:CMB_distribution} in the main text. Our derived expression slightly modifies the original CMB distribution~\cite{shmueli_useful_2005,borges_compoisson_2014,kadane_sums_2016,daly_conway-maxwell-poisson_2016}, as the parameters in the binomial coefficient are not necessarily integers in our case which is due to the ratios $h^*/s_f$ and $h^*/s_g$. Nonetheless, it is possible to justify the use of such a distribution in our context.

The CMB distribution is derived from the sum of two independent Conway–Maxwell–Poisson (CMP) variables (see Eq.~\eqref{eq:CMP_distribution}). The latter were introduced by \citet{conway1961queueing} to represent queuing systems with state-dependent service rates. Reintroduced and studied in Refs.~\cite{shmueli_useful_2005,borges_compoisson_2014,kadane_sums_2016,daly_conway-maxwell-poisson_2016}, the CMP distribution has found applications in modeling data with over- or underdispersion. A CMB variable is interpreted as the sum of identically distributed but not independent Bernoulli random variables. Such situations arise in many domains; for example, in biological systems, a statistical association (i.e., generalization of correlation) between plants or individual cells in a tissue occurs when they compete for nutrient availability~\cite{borges_compoisson_2014,kadane_sums_2016}. Our case fits well within this description, as each Fock state experiences competition between boson raising and lowering processes. These processes are correlated since both occur simultaneously with each application of the effective jump operator.

The normalization constant for bosons distributed according to Eq.~\eqref{eq:CMB_distribution} can be find numerically by summing $\abs{\xi_{k}}^2$ terms for a large enough number of Fock states. Analytically, it can be approximated by
\be
    \cl{N}_\n{CMB}^{-1} = 
    \frac{\theta^{y}\,d}{\left(y!(m-y)!\right)^{2}}\,
    \pFq{3}{2}{1,\,\, y-m,\,\, y-m}{1+y,\,\, 1+y}{\,\theta\,}
\ee
where $\prescript{}{p}{F}_q$ denotes the generalized hypergeometric function, while $m$, $\theta$ and $y=x-k/d$ are the parameters of our CMB distribution defined in Eq.~\eqref{eq:state_distribution_params}. Although this expression differs from the normalization constant of the standard CMB distribution, other characteristics of our CMB distribution can be derived by considering it to be a linear transformation of the standard CMB variable. Then, its expected value and variance follow the expressions in Eq.~\eqref{eq:mean_and_variance_CMB} where~\cite{shmueli_useful_2005,borges_compoisson_2014,kadane_sums_2016,daly_conway-maxwell-poisson_2016}
\be
\begin{split}
    \bb{E}[X] = & \, m^2\,\theta\,
    \frac{\,\prescript{}{2}{F}_1\!\left[1-m, 1-m\,;\,2\,;\,\theta\,\right]\,}
    {\,\prescript{}{2}{F}_1\!\left[-m,-m\,;\,1\,;\,\theta\,\right]\,} \,, \\[5pt]
    \n{Var}(X) = & \, (1 - m^2\,\theta)  m^2\,\theta \,
    \frac{\,\prescript{}{2}{F}_1\!\left[1-m, 1-m\,;\,2\,;\,\theta\,\right]\,}
    {\,\prescript{}{2}{F}_1\!\left[-m,-m\,;\,1\,;\,\theta\,\right]\,} + \\[3pt]
    & \, (1 - m)^2  m^2\,\theta^2\,
    \frac{\,\prescript{}{2}{F}_1\!\left[2-m, 2-m\,;\,3\,;\,\theta\,\right]\,}
    {\,\prescript{}{2}{F}_1\!\left[-m,-m\,;\,1\,;\,\theta\,\right]\,} \,.
\end{split}
\ee
are the expected value and variance of a random variable ${X\sim \n{CMB}(m,\theta,2)}$. Similar to how the binomial distribution is a limiting case of the Poisson distribution, $\n{CMP}(\lambda,\nu)$ follows from $\n{CMB}(m,\theta,\nu)$ when $\lambda=m^\nu p$ and $m\rightarrow\infty$~\cite{shmueli_useful_2005,borges_compoisson_2014,kadane_sums_2016,daly_conway-maxwell-poisson_2016}. This explains the derivation of Eq.~\eqref{eq:CMP_distribution}. Finally, thanks to fact that CMP and CMB distributions are members of the exponential family, other properties such as the skewness, moment generating function, or log-likelihood function can be derived in the same way.

\section{Bessel function approximation of Rabi frequencies} \label{app:bessel_approx}

\begin{figure}
    \centering
    \includegraphics{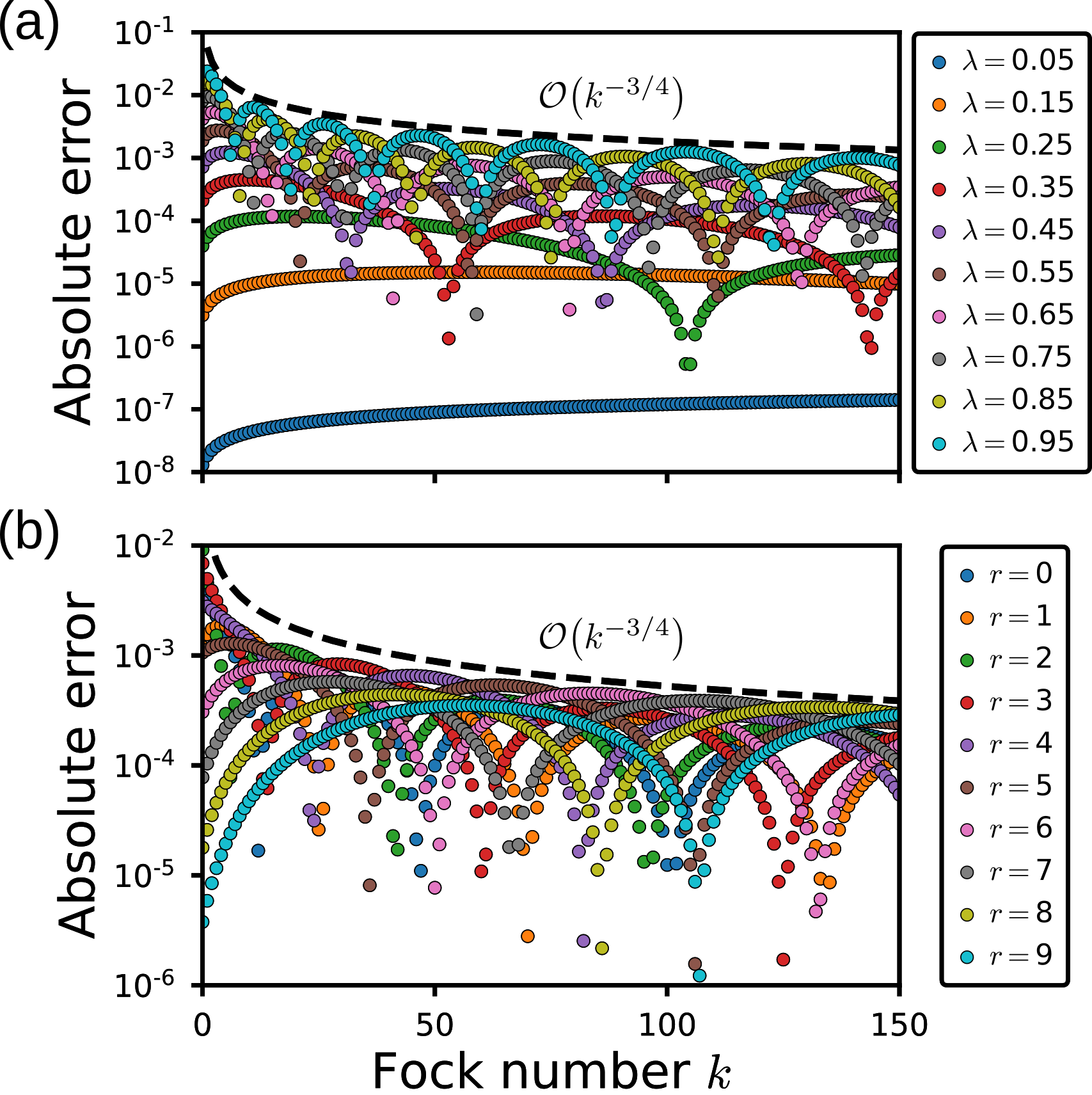}
    \caption{Absolute error between Rabi frequencies defined using Laguerre polynomials (exact expression, Eq.~\eqref{eq:displace_mat_el}) and Bessel functions (approximation, Eq.~\eqref{eq:displace_mat_el_approx}) as a function of scalar parameter $\lambda\in\bb{R}_+$ and integer $r\in\bb{N}$. (a) Error for fixed $r=1$. (b) Error for fixed $\lambda=0.5$.}
    \label{fig_laguerre_vs_bessel}
\end{figure}

We shall now discuss in more details the Bessel function approximation given in Eq.~\eqref{eq:Bessel_approx_rabi_freq} of the Rabi frequencies $\tilde{f}$ and $\tilde{g}$ resulting from the Hamiltonian in Eq.~\eqref{eq:trapped_ions_hamilton}. Let us reformulate the problem as following: consider the time-dependent bosonic operator
\be
    \h{O}(t) = e^{i \lambda (a^{\dagger}e^{-i\omega t} + a e^{i\omega t})} e^{-i r \omega t}
\ee
with $\omega,\lambda\in\bb{R}$ and $r\in\bb{Z}$. Without loss of generality, we can assume $r\geq0$ in which case the time-independent term of the operator reads $\h{O}=\h{a}^{\dagger\,r} f(\h{n},r,\lambda)$ with the function $f$ defined in Eq.~\eqref{eq:f_function_ions}. Then, by our definition of Rabi frequencies, we are interested in the quantity $\tilde{f}(k)=\bra{k+r} \h{O} \ket{k} \equiv {O}_{k+r,k}[\lambda]$. More generally, this function corresponds to the the $r$-th lower diagonal of the displacement operator $\h{D}(i\lambda)$ with $\h{D}(\xi)=\exp\!\left(\xi\h{a}-\xi^*\h{a}^\dagger\right)$ expressed in the Fock basis. It is well known that this function reads~\cite{cahill_ordered_1969}
\be \label{eq:displace_mat_el}
    O_{k+r, k}[\lambda]= \left(\frac{k!}{(k+r)!}\right)^{\!1/2}
    e^{-\lambda^2\!/2} \,\lambda^r \,\,L_k^{(r)}\!\left(\lambda^2\right)
\ee
with $L_k^{(r)}$ denoting generalized Laguerre polynomials. Let us now simplify this expression using two approximations. First, we approximate the Laguerre polynomial using Szeg\"{o}'s Theorem 8.22.4~\cite{szego_orthogonal_1939} which states
\be \label{eq:Laguerre_Bessel_approx}
\begin{split}
     e^{-x/2} \,\, x^{r/2} \,\,L^{(r)}_{k}(x) \approx & 
    \left(k+\frac{r+1}{2}\right)^{\!-r/2}\frac{(k+r)!}{k!} \\
    & \,\,J_{r}\!\left(2 \sqrt{x}\,\sqrt{k+\frac{r+1}{2}}\right)
\end{split}
\ee
where $J_{r}$ is the Bessel function of first kind of order $r$. This expression is valid for $x>0$ and is uniformly bound by $\cl{O}\!\left(k^{\,r/2-3/4}\right)$. Although this bound scales unfavorably for us as $k$ represents the Fock number, we note that for large $k$ (using the Stirling's formula) the prefactor of the Bessel function follows $\cl{O}\!\left(k^{\,r/2}\right)$. Thus, dividing the whole expression by this term solves the scaling issue. The second approximation we require is
\be \label{eq:sqrt_factorial approximation}
    \left(\frac{k!}{(k+r)!}\right)^{\!1/2} \approx 
    \left(k+\frac{r+1}{2}\!\right)^{\!r/2} \, \frac{k!}{(k+r)!}\,.
\ee
This approximation is easily obtained through the Stirling's formula and is uniformly bound by $\cl{O}\!\left(k^{-r-2}\right)$. Thus, combining Eqs.~\eqref{eq:Laguerre_Bessel_approx} and~\eqref{eq:sqrt_factorial approximation}, we obtain the following approximation of Eq.~\eqref{eq:displace_mat_el}
\be \label{eq:displace_mat_el_approx}
    O_{k+r, k}[\lambda] \approx J_{r}\!\left(2 \abs{\lambda}\sqrt{k+\frac{r+1}{2}}\right) + \cl{O}\!\left(k^{-3/4}\right)\,.
\ee
Knowing that in most cases we are interested ${0<\lambda<1.5}$ we numerically compare both expressions using their absolute error. The results are presented in Fig.~\ref{fig_laguerre_vs_bessel}. We see that Eq.~\eqref{eq:displace_mat_el_approx} approximates the desired quantity remarkably well, leading to errors consistently below $1\%$ in the regime we are interested. This makes the formula a promising candidate for numerical applications, such as motional state reconstruction of trapped ions, where the phonon distribution is obtained by fitting the Rabi oscillation of the spin~\cite{meekhof_generation_1996,wineland_experimental_1998}. Notably, the fitting process is often cumbersome due to the complex form of the Rabi frequency in Eq.\eqref{eq:displace_mat_el}. The approximation, however, offers a simpler form (free from factorials and Laguerre terms, thereby avoiding overflow and underflow issues) and can leverage efficient numerical evaluations of Bessel functions. Furthermore, Eq.~\eqref{eq:displace_mat_el_approx} allows us to easily draw a connection between the quantum and classical representations of the Doppler shift induced by atomic motion on emitted radiation. In the classical context, atomic motion modulates the frequency of this radiation, resulting in a spectrum comprising Bessel functions.

\section{Photon recoil superoperator} \label{app:recoil}

In this appendix, we delve into the superoperator representation of the photon recoil effect induced by spontaneous emission, which enables an efficient classical simulation of its impact on the quantum state of an ion's motion. We consider a trapped ion in an excited electronic state $\ket{e}$, susceptible to spontaneous emission into a lower-energy state $\ket{g}$. By tracing out the vacuum modes, we obtain the Liouvillian that governs the ion's dynamics~\cite{stenholm_semiclassical_1986,cirac_laser_1992}
\be \label{eq:liouv_spontan_emission}
    \cl{L}_\n{se}(\h{\rho}) =
    \gamma_{eg} \left(\h{\sigma}_{ge} \,\h{\tilde{\rho}}\,\h{\sigma}_{eg} - 
    \frac{1}{2}\left\{\h{\sigma}_{eg} \h{\sigma}_{ge}, \h{\rho}\right\} \right)
\ee
where $\gamma_{eg}$ is the spontaneous emission rate determined by the natural linewidth of the transition ${\ket{e}\rightarrow\ket{g}}$ and ${\h{\sigma}_{ge}=\h{\sigma}_{eg}^\dagger=\dyad{g}{e}}$. The emitted photon, carrying momentum $\hbar k_{eg}$, imparts a momentum kick to the ion in the opposite direction, which is accounted for in 
\be \label{eq:rho_tilde_spon_em}
    \h{\tilde{\rho}}=\frac{1}{2} \int_{-1}^{1} \n{d}x \,\, W(x) e^{i \eta \h{q} x} \, \h{\rho} \, e^{-i \eta \h{q} x}
\ee
with $\h{q}=\h{a}+\h{a}^\dagger$ being the (unitless) position operator. Here, $\eta=k_{eg}z_0$ correspond to the Lamb-Dicke parameter (see Sec.~\ref{sec:trapped_ions}) with its cosine dependence expressed through the variable $x$. The function ${W(x)=3/4 (1+x^2)}$ represents the angular distribution of the spontaneous emission, assuming a dipole transition from $\ket{e}$ to $\ket{g}$. In the superoperator form, this equation becomes
\be \label{eq:rho_tilde_spon_em_superop}
    |\tilde{\rho}\rangle\!\rangle = \int_{-1}^{1} \n{d}x \,\, W(x)\!\left[
    \big(e^{-i\eta\h{q} x}\big)^{\!\top}
    \!\otimes e^{i \eta\h{q} x}\right]|\rho\rangle\!\rangle
\ee
where $\top$ denotes the transpose operation, and $|\rho\rangle\!\rangle$ -- the vectorized form of the density matrix. We observe that the latter is now decoupled from the integral, and the operator inside the integral is diagonal in the position basis. Consequently, the entries of the diagonal superoperator matrix in this basis are given by the solution of the following integral
\be
\begin{split}
    &\int_{-1}^{1} \frac{3}{4}\left(1+x^{2}\right) e^{i k\left(q_{1}-q_{2}\right) x} = \\
    & =3 \,\n{sinc}\!\left(k\left(q_{1}-q_{2}\right)\right)+3 \frac{\n{cosc}\!\left(k\left(q_{1}-q_{2}\right)\right)}{k\left(q_{1}-q_{2}\right)}
\end{split}
\ee
with $\n{sinc}$ being the unnormalized sinc function and $\n{cosc}$ its derivative. To represent the density matrix in the oscillator's energy basis, we need to perform a change of basis from the position basis to the energy basis. This is achieved using the transformation ${\h{T}_{PF}=\sum_{q,k} \psi_{k}(q)\dyad{q}{k}}$ where
\be
    \psi_{k}(q)=\frac{\pi^{-1 / 4}}{\sqrt{2^{k} k!}}\,e^{-q^{2}/2}\, H_{k}\!\left( q \right)
\ee
is the position representation of Fock state $\ket{k}$ through Hermite polynomials $H_k$. Then, the superoperator transforming $|\rho\rangle\!\rangle$ into $|\tilde{\rho}\rangle\!\rangle$ is in the Fock basis given by ${\cl{W}_{F} = \cl{T}_{FP} \cdot \cl{W}_{P} \cdot \cl{T}_{PF}}$ with ${\cl{T}_{PF}=\h{T}_{PF}\otimes \h{T}_{PF}}$. Finally, we can represent the whole Liouvillian in Eq.~\eqref{eq:liouv_spontan_emission} in a superoperator form taking into account the spin part of the system, too.

This method offers an efficient alternative to the standard numerical evaluations of such Liouvillians, which typically involve performing a Monte Carlo wave function simulation. In this simulation, at every jump, the photon recoil displacement amplitude is sampled according to the dipolar angular distribution $W(x)$. In contrast, the superoperator method presented above circumvents this sampling overhead by exploiting the diagonal form of the superoperator in the position basis, the analytical solution of the integral, and the efficient construction of the transformation matrix $\h{T}_{PF}$.

\medskip

\begin{figure*}
    \includegraphics[]{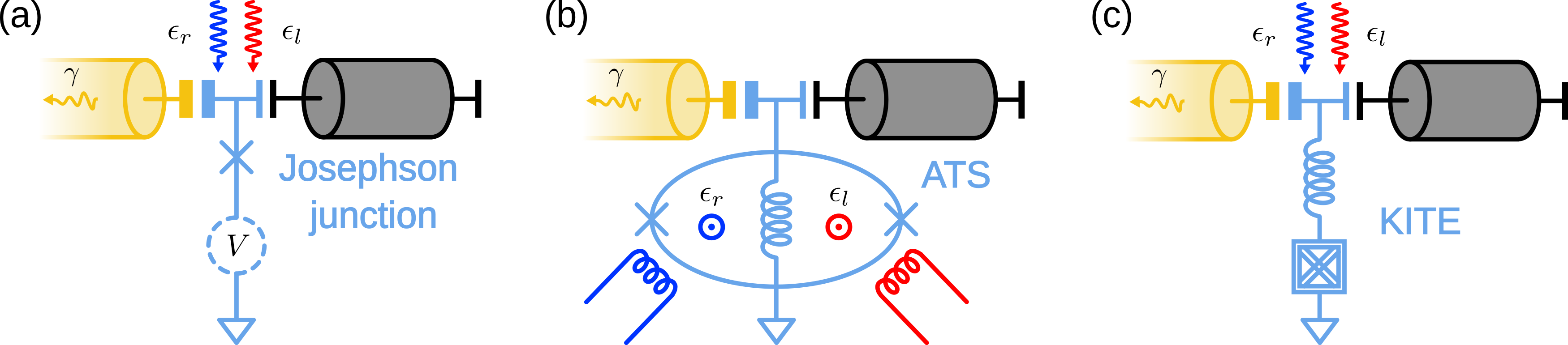}
    {\phantomsubcaption\label{fig_cQED:circuit_JJ}}
    {\phantomsubcaption\label{fig_cQED:circuit_ATS}}
    {\phantomsubcaption\label{fig_cQED:circuit_KITE}}
    \caption{NLRE method for superconducting circuit QED settings. (a-c) Schematic representation of cQED with a high-quality $LC$ oscillator (black) storing the cat-like state manifold and a nonlinear circuit (blue) coupled to a dissipative environment represented here by a lossy $LC$ oscillator (yellow). The nonlinear elements are either (a) a single Josephson junction, (b) an asymmetrically threaded superconducting quantum interference device (ATS), or (c) kinetic interference cotunneling elements (KITE). The toy model of NLRE can be engineered using bichromatic (a,c) microwave or (b) flux drives. In all the circuits we admit an optional DC voltage bias $V$, represented schematically only in (a).}
    \label{fig_cQED}
\end{figure*}

\section{NLRE method for cQED settings} \label{app:nlre_cqed}

We shall now provide a detailed derivation of the NLRE Hamiltonians for three examples of cQED systems that have shown dissipative stabilization of cat codes. The situation we consider is identical to the one presented in the main text: a high-$Q$ quantum $LC$ oscillator coupled to a nonlinear element, which in turn is coupled to a reservoir (a low-$Q$ $LC$ oscillator or transmission line). The three examples differ in the nonlinear element considered, which can be: (a) a single Josephson junction, (b) an asymmetrically threaded superconducting quantum interference device (ATS), or (c) kinetic interference cotunneling elements (KITE). Fig.~\ref{fig_cQED} presents a schematic representation of these systems. 

The first case, utilizing a single Josephson junction, was employed in the initial experimental demonstration of dissipatively stabilized cat states and the first Zeno-type operations on them~\cite{leghtas_confining_2015,touzard_coherent_2018}. The second case, involving an ATS, corresponds to recent experimental realizations of dissipatively stabilized cat qubits showing an exponential suppression of dephasing errors with increasing mean photon number~\cite{lescanne_exponential_2020,berdou_one_2023,reglade_quantum_2024}. The third case, involving KITE, represents a novel approach developed for fault-tolerant readout of cat qubits and is currently under experimental investigation~\cite{cohen_degeneracy-preserving_2017,smith_superconducting_2020,smith_magnifying_2022,smith_spectral_2025}.

As in the example from the main text, we assume the nonlinear element to be in its ground state and the reservoir to be in the vacuum state. Let us first focus on a single Josephson junction as the nonlinear element that we additionally drive using a strong single-tone microwave pump signal $\epsilon(t) = \epsilon_p \cos\!\left( \omega_p t \right)$. Denoting $\h{a}$ (storage) and $\h{c}$ (reservoir) to be the dressed modes of the system and the nonlinear element to be in its ground state, the system Hamiltonian in the non-interacting picture reads
\be \label{eq:hamil_JJ}
    \h{H}_\n{sys} = \omega_a \h{a}^\dagger\h{a} + \omega_c \h{c}^\dagger\h{c} - 
    E_J \cos\!\left( \h{\varphi} \right) + \epsilon(t) \left( \h{c} + \h{c}^\dagger \right)
\ee
with $\omega_a$ and $\omega_c$ being the oscillators frequency, $E_J$ the Josephson junction energy, and $\h{\varphi} = \varphi_a (\h{a}+\h{a}^\dagger) + \varphi_c (\h{c}+\h{c}^\dagger)$ the phase operator across the junction. We place ourselves in the frame displaced by an effective displacement ${\xi(t)\approx\xi_p\exp(-i\omega_p t)}$ with ${\xi_{p}=-i\epsilon_{p}/(\gamma/2+i(\omega_c-\omega_{p}))}$ (obtained from the Langevin equation for the reservoir mode) and rotating at $\omega_a$ and $\omega_c$. Expanding the cosine term around $\varphi_c \abs{\xi_p}$ to first order assuming $\varphi_c\abs{\xi_p}\ll1$ yields the approximate Hamiltonian
\be \label{eq:hamil_JJ_rot}
\begin{split}
    \h{H} \approx &- E_J \cos\!\left( \h{\varphi}(t) \right) + \\
    &+ E_J \varphi_c \left( \xi_p e^{-i\omega_p t} + \xi_p^* e^{i\omega_p t}\right) \sin\!\left( \h{\varphi}(t) \right)
\end{split}
\ee
We consider the general case $\omega_p = p_a \omega_a + p_c \omega_c$ with $p_a,p_c\in\bb{Z}$. Before proceeding with the rotating wave approximation, let us simplify the operator $\exp(i\h{\varphi}(t))$ using the Baker-Campbell-Hausdorff formula and the function operators $f(\h{n},k_\SB,\eta)$ defined in Eq.\eqref{eq:f_function_ions} 
\begin{widetext}
\be \label{eq:exp_expan_cqed}
\begin{split}
    \exp(i\h{\varphi}(t)) &= 
    \exp\!\left(i\varphi_a e^{-i\omega_at}\h{a} + i\varphi_a e^{i\omega_at}\h{a}^\dagger \right) \,\,
    \exp\!\left(i\varphi_c e^{-i\omega_ct}\h{c} + i\varphi_c e^{i\omega_ct}\h{c}^\dagger \right) = \\
    &= e^{-\varphi_a^2/2} \sum_{n,m\in\bb{N}} \frac{(i\varphi_a)^{n+m}}{n!\,m!} 
    e^{i(n-m)\omega_at} \h{a}^{\dagger\,n} \h{a}^{m}  \,\,
    e^{-\varphi_c^2/2} \sum_{k,l\in\bb{N}} \frac{(i\varphi_c)^{k+l}}{k!\,l!} 
    e^{i(k-l)\omega_ct} \h{c}^{\dagger\,k} \h{c}^{l} = \\
    &= \sum_{k_a\in\bb{Z}} e^{i k_a \omega_at} \h{a}^{\dagger\,k_a} f(\h{n},k_a,\varphi_a) \,\,
    \sum_{k_c\in\bb{Z}} e^{i k_c \omega_ct} \h{c}^{\dagger\,k_c} f(\h{n}_c,k_c,\varphi_c)
\end{split}
\ee
with $n_c\equiv\h{c}^\dagger\h{c}$. Using this expression and the fact that $f(\h{n},k_\SB,-\eta)=(-1)^{k_\SB} f(\h{n},k_\SB,\eta)$, we conclude that the trigonometric operator $\cos(\h{\varphi}(t))$ can be expressed as 
\be \label{eq:cosine_op}
    \cos(\h{\varphi}(t)) =\!\! \sum_{k_a,k_c} 
    \h{a}^{\dagger\,k_a} \h{c}^{\dagger\,k_c} f(\h{n},k_a,\varphi_a) f(\h{n}_c,k_c,\varphi_c) 
    e^{i (k_a \omega_a + k_c \omega_c) t }\,\,\delta_{(k_a+k_c)\,\n{mod}\,2,0}\,.
\ee
The expression for $\sin(\h{\varphi}(t))$ is similar, but with the Kronecker delta $\delta_{(k_a+k_c)\,\n{mod}\,2,1}$ and a global phase $-\pi/2$. We can now apply the rotating wave approximation to the previous Hamiltonian by selecting in Eq.~\eqref{eq:cosine_op} only the terms that oscillate with $\omega_p$. For $p_a,p_c>0$, the Hamiltonian reads 
\be \label{eq:hamil_drive_NLRE}
    \h{H} \approx - E_J \, f(\h{n},0,\varphi_a) \, f(\h{n}_c,0,\varphi_c) 
    - i E_J \varphi_c \xi_p \, \h{a}^{\dagger\,p_a} \h{c}^{\dagger\,p_c} f(\h{n},p_a,\varphi_a) f(\h{n}_c,p_c,\varphi_c) 
    \delta_{(p_a+p_c)\,\n{mod}\,2,1} + \n{H.c.}
\ee
\end{widetext}
For $p_a,p_c<0$, replace $\h{a}^{\dagger\,p_a} f(\h{n},p_a,\varphi_a)$ by $f(\h{n},p_a,\varphi_a)\h{a}^{p_a}$ correspondingly. For standard reservoir engineering schemes, one typically requires $p_c=1$, which in this context necessitates that $p_a$ be even. Therefore, we get the Hamiltonian in Eq.~\eqref{eq:hamil_cQED_NLRE} when considering two microwave drives and setting $p_a=r/l$ and $p_c=1$. It comes however with an additional first term that originates from the zeroth order expansion from Eq.~\eqref{eq:hamil_JJ_rot}. This expression corresponds to an always-on resonance condition $k_a=k_c=0$ and gathers various effects such as AC Stark shift, self-Kerr and cross-Kerr terms, all being energy conserving (i.e. it can be diagonalized using Fock states). 

This leads to some restrictive conditions. The first requirement is that this energy-conserving term must be compensated or be negligible compared to the desired stabilizing Hamiltonian. The coefficients of the function $f(\h{n},0,\varphi_a)$ written as power series of $\h{n}$ fall with powers of $\varphi_a$ and factorial denominator. Since the phase fluctuations are typically small, the first few terms of this series dominate the dynamics of which the leading one (AC Stark shift) can be compensated using the drive frequency $\omega_p$~\cite{leghtas_confining_2015,touzard_coherent_2018}. The second requirement is that the Rabi frequencies $\tilde{f}$ and $\tilde{g}$ must be engineered to satisfy the conditions of the NLRE method. The strengths of the microwave drives $\epsilon_{r/l}$ are the only free parameters that can allow an in situ engineering of the crossing points between the Rabi frequencies $\tilde{f}$ and $\tilde{g}$. The range of their tunability is limited by the condition ${\varphi_c\abs{\xi_{r/l}}\ll1}$ (see Eq.~\eqref{eq:hamil_JJ_rot}).

The different constraints imposed by the use of a single Josephson junction to stabilize cat-like state manifolds can be resolved by substituting it with some alternative nonlinear elements. Using for example a superconducting quantum interference device (SQUID) which allows to tune the Josephson energy using an external flux and to add a phase offset to the phase operator in Eq.~\eqref{eq:hamil_JJ}. When this phase offset equals $\pi/2$, the trigonometric term changes from a cosine to a sine function allowing to construct models with odd indices $r,l$. Although SQUIDs may add some additional tunable parameters, the energy-conserving term from Eq.~\eqref{eq:hamil_cQED_NLRE} will persist.

In the scenario when the nonlinear element is an ATS, the system Hamiltonian reads
\be \label{eq:hamil_ATS}
\begin{split}
    \h{H}_\n{sys}\! = \,&\omega_a \h{a}^\dagger\h{a} + \omega_c \h{c}^\dagger\h{c} - 
    2 E_J \cos\!\left(\varphi_\Sigma\right) \cos\!\left( \h{\varphi} + \varphi_\Delta \right) \\
    &+ 2\dif E_J \sin\!\left(\varphi_\Sigma\right) \sin\!\left( \h{\varphi} + \varphi_\Delta \right)
\end{split}
\ee
where $\varphi_\Sigma$ and $\varphi_\Delta$ are the sum and difference of the external fluxes, respectively, and $\dif E_J$ captures the asymmetry between Josephson energies of the junctions composing the ATS. Considering the same parameter regime as in Sec.~\ref{sec:cqed_setting}, the Hamiltonian in the rotating frame takes a similar form as Eq.~\eqref{eq:hamil_ATS_rot} but with $\omega_V=0$. Because of the absence of the zeroth order terms as in single Josephson junction case, the rotating wave approximation reveals only the operators necessary to the NLRE Hamiltonian. Moreover, the expressions $i E_J \varphi_c \xi_p$ and $\delta_{(p_a+p_c)\,\n{mod}\,2,1}$ from Eq.~\eqref{eq:hamil_drive_NLRE} are in this case replaced by $E_J \epsilon_p$ and $\delta_{(p_a+p_c)\,\n{mod}\,2,0}$, preserving the parity constraint on $p_a$ and $p_c$. This approach thus alleviates some of the constraints associated with using a single Josephson junction. However, the two approaches presented so far show another fundamental limitation when one desires to implement models with large orders $r,l$ which is the fact that the normalized Rabi frequencies $\tilde{f}(k)/\Omega_r$ and $\tilde{g}(k)/\Omega_l$ scale as $\cl{O}\left(\varphi_a^\nu/\nu!\right)$ with $\nu$ being $r$ or $l$ (straightforward from Eq.~\eqref{eq:Bessel_approx_rabi_freq}). This means that the normalized Rabi frequencies become exponentially small with the order and must be compensated by some exponentially large driving strengths $\epsilon_{r/l}$.

To solve this issue, one can envisage using instead a KITE device~\cite{smith_superconducting_2020,smith_magnifying_2022} which resembles a SQUID whose Josephson junctions are each connected in series to a large inductance. This creates a generalized Josephson element which when shunted with another large inductance (see~Fig.~\ref{fig_cQED:circuit_KITE}) gives a system Hamiltonian
\be \label{eq:hamil_KITE}
    \h{H}_\n{sys} = \omega_a \h{a}^\dagger\h{a} + \omega_c \h{c}^\dagger\h{c} + 
    (-1)^\mu E_J \cos\!\left(\mu \h{\varphi} \right)\,.
\ee
Here, $\mu\in\bb{N}$ represents the number of Cooper pairs that tunnel through the element. This expression is identical to the Hamiltonian of a single Josephson junction discussed earlier (up to the drive term), and thus yields similar results but with larger zero-point fluctuations. For large $\varphi_{a/c}$, the energy-conserving term in Eq.~\eqref{eq:hamil_drive_NLRE} will become negligible leaving room to the desired Hamiltonian of the NLRE toy model. The NLRE method we propose is then highly relevant to current research and introduces a novel approach to studying the dissipative stabilization of cat-like state manifolds. As shown the method can explain past and present circuit QED experiments, allowing us to identify challenges and conditions for different schemes to work. Notably, all existing schemes are still subject to the restrictive requirement that process orders $r$ and $l$ be either even or odd integers. The novel scheme we presented in Sec.~\ref{sec:cqed_setting} overcomes this limitation using a voltage-biased nonlinear element.

It is important to note that in all the examples discussed above, the rotating wave approximation remains valid as long as the oscillators' frequencies $\omega_a$ and $\omega_b$ are much larger than the coupling strengths (i.e. the Rabi frequencies $\tilde{f}$ and $\tilde{g}$). This condition is typically satisfied in all the presented cases, as we engineer the external drives to allow the use of a small angle approximation. Quantitatively, $\omega_{a/b}$ are of the order of $\unit{\GHz}$ while as seen in Fig.~\ref{fig_ATS:rabi_frequency} the Rabi frequencies are of the order of $\unit{\MHz}$.

\section{Bosonic Hamiltonians using voltage-biased nonlinear circuits} \label{app:dc_bias_cqed}

\begin{figure}
    \centering
    \includegraphics{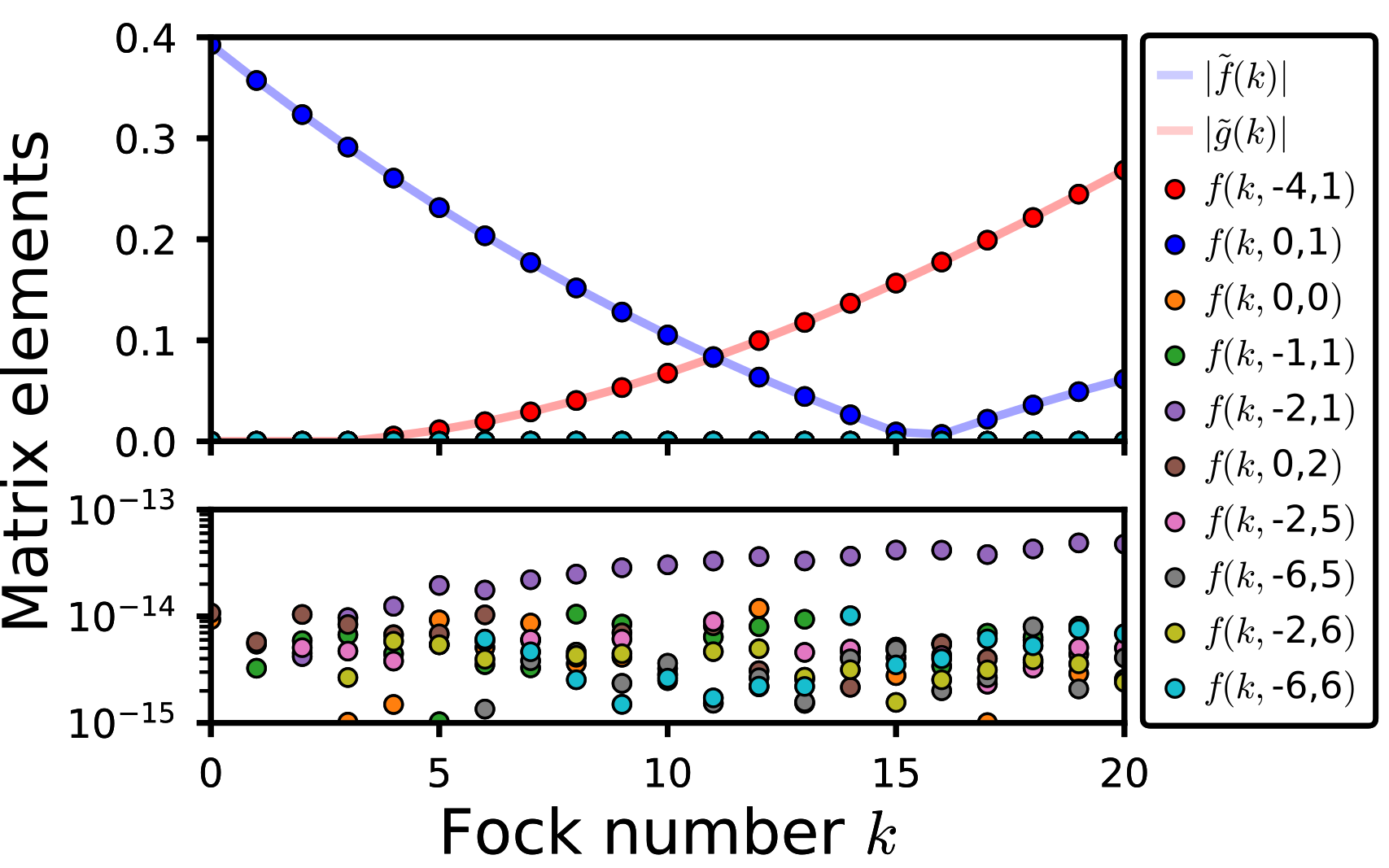}
    \caption{Matrix elements of the Hamiltonian $\h{H}_\n{avg}$ obtained by time averaging the Hamiltonian in Eq.~\eqref{eq:hamil_dc_bias} over $\qty{10}{\ns}$. The parameters of the considered system are identical to those used in Fig.~\ref{fig_ATS:rabi_frequency}. The plotted functions correspond to the matrix off-diagonals $f(k,k_a,k_c)=\abs{\bra{k+k_a,k_c}\h{H}_\n{avg}\ket{k,0}}$ with $k_{a/c}\in\bb{Z}$. We see that the resonant terms of the Hamiltonian follow the analytically derived Rabi frequencies $\tilde{f}$ and $\tilde{g}$.}
    \label{fig_RWA_cQED}
\end{figure}

In this section we present the derivation of the Hamiltonian for a voltage-biased nonlinear circuit in the rotating frame. We consider a nonlinear circuit with an ATS biased by a DC voltage source $V$. For simplicity we assume no asymmetry between the Josephson junctions energies (i.e. ${\dif E_J=0}$) and two external fluxes such that $\varphi_\Delta=\pi/2$ and $\varphi_\Delta=\pi/2+\epsilon(t)$. We modulate the fluxes using $\epsilon(t) = \epsilon_p \cos\!\left( \omega_p t \right)$ with $\epsilon_p\ll1$ and $\omega_p = p_a \omega_a + p_c \omega_c$, where $p_a,p_c\in\bb{Z}$. The Hamiltonian of the system in the oscillator's rotating frame and in the small angle approximation reads
\be \label{eq:hamil_dc_bias}
    \h{H} = - E_J \epsilon_p \left(e^{i\omega_pt} + e^{-i\omega_pt}\right) \sin\!\left( \h{\varphi}(t) + \omega_V t \right)\,.
\ee
Here, $\omega_V\equiv2eV/\hbar=v_a\omega_a+v_c\omega_c$ with $v_a,v_c\in\bb{Z}$ is the frequency associated with the DC voltage source which, as in the main text, we assume to be much smaller than the superconducting gap voltage of the junctions composing the ATS. 

Due to the additional term in the trigonometric function it cannot be rewritten in the form of Eq.~\eqref{eq:cosine_op}. This means each operator $\exp(\pm i(\h{\varphi}(t)+\omega_Vt))$ must be treated separately when performing the rotating wave approximation. Together with $e^{\pm i\omega_pt}$, the two resonant conditions are
\be \label{eq:resonant_conditions}
    \omega_p \pm \omega_V = k_a^{\pm}\omega_a + k_c^{\pm}\omega_c \,\,\,
    \Leftrightarrow \,\,\,
    \left\{
    \begin{matrix}
        p_a \pm v_a = k_a^{\pm} \\ 
        p_c \pm v_c = k_c^{\pm}
    \end{matrix}
    \right.
\ee
where $k_a^{\pm},k_c^{\pm}\in\bb{Z}$ are free variables determining the order of the resonant processes. Explicitly, after the rotating wave approximation the Hamiltonian reads ($k_a^\pm,k_c^\pm>0$)
\be \label{eq:hamil_bias_NLRE}
\begin{split}
    \h{H}&\propto \h{a}^{\dagger\,k_{a}^{+}} \h{c}^{\dagger\,k_{c}^{+}} 
    f(\h{n},k_{a}^{+},\varphi_a) f(\h{n}_c,k_{c}^{+},\varphi_c) \,+ \\
    &+\h{a}^{\dagger\,k_{a}^{-}} \h{c}^{\dagger\,k_{c}^{-}} 
    f(\h{n},k_{a}^{-},\varphi_a) f(\h{n}_c,k_{c}^{-},\varphi_c) + \n{H.c.}
\end{split}
\ee
As before, for $k_a^\pm,k_c^\pm<0$, replace $\h{a}^{\dagger\,k_a^\pm} f(\h{n},k_a^\pm,\varphi_a)$ by $f(\h{n},k_a^\pm,\varphi_a)\h{a}^{k_a^\pm}$ correspondingly. The bias voltage thus introduces an additional degree of freedom and helps lift the parity constraint on $p_a$ and $p_c$ observed in previous examples. We choose $k_{a/c}^{+}$ to be the orders of the desired processes. Therefore, the objective is to engineer $\omega_p$ and $\omega_V$ such that either their difference prevents the resonant condition from being satisfied (i.e., there exist no ${k_{a/c}^{-}\in\mathbb{Z}}$ that fulfill Eq.~\eqref{eq:resonant_conditions}), or such that the orders $k_{a/c}^{-}$ of the undesired processes are large enough to be negligible due to the exponential suppression of the Rabi frequencies for large orders. We opt for the first solution as it is more versatile and straightforward to engineer as we can simply impose 
\be \label{eq:off_resonant_condition}
    \omega_p - \omega_V = \frac{1}{2} \left(k_a^{-}\omega_a + k_c^{-}\omega_c\right) 
    \quad\n{for}\quad k_{a,c}^{-}\in\bb{Z}\,.
\ee

As an example let us take the situation presented in the main text and depicted in Fig.~\ref{fig_ATS:rabi_frequency}. We consider the same set of parameters and numerically average the Hamiltonian in Eq.~\eqref{eq:hamil_dc_bias} over a time span of $\qty{10}{\ns}$ (integration step $\qty{1}{\ps}$). The result $\h{H}_\n{avg}$ corresponds to the first term in the Magnus expansion of the time evolution operator and effectively to the rotating wave approximation Hamiltonian. We extract the off-diagonal elements of this operator in the Fock basis. Specifically, we evaluate the function $f(k,k_a,k_c)=\abs{\bra{k+k_a,k_c}\h{H}_\n{avg}\ket{k,0}}$ for various integers $k_{a/c}\in\bb{Z}$ and plot them over the Fock number $k$. The results are shown in Fig.~\ref{fig_RWA_cQED}. We observe that the resonant terms of the Hamiltonian follow the analytically derived Rabi frequencies $\tilde{f}$ and $\tilde{g}$, confirming the validity of the rotating wave approximation. The off-resonant terms are average out to values close to zero and thus can be neglected in the analysis of the system's dynamics. In particular, this is also the case for the off-diagonals that are the closest to our engineered off-resonant condition from Eq.~\eqref{eq:off_resonant_condition}, namely $\omega_r - \omega_V = 6.5\omega_a-5.5\omega_c$ and $\omega_l - \omega_V = 2.5\omega_a-5.5\omega_c$. 

\section{Rabi frequencies in the generalized NLRE model} \label{app:fun_in_general_NLRE}

In this section, we derive the expressions for the functions $\tilde{f}$ and $\tilde{g}$ in the generalized NLRE model. As described in the Sec.~\ref{sec:bogoliubov_transf}, the NLRE method can be extended to stabilize bosonic state that are linked to nonlinear cat states presented in this work as long as there exist an isomorphic transformation $\h{U}$ between the two. The stabilization operator in the standard ladder picture then reads
\be \label{eq:hamil_bogoliubov}
    \h{K} = \h{U} \left(\h{a}^{\dagger\,r} f(\h{n}) - g(\h{n}) \h{a}^{l} \right) \h{U}^\dagger
\ee
with the expression in the parentheses being the NLRE toy model from Eq.~\eqref{eq:hamilton_modif}. After performing the transformation the stabilization operator $\h{K}$ will have the general form presented in Eq.~\eqref{eq:hamilton_bogoliubov_a_at} where we extend the definition of the Rabi frequencies to
\be
    \tilde{f}_j(k)\!=\! \bra{k+j}\!\h{K}\!\ket{k}
    \quad\n{and}\quad
    \tilde{g}_j(k)\!=\! \bra{k}\!\h{K}\!\ket{k+j}.
\ee
In the Fock basis, these functions are defined using the matrix elements of the transformation $\h{U}$ and the functions $\tilde{f}$ and $\tilde{g}$ of the underlying NLRE toy model. Explicitly, we have 
\be \label{eq:tilde_f_bogol}
\begin{split}
    \tilde{f}_j(k) = \sum_{k'=0}^\infty  
    &\quad\bra{k+j}\!\h{U}\!\ket{k'+r}\, \bra{k'}\!\h{U}^\dagger\!\ket{k}\, \tilde{f}(k') \, - \\
    &-\bra{k+j}\!\h{U}\!\ket{k'}\, \bra{k'+l}\!\h{U}^\dagger\!\ket{k}\, \tilde{g}(k')
\end{split}
\ee
and similarly for $\tilde{g}_j(k)$. The matrix elements can be calculated numerically for an arbitrary transformation $\h{U}$. However, the analytical form of some interesting transformations are know and can thus be utilized for designing new models.

For the specific example from the main text, the transformation $\h{U}$ corresponds to a single-mode unitary Bogoliubov transformation decomposed in general into squeezing and rotation of phase space. Since the orientation of the cat states can be adjusted using the relative phase of the boson raising and lowering processes (see Sec.~\ref{sec:simplest_nlre_setting}), we focus exclusively on squeezing transformations $\h{S}(\zeta)$ with $\zeta\in\bb{C}$. The matrix elements of the squeezing operator in the Fock basis are given by~\cite{varro_coherent_2022}
\be \label{eq:squeezing_matrix_elements}
    \bra{k}\!\h{S}\!\left(\abs{\zeta}e^{i\phi}\right)\!\ket{k'} = \sqrt{\cos(\theta)}\, X_n^m(\theta,\phi)
\ee
where ${\cos(\theta)=\n{cosh}(\abs{\zeta})^{-1}}$, ${n=(k+k')/2}$ and ${m=(k-k')/2}$. The function $X_n^m$ is the non-normalized spherical harmonics defined as 
\be
    X_n^m(\theta,\phi) = \sqrt{\frac{4\pi}{2n+1}}\, Y_n^m(\theta,\phi)
\ee
and captures the peculiar property of the squeezing operator that it only couples states with the same parity. In other words, the matrix elements in Eq.~\eqref{eq:squeezing_matrix_elements} are non-zero only if $k$ and $k'$ have the same parity. Given this and Eq.~\eqref{eq:tilde_f_bogol}, we conclude that the Rabi frequencies $\tilde{f}_j$ and $\tilde{g}_j$ will be non-zero only if $j$ shares the same parity than $r$ or $l$. In particular, schemes stabilizing two dimensional manifolds will be composed of Rabi frequencies with either even $j$ if $(r,l)=(0,2)$ or $(2,0)$; or odd $j$ for $(1,1)$ schemes.

{
Let us illustrate this with the examples presented in the main text and show how previously proposed schemes for stabilization and autonomous correction of squeezed cat codes fit in our generalized NLRE framework. First, consider the scheme proposed in Ref.~\cite{schlegel_quantum_2022,hillmann_quantum_2023}. The stabilization operator reads
\be \label{eq:schlegel_stabilization}
\begin{split}
    \h{K}&=\h{S}(\zeta)\big(\h{a}^2-\alpha^2\big)\h{S}(\zeta)^\dagger \\ 
    &=\h{a}^{\dagger\,2}\,\sz^2 - \Big[\alpha^2 + \cz \sz + 2 \,\cz \sz\, \h{n}\Big] + \cz^2\,\h{a}^2
\end{split}
\ee
where $\cz=\cosh(\zeta)$ and $\sz=\sinh(\zeta)$ and we assume $\zeta$ to be real for simplicity. From this explicit notation, it is straightforward to see that the operator involves three even-order boson processes and we can easily identify the functions $f_0$, $f_2$ and $g_2$. Similarly, the stabilization operator proposed in Ref.~\cite{xu_autonomous_2023} reads   
\be \label{eq:xu_stabilization}
\begin{split}
    &\h{K}=\big(c_1\h{a}^\dagger+c_2\h{a}\big)\h{S}(\zeta)\big(\h{a}^2-\alpha^2\big)\h{S}(\zeta)^\dagger \\
    &= \h{a}^{\dagger\,3}\,c_2\sz^2 + c_1\cz^2\,\h{a}^3 \\
    &- \h{a}^\dagger\,\Big[ c_2 \alpha^2 + c_2\cz\sz - 2 c_1\sz^2 + \big(2 c_2\cz\sz - c_1\sz^2\big)\h{n}\Big]\\
    &- \Big[c_1 \alpha^2 + 3 c_1\cz\sz + \big(2 c_1\cz\sz - c_2\cz^2\big)\h{n} \Big]\,\h{a}
\end{split}
\ee
where $c_{1,2}$ are constants satisfying $c_1 + c_2 = 1$ (see Ref.~\cite{xu_autonomous_2023} for details). In this case, the operator involves four odd-order boson processes with clearly identifiable functions $f_1$, $f_3$, $g_1$ and $g_3$. Both examples thus fit perfectly in our generalized NLRE framework. To conclude, let us show that Eqs.~\eqref{eq:schlegel_stabilization} and~\eqref{eq:xu_stabilization} are equivalent to the squeezed versions of the $(0,2)$ and $(1,1)$ NLRE toy models, respectively. We start from Eq.~\eqref{eq:hamil_bogoliubov} and consider the minimal case of first-order polynomial functions $f(\h{n})=\alpha_f+\beta_f\h{n}$ and $g(n)=\alpha_g+\beta_g\h{n}$ with $\alpha_f,\beta_g\in\bb{R}$ being free parameters. We then get for both cases
\begin{widetext}
\be \label{eq:bogol_02}
\begin{split}
    \h{K} = \,&\h{S}(\zeta)\big(f(\h{n}) - g(\h{n})\h{a}^2\big)\h{S}(\zeta)^\dagger
    = \h{a}^{\dagger\,4}\,\beta_g\cz\sz^3
    - \h{a}^{\dagger\,2} \Big[\alpha_g\sz^2 + \beta_f\cz\sz + 3\beta_g\cz^2\sz^2 + 3\beta_g\sz^4 
    + \big(\beta_g\sz^4 + 3\beta_g\cz^2\sz^2\big)\,\h{n} \Big] \\
    &+ \Big[\alpha_f + \alpha_g\cz\sz + \beta_f\sz^2 + 3\beta_g\cz\sz^3 
    + \big(2\alpha_g\cz\sz + \beta_f\cz^2 + \beta_f\sz^2 + 6\beta_g\cz\sz^3\big)\,\h{n}
    + \big(3\beta_g\cz^3\sz + 3\beta_g\cz\sz^3\big)\,\h{n}^2 \Big] \\
    &- \Big[\alpha_g\cz^2 + \beta_f\cz\sz + 6\beta_g\cz^2\sz^2 
    + \big(\beta_g\cz^4 + 3\beta_g\cz^2\sz^2\big)\,\h{n} \Big]\,\h{a}^2
    + \beta_g\cz^3\sz\,\h{a}^4 \,,
\end{split}
\ee
\be \label{eq:bogol_11}
\begin{split}
    \h{K} =\, &\h{S}(\zeta)\big(f(\h{n})\h{a}^\dagger - g(\h{n})\h{a}\big)\h{S}(\zeta)^\dagger 
    = - \h{a}^{\dagger 3} \Big[ \beta_f\cz^2\sz + \beta_g\cz \sz^2\Big]
    + \Big[ \beta_f \cz \sz^2  + \beta_g \cz^2 \sz \Big] \h{a}^3 \\
    &- \Big[ \alpha_f\cz + \alpha_f\sz + \beta_f\cz^2\sz + 2 \beta_f\sz^3 + 3\beta_g\cz\sz^2 + \big(2\beta_f\cz^2\sz + \beta_f\sz^3 + \beta_g\cz^3 + 2\beta_g\cz\sz^2\big)\,\h{n} \Big] \h{a} \\
    &+ \h{a}^\dagger \Big[ \alpha_f\cz + \alpha_g\sz + 3\beta_f\cz\sz^2 + \beta_g\cz^2\sz  + 2 \beta_g\sz^3 + \big(\beta_f\cz^3  + 2\beta_f\cz\sz^2 + 2\beta_g\cz^2\sz + \beta_g\sz^3\big) \h{n} \Big]\,.
\end{split}
\ee
\end{widetext}
From these expressions, we can identify the coefficients of Eqs.~\eqref{eq:schlegel_stabilization} and~\eqref{eq:xu_stabilization} and find that they are equivalent up to a global prefactor. Beyond the equivalence, this also shows that the squeezed-cat stabilization operator proposed in Refs.~\cite{schlegel_quantum_2022,hillmann_quantum_2023} is the squeezed version of a particular $(0,2)$ NLRE scheme, while the one from Ref.~\cite{xu_autonomous_2023} corresponds to a squeezed $(1,1)$ scheme. 

}

This opens the way towards the design of novel stabilization schemes through reverse engineering. Let us consider a system where jump operators $\h{K}$ with two pairs of boson raising and lowering processes can be implemented (e.g., using two bichromatic laser beams in a trapped ions setup or a four-tones drive in cavity QED systems). Then fixing the orders of the drives to be $(r_1,r_2,l_1,l_2)=(1,3,1,3)$ and carefully tuning the system parameters (drive strengths, LD parameters,...), one would be able to stabilize a squeezed cat manifold which in the transformed basis would be similar to a $(1,1)$ scheme. The reverse engineering consists then in finding numerically the right parameters such that $\tilde{f}$ and $\tilde{g}$ cross adequately. The analytical form of these functions will be similar to Eq.~\eqref{eq:tilde_f_bogol}.

\clearpage
\bibliography{references}

\begin{thebibliography}{158}%
\makeatletter
\providecommand \@ifxundefined [1]{%
 \@ifx{#1\undefined}
}%
\providecommand \@ifnum [1]{%
 \ifnum #1\expandafter \@firstoftwo
 \else \expandafter \@secondoftwo
 \fi
}%
\providecommand \@ifx [1]{%
 \ifx #1\expandafter \@firstoftwo
 \else \expandafter \@secondoftwo
 \fi
}%
\providecommand \natexlab [1]{#1}%
\providecommand \enquote  [1]{``#1''}%
\providecommand \bibnamefont  [1]{#1}%
\providecommand \bibfnamefont [1]{#1}%
\providecommand \citenamefont [1]{#1}%
\providecommand \href@noop [0]{\@secondoftwo}%
\providecommand \href [0]{\begingroup \@sanitize@url \@href}%
\providecommand \@href[1]{\@@startlink{#1}\@@href}%
\providecommand \@@href[1]{\endgroup#1\@@endlink}%
\providecommand \@sanitize@url [0]{\catcode `\\12\catcode `\$12\catcode `\&12\catcode `\#12\catcode `\^12\catcode `\_12\catcode `\%12\relax}%
\providecommand \@@startlink[1]{}%
\providecommand \@@endlink[0]{}%
\providecommand \url  [0]{\begingroup\@sanitize@url \@url }%
\providecommand \@url [1]{\endgroup\@href {#1}{\urlprefix }}%
\providecommand \urlprefix  [0]{URL }%
\providecommand \Eprint [0]{\href }%
\providecommand \doibase [0]{https://doi.org/}%
\providecommand \selectlanguage [0]{\@gobble}%
\providecommand \bibinfo  [0]{\@secondoftwo}%
\providecommand \bibfield  [0]{\@secondoftwo}%
\providecommand \translation [1]{[#1]}%
\providecommand \BibitemOpen [0]{}%
\providecommand \bibitemStop [0]{}%
\providecommand \bibitemNoStop [0]{.\EOS\space}%
\providecommand \EOS [0]{\spacefactor3000\relax}%
\providecommand \BibitemShut  [1]{\csname bibitem#1\endcsname}%
\let\auto@bib@innerbib\@empty
\bibitem [{\citenamefont {Plenio}\ \emph {et~al.}(1999)\citenamefont {Plenio}, \citenamefont {Huelga}, \citenamefont {Beige},\ and\ \citenamefont {Knight}}]{plenio_cavity_1999}%
  \BibitemOpen
  \bibfield  {author} {\bibinfo {author} {\bibfnamefont {M.~B.}\ \bibnamefont {Plenio}}, \bibinfo {author} {\bibfnamefont {S.~F.}\ \bibnamefont {Huelga}}, \bibinfo {author} {\bibfnamefont {A.}~\bibnamefont {Beige}},\ and\ \bibinfo {author} {\bibfnamefont {P.~L.}\ \bibnamefont {Knight}},\ }\bibfield  {title} {\bibinfo {title} {Cavity-loss-induced generation of entangled atoms},\ }\href {https://doi.org/10.1103/PhysRevA.59.2468} {\bibfield  {journal} {\bibinfo  {journal} {Phys. Rev. A}\ }\textbf {\bibinfo {volume} {59}},\ \bibinfo {pages} {2468} (\bibinfo {year} {1999})}\BibitemShut {NoStop}%
\bibitem [{\citenamefont {Diehl}\ \emph {et~al.}(2008)\citenamefont {Diehl}, \citenamefont {Micheli}, \citenamefont {Kantian}, \citenamefont {Kraus}, \citenamefont {B{\"u}chler},\ and\ \citenamefont {Zoller}}]{diehl_quantum_2008}%
  \BibitemOpen
  \bibfield  {author} {\bibinfo {author} {\bibfnamefont {S.}~\bibnamefont {Diehl}}, \bibinfo {author} {\bibfnamefont {A.}~\bibnamefont {Micheli}}, \bibinfo {author} {\bibfnamefont {A.}~\bibnamefont {Kantian}}, \bibinfo {author} {\bibfnamefont {B.}~\bibnamefont {Kraus}}, \bibinfo {author} {\bibfnamefont {H.~P.}\ \bibnamefont {B{\"u}chler}},\ and\ \bibinfo {author} {\bibfnamefont {P.}~\bibnamefont {Zoller}},\ }\bibfield  {title} {\bibinfo {title} {Quantum states and phases in driven open quantum systems with cold atoms},\ }\href {https://doi.org/10.1038/nphys1073} {\bibfield  {journal} {\bibinfo  {journal} {Nat. Phys.}\ }\textbf {\bibinfo {volume} {4}},\ \bibinfo {pages} {878} (\bibinfo {year} {2008})}\BibitemShut {NoStop}%
\bibitem [{\citenamefont {Vacanti}\ and\ \citenamefont {Beige}(2009)}]{vacanti_cooling_2009}%
  \BibitemOpen
  \bibfield  {author} {\bibinfo {author} {\bibfnamefont {G.}~\bibnamefont {Vacanti}}\ and\ \bibinfo {author} {\bibfnamefont {A.}~\bibnamefont {Beige}},\ }\bibfield  {title} {\bibinfo {title} {Cooling atoms into entangled states},\ }\href {https://doi.org/10.1088/1367-2630/11/8/083008} {\bibfield  {journal} {\bibinfo  {journal} {New J. Phys.}\ }\textbf {\bibinfo {volume} {11}},\ \bibinfo {pages} {083008} (\bibinfo {year} {2009})}\BibitemShut {NoStop}%
\bibitem [{\citenamefont {Krauter}\ \emph {et~al.}(2011)\citenamefont {Krauter}, \citenamefont {Muschik}, \citenamefont {Jensen}, \citenamefont {Wasilewski}, \citenamefont {Petersen}, \citenamefont {Cirac},\ and\ \citenamefont {Polzik}}]{krauter_entanglement_2011}%
  \BibitemOpen
  \bibfield  {author} {\bibinfo {author} {\bibfnamefont {H.}~\bibnamefont {Krauter}}, \bibinfo {author} {\bibfnamefont {C.~A.}\ \bibnamefont {Muschik}}, \bibinfo {author} {\bibfnamefont {K.}~\bibnamefont {Jensen}}, \bibinfo {author} {\bibfnamefont {W.}~\bibnamefont {Wasilewski}}, \bibinfo {author} {\bibfnamefont {J.~M.}\ \bibnamefont {Petersen}}, \bibinfo {author} {\bibfnamefont {J.~I.}\ \bibnamefont {Cirac}},\ and\ \bibinfo {author} {\bibfnamefont {E.~S.}\ \bibnamefont {Polzik}},\ }\bibfield  {title} {\bibinfo {title} {Entanglement {Generated} by {Dissipation} and {Steady} {State} {Entanglement} of {Two} {Macroscopic} {Objects}},\ }\href {https://doi.org/10.1103/PhysRevLett.107.080503} {\bibfield  {journal} {\bibinfo  {journal} {Phys. Rev. Lett.}\ }\textbf {\bibinfo {volume} {107}},\ \bibinfo {pages} {080503} (\bibinfo {year} {2011})}\BibitemShut {NoStop}%
\bibitem [{\citenamefont {Cho}\ \emph {et~al.}(2011)\citenamefont {Cho}, \citenamefont {Bose},\ and\ \citenamefont {Kim}}]{cho_optical_2011}%
  \BibitemOpen
  \bibfield  {author} {\bibinfo {author} {\bibfnamefont {J.}~\bibnamefont {Cho}}, \bibinfo {author} {\bibfnamefont {S.}~\bibnamefont {Bose}},\ and\ \bibinfo {author} {\bibfnamefont {M.~S.}\ \bibnamefont {Kim}},\ }\bibfield  {title} {\bibinfo {title} {Optical {Pumping} into {Many}-{Body} {Entanglement}},\ }\href {https://doi.org/10.1103/PhysRevLett.106.020504} {\bibfield  {journal} {\bibinfo  {journal} {Phys. Rev. Lett.}\ }\textbf {\bibinfo {volume} {106}},\ \bibinfo {pages} {020504} (\bibinfo {year} {2011})}\BibitemShut {NoStop}%
\bibitem [{\citenamefont {Lin}\ \emph {et~al.}(2013)\citenamefont {Lin}, \citenamefont {Gaebler}, \citenamefont {Reiter}, \citenamefont {Tan}, \citenamefont {Bowler}, \citenamefont {Sørensen}, \citenamefont {Leibfried},\ and\ \citenamefont {Wineland}}]{lin_dissipative_2013}%
  \BibitemOpen
  \bibfield  {author} {\bibinfo {author} {\bibfnamefont {Y.}~\bibnamefont {Lin}}, \bibinfo {author} {\bibfnamefont {J.~P.}\ \bibnamefont {Gaebler}}, \bibinfo {author} {\bibfnamefont {F.}~\bibnamefont {Reiter}}, \bibinfo {author} {\bibfnamefont {T.~R.}\ \bibnamefont {Tan}}, \bibinfo {author} {\bibfnamefont {R.}~\bibnamefont {Bowler}}, \bibinfo {author} {\bibfnamefont {A.~S.}\ \bibnamefont {Sørensen}}, \bibinfo {author} {\bibfnamefont {D.}~\bibnamefont {Leibfried}},\ and\ \bibinfo {author} {\bibfnamefont {D.~J.}\ \bibnamefont {Wineland}},\ }\bibfield  {title} {\bibinfo {title} {Dissipative production of a maximally entangled steady state of two quantum bits},\ }\href {https://doi.org/10.1038/nature12801} {\bibfield  {journal} {\bibinfo  {journal} {Nature}\ }\textbf {\bibinfo {volume} {504}},\ \bibinfo {pages} {415} (\bibinfo {year} {2013})}\BibitemShut {NoStop}%
\bibitem [{\citenamefont {Shankar}\ \emph {et~al.}(2013)\citenamefont {Shankar}, \citenamefont {Hatridge}, \citenamefont {Leghtas}, \citenamefont {Sliwa}, \citenamefont {Narla}, \citenamefont {Vool}, \citenamefont {Girvin}, \citenamefont {Frunzio}, \citenamefont {Mirrahimi},\ and\ \citenamefont {Devoret}}]{shankar_autonomously_2013}%
  \BibitemOpen
  \bibfield  {author} {\bibinfo {author} {\bibfnamefont {S.}~\bibnamefont {Shankar}}, \bibinfo {author} {\bibfnamefont {M.}~\bibnamefont {Hatridge}}, \bibinfo {author} {\bibfnamefont {Z.}~\bibnamefont {Leghtas}}, \bibinfo {author} {\bibfnamefont {K.~M.}\ \bibnamefont {Sliwa}}, \bibinfo {author} {\bibfnamefont {A.}~\bibnamefont {Narla}}, \bibinfo {author} {\bibfnamefont {U.}~\bibnamefont {Vool}}, \bibinfo {author} {\bibfnamefont {S.~M.}\ \bibnamefont {Girvin}}, \bibinfo {author} {\bibfnamefont {L.}~\bibnamefont {Frunzio}}, \bibinfo {author} {\bibfnamefont {M.}~\bibnamefont {Mirrahimi}},\ and\ \bibinfo {author} {\bibfnamefont {M.~H.}\ \bibnamefont {Devoret}},\ }\bibfield  {title} {\bibinfo {title} {Autonomously stabilized entanglement between two superconducting quantum bits},\ }\href {https://doi.org/10.1038/nature12802} {\bibfield  {journal} {\bibinfo  {journal} {Nature}\ }\textbf {\bibinfo {volume} {504}},\ \bibinfo {pages} {419} (\bibinfo {year} {2013})}\BibitemShut {NoStop}%
\bibitem [{\citenamefont {Navarrete-Benlloch}\ \emph {et~al.}(2014)\citenamefont {Navarrete-Benlloch}, \citenamefont {Garc\'{\i}a-Ripoll},\ and\ \citenamefont {Porras}}]{navarrete_inducing_2014}%
  \BibitemOpen
  \bibfield  {author} {\bibinfo {author} {\bibfnamefont {C.}~\bibnamefont {Navarrete-Benlloch}}, \bibinfo {author} {\bibfnamefont {J.~J.}\ \bibnamefont {Garc\'{\i}a-Ripoll}},\ and\ \bibinfo {author} {\bibfnamefont {D.}~\bibnamefont {Porras}},\ }\bibfield  {title} {\bibinfo {title} {Inducing {Nonclassical} {Lasing} via {Periodic} {Drivings} in {Circuit} {Quantum} {Electrodynamics}},\ }\href {https://doi.org/10.1103/PhysRevLett.113.193601} {\bibfield  {journal} {\bibinfo  {journal} {Phys. Rev. Lett.}\ }\textbf {\bibinfo {volume} {113}},\ \bibinfo {pages} {193601} (\bibinfo {year} {2014})}\BibitemShut {NoStop}%
\bibitem [{\citenamefont {Reiter}\ \emph {et~al.}(2016)\citenamefont {Reiter}, \citenamefont {Reeb},\ and\ \citenamefont {S\o{}rensen}}]{reiter_scalable_2016}%
  \BibitemOpen
  \bibfield  {author} {\bibinfo {author} {\bibfnamefont {F.}~\bibnamefont {Reiter}}, \bibinfo {author} {\bibfnamefont {D.}~\bibnamefont {Reeb}},\ and\ \bibinfo {author} {\bibfnamefont {A.~S.}\ \bibnamefont {S\o{}rensen}},\ }\bibfield  {title} {\bibinfo {title} {Scalable {Dissipative} {Preparation} of {Many}-{Body} {Entanglement}},\ }\href {https://doi.org/10.1103/PhysRevLett.117.040501} {\bibfield  {journal} {\bibinfo  {journal} {Phys. Rev. Lett.}\ }\textbf {\bibinfo {volume} {117}},\ \bibinfo {pages} {040501} (\bibinfo {year} {2016})}\BibitemShut {NoStop}%
\bibitem [{\citenamefont {Cole}\ \emph {et~al.}(2022)\citenamefont {Cole}, \citenamefont {Erickson}, \citenamefont {Zarantonello}, \citenamefont {Horn}, \citenamefont {Hou}, \citenamefont {Wu}, \citenamefont {Slichter}, \citenamefont {Reiter}, \citenamefont {Koch},\ and\ \citenamefont {Leibfried}}]{cole_resource_efficient_2022}%
  \BibitemOpen
  \bibfield  {author} {\bibinfo {author} {\bibfnamefont {D.~C.}\ \bibnamefont {Cole}}, \bibinfo {author} {\bibfnamefont {S.~D.}\ \bibnamefont {Erickson}}, \bibinfo {author} {\bibfnamefont {G.}~\bibnamefont {Zarantonello}}, \bibinfo {author} {\bibfnamefont {K.~P.}\ \bibnamefont {Horn}}, \bibinfo {author} {\bibfnamefont {P.-Y.}\ \bibnamefont {Hou}}, \bibinfo {author} {\bibfnamefont {J.~J.}\ \bibnamefont {Wu}}, \bibinfo {author} {\bibfnamefont {D.~H.}\ \bibnamefont {Slichter}}, \bibinfo {author} {\bibfnamefont {F.}~\bibnamefont {Reiter}}, \bibinfo {author} {\bibfnamefont {C.~P.}\ \bibnamefont {Koch}},\ and\ \bibinfo {author} {\bibfnamefont {D.}~\bibnamefont {Leibfried}},\ }\bibfield  {title} {\bibinfo {title} {Resource-{Efficient} {Dissipative} {Entanglement} of {Two} {Trapped}-{Ion} {Qubits}},\ }\href {https://doi.org/10.1103/PhysRevLett.128.080502} {\bibfield  {journal} {\bibinfo  {journal} {Phys. Rev. Lett.}\ }\textbf {\bibinfo {volume} {128}},\ \bibinfo {pages} {080502} (\bibinfo {year} {2022})}\BibitemShut {NoStop}%
\bibitem [{\citenamefont {Malinowski}\ \emph {et~al.}(2022)\citenamefont {Malinowski}, \citenamefont {Zhang}, \citenamefont {Negnevitsky}, \citenamefont {Rojkov}, \citenamefont {Reiter}, \citenamefont {Nguyen}, \citenamefont {Stadler}, \citenamefont {Kienzler}, \citenamefont {Mehta},\ and\ \citenamefont {Home}}]{malinowski_generation_2022}%
  \BibitemOpen
  \bibfield  {author} {\bibinfo {author} {\bibfnamefont {M.}~\bibnamefont {Malinowski}}, \bibinfo {author} {\bibfnamefont {C.}~\bibnamefont {Zhang}}, \bibinfo {author} {\bibfnamefont {V.}~\bibnamefont {Negnevitsky}}, \bibinfo {author} {\bibfnamefont {I.}~\bibnamefont {Rojkov}}, \bibinfo {author} {\bibfnamefont {F.}~\bibnamefont {Reiter}}, \bibinfo {author} {\bibfnamefont {T.-L.}\ \bibnamefont {Nguyen}}, \bibinfo {author} {\bibfnamefont {M.}~\bibnamefont {Stadler}}, \bibinfo {author} {\bibfnamefont {D.}~\bibnamefont {Kienzler}}, \bibinfo {author} {\bibfnamefont {K.~K.}\ \bibnamefont {Mehta}},\ and\ \bibinfo {author} {\bibfnamefont {J.~P.}\ \bibnamefont {Home}},\ }\bibfield  {title} {\bibinfo {title} {Generation of a {Maximally} {Entangled} {State} {Using} {Collective} {Optical} {Pumping}},\ }\href {https://doi.org/10.1103/PhysRevLett.128.080503} {\bibfield  {journal} {\bibinfo  {journal} {Phys. Rev. Lett.}\ }\textbf {\bibinfo {volume} {128}},\ \bibinfo {pages} {080503} (\bibinfo {year} {2022})}\BibitemShut {NoStop}%
\bibitem [{\citenamefont {Paz}\ and\ \citenamefont {Zurek}(1998)}]{paz_continuous_1998}%
  \BibitemOpen
  \bibfield  {author} {\bibinfo {author} {\bibfnamefont {J.~P.}\ \bibnamefont {Paz}}\ and\ \bibinfo {author} {\bibfnamefont {W.~H.}\ \bibnamefont {Zurek}},\ }\bibfield  {title} {\bibinfo {title} {Continuous error correction},\ }\href {https://doi.org/10.1098/rspa.1998.0165} {\bibfield  {journal} {\bibinfo  {journal} {Proceedings of the Royal Society of London. Series A: Mathematical, Physical and Engineering Sciences}\ }\textbf {\bibinfo {volume} {454}},\ \bibinfo {pages} {355} (\bibinfo {year} {1998})}\BibitemShut {NoStop}%
\bibitem [{\citenamefont {Barnes}\ and\ \citenamefont {Warren}(2000)}]{barnes_automatic_2000}%
  \BibitemOpen
  \bibfield  {author} {\bibinfo {author} {\bibfnamefont {J.~P.}\ \bibnamefont {Barnes}}\ and\ \bibinfo {author} {\bibfnamefont {W.~S.}\ \bibnamefont {Warren}},\ }\bibfield  {title} {\bibinfo {title} {Automatic {Quantum} {Error} {Correction}},\ }\href {https://doi.org/10.1103/PhysRevLett.85.856} {\bibfield  {journal} {\bibinfo  {journal} {Phys. Rev. Lett.}\ }\textbf {\bibinfo {volume} {85}},\ \bibinfo {pages} {856} (\bibinfo {year} {2000})}\BibitemShut {NoStop}%
\bibitem [{\citenamefont {Sarovar}\ and\ \citenamefont {Milburn}(2005)}]{sarovar_continuous_2005}%
  \BibitemOpen
  \bibfield  {author} {\bibinfo {author} {\bibfnamefont {M.}~\bibnamefont {Sarovar}}\ and\ \bibinfo {author} {\bibfnamefont {G.~J.}\ \bibnamefont {Milburn}},\ }\bibfield  {title} {\bibinfo {title} {Continuous quantum error correction by cooling},\ }\href {https://doi.org/10.1103/PhysRevA.72.012306} {\bibfield  {journal} {\bibinfo  {journal} {Phys. Rev. A}\ }\textbf {\bibinfo {volume} {72}},\ \bibinfo {pages} {012306} (\bibinfo {year} {2005})}\BibitemShut {NoStop}%
\bibitem [{\citenamefont {Pastawski}\ \emph {et~al.}(2011)\citenamefont {Pastawski}, \citenamefont {Clemente},\ and\ \citenamefont {Cirac}}]{pastawski_quantum_2011}%
  \BibitemOpen
  \bibfield  {author} {\bibinfo {author} {\bibfnamefont {F.}~\bibnamefont {Pastawski}}, \bibinfo {author} {\bibfnamefont {L.}~\bibnamefont {Clemente}},\ and\ \bibinfo {author} {\bibfnamefont {J.~I.}\ \bibnamefont {Cirac}},\ }\bibfield  {title} {\bibinfo {title} {Quantum memories based on engineered dissipation},\ }\href {https://doi.org/10.1103/PhysRevA.83.012304} {\bibfield  {journal} {\bibinfo  {journal} {Phys. Rev. A}\ }\textbf {\bibinfo {volume} {83}},\ \bibinfo {pages} {012304} (\bibinfo {year} {2011})}\BibitemShut {NoStop}%
\bibitem [{\citenamefont {Kapit}(2016)}]{kapit_hardware_2016}%
  \BibitemOpen
  \bibfield  {author} {\bibinfo {author} {\bibfnamefont {E.}~\bibnamefont {Kapit}},\ }\bibfield  {title} {\bibinfo {title} {Hardware-{Efficient} and {Fully} {Autonomous} {Quantum} {Error} {Correction} in {Superconducting} {Circuits}},\ }\href {https://doi.org/10.1103/PhysRevLett.116.150501} {\bibfield  {journal} {\bibinfo  {journal} {Phys. Rev. Lett.}\ }\textbf {\bibinfo {volume} {116}},\ \bibinfo {pages} {150501} (\bibinfo {year} {2016})}\BibitemShut {NoStop}%
\bibitem [{\citenamefont {Reiter}\ \emph {et~al.}(2017)\citenamefont {Reiter}, \citenamefont {Sørensen}, \citenamefont {Zoller},\ and\ \citenamefont {Muschik}}]{reiter_dissipative_2017}%
  \BibitemOpen
  \bibfield  {author} {\bibinfo {author} {\bibfnamefont {F.}~\bibnamefont {Reiter}}, \bibinfo {author} {\bibfnamefont {A.~S.}\ \bibnamefont {Sørensen}}, \bibinfo {author} {\bibfnamefont {P.}~\bibnamefont {Zoller}},\ and\ \bibinfo {author} {\bibfnamefont {C.~A.}\ \bibnamefont {Muschik}},\ }\bibfield  {title} {\bibinfo {title} {Dissipative quantum error correction and application to quantum sensing with trapped ions},\ }\href {https://doi.org/10.1038/s41467-017-01895-5} {\bibfield  {journal} {\bibinfo  {journal} {Nat. Commun.}\ }\textbf {\bibinfo {volume} {8}},\ \bibinfo {pages} {1822} (\bibinfo {year} {2017})}\BibitemShut {NoStop}%
\bibitem [{\citenamefont {Lihm}\ \emph {et~al.}(2018)\citenamefont {Lihm}, \citenamefont {Noh},\ and\ \citenamefont {Fischer}}]{lihm_implementation_2018}%
  \BibitemOpen
  \bibfield  {author} {\bibinfo {author} {\bibfnamefont {J.-M.}\ \bibnamefont {Lihm}}, \bibinfo {author} {\bibfnamefont {K.}~\bibnamefont {Noh}},\ and\ \bibinfo {author} {\bibfnamefont {U.~R.}\ \bibnamefont {Fischer}},\ }\bibfield  {title} {\bibinfo {title} {Implementation-independent sufficient condition of the {Knill}-{Laflamme} type for the autonomous protection of logical qudits by strong engineered dissipation},\ }\href {https://doi.org/10.1103/PhysRevA.98.012317} {\bibfield  {journal} {\bibinfo  {journal} {Phys. Rev. A}\ }\textbf {\bibinfo {volume} {98}},\ \bibinfo {pages} {012317} (\bibinfo {year} {2018})}\BibitemShut {NoStop}%
\bibitem [{\citenamefont {de~Neeve}\ \emph {et~al.}(2022)\citenamefont {de~Neeve}, \citenamefont {Nguyen}, \citenamefont {Behrle},\ and\ \citenamefont {Home}}]{de_neeve_error_2022}%
  \BibitemOpen
  \bibfield  {author} {\bibinfo {author} {\bibfnamefont {B.}~\bibnamefont {de~Neeve}}, \bibinfo {author} {\bibfnamefont {T.-L.}\ \bibnamefont {Nguyen}}, \bibinfo {author} {\bibfnamefont {T.}~\bibnamefont {Behrle}},\ and\ \bibinfo {author} {\bibfnamefont {J.~P.}\ \bibnamefont {Home}},\ }\bibfield  {title} {\bibinfo {title} {Error correction of a logical grid state qubit by dissipative pumping},\ }\href {https://www.nature.com/articles/s41567-021-01487-7} {\bibfield  {journal} {\bibinfo  {journal} {Nat. Phys.}\ }\textbf {\bibinfo {volume} {18}},\ \bibinfo {pages} {296} (\bibinfo {year} {2022})}\BibitemShut {NoStop}%
\bibitem [{\citenamefont {Poyatos}\ \emph {et~al.}(1996)\citenamefont {Poyatos}, \citenamefont {Cirac},\ and\ \citenamefont {Zoller}}]{poyatos_quantum_1996}%
  \BibitemOpen
  \bibfield  {author} {\bibinfo {author} {\bibfnamefont {J.~F.}\ \bibnamefont {Poyatos}}, \bibinfo {author} {\bibfnamefont {J.~I.}\ \bibnamefont {Cirac}},\ and\ \bibinfo {author} {\bibfnamefont {P.}~\bibnamefont {Zoller}},\ }\bibfield  {title} {\bibinfo {title} {Quantum {Reservoir} {Engineering} with {Laser} {Cooled} {Trapped} {Ions}},\ }\href {https://doi.org/10.1103/PhysRevLett.77.4728} {\bibfield  {journal} {\bibinfo  {journal} {Phys. Rev. Lett.}\ }\textbf {\bibinfo {volume} {77}},\ \bibinfo {pages} {4728} (\bibinfo {year} {1996})}\BibitemShut {NoStop}%
\bibitem [{\citenamefont {Barreiro}\ \emph {et~al.}(2011)\citenamefont {Barreiro}, \citenamefont {M{\"u}ller}, \citenamefont {Schindler}, \citenamefont {Nigg}, \citenamefont {Monz}, \citenamefont {Chwalla}, \citenamefont {Hennrich}, \citenamefont {Roos}, \citenamefont {Zoller},\ and\ \citenamefont {Blatt}}]{barreiro_open_2011}%
  \BibitemOpen
  \bibfield  {author} {\bibinfo {author} {\bibfnamefont {J.~T.}\ \bibnamefont {Barreiro}}, \bibinfo {author} {\bibfnamefont {M.}~\bibnamefont {M{\"u}ller}}, \bibinfo {author} {\bibfnamefont {P.}~\bibnamefont {Schindler}}, \bibinfo {author} {\bibfnamefont {D.}~\bibnamefont {Nigg}}, \bibinfo {author} {\bibfnamefont {T.}~\bibnamefont {Monz}}, \bibinfo {author} {\bibfnamefont {M.}~\bibnamefont {Chwalla}}, \bibinfo {author} {\bibfnamefont {M.}~\bibnamefont {Hennrich}}, \bibinfo {author} {\bibfnamefont {C.~F.}\ \bibnamefont {Roos}}, \bibinfo {author} {\bibfnamefont {P.}~\bibnamefont {Zoller}},\ and\ \bibinfo {author} {\bibfnamefont {R.}~\bibnamefont {Blatt}},\ }\bibfield  {title} {\bibinfo {title} {An open-system quantum simulator with trapped ions},\ }\href {https://doi.org/10.1038/nature09801} {\bibfield  {journal} {\bibinfo  {journal} {Nature}\ }\textbf {\bibinfo {volume} {470}},\ \bibinfo {pages} {486} (\bibinfo {year} {2011})}\BibitemShut {NoStop}%
\bibitem [{\citenamefont {Raghunandan}\ \emph {et~al.}(2020)\citenamefont {Raghunandan}, \citenamefont {Wolf}, \citenamefont {Ospelkaus}, \citenamefont {Schmidt},\ and\ \citenamefont {Weimer}}]{raghunandan_initialization_2020}%
  \BibitemOpen
  \bibfield  {author} {\bibinfo {author} {\bibfnamefont {M.}~\bibnamefont {Raghunandan}}, \bibinfo {author} {\bibfnamefont {F.}~\bibnamefont {Wolf}}, \bibinfo {author} {\bibfnamefont {C.}~\bibnamefont {Ospelkaus}}, \bibinfo {author} {\bibfnamefont {P.~O.}\ \bibnamefont {Schmidt}},\ and\ \bibinfo {author} {\bibfnamefont {H.}~\bibnamefont {Weimer}},\ }\bibfield  {title} {\bibinfo {title} {Initialization of quantum simulators by sympathetic cooling},\ }\href {https://doi.org/10.1126/sciadv.aaw9268} {\bibfield  {journal} {\bibinfo  {journal} {Sci. Adv.}\ }\textbf {\bibinfo {volume} {6}},\ \bibinfo {pages} {eaaw9268} (\bibinfo {year} {2020})}\BibitemShut {NoStop}%
\bibitem [{\citenamefont {Verstraete}\ \emph {et~al.}(2009)\citenamefont {Verstraete}, \citenamefont {Wolf},\ and\ \citenamefont {Ignacio~Cirac}}]{verstraete_quantum_2009}%
  \BibitemOpen
  \bibfield  {author} {\bibinfo {author} {\bibfnamefont {F.}~\bibnamefont {Verstraete}}, \bibinfo {author} {\bibfnamefont {M.~M.}\ \bibnamefont {Wolf}},\ and\ \bibinfo {author} {\bibfnamefont {J.}~\bibnamefont {Ignacio~Cirac}},\ }\bibfield  {title} {\bibinfo {title} {Quantum computation and quantum-state engineering driven by dissipation},\ }\href {https://doi.org/10.1038/nphys1342} {\bibfield  {journal} {\bibinfo  {journal} {Nat. Phys.}\ }\textbf {\bibinfo {volume} {5}},\ \bibinfo {pages} {633} (\bibinfo {year} {2009})}\BibitemShut {NoStop}%
\bibitem [{\citenamefont {Kienzler}\ \emph {et~al.}(2015)\citenamefont {Kienzler}, \citenamefont {Lo}, \citenamefont {Keitch}, \citenamefont {de~Clercq}, \citenamefont {Leupold}, \citenamefont {Lindenfelser}, \citenamefont {Marinelli}, \citenamefont {Negnevitsky},\ and\ \citenamefont {Home}}]{kienzler_quantum_2015}%
  \BibitemOpen
  \bibfield  {author} {\bibinfo {author} {\bibfnamefont {D.}~\bibnamefont {Kienzler}}, \bibinfo {author} {\bibfnamefont {H.-Y.}\ \bibnamefont {Lo}}, \bibinfo {author} {\bibfnamefont {B.}~\bibnamefont {Keitch}}, \bibinfo {author} {\bibfnamefont {L.}~\bibnamefont {de~Clercq}}, \bibinfo {author} {\bibfnamefont {F.}~\bibnamefont {Leupold}}, \bibinfo {author} {\bibfnamefont {F.}~\bibnamefont {Lindenfelser}}, \bibinfo {author} {\bibfnamefont {M.}~\bibnamefont {Marinelli}}, \bibinfo {author} {\bibfnamefont {V.}~\bibnamefont {Negnevitsky}},\ and\ \bibinfo {author} {\bibfnamefont {J.~P.}\ \bibnamefont {Home}},\ }\bibfield  {title} {\bibinfo {title} {Quantum harmonic oscillator state synthesis by reservoir engineering},\ }\href {https://doi.org/10.1126/science.1261033} {\bibfield  {journal} {\bibinfo  {journal} {Science}\ }\textbf {\bibinfo {volume} {347}},\ \bibinfo {pages} {53} (\bibinfo {year} {2015})}\BibitemShut {NoStop}%
\bibitem [{\citenamefont {Cirac}\ \emph {et~al.}(1993)\citenamefont {Cirac}, \citenamefont {Parkins}, \citenamefont {Blatt},\ and\ \citenamefont {Zoller}}]{cirac_dark_1993}%
  \BibitemOpen
  \bibfield  {author} {\bibinfo {author} {\bibfnamefont {J.~I.}\ \bibnamefont {Cirac}}, \bibinfo {author} {\bibfnamefont {A.~S.}\ \bibnamefont {Parkins}}, \bibinfo {author} {\bibfnamefont {R.}~\bibnamefont {Blatt}},\ and\ \bibinfo {author} {\bibfnamefont {P.}~\bibnamefont {Zoller}},\ }\bibfield  {title} {\bibinfo {title} {``{Dark}'' squeezed states of the motion of a trapped ion},\ }\href {https://doi.org/10.1103/PhysRevLett.70.556} {\bibfield  {journal} {\bibinfo  {journal} {Phys. Rev. Lett.}\ }\textbf {\bibinfo {volume} {70}},\ \bibinfo {pages} {556} (\bibinfo {year} {1993})}\BibitemShut {NoStop}%
\bibitem [{\citenamefont {Kronwald}\ \emph {et~al.}(2013)\citenamefont {Kronwald}, \citenamefont {Marquardt},\ and\ \citenamefont {Clerk}}]{kronwald_arbitrarily_2013}%
  \BibitemOpen
  \bibfield  {author} {\bibinfo {author} {\bibfnamefont {A.}~\bibnamefont {Kronwald}}, \bibinfo {author} {\bibfnamefont {F.}~\bibnamefont {Marquardt}},\ and\ \bibinfo {author} {\bibfnamefont {A.~A.}\ \bibnamefont {Clerk}},\ }\bibfield  {title} {\bibinfo {title} {Arbitrarily large steady-state bosonic squeezing via dissipation},\ }\href {https://doi.org/10.1103/PhysRevA.88.063833} {\bibfield  {journal} {\bibinfo  {journal} {Phys. Rev. A}\ }\textbf {\bibinfo {volume} {88}},\ \bibinfo {pages} {063833} (\bibinfo {year} {2013})}\BibitemShut {NoStop}%
\bibitem [{\citenamefont {Lo}\ \emph {et~al.}(2015)\citenamefont {Lo}, \citenamefont {Kienzler}, \citenamefont {de~Clercq}, \citenamefont {Marinelli}, \citenamefont {Negnevitsky}, \citenamefont {Keitch},\ and\ \citenamefont {Home}}]{lo_spinmotion_2015}%
  \BibitemOpen
  \bibfield  {author} {\bibinfo {author} {\bibfnamefont {H.-Y.}\ \bibnamefont {Lo}}, \bibinfo {author} {\bibfnamefont {D.}~\bibnamefont {Kienzler}}, \bibinfo {author} {\bibfnamefont {L.}~\bibnamefont {de~Clercq}}, \bibinfo {author} {\bibfnamefont {M.}~\bibnamefont {Marinelli}}, \bibinfo {author} {\bibfnamefont {V.}~\bibnamefont {Negnevitsky}}, \bibinfo {author} {\bibfnamefont {B.~C.}\ \bibnamefont {Keitch}},\ and\ \bibinfo {author} {\bibfnamefont {J.~P.}\ \bibnamefont {Home}},\ }\bibfield  {title} {\bibinfo {title} {Spin–motion entanglement and state diagnosis with squeezed oscillator wavepackets},\ }\href {https://doi.org/10.1038/nature14458} {\bibfield  {journal} {\bibinfo  {journal} {Nature}\ }\textbf {\bibinfo {volume} {521}},\ \bibinfo {pages} {336} (\bibinfo {year} {2015})}\BibitemShut {NoStop}%
\bibitem [{\citenamefont {Wollman}\ \emph {et~al.}(2015)\citenamefont {Wollman}, \citenamefont {Lei}, \citenamefont {Weinstein}, \citenamefont {Suh}, \citenamefont {Kronwald}, \citenamefont {Marquardt}, \citenamefont {Clerk},\ and\ \citenamefont {Schwab}}]{wollman_quantum_2015}%
  \BibitemOpen
  \bibfield  {author} {\bibinfo {author} {\bibfnamefont {E.~E.}\ \bibnamefont {Wollman}}, \bibinfo {author} {\bibfnamefont {C.~U.}\ \bibnamefont {Lei}}, \bibinfo {author} {\bibfnamefont {A.~J.}\ \bibnamefont {Weinstein}}, \bibinfo {author} {\bibfnamefont {J.}~\bibnamefont {Suh}}, \bibinfo {author} {\bibfnamefont {A.}~\bibnamefont {Kronwald}}, \bibinfo {author} {\bibfnamefont {F.}~\bibnamefont {Marquardt}}, \bibinfo {author} {\bibfnamefont {A.~A.}\ \bibnamefont {Clerk}},\ and\ \bibinfo {author} {\bibfnamefont {K.~C.}\ \bibnamefont {Schwab}},\ }\bibfield  {title} {\bibinfo {title} {Quantum squeezing of motion in a mechanical resonator},\ }\href {https://doi.org/10.1126/science.aac5138} {\bibfield  {journal} {\bibinfo  {journal} {Science}\ }\textbf {\bibinfo {volume} {349}},\ \bibinfo {pages} {952} (\bibinfo {year} {2015})}\BibitemShut {NoStop}%
\bibitem [{\citenamefont {Kienzler}\ \emph {et~al.}(2017)\citenamefont {Kienzler}, \citenamefont {Lo}, \citenamefont {Negnevitsky}, \citenamefont {Fl\"uhmann}, \citenamefont {Marinelli},\ and\ \citenamefont {Home}}]{kienzler_quantum_2017}%
  \BibitemOpen
  \bibfield  {author} {\bibinfo {author} {\bibfnamefont {D.}~\bibnamefont {Kienzler}}, \bibinfo {author} {\bibfnamefont {H.-Y.}\ \bibnamefont {Lo}}, \bibinfo {author} {\bibfnamefont {V.}~\bibnamefont {Negnevitsky}}, \bibinfo {author} {\bibfnamefont {C.}~\bibnamefont {Fl\"uhmann}}, \bibinfo {author} {\bibfnamefont {M.}~\bibnamefont {Marinelli}},\ and\ \bibinfo {author} {\bibfnamefont {J.~P.}\ \bibnamefont {Home}},\ }\bibfield  {title} {\bibinfo {title} {Quantum {Harmonic} {Oscillator} {State} {Control} in a {Squeezed} {Fock} {Basis}},\ }\href {https://doi.org/10.1103/PhysRevLett.119.033602} {\bibfield  {journal} {\bibinfo  {journal} {Phys. Rev. Lett.}\ }\textbf {\bibinfo {volume} {119}},\ \bibinfo {pages} {033602} (\bibinfo {year} {2017})}\BibitemShut {NoStop}%
\bibitem [{\citenamefont {Behrle}\ \emph {et~al.}(2023)\citenamefont {Behrle}, \citenamefont {Nguyen}, \citenamefont {Reiter}, \citenamefont {Baur}, \citenamefont {de~Neeve}, \citenamefont {Stadler}, \citenamefont {Marinelli}, \citenamefont {Lancellotti}, \citenamefont {Yelin},\ and\ \citenamefont {Home}}]{behrle_phonon_2023}%
  \BibitemOpen
  \bibfield  {author} {\bibinfo {author} {\bibfnamefont {T.}~\bibnamefont {Behrle}}, \bibinfo {author} {\bibfnamefont {T.~L.}\ \bibnamefont {Nguyen}}, \bibinfo {author} {\bibfnamefont {F.}~\bibnamefont {Reiter}}, \bibinfo {author} {\bibfnamefont {D.}~\bibnamefont {Baur}}, \bibinfo {author} {\bibfnamefont {B.}~\bibnamefont {de~Neeve}}, \bibinfo {author} {\bibfnamefont {M.}~\bibnamefont {Stadler}}, \bibinfo {author} {\bibfnamefont {M.}~\bibnamefont {Marinelli}}, \bibinfo {author} {\bibfnamefont {F.}~\bibnamefont {Lancellotti}}, \bibinfo {author} {\bibfnamefont {S.~F.}\ \bibnamefont {Yelin}},\ and\ \bibinfo {author} {\bibfnamefont {J.~P.}\ \bibnamefont {Home}},\ }\bibfield  {title} {\bibinfo {title} {Phonon {Laser} in the {Quantum} {Regime}},\ }\href {https://doi.org/10.1103/PhysRevLett.131.043605} {\bibfield  {journal} {\bibinfo  {journal} {Phys. Rev. Lett.}\ }\textbf {\bibinfo {volume} {131}},\ \bibinfo {pages} {043605} (\bibinfo {year} {2023})}\BibitemShut {NoStop}%
\bibitem [{\citenamefont {Ma}\ \emph {et~al.}(2019)\citenamefont {Ma}, \citenamefont {Saxberg}, \citenamefont {Owens}, \citenamefont {Leung}, \citenamefont {Lu}, \citenamefont {Simon},\ and\ \citenamefont {Schuster}}]{ma_dissipatively_2019}%
  \BibitemOpen
  \bibfield  {author} {\bibinfo {author} {\bibfnamefont {R.}~\bibnamefont {Ma}}, \bibinfo {author} {\bibfnamefont {B.}~\bibnamefont {Saxberg}}, \bibinfo {author} {\bibfnamefont {C.}~\bibnamefont {Owens}}, \bibinfo {author} {\bibfnamefont {N.}~\bibnamefont {Leung}}, \bibinfo {author} {\bibfnamefont {Y.}~\bibnamefont {Lu}}, \bibinfo {author} {\bibfnamefont {J.}~\bibnamefont {Simon}},\ and\ \bibinfo {author} {\bibfnamefont {D.~I.}\ \bibnamefont {Schuster}},\ }\bibfield  {title} {\bibinfo {title} {A dissipatively stabilized {Mott} insulator of photons},\ }\href {https://doi.org/10.1038/s41586-019-0897-9} {\bibfield  {journal} {\bibinfo  {journal} {Nature}\ }\textbf {\bibinfo {volume} {566}},\ \bibinfo {pages} {51} (\bibinfo {year} {2019})}\BibitemShut {NoStop}%
\bibitem [{\citenamefont {Grimsmo}\ \emph {et~al.}(2020)\citenamefont {Grimsmo}, \citenamefont {Combes},\ and\ \citenamefont {Baragiola}}]{grimso_rotational_2020}%
  \BibitemOpen
  \bibfield  {author} {\bibinfo {author} {\bibfnamefont {A.~L.}\ \bibnamefont {Grimsmo}}, \bibinfo {author} {\bibfnamefont {J.}~\bibnamefont {Combes}},\ and\ \bibinfo {author} {\bibfnamefont {B.~Q.}\ \bibnamefont {Baragiola}},\ }\bibfield  {title} {\bibinfo {title} {Quantum computing with rotation-symmetric bosonic codes},\ }\href {https://doi.org/10.1103/PhysRevX.10.011058} {\bibfield  {journal} {\bibinfo  {journal} {Phys. Rev. X}\ }\textbf {\bibinfo {volume} {10}},\ \bibinfo {pages} {011058} (\bibinfo {year} {2020})}\BibitemShut {NoStop}%
\bibitem [{\citenamefont {Gottesman}\ \emph {et~al.}(2001)\citenamefont {Gottesman}, \citenamefont {Kitaev},\ and\ \citenamefont {Preskill}}]{gottesman_encoding_2001}%
  \BibitemOpen
  \bibfield  {author} {\bibinfo {author} {\bibfnamefont {D.}~\bibnamefont {Gottesman}}, \bibinfo {author} {\bibfnamefont {A.}~\bibnamefont {Kitaev}},\ and\ \bibinfo {author} {\bibfnamefont {J.}~\bibnamefont {Preskill}},\ }\bibfield  {title} {\bibinfo {title} {Encoding a qubit in an oscillator},\ }\href {https://doi.org/10.1103/PhysRevA.64.012310} {\bibfield  {journal} {\bibinfo  {journal} {Phys. Rev. A}\ }\textbf {\bibinfo {volume} {64}},\ \bibinfo {pages} {012310} (\bibinfo {year} {2001})}\BibitemShut {NoStop}%
\bibitem [{\citenamefont {Cochrane}\ \emph {et~al.}(1999)\citenamefont {Cochrane}, \citenamefont {Milburn},\ and\ \citenamefont {Munro}}]{cochrane_macroscopically_1999}%
  \BibitemOpen
  \bibfield  {author} {\bibinfo {author} {\bibfnamefont {P.~T.}\ \bibnamefont {Cochrane}}, \bibinfo {author} {\bibfnamefont {G.~J.}\ \bibnamefont {Milburn}},\ and\ \bibinfo {author} {\bibfnamefont {W.~J.}\ \bibnamefont {Munro}},\ }\bibfield  {title} {\bibinfo {title} {Macroscopically distinct quantum-superposition states as a bosonic code for amplitude damping},\ }\href {https://doi.org/10.1103/PhysRevA.59.2631} {\bibfield  {journal} {\bibinfo  {journal} {Phys. Rev. A}\ }\textbf {\bibinfo {volume} {59}},\ \bibinfo {pages} {2631} (\bibinfo {year} {1999})}\BibitemShut {NoStop}%
\bibitem [{\citenamefont {Mirrahimi}\ \emph {et~al.}(2014)\citenamefont {Mirrahimi}, \citenamefont {Leghtas}, \citenamefont {Albert}, \citenamefont {Touzard}, \citenamefont {Schoelkopf}, \citenamefont {Jiang},\ and\ \citenamefont {Devoret}}]{mirrahimi_dynamically_2014}%
  \BibitemOpen
  \bibfield  {author} {\bibinfo {author} {\bibfnamefont {M.}~\bibnamefont {Mirrahimi}}, \bibinfo {author} {\bibfnamefont {Z.}~\bibnamefont {Leghtas}}, \bibinfo {author} {\bibfnamefont {V.~V.}\ \bibnamefont {Albert}}, \bibinfo {author} {\bibfnamefont {S.}~\bibnamefont {Touzard}}, \bibinfo {author} {\bibfnamefont {R.~J.}\ \bibnamefont {Schoelkopf}}, \bibinfo {author} {\bibfnamefont {L.}~\bibnamefont {Jiang}},\ and\ \bibinfo {author} {\bibfnamefont {M.~H.}\ \bibnamefont {Devoret}},\ }\bibfield  {title} {\bibinfo {title} {Dynamically protected cat-qubits: a new paradigm for universal quantum computation},\ }\href {https://doi.org/10.1088/1367-2630/16/4/045014} {\bibfield  {journal} {\bibinfo  {journal} {New J. Phys.}\ }\textbf {\bibinfo {volume} {16}},\ \bibinfo {pages} {045014} (\bibinfo {year} {2014})}\BibitemShut {NoStop}%
\bibitem [{\citenamefont {Leghtas}\ \emph {et~al.}(2015)\citenamefont {Leghtas}, \citenamefont {Touzard}, \citenamefont {Pop}, \citenamefont {Kou}, \citenamefont {Vlastakis}, \citenamefont {Petrenko}, \citenamefont {Sliwa}, \citenamefont {Narla}, \citenamefont {Shankar}, \citenamefont {Hatridge}, \citenamefont {Reagor}, \citenamefont {Frunzio}, \citenamefont {Schoelkopf}, \citenamefont {Mirrahimi},\ and\ \citenamefont {Devoret}}]{leghtas_confining_2015}%
  \BibitemOpen
  \bibfield  {author} {\bibinfo {author} {\bibfnamefont {Z.}~\bibnamefont {Leghtas}}, \bibinfo {author} {\bibfnamefont {S.}~\bibnamefont {Touzard}}, \bibinfo {author} {\bibfnamefont {I.~M.}\ \bibnamefont {Pop}}, \bibinfo {author} {\bibfnamefont {A.}~\bibnamefont {Kou}}, \bibinfo {author} {\bibfnamefont {B.}~\bibnamefont {Vlastakis}}, \bibinfo {author} {\bibfnamefont {A.}~\bibnamefont {Petrenko}}, \bibinfo {author} {\bibfnamefont {K.~M.}\ \bibnamefont {Sliwa}}, \bibinfo {author} {\bibfnamefont {A.}~\bibnamefont {Narla}}, \bibinfo {author} {\bibfnamefont {S.}~\bibnamefont {Shankar}}, \bibinfo {author} {\bibfnamefont {M.~J.}\ \bibnamefont {Hatridge}}, \bibinfo {author} {\bibfnamefont {M.}~\bibnamefont {Reagor}}, \bibinfo {author} {\bibfnamefont {L.}~\bibnamefont {Frunzio}}, \bibinfo {author} {\bibfnamefont {R.~J.}\ \bibnamefont {Schoelkopf}}, \bibinfo {author} {\bibfnamefont {M.}~\bibnamefont {Mirrahimi}},\ and\ \bibinfo {author} {\bibfnamefont {M.~H.}\ \bibnamefont {Devoret}},\ }\bibfield  {title} {\bibinfo {title} {Confining the state of light to a quantum manifold by engineered two-photon loss},\ }\href {https://doi.org/10.1126/science.aaa2085} {\bibfield  {journal} {\bibinfo  {journal} {Science}\ }\textbf {\bibinfo {volume} {347}},\ \bibinfo {pages} {853} (\bibinfo {year} {2015})}\BibitemShut {NoStop}%
\bibitem [{\citenamefont {Touzard}\ \emph {et~al.}(2018)\citenamefont {Touzard}, \citenamefont {Grimm}, \citenamefont {Leghtas}, \citenamefont {Mundhada}, \citenamefont {Reinhold}, \citenamefont {Axline}, \citenamefont {Reagor}, \citenamefont {Chou}, \citenamefont {Blumoff}, \citenamefont {Sliwa}, \citenamefont {Shankar}, \citenamefont {Frunzio}, \citenamefont {Schoelkopf}, \citenamefont {Mirrahimi},\ and\ \citenamefont {Devoret}}]{touzard_coherent_2018}%
  \BibitemOpen
  \bibfield  {author} {\bibinfo {author} {\bibfnamefont {S.}~\bibnamefont {Touzard}}, \bibinfo {author} {\bibfnamefont {A.}~\bibnamefont {Grimm}}, \bibinfo {author} {\bibfnamefont {Z.}~\bibnamefont {Leghtas}}, \bibinfo {author} {\bibfnamefont {S.~O.}\ \bibnamefont {Mundhada}}, \bibinfo {author} {\bibfnamefont {P.}~\bibnamefont {Reinhold}}, \bibinfo {author} {\bibfnamefont {C.}~\bibnamefont {Axline}}, \bibinfo {author} {\bibfnamefont {M.}~\bibnamefont {Reagor}}, \bibinfo {author} {\bibfnamefont {K.}~\bibnamefont {Chou}}, \bibinfo {author} {\bibfnamefont {J.}~\bibnamefont {Blumoff}}, \bibinfo {author} {\bibfnamefont {K.~M.}\ \bibnamefont {Sliwa}}, \bibinfo {author} {\bibfnamefont {S.}~\bibnamefont {Shankar}}, \bibinfo {author} {\bibfnamefont {L.}~\bibnamefont {Frunzio}}, \bibinfo {author} {\bibfnamefont {R.~J.}\ \bibnamefont {Schoelkopf}}, \bibinfo {author} {\bibfnamefont {M.}~\bibnamefont {Mirrahimi}},\ and\ \bibinfo {author} {\bibfnamefont {M.~H.}\ \bibnamefont {Devoret}},\ }\bibfield  {title} {\bibinfo {title} {Coherent {Oscillations} inside a {Quantum} {Manifold} {Stabilized} by {Dissipation}},\ }\href {https://doi.org/10.1103/PhysRevX.8.021005} {\bibfield  {journal} {\bibinfo  {journal} {Phys. Rev. X}\ }\textbf {\bibinfo {volume} {8}},\ \bibinfo {pages} {021005} (\bibinfo {year} {2018})}\BibitemShut {NoStop}%
\bibitem [{\citenamefont {Grimm}\ \emph {et~al.}(2020)\citenamefont {Grimm}, \citenamefont {Frattini}, \citenamefont {Puri}, \citenamefont {Mundhada}, \citenamefont {Touzard}, \citenamefont {Mirrahimi}, \citenamefont {Girvin}, \citenamefont {Shankar},\ and\ \citenamefont {Devoret}}]{grimm_stabilization_2020}%
  \BibitemOpen
  \bibfield  {author} {\bibinfo {author} {\bibfnamefont {A.}~\bibnamefont {Grimm}}, \bibinfo {author} {\bibfnamefont {N.~E.}\ \bibnamefont {Frattini}}, \bibinfo {author} {\bibfnamefont {S.}~\bibnamefont {Puri}}, \bibinfo {author} {\bibfnamefont {S.~O.}\ \bibnamefont {Mundhada}}, \bibinfo {author} {\bibfnamefont {S.}~\bibnamefont {Touzard}}, \bibinfo {author} {\bibfnamefont {M.}~\bibnamefont {Mirrahimi}}, \bibinfo {author} {\bibfnamefont {S.~M.}\ \bibnamefont {Girvin}}, \bibinfo {author} {\bibfnamefont {S.}~\bibnamefont {Shankar}},\ and\ \bibinfo {author} {\bibfnamefont {M.~H.}\ \bibnamefont {Devoret}},\ }\bibfield  {title} {\bibinfo {title} {Stabilization and operation of a {Kerr}-cat qubit},\ }\href {https://doi.org/10.1038/s41586-020-2587-z} {\bibfield  {journal} {\bibinfo  {journal} {Nature}\ }\textbf {\bibinfo {volume} {584}},\ \bibinfo {pages} {205} (\bibinfo {year} {2020})}\BibitemShut {NoStop}%
\bibitem [{\citenamefont {Campagne-Ibarcq}\ \emph {et~al.}(2020)\citenamefont {Campagne-Ibarcq}, \citenamefont {Eickbusch}, \citenamefont {Touzard}, \citenamefont {Zalys-Geller}, \citenamefont {Frattini}, \citenamefont {Sivak}, \citenamefont {Reinhold}, \citenamefont {Puri}, \citenamefont {Shankar}, \citenamefont {Schoelkopf}, \citenamefont {Frunzio}, \citenamefont {Mirrahimi},\ and\ \citenamefont {Devoret}}]{campagne-ibarcq_quantum_2020}%
  \BibitemOpen
  \bibfield  {author} {\bibinfo {author} {\bibfnamefont {P.}~\bibnamefont {Campagne-Ibarcq}}, \bibinfo {author} {\bibfnamefont {A.}~\bibnamefont {Eickbusch}}, \bibinfo {author} {\bibfnamefont {S.}~\bibnamefont {Touzard}}, \bibinfo {author} {\bibfnamefont {E.}~\bibnamefont {Zalys-Geller}}, \bibinfo {author} {\bibfnamefont {N.~E.}\ \bibnamefont {Frattini}}, \bibinfo {author} {\bibfnamefont {V.~V.}\ \bibnamefont {Sivak}}, \bibinfo {author} {\bibfnamefont {P.}~\bibnamefont {Reinhold}}, \bibinfo {author} {\bibfnamefont {S.}~\bibnamefont {Puri}}, \bibinfo {author} {\bibfnamefont {S.}~\bibnamefont {Shankar}}, \bibinfo {author} {\bibfnamefont {R.~J.}\ \bibnamefont {Schoelkopf}}, \bibinfo {author} {\bibfnamefont {L.}~\bibnamefont {Frunzio}}, \bibinfo {author} {\bibfnamefont {M.}~\bibnamefont {Mirrahimi}},\ and\ \bibinfo {author} {\bibfnamefont {M.~H.}\ \bibnamefont {Devoret}},\ }\bibfield  {title} {\bibinfo {title} {Quantum error correction of a qubit encoded in grid states of an oscillator},\ }\href {https://doi.org/10.1038/s41586-020-2603-3} {\bibfield  {journal} {\bibinfo  {journal} {Nature}\ }\textbf {\bibinfo {volume} {584}},\ \bibinfo {pages} {368} (\bibinfo {year} {2020})}\BibitemShut {NoStop}%
\bibitem [{\citenamefont {Lescanne}\ \emph {et~al.}(2020)\citenamefont {Lescanne}, \citenamefont {Villiers}, \citenamefont {Peronnin}, \citenamefont {Sarlette}, \citenamefont {Delbecq}, \citenamefont {Huard}, \citenamefont {Kontos}, \citenamefont {Mirrahimi},\ and\ \citenamefont {Leghtas}}]{lescanne_exponential_2020}%
  \BibitemOpen
  \bibfield  {author} {\bibinfo {author} {\bibfnamefont {R.}~\bibnamefont {Lescanne}}, \bibinfo {author} {\bibfnamefont {M.}~\bibnamefont {Villiers}}, \bibinfo {author} {\bibfnamefont {T.}~\bibnamefont {Peronnin}}, \bibinfo {author} {\bibfnamefont {A.}~\bibnamefont {Sarlette}}, \bibinfo {author} {\bibfnamefont {M.}~\bibnamefont {Delbecq}}, \bibinfo {author} {\bibfnamefont {B.}~\bibnamefont {Huard}}, \bibinfo {author} {\bibfnamefont {T.}~\bibnamefont {Kontos}}, \bibinfo {author} {\bibfnamefont {M.}~\bibnamefont {Mirrahimi}},\ and\ \bibinfo {author} {\bibfnamefont {Z.}~\bibnamefont {Leghtas}},\ }\bibfield  {title} {\bibinfo {title} {Exponential suppression of bit-flips in a qubit encoded in an oscillator},\ }\href {https://doi.org/10.1038/s41567-020-0824-x} {\bibfield  {journal} {\bibinfo  {journal} {Nat. Phys.}\ }\textbf {\bibinfo {volume} {16}},\ \bibinfo {pages} {509} (\bibinfo {year} {2020})}\BibitemShut {NoStop}%
\bibitem [{\citenamefont {Gertler}\ \emph {et~al.}(2021)\citenamefont {Gertler}, \citenamefont {Baker}, \citenamefont {Li}, \citenamefont {Shirol}, \citenamefont {Koch},\ and\ \citenamefont {Wang}}]{gertler_protecting_2021}%
  \BibitemOpen
  \bibfield  {author} {\bibinfo {author} {\bibfnamefont {J.~M.}\ \bibnamefont {Gertler}}, \bibinfo {author} {\bibfnamefont {B.}~\bibnamefont {Baker}}, \bibinfo {author} {\bibfnamefont {J.}~\bibnamefont {Li}}, \bibinfo {author} {\bibfnamefont {S.}~\bibnamefont {Shirol}}, \bibinfo {author} {\bibfnamefont {J.}~\bibnamefont {Koch}},\ and\ \bibinfo {author} {\bibfnamefont {C.}~\bibnamefont {Wang}},\ }\bibfield  {title} {\bibinfo {title} {Protecting a bosonic qubit with autonomous quantum error correction},\ }\href {https://doi.org/10.1038/s41586-021-03257-0} {\bibfield  {journal} {\bibinfo  {journal} {Nature}\ }\textbf {\bibinfo {volume} {590}},\ \bibinfo {pages} {243} (\bibinfo {year} {2021})}\BibitemShut {NoStop}%
\bibitem [{\citenamefont {Kwon}\ \emph {et~al.}(2022)\citenamefont {Kwon}, \citenamefont {Watabe},\ and\ \citenamefont {Tsai}}]{kwon_autonomous_2022}%
  \BibitemOpen
  \bibfield  {author} {\bibinfo {author} {\bibfnamefont {S.}~\bibnamefont {Kwon}}, \bibinfo {author} {\bibfnamefont {S.}~\bibnamefont {Watabe}},\ and\ \bibinfo {author} {\bibfnamefont {J.-S.}\ \bibnamefont {Tsai}},\ }\bibfield  {title} {\bibinfo {title} {Autonomous quantum error correction in a four-photon {Kerr} parametric oscillator},\ }\href {https://doi.org/10.1038/s41534-022-00553-z} {\bibfield  {journal} {\bibinfo  {journal} {npj Quantum Inf.}\ }\textbf {\bibinfo {volume} {8}},\ \bibinfo {pages} {1} (\bibinfo {year} {2022})}\BibitemShut {NoStop}%
\bibitem [{\citenamefont {Sivak}\ \emph {et~al.}(2023)\citenamefont {Sivak}, \citenamefont {Eickbusch}, \citenamefont {Royer}, \citenamefont {Singh}, \citenamefont {Tsioutsios}, \citenamefont {Ganjam}, \citenamefont {Miano}, \citenamefont {Brock}, \citenamefont {Ding}, \citenamefont {Frunzio}, \citenamefont {Girvin}, \citenamefont {Schoelkopf},\ and\ \citenamefont {Devoret}}]{sivak_real-time_2023}%
  \BibitemOpen
  \bibfield  {author} {\bibinfo {author} {\bibfnamefont {V.~V.}\ \bibnamefont {Sivak}}, \bibinfo {author} {\bibfnamefont {A.}~\bibnamefont {Eickbusch}}, \bibinfo {author} {\bibfnamefont {B.}~\bibnamefont {Royer}}, \bibinfo {author} {\bibfnamefont {S.}~\bibnamefont {Singh}}, \bibinfo {author} {\bibfnamefont {I.}~\bibnamefont {Tsioutsios}}, \bibinfo {author} {\bibfnamefont {S.}~\bibnamefont {Ganjam}}, \bibinfo {author} {\bibfnamefont {A.}~\bibnamefont {Miano}}, \bibinfo {author} {\bibfnamefont {B.~L.}\ \bibnamefont {Brock}}, \bibinfo {author} {\bibfnamefont {A.~Z.}\ \bibnamefont {Ding}}, \bibinfo {author} {\bibfnamefont {L.}~\bibnamefont {Frunzio}}, \bibinfo {author} {\bibfnamefont {S.~M.}\ \bibnamefont {Girvin}}, \bibinfo {author} {\bibfnamefont {R.~J.}\ \bibnamefont {Schoelkopf}},\ and\ \bibinfo {author} {\bibfnamefont {M.~H.}\ \bibnamefont {Devoret}},\ }\bibfield  {title} {\bibinfo {title} {Real-time quantum error correction beyond break-even},\ }\href {https://doi.org/10.1038/s41586-023-05782-6} {\bibfield  {journal} {\bibinfo  {journal} {Nature}\ }\textbf {\bibinfo {volume} {616}},\ \bibinfo {pages} {50} (\bibinfo {year} {2023})}\BibitemShut {NoStop}%
\bibitem [{\citenamefont {Ni}\ \emph {et~al.}(2023)\citenamefont {Ni}, \citenamefont {Li}, \citenamefont {Deng}, \citenamefont {Cai}, \citenamefont {Zhang}, \citenamefont {Wang}, \citenamefont {Yang}, \citenamefont {Yu}, \citenamefont {Yan}, \citenamefont {Liu}, \citenamefont {Zou}, \citenamefont {Sun}, \citenamefont {Zheng}, \citenamefont {Xu},\ and\ \citenamefont {Yu}}]{ni_beating_2023}%
  \BibitemOpen
  \bibfield  {author} {\bibinfo {author} {\bibfnamefont {Z.}~\bibnamefont {Ni}}, \bibinfo {author} {\bibfnamefont {S.}~\bibnamefont {Li}}, \bibinfo {author} {\bibfnamefont {X.}~\bibnamefont {Deng}}, \bibinfo {author} {\bibfnamefont {Y.}~\bibnamefont {Cai}}, \bibinfo {author} {\bibfnamefont {L.}~\bibnamefont {Zhang}}, \bibinfo {author} {\bibfnamefont {W.}~\bibnamefont {Wang}}, \bibinfo {author} {\bibfnamefont {Z.-B.}\ \bibnamefont {Yang}}, \bibinfo {author} {\bibfnamefont {H.}~\bibnamefont {Yu}}, \bibinfo {author} {\bibfnamefont {F.}~\bibnamefont {Yan}}, \bibinfo {author} {\bibfnamefont {S.}~\bibnamefont {Liu}}, \bibinfo {author} {\bibfnamefont {C.-L.}\ \bibnamefont {Zou}}, \bibinfo {author} {\bibfnamefont {L.}~\bibnamefont {Sun}}, \bibinfo {author} {\bibfnamefont {S.-B.}\ \bibnamefont {Zheng}}, \bibinfo {author} {\bibfnamefont {Y.}~\bibnamefont {Xu}},\ and\ \bibinfo {author} {\bibfnamefont {D.}~\bibnamefont {Yu}},\ }\bibfield  {title} {\bibinfo {title} {Beating the break-even point with a discrete-variable-encoded logical qubit},\ }\href {https://doi.org/10.1038/s41586-023-05784-4} {\bibfield  {journal} {\bibinfo  {journal} {Nature}\ }\textbf {\bibinfo {volume} {616}},\ \bibinfo {pages} {56} (\bibinfo {year} {2023})}\BibitemShut {NoStop}%
\bibitem [{\citenamefont {Berdou}\ \emph {et~al.}(2023)\citenamefont {Berdou}, \citenamefont {Murani}, \citenamefont {R\'eglade}, \citenamefont {Smith}, \citenamefont {Villiers}, \citenamefont {Palomo}, \citenamefont {Rosticher}, \citenamefont {Denis}, \citenamefont {Morfin}, \citenamefont {Delbecq}, \citenamefont {Kontos}, \citenamefont {Pankratova}, \citenamefont {Rautschke}, \citenamefont {Peronnin}, \citenamefont {Sellem}, \citenamefont {Rouchon}, \citenamefont {Sarlette}, \citenamefont {Mirrahimi}, \citenamefont {Campagne-Ibarcq}, \citenamefont {Jezouin}, \citenamefont {Lescanne},\ and\ \citenamefont {Leghtas}}]{berdou_one_2023}%
  \BibitemOpen
  \bibfield  {author} {\bibinfo {author} {\bibfnamefont {C.}~\bibnamefont {Berdou}}, \bibinfo {author} {\bibfnamefont {A.}~\bibnamefont {Murani}}, \bibinfo {author} {\bibfnamefont {U.}~\bibnamefont {R\'eglade}}, \bibinfo {author} {\bibfnamefont {W.~C.}\ \bibnamefont {Smith}}, \bibinfo {author} {\bibfnamefont {M.}~\bibnamefont {Villiers}}, \bibinfo {author} {\bibfnamefont {J.}~\bibnamefont {Palomo}}, \bibinfo {author} {\bibfnamefont {M.}~\bibnamefont {Rosticher}}, \bibinfo {author} {\bibfnamefont {A.}~\bibnamefont {Denis}}, \bibinfo {author} {\bibfnamefont {P.}~\bibnamefont {Morfin}}, \bibinfo {author} {\bibfnamefont {M.}~\bibnamefont {Delbecq}}, \bibinfo {author} {\bibfnamefont {T.}~\bibnamefont {Kontos}}, \bibinfo {author} {\bibfnamefont {N.}~\bibnamefont {Pankratova}}, \bibinfo {author} {\bibfnamefont {F.}~\bibnamefont {Rautschke}}, \bibinfo {author} {\bibfnamefont {T.}~\bibnamefont {Peronnin}}, \bibinfo {author} {\bibfnamefont {L.-A.}\ \bibnamefont {Sellem}}, \bibinfo {author} {\bibfnamefont {P.}~\bibnamefont {Rouchon}}, \bibinfo {author} {\bibfnamefont {A.}~\bibnamefont {Sarlette}}, \bibinfo {author} {\bibfnamefont {M.}~\bibnamefont {Mirrahimi}}, \bibinfo {author} {\bibfnamefont {P.}~\bibnamefont {Campagne-Ibarcq}}, \bibinfo {author} {\bibfnamefont {S.}~\bibnamefont {Jezouin}}, \bibinfo {author} {\bibfnamefont {R.}~\bibnamefont {Lescanne}},\ and\ \bibinfo {author} {\bibfnamefont {Z.}~\bibnamefont {Leghtas}},\ }\bibfield  {title} {\bibinfo {title} {One {Hundred} {Second} {Bit}-{Flip} {Time} in a {Two}-{Photon} {Dissipative} {Oscillator}},\ }\href {https://doi.org/10.1103/PRXQuantum.4.020350} {\bibfield  {journal} {\bibinfo  {journal} {PRX Quantum}\ }\textbf {\bibinfo {volume} {4}},\ \bibinfo {pages} {020350} (\bibinfo {year} {2023})}\BibitemShut {NoStop}%
\bibitem [{\citenamefont {R{\'e}glade}\ \emph {et~al.}(2024)\citenamefont {R{\'e}glade}, \citenamefont {Bocquet}, \citenamefont {Gautier}, \citenamefont {Cohen}, \citenamefont {Marquet}, \citenamefont {Albertinale}, \citenamefont {Pankratova}, \citenamefont {Hall{\'e}n}, \citenamefont {Rautschke}, \citenamefont {Sellem}, \citenamefont {Rouchon}, \citenamefont {Sarlette}, \citenamefont {Mirrahimi}, \citenamefont {Campagne-Ibarcq}, \citenamefont {Lescanne}, \citenamefont {Jezouin},\ and\ \citenamefont {Leghtas}}]{reglade_quantum_2024}%
  \BibitemOpen
  \bibfield  {author} {\bibinfo {author} {\bibfnamefont {U.}~\bibnamefont {R{\'e}glade}}, \bibinfo {author} {\bibfnamefont {A.}~\bibnamefont {Bocquet}}, \bibinfo {author} {\bibfnamefont {R.}~\bibnamefont {Gautier}}, \bibinfo {author} {\bibfnamefont {J.}~\bibnamefont {Cohen}}, \bibinfo {author} {\bibfnamefont {A.}~\bibnamefont {Marquet}}, \bibinfo {author} {\bibfnamefont {E.}~\bibnamefont {Albertinale}}, \bibinfo {author} {\bibfnamefont {N.}~\bibnamefont {Pankratova}}, \bibinfo {author} {\bibfnamefont {M.}~\bibnamefont {Hall{\'e}n}}, \bibinfo {author} {\bibfnamefont {F.}~\bibnamefont {Rautschke}}, \bibinfo {author} {\bibfnamefont {L.-A.}\ \bibnamefont {Sellem}}, \bibinfo {author} {\bibfnamefont {P.}~\bibnamefont {Rouchon}}, \bibinfo {author} {\bibfnamefont {A.}~\bibnamefont {Sarlette}}, \bibinfo {author} {\bibfnamefont {M.}~\bibnamefont {Mirrahimi}}, \bibinfo {author} {\bibfnamefont {P.}~\bibnamefont {Campagne-Ibarcq}}, \bibinfo {author} {\bibfnamefont {R.}~\bibnamefont {Lescanne}}, \bibinfo {author} {\bibfnamefont {S.}~\bibnamefont {Jezouin}},\ and\ \bibinfo {author} {\bibfnamefont {Z.}~\bibnamefont {Leghtas}},\ }\bibfield  {title} {\bibinfo {title} {Quantum control of a cat qubit with bit-flip times exceeding ten seconds},\ }\href {https://doi.org/10.1038/s41586-024-07294-3} {\bibfield  {journal} {\bibinfo  {journal} {Nature}\ }\textbf {\bibinfo {volume} {629}},\ \bibinfo {pages} {778} (\bibinfo {year} {2024})}\BibitemShut {NoStop}%
\bibitem [{\citenamefont {Fl\"uhmann}\ \emph {et~al.}(2019)\citenamefont {Fl\"uhmann}, \citenamefont {Nguyen}, \citenamefont {Marinelli}, \citenamefont {Negnevitsky}, \citenamefont {Mehta},\ and\ \citenamefont {Home}}]{fluhmann_encoding_2019}%
  \BibitemOpen
  \bibfield  {author} {\bibinfo {author} {\bibfnamefont {C.}~\bibnamefont {Fl\"uhmann}}, \bibinfo {author} {\bibfnamefont {T.~L.}\ \bibnamefont {Nguyen}}, \bibinfo {author} {\bibfnamefont {M.}~\bibnamefont {Marinelli}}, \bibinfo {author} {\bibfnamefont {V.}~\bibnamefont {Negnevitsky}}, \bibinfo {author} {\bibfnamefont {K.}~\bibnamefont {Mehta}},\ and\ \bibinfo {author} {\bibfnamefont {J.~P.}\ \bibnamefont {Home}},\ }\bibfield  {title} {\bibinfo {title} {Encoding a qubit in a trapped-ion mechanical oscillator},\ }\href {https://doi.org/10.1038/s41586-019-0960-6} {\bibfield  {journal} {\bibinfo  {journal} {Nature}\ }\textbf {\bibinfo {volume} {566}},\ \bibinfo {pages} {513} (\bibinfo {year} {2019})}\BibitemShut {NoStop}%
\bibitem [{\citenamefont {Hach~III}\ and\ \citenamefont {Gerry}(1994)}]{hach_iii_generation_1994}%
  \BibitemOpen
  \bibfield  {author} {\bibinfo {author} {\bibfnamefont {E.~E.}\ \bibnamefont {Hach~III}}\ and\ \bibinfo {author} {\bibfnamefont {C.~C.}\ \bibnamefont {Gerry}},\ }\bibfield  {title} {\bibinfo {title} {Generation of mixtures of {Schr}{\textbackslash}"odinger-cat states from a competitive two-photon process},\ }\href {https://doi.org/10.1103/PhysRevA.49.490} {\bibfield  {journal} {\bibinfo  {journal} {Phys. Rev. A}\ }\textbf {\bibinfo {volume} {49}},\ \bibinfo {pages} {490} (\bibinfo {year} {1994})}\BibitemShut {NoStop}%
\bibitem [{\citenamefont {Gilles}\ \emph {et~al.}(1994)\citenamefont {Gilles}, \citenamefont {Garraway},\ and\ \citenamefont {Knight}}]{gilles_generation_1994}%
  \BibitemOpen
  \bibfield  {author} {\bibinfo {author} {\bibfnamefont {L.}~\bibnamefont {Gilles}}, \bibinfo {author} {\bibfnamefont {B.~M.}\ \bibnamefont {Garraway}},\ and\ \bibinfo {author} {\bibfnamefont {P.~L.}\ \bibnamefont {Knight}},\ }\bibfield  {title} {\bibinfo {title} {Generation of nonclassical light by dissipative two-photon processes},\ }\href {https://doi.org/10.1103/PhysRevA.49.2785} {\bibfield  {journal} {\bibinfo  {journal} {Phys. Rev. A}\ }\textbf {\bibinfo {volume} {49}},\ \bibinfo {pages} {2785} (\bibinfo {year} {1994})}\BibitemShut {NoStop}%
\bibitem [{\citenamefont {de~Matos~Filho}\ and\ \citenamefont {Vogel}(1996{\natexlab{a}})}]{de_matos_filho_even_1996}%
  \BibitemOpen
  \bibfield  {author} {\bibinfo {author} {\bibfnamefont {R.~L.}\ \bibnamefont {de~Matos~Filho}}\ and\ \bibinfo {author} {\bibfnamefont {W.}~\bibnamefont {Vogel}},\ }\bibfield  {title} {\bibinfo {title} {Even and {Odd} {Coherent} {States} of the {Motion} of a {Trapped} {Ion}},\ }\href {https://doi.org/10.1103/PhysRevLett.76.608} {\bibfield  {journal} {\bibinfo  {journal} {Phys. Rev. Lett.}\ }\textbf {\bibinfo {volume} {76}},\ \bibinfo {pages} {608} (\bibinfo {year} {1996}{\natexlab{a}})}\BibitemShut {NoStop}%
\bibitem [{\citenamefont {Sarlette}\ \emph {et~al.}(2011)\citenamefont {Sarlette}, \citenamefont {Raimond}, \citenamefont {Brune},\ and\ \citenamefont {Rouchon}}]{sarlette_stabilization_2011}%
  \BibitemOpen
  \bibfield  {author} {\bibinfo {author} {\bibfnamefont {A.}~\bibnamefont {Sarlette}}, \bibinfo {author} {\bibfnamefont {J.~M.}\ \bibnamefont {Raimond}}, \bibinfo {author} {\bibfnamefont {M.}~\bibnamefont {Brune}},\ and\ \bibinfo {author} {\bibfnamefont {P.}~\bibnamefont {Rouchon}},\ }\bibfield  {title} {\bibinfo {title} {Stabilization of {Nonclassical} {States} of the {Radiation} {Field} in a {Cavity} by {Reservoir} {Engineering}},\ }\href {https://doi.org/10.1103/PhysRevLett.107.010402} {\bibfield  {journal} {\bibinfo  {journal} {Phys. Rev. Lett.}\ }\textbf {\bibinfo {volume} {107}},\ \bibinfo {pages} {010402} (\bibinfo {year} {2011})}\BibitemShut {NoStop}%
\bibitem [{\citenamefont {Mundhada}\ \emph {et~al.}(2017)\citenamefont {Mundhada}, \citenamefont {Grimm}, \citenamefont {Touzard}, \citenamefont {Vool}, \citenamefont {Shankar}, \citenamefont {Devoret},\ and\ \citenamefont {Mirrahimi}}]{mundhada_generating_2017}%
  \BibitemOpen
  \bibfield  {author} {\bibinfo {author} {\bibfnamefont {S.~O.}\ \bibnamefont {Mundhada}}, \bibinfo {author} {\bibfnamefont {A.}~\bibnamefont {Grimm}}, \bibinfo {author} {\bibfnamefont {S.}~\bibnamefont {Touzard}}, \bibinfo {author} {\bibfnamefont {U.}~\bibnamefont {Vool}}, \bibinfo {author} {\bibfnamefont {S.}~\bibnamefont {Shankar}}, \bibinfo {author} {\bibfnamefont {M.~H.}\ \bibnamefont {Devoret}},\ and\ \bibinfo {author} {\bibfnamefont {M.}~\bibnamefont {Mirrahimi}},\ }\bibfield  {title} {\bibinfo {title} {Generating higher-order quantum dissipation from lower-order parametric processes},\ }\href {https://doi.org/10.1088/2058-9565/aa6e9d} {\bibfield  {journal} {\bibinfo  {journal} {Quantum Sci. Technol.}\ }\textbf {\bibinfo {volume} {2}},\ \bibinfo {pages} {024005} (\bibinfo {year} {2017})}\BibitemShut {NoStop}%
\bibitem [{\citenamefont {Mamaev}\ \emph {et~al.}(2018)\citenamefont {Mamaev}, \citenamefont {Govia},\ and\ \citenamefont {Clerk}}]{mamaev_dissipative_2018}%
  \BibitemOpen
  \bibfield  {author} {\bibinfo {author} {\bibfnamefont {M.}~\bibnamefont {Mamaev}}, \bibinfo {author} {\bibfnamefont {L.~C.~G.}\ \bibnamefont {Govia}},\ and\ \bibinfo {author} {\bibfnamefont {A.~A.}\ \bibnamefont {Clerk}},\ }\bibfield  {title} {\bibinfo {title} {Dissipative stabilization of entangled cat states using a driven {Bose}-{Hubbard} dimer},\ }\href {https://doi.org/10.22331/q-2018-03-27-58} {\bibfield  {journal} {\bibinfo  {journal} {Quantum}\ }\textbf {\bibinfo {volume} {2}},\ \bibinfo {pages} {58} (\bibinfo {year} {2018})}\BibitemShut {NoStop}%
\bibitem [{\citenamefont {Putterman}\ \emph {et~al.}(2022)\citenamefont {Putterman}, \citenamefont {Iverson}, \citenamefont {Xu}, \citenamefont {Jiang}, \citenamefont {Painter}, \citenamefont {Brand\~ao},\ and\ \citenamefont {Noh}}]{putterman_stabilizing_2022}%
  \BibitemOpen
  \bibfield  {author} {\bibinfo {author} {\bibfnamefont {H.}~\bibnamefont {Putterman}}, \bibinfo {author} {\bibfnamefont {J.}~\bibnamefont {Iverson}}, \bibinfo {author} {\bibfnamefont {Q.}~\bibnamefont {Xu}}, \bibinfo {author} {\bibfnamefont {L.}~\bibnamefont {Jiang}}, \bibinfo {author} {\bibfnamefont {O.}~\bibnamefont {Painter}}, \bibinfo {author} {\bibfnamefont {F.~G. S.~L.}\ \bibnamefont {Brand\~ao}},\ and\ \bibinfo {author} {\bibfnamefont {K.}~\bibnamefont {Noh}},\ }\bibfield  {title} {\bibinfo {title} {Stabilizing a {Bosonic} {Qubit} {Using} {Colored} {Dissipation}},\ }\href {https://doi.org/10.1103/PhysRevLett.128.110502} {\bibfield  {journal} {\bibinfo  {journal} {Phys. Rev. Lett.}\ }\textbf {\bibinfo {volume} {128}},\ \bibinfo {pages} {110502} (\bibinfo {year} {2022})}\BibitemShut {NoStop}%
\bibitem [{\citenamefont {Albert}\ and\ \citenamefont {Jiang}(2014)}]{albert_symmetries_2014}%
  \BibitemOpen
  \bibfield  {author} {\bibinfo {author} {\bibfnamefont {V.~V.}\ \bibnamefont {Albert}}\ and\ \bibinfo {author} {\bibfnamefont {L.}~\bibnamefont {Jiang}},\ }\bibfield  {title} {\bibinfo {title} {Symmetries and conserved quantities in {Lindblad} master equations},\ }\href {https://doi.org/10.1103/PhysRevA.89.022118} {\bibfield  {journal} {\bibinfo  {journal} {Phys. Rev. A}\ }\textbf {\bibinfo {volume} {89}},\ \bibinfo {pages} {022118} (\bibinfo {year} {2014})}\BibitemShut {NoStop}%
\bibitem [{\citenamefont {Albert}\ \emph {et~al.}(2019)\citenamefont {Albert}, \citenamefont {Mundhada}, \citenamefont {Grimm}, \citenamefont {Touzard}, \citenamefont {Devoret},\ and\ \citenamefont {Jiang}}]{albert_pair_cat_2019}%
  \BibitemOpen
  \bibfield  {author} {\bibinfo {author} {\bibfnamefont {V.~V.}\ \bibnamefont {Albert}}, \bibinfo {author} {\bibfnamefont {S.~O.}\ \bibnamefont {Mundhada}}, \bibinfo {author} {\bibfnamefont {A.}~\bibnamefont {Grimm}}, \bibinfo {author} {\bibfnamefont {S.}~\bibnamefont {Touzard}}, \bibinfo {author} {\bibfnamefont {M.~H.}\ \bibnamefont {Devoret}},\ and\ \bibinfo {author} {\bibfnamefont {L.}~\bibnamefont {Jiang}},\ }\bibfield  {title} {\bibinfo {title} {Pair-cat codes: autonomous error-correction with low-order nonlinearity},\ }\href {https://doi.org/10.1088/2058-9565/ab1e69} {\bibfield  {journal} {\bibinfo  {journal} {Quantum Sci. Technol.}\ }\textbf {\bibinfo {volume} {4}},\ \bibinfo {pages} {035007} (\bibinfo {year} {2019})}\BibitemShut {NoStop}%
\bibitem [{\citenamefont {Goto}(2016)}]{goto2016universal}%
  \BibitemOpen
  \bibfield  {author} {\bibinfo {author} {\bibfnamefont {H.}~\bibnamefont {Goto}},\ }\bibfield  {title} {\bibinfo {title} {Universal quantum computation with a nonlinear oscillator network},\ }\href {https://doi.org/10.1103/PhysRevA.93.050301} {\bibfield  {journal} {\bibinfo  {journal} {Phys. Rev. A}\ }\textbf {\bibinfo {volume} {93}},\ \bibinfo {pages} {050301(R)} (\bibinfo {year} {2016})}\BibitemShut {NoStop}%
\bibitem [{\citenamefont {Albert}\ \emph {et~al.}(2016)\citenamefont {Albert}, \citenamefont {Shu}, \citenamefont {Krastanov}, \citenamefont {Shen}, \citenamefont {Liu}, \citenamefont {Yang}, \citenamefont {Schoelkopf}, \citenamefont {Mirrahimi}, \citenamefont {Devoret},\ and\ \citenamefont {Jiang}}]{albert_holonomic_2016}%
  \BibitemOpen
  \bibfield  {author} {\bibinfo {author} {\bibfnamefont {V.~V.}\ \bibnamefont {Albert}}, \bibinfo {author} {\bibfnamefont {C.}~\bibnamefont {Shu}}, \bibinfo {author} {\bibfnamefont {S.}~\bibnamefont {Krastanov}}, \bibinfo {author} {\bibfnamefont {C.}~\bibnamefont {Shen}}, \bibinfo {author} {\bibfnamefont {R.-B.}\ \bibnamefont {Liu}}, \bibinfo {author} {\bibfnamefont {Z.-B.}\ \bibnamefont {Yang}}, \bibinfo {author} {\bibfnamefont {R.~J.}\ \bibnamefont {Schoelkopf}}, \bibinfo {author} {\bibfnamefont {M.}~\bibnamefont {Mirrahimi}}, \bibinfo {author} {\bibfnamefont {M.~H.}\ \bibnamefont {Devoret}},\ and\ \bibinfo {author} {\bibfnamefont {L.}~\bibnamefont {Jiang}},\ }\bibfield  {title} {\bibinfo {title} {Holonomic {Quantum} {Control} with {Continuous} {Variable} {Systems}},\ }\href {https://doi.org/10.1103/PhysRevLett.116.140502} {\bibfield  {journal} {\bibinfo  {journal} {Phys. Rev. Lett.}\ }\textbf {\bibinfo {volume} {116}},\ \bibinfo {pages} {140502} (\bibinfo {year} {2016})}\BibitemShut {NoStop}%
\bibitem [{\citenamefont {Puri}\ \emph {et~al.}(2020)\citenamefont {Puri}, \citenamefont {St-Jean}, \citenamefont {Gross}, \citenamefont {Grimm}, \citenamefont {Frattini}, \citenamefont {Iyer}, \citenamefont {Krishna}, \citenamefont {Touzard}, \citenamefont {Jiang}, \citenamefont {Blais}, \citenamefont {Flammia},\ and\ \citenamefont {Girvin}}]{puri_bias_preserving_2020}%
  \BibitemOpen
  \bibfield  {author} {\bibinfo {author} {\bibfnamefont {S.}~\bibnamefont {Puri}}, \bibinfo {author} {\bibfnamefont {L.}~\bibnamefont {St-Jean}}, \bibinfo {author} {\bibfnamefont {J.~A.}\ \bibnamefont {Gross}}, \bibinfo {author} {\bibfnamefont {A.}~\bibnamefont {Grimm}}, \bibinfo {author} {\bibfnamefont {N.~E.}\ \bibnamefont {Frattini}}, \bibinfo {author} {\bibfnamefont {P.~S.}\ \bibnamefont {Iyer}}, \bibinfo {author} {\bibfnamefont {A.}~\bibnamefont {Krishna}}, \bibinfo {author} {\bibfnamefont {S.}~\bibnamefont {Touzard}}, \bibinfo {author} {\bibfnamefont {L.}~\bibnamefont {Jiang}}, \bibinfo {author} {\bibfnamefont {A.}~\bibnamefont {Blais}}, \bibinfo {author} {\bibfnamefont {S.~T.}\ \bibnamefont {Flammia}},\ and\ \bibinfo {author} {\bibfnamefont {S.~M.}\ \bibnamefont {Girvin}},\ }\bibfield  {title} {\bibinfo {title} {Bias-preserving gates with stabilized cat qubits},\ }\href {https://doi.org/10.1126/sciadv.aay5901} {\bibfield  {journal} {\bibinfo  {journal} {Sci. Adv.}\ }\textbf {\bibinfo {volume} {6}},\ \bibinfo {pages} {eaay5901} (\bibinfo {year} {2020})}\BibitemShut {NoStop}%
\bibitem [{\citenamefont {Gautier}\ \emph {et~al.}(2023)\citenamefont {Gautier}, \citenamefont {Mirrahimi},\ and\ \citenamefont {Sarlette}}]{gautier_designing_2023}%
  \BibitemOpen
  \bibfield  {author} {\bibinfo {author} {\bibfnamefont {R.}~\bibnamefont {Gautier}}, \bibinfo {author} {\bibfnamefont {M.}~\bibnamefont {Mirrahimi}},\ and\ \bibinfo {author} {\bibfnamefont {A.}~\bibnamefont {Sarlette}},\ }\bibfield  {title} {\bibinfo {title} {Designing {High}-{Fidelity} {Zeno} {Gates} for {Dissipative} {Cat} {Qubits}},\ }\href {https://doi.org/10.1103/PRXQuantum.4.040316} {\bibfield  {journal} {\bibinfo  {journal} {PRX Quantum}\ }\textbf {\bibinfo {volume} {4}},\ \bibinfo {pages} {040316} (\bibinfo {year} {2023})}\BibitemShut {NoStop}%
\bibitem [{\citenamefont {Schlegel}\ \emph {et~al.}(2022)\citenamefont {Schlegel}, \citenamefont {Minganti},\ and\ \citenamefont {Savona}}]{schlegel_quantum_2022}%
  \BibitemOpen
  \bibfield  {author} {\bibinfo {author} {\bibfnamefont {D.~S.}\ \bibnamefont {Schlegel}}, \bibinfo {author} {\bibfnamefont {F.}~\bibnamefont {Minganti}},\ and\ \bibinfo {author} {\bibfnamefont {V.}~\bibnamefont {Savona}},\ }\bibfield  {title} {\bibinfo {title} {Quantum error correction using squeezed {Schr}{\textbackslash}"odinger cat states},\ }\href {https://doi.org/10.1103/PhysRevA.106.022431} {\bibfield  {journal} {\bibinfo  {journal} {Phys. Rev. A}\ }\textbf {\bibinfo {volume} {106}},\ \bibinfo {pages} {022431} (\bibinfo {year} {2022})}\BibitemShut {NoStop}%
\bibitem [{\citenamefont {Hillmann}\ and\ \citenamefont {Quijandr\'{\i}a}(2023)}]{hillmann_quantum_2023}%
  \BibitemOpen
  \bibfield  {author} {\bibinfo {author} {\bibfnamefont {T.}~\bibnamefont {Hillmann}}\ and\ \bibinfo {author} {\bibfnamefont {F.}~\bibnamefont {Quijandr\'{\i}a}},\ }\bibfield  {title} {\bibinfo {title} {Quantum error correction with dissipatively stabilized squeezed-cat qubits},\ }\href {https://doi.org/10.1103/PhysRevA.107.032423} {\bibfield  {journal} {\bibinfo  {journal} {Phys. Rev. A}\ }\textbf {\bibinfo {volume} {107}},\ \bibinfo {pages} {032423} (\bibinfo {year} {2023})}\BibitemShut {NoStop}%
\bibitem [{\citenamefont {Xu}\ \emph {et~al.}(2023)\citenamefont {Xu}, \citenamefont {Zheng}, \citenamefont {Wang}, \citenamefont {Zoller}, \citenamefont {Clerk},\ and\ \citenamefont {Jiang}}]{xu_autonomous_2023}%
  \BibitemOpen
  \bibfield  {author} {\bibinfo {author} {\bibfnamefont {Q.}~\bibnamefont {Xu}}, \bibinfo {author} {\bibfnamefont {G.}~\bibnamefont {Zheng}}, \bibinfo {author} {\bibfnamefont {Y.-X.}\ \bibnamefont {Wang}}, \bibinfo {author} {\bibfnamefont {P.}~\bibnamefont {Zoller}}, \bibinfo {author} {\bibfnamefont {A.~A.}\ \bibnamefont {Clerk}},\ and\ \bibinfo {author} {\bibfnamefont {L.}~\bibnamefont {Jiang}},\ }\bibfield  {title} {\bibinfo {title} {Autonomous quantum error correction and fault-tolerant quantum computation with squeezed cat qubits},\ }\href {https://doi.org/10.1038/s41534-023-00746-0} {\bibfield  {journal} {\bibinfo  {journal} {npj Quantum Inf.}\ }\textbf {\bibinfo {volume} {9}},\ \bibinfo {pages} {1} (\bibinfo {year} {2023})}\BibitemShut {NoStop}%
\bibitem [{\citenamefont {Wineland}\ \emph {et~al.}(1998)\citenamefont {Wineland}, \citenamefont {Monroe}, \citenamefont {Itano}, \citenamefont {Leibfried}, \citenamefont {King},\ and\ \citenamefont {Meekhof}}]{wineland_experimental_1998}%
  \BibitemOpen
  \bibfield  {author} {\bibinfo {author} {\bibfnamefont {D.~J.}\ \bibnamefont {Wineland}}, \bibinfo {author} {\bibfnamefont {C.}~\bibnamefont {Monroe}}, \bibinfo {author} {\bibfnamefont {W.~M.}\ \bibnamefont {Itano}}, \bibinfo {author} {\bibfnamefont {D.}~\bibnamefont {Leibfried}}, \bibinfo {author} {\bibfnamefont {B.~E.}\ \bibnamefont {King}},\ and\ \bibinfo {author} {\bibfnamefont {D.~M.}\ \bibnamefont {Meekhof}},\ }\bibfield  {title} {\bibinfo {title} {Experimental {Issues} in {Coherent} {Quantum}-{State} {Manipulation} of {Trapped} {Atomic} {Ions}},\ }\href {https://doi.org/10.6028/jres.103.019} {\bibfield  {journal} {\bibinfo  {journal} {J. Res. Nat. Inst. Stand. Technol.}\ }\textbf {\bibinfo {volume} {103}},\ \bibinfo {pages} {259} (\bibinfo {year} {1998})}\BibitemShut {NoStop}%
\bibitem [{\citenamefont {Blais}\ \emph {et~al.}(2021)\citenamefont {Blais}, \citenamefont {Grimsmo}, \citenamefont {Girvin},\ and\ \citenamefont {Wallraff}}]{blais_circuit_2021}%
  \BibitemOpen
  \bibfield  {author} {\bibinfo {author} {\bibfnamefont {A.}~\bibnamefont {Blais}}, \bibinfo {author} {\bibfnamefont {A.~L.}\ \bibnamefont {Grimsmo}}, \bibinfo {author} {\bibfnamefont {S.~M.}\ \bibnamefont {Girvin}},\ and\ \bibinfo {author} {\bibfnamefont {A.}~\bibnamefont {Wallraff}},\ }\bibfield  {title} {\bibinfo {title} {Circuit quantum electrodynamics},\ }\href {https://doi.org/10.1103/RevModPhys.93.025005} {\bibfield  {journal} {\bibinfo  {journal} {Rev. Mod. Phys.}\ }\textbf {\bibinfo {volume} {93}},\ \bibinfo {pages} {025005} (\bibinfo {year} {2021})}\BibitemShut {NoStop}%
\bibitem [{\citenamefont {Leghtas}\ \emph {et~al.}(2013)\citenamefont {Leghtas}, \citenamefont {Kirchmair}, \citenamefont {Vlastakis}, \citenamefont {Schoelkopf}, \citenamefont {Devoret},\ and\ \citenamefont {Mirrahimi}}]{leghtas_hardware_2013}%
  \BibitemOpen
  \bibfield  {author} {\bibinfo {author} {\bibfnamefont {Z.}~\bibnamefont {Leghtas}}, \bibinfo {author} {\bibfnamefont {G.}~\bibnamefont {Kirchmair}}, \bibinfo {author} {\bibfnamefont {B.}~\bibnamefont {Vlastakis}}, \bibinfo {author} {\bibfnamefont {R.~J.}\ \bibnamefont {Schoelkopf}}, \bibinfo {author} {\bibfnamefont {M.~H.}\ \bibnamefont {Devoret}},\ and\ \bibinfo {author} {\bibfnamefont {M.}~\bibnamefont {Mirrahimi}},\ }\bibfield  {title} {\bibinfo {title} {Hardware-{Efficient} {Autonomous} {Quantum} {Memory} {Protection}},\ }\href {https://doi.org/10.1103/PhysRevLett.111.120501} {\bibfield  {journal} {\bibinfo  {journal} {Phys. Rev. Lett.}\ }\textbf {\bibinfo {volume} {111}},\ \bibinfo {pages} {120501} (\bibinfo {year} {2013})}\BibitemShut {NoStop}%
\bibitem [{\citenamefont {McDonnell}\ \emph {et~al.}(2007)\citenamefont {McDonnell}, \citenamefont {Home}, \citenamefont {Lucas}, \citenamefont {Imreh}, \citenamefont {Keitch}, \citenamefont {Szwer}, \citenamefont {Thomas}, \citenamefont {Webster}, \citenamefont {Stacey},\ and\ \citenamefont {Steane}}]{mcdonnell_long-lived_2007}%
  \BibitemOpen
  \bibfield  {author} {\bibinfo {author} {\bibfnamefont {M.~J.}\ \bibnamefont {McDonnell}}, \bibinfo {author} {\bibfnamefont {J.~P.}\ \bibnamefont {Home}}, \bibinfo {author} {\bibfnamefont {D.~M.}\ \bibnamefont {Lucas}}, \bibinfo {author} {\bibfnamefont {G.}~\bibnamefont {Imreh}}, \bibinfo {author} {\bibfnamefont {B.~C.}\ \bibnamefont {Keitch}}, \bibinfo {author} {\bibfnamefont {D.~J.}\ \bibnamefont {Szwer}}, \bibinfo {author} {\bibfnamefont {N.~R.}\ \bibnamefont {Thomas}}, \bibinfo {author} {\bibfnamefont {S.~C.}\ \bibnamefont {Webster}}, \bibinfo {author} {\bibfnamefont {D.~N.}\ \bibnamefont {Stacey}},\ and\ \bibinfo {author} {\bibfnamefont {A.~M.}\ \bibnamefont {Steane}},\ }\bibfield  {title} {\bibinfo {title} {Long-{Lived} {Mesoscopic} {Entanglement} outside the {Lamb}-{Dicke} {Regime}},\ }\href {https://doi.org/10.1103/PhysRevLett.98.063603} {\bibfield  {journal} {\bibinfo  {journal} {Phys. Rev. Lett.}\ }\textbf {\bibinfo {volume} {98}},\ \bibinfo {pages} {063603} (\bibinfo {year} {2007})}\BibitemShut {NoStop}%
\bibitem [{\citenamefont {Stutter}\ \emph {et~al.}(2018)\citenamefont {Stutter}, \citenamefont {Hrmo}, \citenamefont {Jarlaud}, \citenamefont {Joshi}, \citenamefont {Goodwin},\ and\ \citenamefont {Thompson}}]{stutter_sideband_2018}%
  \BibitemOpen
  \bibfield  {author} {\bibinfo {author} {\bibfnamefont {G.}~\bibnamefont {Stutter}}, \bibinfo {author} {\bibfnamefont {P.}~\bibnamefont {Hrmo}}, \bibinfo {author} {\bibfnamefont {V.}~\bibnamefont {Jarlaud}}, \bibinfo {author} {\bibfnamefont {M.~K.}\ \bibnamefont {Joshi}}, \bibinfo {author} {\bibfnamefont {J.~F.}\ \bibnamefont {Goodwin}},\ and\ \bibinfo {author} {\bibfnamefont {R.~C.}\ \bibnamefont {Thompson}},\ }\bibfield  {title} {\bibinfo {title} {Sideband cooling of small ion {Coulomb} crystals in a {Penning} trap},\ }\href {https://doi.org/10.1080/09500340.2017.1376719} {\bibfield  {journal} {\bibinfo  {journal} {J. Mod. Opt.}\ }\textbf {\bibinfo {volume} {65}},\ \bibinfo {pages} {549} (\bibinfo {year} {2018})}\BibitemShut {NoStop}%
\bibitem [{\citenamefont {Hrmo}\ \emph {et~al.}(2019)\citenamefont {Hrmo}, \citenamefont {Joshi}, \citenamefont {Jarlaud}, \citenamefont {Corfield},\ and\ \citenamefont {Thompson}}]{hrmo_sideband_2019}%
  \BibitemOpen
  \bibfield  {author} {\bibinfo {author} {\bibfnamefont {P.}~\bibnamefont {Hrmo}}, \bibinfo {author} {\bibfnamefont {M.~K.}\ \bibnamefont {Joshi}}, \bibinfo {author} {\bibfnamefont {V.}~\bibnamefont {Jarlaud}}, \bibinfo {author} {\bibfnamefont {O.}~\bibnamefont {Corfield}},\ and\ \bibinfo {author} {\bibfnamefont {R.~C.}\ \bibnamefont {Thompson}},\ }\bibfield  {title} {\bibinfo {title} {Sideband cooling of the radial modes of motion of a single ion in a {Penning} trap},\ }\href {https://doi.org/10.1103/PhysRevA.100.043414} {\bibfield  {journal} {\bibinfo  {journal} {Phys. Rev. A}\ }\textbf {\bibinfo {volume} {100}},\ \bibinfo {pages} {043414} (\bibinfo {year} {2019})}\BibitemShut {NoStop}%
\bibitem [{\citenamefont {Jarlaud}\ \emph {et~al.}(2020)\citenamefont {Jarlaud}, \citenamefont {Hrmo}, \citenamefont {Joshi},\ and\ \citenamefont {Thompson}}]{jarlaud_coherence_2020}%
  \BibitemOpen
  \bibfield  {author} {\bibinfo {author} {\bibfnamefont {V.}~\bibnamefont {Jarlaud}}, \bibinfo {author} {\bibfnamefont {P.}~\bibnamefont {Hrmo}}, \bibinfo {author} {\bibfnamefont {M.~K.}\ \bibnamefont {Joshi}},\ and\ \bibinfo {author} {\bibfnamefont {R.~C.}\ \bibnamefont {Thompson}},\ }\bibfield  {title} {\bibinfo {title} {Coherence properties of highly-excited motional states of a trapped ion},\ }\href {https://doi.org/10.1088/1361-6455/abc271} {\bibfield  {journal} {\bibinfo  {journal} {J. Phys. B: At., Mol. Opt. Phys.}\ }\textbf {\bibinfo {volume} {54}},\ \bibinfo {pages} {015501} (\bibinfo {year} {2020})}\BibitemShut {NoStop}%
\bibitem [{\citenamefont {Hofheinz}\ \emph {et~al.}(2011)\citenamefont {Hofheinz}, \citenamefont {Portier}, \citenamefont {Baudouin}, \citenamefont {Joyez}, \citenamefont {Vion}, \citenamefont {Bertet}, \citenamefont {Roche},\ and\ \citenamefont {Esteve}}]{hofheinz_bright_2011}%
  \BibitemOpen
  \bibfield  {author} {\bibinfo {author} {\bibfnamefont {M.}~\bibnamefont {Hofheinz}}, \bibinfo {author} {\bibfnamefont {F.}~\bibnamefont {Portier}}, \bibinfo {author} {\bibfnamefont {Q.}~\bibnamefont {Baudouin}}, \bibinfo {author} {\bibfnamefont {P.}~\bibnamefont {Joyez}}, \bibinfo {author} {\bibfnamefont {D.}~\bibnamefont {Vion}}, \bibinfo {author} {\bibfnamefont {P.}~\bibnamefont {Bertet}}, \bibinfo {author} {\bibfnamefont {P.}~\bibnamefont {Roche}},\ and\ \bibinfo {author} {\bibfnamefont {D.}~\bibnamefont {Esteve}},\ }\bibfield  {title} {\bibinfo {title} {Bright {Side} of the {Coulomb} {Blockade}},\ }\href {https://doi.org/10.1103/PhysRevLett.106.217005} {\bibfield  {journal} {\bibinfo  {journal} {Phys. Rev. Lett.}\ }\textbf {\bibinfo {volume} {106}},\ \bibinfo {pages} {217005} (\bibinfo {year} {2011})}\BibitemShut {NoStop}%
\bibitem [{\citenamefont {Rolland}\ \emph {et~al.}(2019)\citenamefont {Rolland}, \citenamefont {Peugeot}, \citenamefont {Dambach}, \citenamefont {Westig}, \citenamefont {Kubala}, \citenamefont {Mukharsky}, \citenamefont {Altimiras}, \citenamefont {le~Sueur}, \citenamefont {Joyez}, \citenamefont {Vion}, \citenamefont {Roche}, \citenamefont {Esteve}, \citenamefont {Ankerhold},\ and\ \citenamefont {Portier}}]{rolland_antibunched_2019}%
  \BibitemOpen
  \bibfield  {author} {\bibinfo {author} {\bibfnamefont {C.}~\bibnamefont {Rolland}}, \bibinfo {author} {\bibfnamefont {A.}~\bibnamefont {Peugeot}}, \bibinfo {author} {\bibfnamefont {S.}~\bibnamefont {Dambach}}, \bibinfo {author} {\bibfnamefont {M.}~\bibnamefont {Westig}}, \bibinfo {author} {\bibfnamefont {B.}~\bibnamefont {Kubala}}, \bibinfo {author} {\bibfnamefont {Y.}~\bibnamefont {Mukharsky}}, \bibinfo {author} {\bibfnamefont {C.}~\bibnamefont {Altimiras}}, \bibinfo {author} {\bibfnamefont {H.}~\bibnamefont {le~Sueur}}, \bibinfo {author} {\bibfnamefont {P.}~\bibnamefont {Joyez}}, \bibinfo {author} {\bibfnamefont {D.}~\bibnamefont {Vion}}, \bibinfo {author} {\bibfnamefont {P.}~\bibnamefont {Roche}}, \bibinfo {author} {\bibfnamefont {D.}~\bibnamefont {Esteve}}, \bibinfo {author} {\bibfnamefont {J.}~\bibnamefont {Ankerhold}},\ and\ \bibinfo {author} {\bibfnamefont {F.}~\bibnamefont {Portier}},\ }\bibfield  {title} {\bibinfo {title} {Antibunched {Photons} {Emitted} by a dc-{Biased} {Josephson} {Junction}},\ }\href {https://doi.org/10.1103/PhysRevLett.122.186804} {\bibfield  {journal} {\bibinfo  {journal} {Phys. Rev. Lett.}\ }\textbf {\bibinfo {volume} {122}},\ \bibinfo {pages} {186804} (\bibinfo {year} {2019})}\BibitemShut {NoStop}%
\bibitem [{\citenamefont {Peugeot}\ \emph {et~al.}(2021)\citenamefont {Peugeot}, \citenamefont {M\'enard}, \citenamefont {Dambach}, \citenamefont {Westig}, \citenamefont {Kubala}, \citenamefont {Mukharsky}, \citenamefont {Altimiras}, \citenamefont {Joyez}, \citenamefont {Vion}, \citenamefont {Roche}, \citenamefont {Esteve}, \citenamefont {Milman}, \citenamefont {Lepp\"akangas}, \citenamefont {Johansson}, \citenamefont {Hofheinz}, \citenamefont {Ankerhold},\ and\ \citenamefont {Portier}}]{peugeot_generating_2021}%
  \BibitemOpen
  \bibfield  {author} {\bibinfo {author} {\bibfnamefont {A.}~\bibnamefont {Peugeot}}, \bibinfo {author} {\bibfnamefont {G.}~\bibnamefont {M\'enard}}, \bibinfo {author} {\bibfnamefont {S.}~\bibnamefont {Dambach}}, \bibinfo {author} {\bibfnamefont {M.}~\bibnamefont {Westig}}, \bibinfo {author} {\bibfnamefont {B.}~\bibnamefont {Kubala}}, \bibinfo {author} {\bibfnamefont {Y.}~\bibnamefont {Mukharsky}}, \bibinfo {author} {\bibfnamefont {C.}~\bibnamefont {Altimiras}}, \bibinfo {author} {\bibfnamefont {P.}~\bibnamefont {Joyez}}, \bibinfo {author} {\bibfnamefont {D.}~\bibnamefont {Vion}}, \bibinfo {author} {\bibfnamefont {P.}~\bibnamefont {Roche}}, \bibinfo {author} {\bibfnamefont {D.}~\bibnamefont {Esteve}}, \bibinfo {author} {\bibfnamefont {P.}~\bibnamefont {Milman}}, \bibinfo {author} {\bibfnamefont {J.}~\bibnamefont {Lepp\"akangas}}, \bibinfo {author} {\bibfnamefont {G.}~\bibnamefont {Johansson}}, \bibinfo {author} {\bibfnamefont {M.}~\bibnamefont {Hofheinz}}, \bibinfo {author} {\bibfnamefont {J.}~\bibnamefont {Ankerhold}},\ and\ \bibinfo {author} {\bibfnamefont {F.}~\bibnamefont {Portier}},\ }\bibfield  {title} {\bibinfo {title} {Generating {Two} {Continuous} {Entangled} {Microwave} {Beams} {Using} a dc-{Biased} {Josephson} {Junction}},\ }\href {https://doi.org/10.1103/PhysRevX.11.031008} {\bibfield  {journal} {\bibinfo  {journal} {Phys. Rev. X}\ }\textbf {\bibinfo {volume} {11}},\ \bibinfo {pages} {031008} (\bibinfo {year} {2021})}\BibitemShut {NoStop}%
\bibitem [{\citenamefont {M\'enard}\ \emph {et~al.}(2022)\citenamefont {M\'enard}, \citenamefont {Peugeot}, \citenamefont {Padurariu}, \citenamefont {Rolland}, \citenamefont {Kubala}, \citenamefont {Mukharsky}, \citenamefont {Iftikhar}, \citenamefont {Altimiras}, \citenamefont {Roche}, \citenamefont {le~Sueur}, \citenamefont {Joyez}, \citenamefont {Vion}, \citenamefont {Esteve}, \citenamefont {Ankerhold},\ and\ \citenamefont {Portier}}]{menard_emission_2022}%
  \BibitemOpen
  \bibfield  {author} {\bibinfo {author} {\bibfnamefont {G.~C.}\ \bibnamefont {M\'enard}}, \bibinfo {author} {\bibfnamefont {A.}~\bibnamefont {Peugeot}}, \bibinfo {author} {\bibfnamefont {C.}~\bibnamefont {Padurariu}}, \bibinfo {author} {\bibfnamefont {C.}~\bibnamefont {Rolland}}, \bibinfo {author} {\bibfnamefont {B.}~\bibnamefont {Kubala}}, \bibinfo {author} {\bibfnamefont {Y.}~\bibnamefont {Mukharsky}}, \bibinfo {author} {\bibfnamefont {Z.}~\bibnamefont {Iftikhar}}, \bibinfo {author} {\bibfnamefont {C.}~\bibnamefont {Altimiras}}, \bibinfo {author} {\bibfnamefont {P.}~\bibnamefont {Roche}}, \bibinfo {author} {\bibfnamefont {H.}~\bibnamefont {le~Sueur}}, \bibinfo {author} {\bibfnamefont {P.}~\bibnamefont {Joyez}}, \bibinfo {author} {\bibfnamefont {D.}~\bibnamefont {Vion}}, \bibinfo {author} {\bibfnamefont {D.}~\bibnamefont {Esteve}}, \bibinfo {author} {\bibfnamefont {J.}~\bibnamefont {Ankerhold}},\ and\ \bibinfo {author} {\bibfnamefont {F.}~\bibnamefont {Portier}},\ }\bibfield  {title} {\bibinfo {title} {Emission of {Photon} {Multiplets} by a dc-{Biased} {Superconducting} {Circuit}},\ }\href {https://doi.org/10.1103/PhysRevX.12.021006} {\bibfield  {journal} {\bibinfo  {journal} {Phys. Rev. X}\ }\textbf {\bibinfo {volume} {12}},\ \bibinfo {pages} {021006} (\bibinfo {year} {2022})}\BibitemShut {NoStop}%
\bibitem [{\citenamefont {Cohen}\ \emph {et~al.}(2017)\citenamefont {Cohen}, \citenamefont {Smith}, \citenamefont {Devoret},\ and\ \citenamefont {Mirrahimi}}]{cohen_degeneracy-preserving_2017}%
  \BibitemOpen
  \bibfield  {author} {\bibinfo {author} {\bibfnamefont {J.}~\bibnamefont {Cohen}}, \bibinfo {author} {\bibfnamefont {W.~C.}\ \bibnamefont {Smith}}, \bibinfo {author} {\bibfnamefont {M.~H.}\ \bibnamefont {Devoret}},\ and\ \bibinfo {author} {\bibfnamefont {M.}~\bibnamefont {Mirrahimi}},\ }\bibfield  {title} {\bibinfo {title} {Degeneracy-{Preserving} {Quantum} {Nondemolition} {Measurement} of {Parity}-{Type} {Observables} for {Cat} {Qubits}},\ }\href {https://doi.org/10.1103/PhysRevLett.119.060503} {\bibfield  {journal} {\bibinfo  {journal} {Phys. Rev. Lett.}\ }\textbf {\bibinfo {volume} {119}},\ \bibinfo {pages} {060503} (\bibinfo {year} {2017})}\BibitemShut {NoStop}%
\bibitem [{\citenamefont {Smith}\ \emph {et~al.}(2020)\citenamefont {Smith}, \citenamefont {Kou}, \citenamefont {Xiao}, \citenamefont {Vool},\ and\ \citenamefont {Devoret}}]{smith_superconducting_2020}%
  \BibitemOpen
  \bibfield  {author} {\bibinfo {author} {\bibfnamefont {W.~C.}\ \bibnamefont {Smith}}, \bibinfo {author} {\bibfnamefont {A.}~\bibnamefont {Kou}}, \bibinfo {author} {\bibfnamefont {X.}~\bibnamefont {Xiao}}, \bibinfo {author} {\bibfnamefont {U.}~\bibnamefont {Vool}},\ and\ \bibinfo {author} {\bibfnamefont {M.~H.}\ \bibnamefont {Devoret}},\ }\bibfield  {title} {\bibinfo {title} {Superconducting circuit protected by two-{Cooper}-pair tunneling},\ }\href {https://doi.org/10.1038/s41534-019-0231-2} {\bibfield  {journal} {\bibinfo  {journal} {npj Quantum Inf.}\ }\textbf {\bibinfo {volume} {6}},\ \bibinfo {pages} {1} (\bibinfo {year} {2020})}\BibitemShut {NoStop}%
\bibitem [{\citenamefont {Smith}\ \emph {et~al.}(2022)\citenamefont {Smith}, \citenamefont {Villiers}, \citenamefont {Marquet}, \citenamefont {Palomo}, \citenamefont {Delbecq}, \citenamefont {Kontos}, \citenamefont {Campagne-Ibarcq}, \citenamefont {Dou\ifmmode~\mbox{\c{c}}\else \c{c}\fi{}ot},\ and\ \citenamefont {Leghtas}}]{smith_magnifying_2022}%
  \BibitemOpen
  \bibfield  {author} {\bibinfo {author} {\bibfnamefont {W.~C.}\ \bibnamefont {Smith}}, \bibinfo {author} {\bibfnamefont {M.}~\bibnamefont {Villiers}}, \bibinfo {author} {\bibfnamefont {A.}~\bibnamefont {Marquet}}, \bibinfo {author} {\bibfnamefont {J.}~\bibnamefont {Palomo}}, \bibinfo {author} {\bibfnamefont {M.~R.}\ \bibnamefont {Delbecq}}, \bibinfo {author} {\bibfnamefont {T.}~\bibnamefont {Kontos}}, \bibinfo {author} {\bibfnamefont {P.}~\bibnamefont {Campagne-Ibarcq}}, \bibinfo {author} {\bibfnamefont {B.}~\bibnamefont {Dou\ifmmode~\mbox{\c{c}}\else \c{c}\fi{}ot}},\ and\ \bibinfo {author} {\bibfnamefont {Z.}~\bibnamefont {Leghtas}},\ }\bibfield  {title} {\bibinfo {title} {Magnifying {Quantum} {Phase} {Fluctuations} with {Cooper}-{Pair} {Pairing}},\ }\href {https://doi.org/10.1103/PhysRevX.12.021002} {\bibfield  {journal} {\bibinfo  {journal} {Phys. Rev. X}\ }\textbf {\bibinfo {volume} {12}},\ \bibinfo {pages} {021002} (\bibinfo {year} {2022})}\BibitemShut {NoStop}%
\bibitem [{\citenamefont {Smith}\ \emph {et~al.}(2025)\citenamefont {Smith}, \citenamefont {Borgognoni}, \citenamefont {Villiers}, \citenamefont {Roverc’h}, \citenamefont {Palomo}, \citenamefont {Delbecq}, \citenamefont {Kontos}, \citenamefont {Campagne-Ibarcq}, \citenamefont {Douçot},\ and\ \citenamefont {Leghtas}}]{smith_spectral_2025}%
  \BibitemOpen
  \bibfield  {author} {\bibinfo {author} {\bibfnamefont {W.~C.}\ \bibnamefont {Smith}}, \bibinfo {author} {\bibfnamefont {A.}~\bibnamefont {Borgognoni}}, \bibinfo {author} {\bibfnamefont {M.}~\bibnamefont {Villiers}}, \bibinfo {author} {\bibfnamefont {E.}~\bibnamefont {Roverc’h}}, \bibinfo {author} {\bibfnamefont {J.}~\bibnamefont {Palomo}}, \bibinfo {author} {\bibfnamefont {M.~R.}\ \bibnamefont {Delbecq}}, \bibinfo {author} {\bibfnamefont {T.}~\bibnamefont {Kontos}}, \bibinfo {author} {\bibfnamefont {P.}~\bibnamefont {Campagne-Ibarcq}}, \bibinfo {author} {\bibfnamefont {B.}~\bibnamefont {Douçot}},\ and\ \bibinfo {author} {\bibfnamefont {Z.}~\bibnamefont {Leghtas}},\ }\bibfield  {title} {\bibinfo {title} {Spectral signature of high-order photon processes enhanced by {Cooper}-pair pairing},\ }\href {https://doi.org/10.1038/s41467-025-62047-8} {\bibfield  {journal} {\bibinfo  {journal} {Nat. Commun.}\ }\textbf {\bibinfo {volume} {16}},\ \bibinfo {pages} {8359} (\bibinfo {year} {2025})}\BibitemShut {NoStop}%
\bibitem [{Note1()}]{Note1}%
  \BibitemOpen
  \bibinfo {note} {{The term originates from the description of coherent driving in closed systems without dissipation, and we retain it here for historical consistency.}}\BibitemShut {Stop}%
\bibitem [{\citenamefont {Chamberland}\ \emph {et~al.}(2022)\citenamefont {Chamberland}, \citenamefont {Noh}, \citenamefont {Arrangoiz-Arriola}, \citenamefont {Campbell}, \citenamefont {Hann}, \citenamefont {Iverson}, \citenamefont {Putterman}, \citenamefont {Bohdanowicz}, \citenamefont {Flammia}, \citenamefont {Keller}, \citenamefont {Refael}, \citenamefont {Preskill}, \citenamefont {Jiang}, \citenamefont {Safavi-Naeini}, \citenamefont {Painter},\ and\ \citenamefont {Brand\~ao}}]{chamberland_building_2022}%
  \BibitemOpen
  \bibfield  {author} {\bibinfo {author} {\bibfnamefont {C.}~\bibnamefont {Chamberland}}, \bibinfo {author} {\bibfnamefont {K.}~\bibnamefont {Noh}}, \bibinfo {author} {\bibfnamefont {P.}~\bibnamefont {Arrangoiz-Arriola}}, \bibinfo {author} {\bibfnamefont {E.~T.}\ \bibnamefont {Campbell}}, \bibinfo {author} {\bibfnamefont {C.~T.}\ \bibnamefont {Hann}}, \bibinfo {author} {\bibfnamefont {J.}~\bibnamefont {Iverson}}, \bibinfo {author} {\bibfnamefont {H.}~\bibnamefont {Putterman}}, \bibinfo {author} {\bibfnamefont {T.~C.}\ \bibnamefont {Bohdanowicz}}, \bibinfo {author} {\bibfnamefont {S.~T.}\ \bibnamefont {Flammia}}, \bibinfo {author} {\bibfnamefont {A.}~\bibnamefont {Keller}}, \bibinfo {author} {\bibfnamefont {G.}~\bibnamefont {Refael}}, \bibinfo {author} {\bibfnamefont {J.}~\bibnamefont {Preskill}}, \bibinfo {author} {\bibfnamefont {L.}~\bibnamefont {Jiang}}, \bibinfo {author} {\bibfnamefont {A.~H.}\ \bibnamefont {Safavi-Naeini}}, \bibinfo {author} {\bibfnamefont {O.}~\bibnamefont {Painter}},\ and\ \bibinfo {author} {\bibfnamefont {F.~G. S.~L.}\ \bibnamefont {Brand\~ao}},\ }\bibfield  {title} {\bibinfo {title} {Building a {Fault}-{Tolerant} {Quantum} {Computer} {Using} {Concatenated} {Cat} {Codes}},\ }\href {https://doi.org/10.1103/PRXQuantum.3.010329} {\bibfield  {journal} {\bibinfo  {journal} {PRX Quantum}\ }\textbf {\bibinfo {volume} {3}},\ \bibinfo {pages} {010329} (\bibinfo {year} {2022})}\BibitemShut {NoStop}%
\bibitem [{\citenamefont {Gautier}\ \emph {et~al.}(2022)\citenamefont {Gautier}, \citenamefont {Sarlette},\ and\ \citenamefont {Mirrahimi}}]{gautier_combined_2022}%
  \BibitemOpen
  \bibfield  {author} {\bibinfo {author} {\bibfnamefont {R.}~\bibnamefont {Gautier}}, \bibinfo {author} {\bibfnamefont {A.}~\bibnamefont {Sarlette}},\ and\ \bibinfo {author} {\bibfnamefont {M.}~\bibnamefont {Mirrahimi}},\ }\bibfield  {title} {\bibinfo {title} {Combined {Dissipative} and {Hamiltonian} {Confinement} of {Cat} {Qubits}},\ }\href {https://doi.org/10.1103/PRXQuantum.3.020339} {\bibfield  {journal} {\bibinfo  {journal} {PRX Quantum}\ }\textbf {\bibinfo {volume} {3}},\ \bibinfo {pages} {020339} (\bibinfo {year} {2022})}\BibitemShut {NoStop}%
\bibitem [{\citenamefont {de~Matos~Filho}\ and\ \citenamefont {Vogel}(1996{\natexlab{b}})}]{de_matos_filho_nonlinear_1996}%
  \BibitemOpen
  \bibfield  {author} {\bibinfo {author} {\bibfnamefont {R.~L.}\ \bibnamefont {de~Matos~Filho}}\ and\ \bibinfo {author} {\bibfnamefont {W.}~\bibnamefont {Vogel}},\ }\bibfield  {title} {\bibinfo {title} {Nonlinear coherent states},\ }\href {https://doi.org/10.1103/PhysRevA.54.4560} {\bibfield  {journal} {\bibinfo  {journal} {Phys. Rev. A}\ }\textbf {\bibinfo {volume} {54}},\ \bibinfo {pages} {4560} (\bibinfo {year} {1996}{\natexlab{b}})}\BibitemShut {NoStop}%
\bibitem [{\citenamefont {Dodonov}(2002)}]{dodonov_nonclassical_2002}%
  \BibitemOpen
  \bibfield  {author} {\bibinfo {author} {\bibfnamefont {V.~V.}\ \bibnamefont {Dodonov}},\ }\bibfield  {title} {\bibinfo {title} {`{Nonclassical}' states in quantum optics: a `squeezed' review of the first 75 years},\ }\href {https://doi.org/10.1088/1464-4266/4/1/201} {\bibfield  {journal} {\bibinfo  {journal} {J. Opt. B: Quantum Semiclassical Opt.}\ }\textbf {\bibinfo {volume} {4}},\ \bibinfo {pages} {R1} (\bibinfo {year} {2002})}\BibitemShut {NoStop}%
\bibitem [{\citenamefont {Man’ko}\ \emph {et~al.}(2000)\citenamefont {Man’ko}, \citenamefont {Marmo}, \citenamefont {Porzio}, \citenamefont {Solimeno},\ and\ \citenamefont {Zaccaria}}]{manko_trapped_2000}%
  \BibitemOpen
  \bibfield  {author} {\bibinfo {author} {\bibfnamefont {V.}~\bibnamefont {Man’ko}}, \bibinfo {author} {\bibfnamefont {G.}~\bibnamefont {Marmo}}, \bibinfo {author} {\bibfnamefont {A.}~\bibnamefont {Porzio}}, \bibinfo {author} {\bibfnamefont {S.}~\bibnamefont {Solimeno}},\ and\ \bibinfo {author} {\bibfnamefont {F.}~\bibnamefont {Zaccaria}},\ }\bibfield  {title} {\bibinfo {title} {Trapped ions in laser fields: {A} benchmark for deformed quantum oscillators},\ }\href {https://doi.org/10.1103/PhysRevA.62.053407} {\bibfield  {journal} {\bibinfo  {journal} {Phys. Rev. A}\ }\textbf {\bibinfo {volume} {62}},\ \bibinfo {pages} {053407} (\bibinfo {year} {2000})}\BibitemShut {NoStop}%
\bibitem [{\citenamefont {Kraus}\ \emph {et~al.}(2008)\citenamefont {Kraus}, \citenamefont {B\"uchler}, \citenamefont {Diehl}, \citenamefont {Kantian}, \citenamefont {Micheli},\ and\ \citenamefont {Zoller}}]{kraus_preparation_2008}%
  \BibitemOpen
  \bibfield  {author} {\bibinfo {author} {\bibfnamefont {B.}~\bibnamefont {Kraus}}, \bibinfo {author} {\bibfnamefont {H.~P.}\ \bibnamefont {B\"uchler}}, \bibinfo {author} {\bibfnamefont {S.}~\bibnamefont {Diehl}}, \bibinfo {author} {\bibfnamefont {A.}~\bibnamefont {Kantian}}, \bibinfo {author} {\bibfnamefont {A.}~\bibnamefont {Micheli}},\ and\ \bibinfo {author} {\bibfnamefont {P.}~\bibnamefont {Zoller}},\ }\bibfield  {title} {\bibinfo {title} {Preparation of entangled states by quantum {Markov} processes},\ }\href {https://doi.org/10.1103/PhysRevA.78.042307} {\bibfield  {journal} {\bibinfo  {journal} {Phys. Rev. A}\ }\textbf {\bibinfo {volume} {78}},\ \bibinfo {pages} {042307} (\bibinfo {year} {2008})}\BibitemShut {NoStop}%
\bibitem [{Note2()}]{Note2}%
  \BibitemOpen
  \bibinfo {note} {We keep this assumption for the remaining of the paper.}\BibitemShut {Stop}%
\bibitem [{\citenamefont {Minganti}\ \emph {et~al.}(2018)\citenamefont {Minganti}, \citenamefont {Biella}, \citenamefont {Bartolo},\ and\ \citenamefont {Ciuti}}]{minganti_spectral_2018}%
  \BibitemOpen
  \bibfield  {author} {\bibinfo {author} {\bibfnamefont {F.}~\bibnamefont {Minganti}}, \bibinfo {author} {\bibfnamefont {A.}~\bibnamefont {Biella}}, \bibinfo {author} {\bibfnamefont {N.}~\bibnamefont {Bartolo}},\ and\ \bibinfo {author} {\bibfnamefont {C.}~\bibnamefont {Ciuti}},\ }\bibfield  {title} {\bibinfo {title} {Spectral theory of {Liouvillians} for dissipative phase transitions},\ }\href {https://doi.org/10.1103/PhysRevA.98.042118} {\bibfield  {journal} {\bibinfo  {journal} {Phys. Rev. A}\ }\textbf {\bibinfo {volume} {98}},\ \bibinfo {pages} {042118} (\bibinfo {year} {2018})}\BibitemShut {NoStop}%
\bibitem [{Note3()}]{Note3}%
  \BibitemOpen
  \bibinfo {note} {We can in fact identify two processes: when $l>0$, the population in $\ket {\Xi _\mu }$ with ${l \leq \mu < d}$ will leak into the other dark states; when $l=0$, an incoherent dephasing will happen.}\BibitemShut {Stop}%
\bibitem [{\citenamefont {Labay-Mora}\ \emph {et~al.}(2023)\citenamefont {Labay-Mora}, \citenamefont {Zambrini},\ and\ \citenamefont {Giorgi}}]{labay-mora_quantum_2023}%
  \BibitemOpen
  \bibfield  {author} {\bibinfo {author} {\bibfnamefont {A.}~\bibnamefont {Labay-Mora}}, \bibinfo {author} {\bibfnamefont {R.}~\bibnamefont {Zambrini}},\ and\ \bibinfo {author} {\bibfnamefont {G.~L.}\ \bibnamefont {Giorgi}},\ }\bibfield  {title} {\bibinfo {title} {Quantum {Associative} {Memory} with a {Single} {Driven}-{Dissipative} {Nonlinear} {Oscillator}},\ }\href {https://doi.org/10.1103/PhysRevLett.130.190602} {\bibfield  {journal} {\bibinfo  {journal} {Phys. Rev. Lett.}\ }\textbf {\bibinfo {volume} {130}},\ \bibinfo {pages} {190602} (\bibinfo {year} {2023})}\BibitemShut {NoStop}%
\bibitem [{\citenamefont {Labay-Mora}\ \emph {et~al.}(2024)\citenamefont {Labay-Mora}, \citenamefont {Zambrini},\ and\ \citenamefont {Giorgi}}]{labay-mora_quantum_2024}%
  \BibitemOpen
  \bibfield  {author} {\bibinfo {author} {\bibfnamefont {A.}~\bibnamefont {Labay-Mora}}, \bibinfo {author} {\bibfnamefont {R.}~\bibnamefont {Zambrini}},\ and\ \bibinfo {author} {\bibfnamefont {G.~L.}\ \bibnamefont {Giorgi}},\ }\bibfield  {title} {\bibinfo {title} {Quantum memories for squeezed and coherent superpositions in a driven-dissipative nonlinear oscillator},\ }\href {https://doi.org/10.1103/PhysRevA.109.032407} {\bibfield  {journal} {\bibinfo  {journal} {Phys. Rev. A}\ }\textbf {\bibinfo {volume} {109}},\ \bibinfo {pages} {032407} (\bibinfo {year} {2024})}\BibitemShut {NoStop}%
\bibitem [{\citenamefont {Labay-Mora}\ \emph {et~al.}(2025)\citenamefont {Labay-Mora}, \citenamefont {Fiorelli}, \citenamefont {Zambrini},\ and\ \citenamefont {Giorgi}}]{labay-mora_theoretical_2025}%
  \BibitemOpen
  \bibfield  {author} {\bibinfo {author} {\bibfnamefont {A.}~\bibnamefont {Labay-Mora}}, \bibinfo {author} {\bibfnamefont {E.}~\bibnamefont {Fiorelli}}, \bibinfo {author} {\bibfnamefont {R.}~\bibnamefont {Zambrini}},\ and\ \bibinfo {author} {\bibfnamefont {G.~L.}\ \bibnamefont {Giorgi}},\ }\bibfield  {title} {\bibinfo {title} {Theoretical framework for quantum associative memories},\ }\href {https://doi.org/10.1088/2058-9565/ade184} {\bibfield  {journal} {\bibinfo  {journal} {Quantum Science and Technology}\ }\textbf {\bibinfo {volume} {10}},\ \bibinfo {pages} {035050} (\bibinfo {year} {2025})}\BibitemShut {NoStop}%
\bibitem [{\citenamefont {Mandel}(1979)}]{mandel1979sub}%
  \BibitemOpen
  \bibfield  {author} {\bibinfo {author} {\bibfnamefont {L.}~\bibnamefont {Mandel}},\ }\bibfield  {title} {\bibinfo {title} {Sub-poissonian photon statistics in resonance fluorescence},\ }\href@noop {} {\bibfield  {journal} {\bibinfo  {journal} {Opt. Lett.}\ }\textbf {\bibinfo {volume} {4}},\ \bibinfo {pages} {205} (\bibinfo {year} {1979})}\BibitemShut {NoStop}%
\bibitem [{Note4()}]{Note4}%
  \BibitemOpen
  \bibinfo {note} {When $\protect \tilde {f}$ and $\protect \tilde {g}$ are nonlinear, the slopes $s_f$ and $s_g$ are defined as $\protect \frac {\partial }{\partial k}\protect \,\protect \tilde {f} (k^*)$ and $\protect \frac {\partial }{\partial k}\protect \,\protect \tilde {g}(k^*+r)$, respectively.}\BibitemShut {Stop}%
\bibitem [{\citenamefont {Shmueli}\ \emph {et~al.}(2005)\citenamefont {Shmueli}, \citenamefont {Minka}, \citenamefont {Kadane}, \citenamefont {Borle},\ and\ \citenamefont {Boatwright}}]{shmueli_useful_2005}%
  \BibitemOpen
  \bibfield  {author} {\bibinfo {author} {\bibfnamefont {G.}~\bibnamefont {Shmueli}}, \bibinfo {author} {\bibfnamefont {T.~P.}\ \bibnamefont {Minka}}, \bibinfo {author} {\bibfnamefont {J.~B.}\ \bibnamefont {Kadane}}, \bibinfo {author} {\bibfnamefont {S.}~\bibnamefont {Borle}},\ and\ \bibinfo {author} {\bibfnamefont {P.}~\bibnamefont {Boatwright}},\ }\bibfield  {title} {\bibinfo {title} {A {Useful} {Distribution} for {Fitting} {Discrete} {Data}: {Revival} of the {Conway}–{Maxwell}–{Poisson} {Distribution}},\ }\href {https://doi.org/10.1111/j.1467-9876.2005.00474.x} {\bibfield  {journal} {\bibinfo  {journal} {J. R. Stat. Soc., C: Appl. Stat.}\ }\textbf {\bibinfo {volume} {54}},\ \bibinfo {pages} {127} (\bibinfo {year} {2005})}\BibitemShut {NoStop}%
\bibitem [{\citenamefont {Borges}\ \emph {et~al.}(2014)\citenamefont {Borges}, \citenamefont {Rodrigues}, \citenamefont {Balakrishnan},\ and\ \citenamefont {Bazán}}]{borges_compoisson_2014}%
  \BibitemOpen
  \bibfield  {author} {\bibinfo {author} {\bibfnamefont {P.}~\bibnamefont {Borges}}, \bibinfo {author} {\bibfnamefont {J.}~\bibnamefont {Rodrigues}}, \bibinfo {author} {\bibfnamefont {N.}~\bibnamefont {Balakrishnan}},\ and\ \bibinfo {author} {\bibfnamefont {J.}~\bibnamefont {Bazán}},\ }\bibfield  {title} {\bibinfo {title} {A {COM}–{Poisson} type generalization of the binomial distribution and its properties and applications},\ }\href {https://doi.org/10.1016/j.spl.2014.01.019} {\bibfield  {journal} {\bibinfo  {journal} {Statistics \& Probability Letters}\ }\textbf {\bibinfo {volume} {87}},\ \bibinfo {pages} {158} (\bibinfo {year} {2014})}\BibitemShut {NoStop}%
\bibitem [{\citenamefont {Kadane}(2016)}]{kadane_sums_2016}%
  \BibitemOpen
  \bibfield  {author} {\bibinfo {author} {\bibfnamefont {J.~B.}\ \bibnamefont {Kadane}},\ }\bibfield  {title} {\bibinfo {title} {Sums of {Possibly} {Associated} {Bernoulli} {Variables}: {The} {Conway}–{Maxwell}-{Binomial} {Distribution}},\ }\href {https://doi.org/10.1214/15-BA955} {\bibfield  {journal} {\bibinfo  {journal} {Bayesian Anal.}\ }\textbf {\bibinfo {volume} {11}},\ \bibinfo {pages} {403} (\bibinfo {year} {2016})}\BibitemShut {NoStop}%
\bibitem [{\citenamefont {Daly}\ and\ \citenamefont {Gaunt}(2016)}]{daly_conway-maxwell-poisson_2016}%
  \BibitemOpen
  \bibfield  {author} {\bibinfo {author} {\bibfnamefont {F.}~\bibnamefont {Daly}}\ and\ \bibinfo {author} {\bibfnamefont {R.~E.}\ \bibnamefont {Gaunt}},\ }\bibfield  {title} {\bibinfo {title} {The {Conway}-{Maxwell}-{Poisson} distribution:{Distributional} theory and approximation},\ }\href {https://doi.org/10.30757/ALEA.v13-25} {\bibfield  {journal} {\bibinfo  {journal} {ALEA, Lat. Am. J. Probab. Math. Stat.}\ }\textbf {\bibinfo {volume} {13}},\ \bibinfo {pages} {635} (\bibinfo {year} {2016})}\BibitemShut {NoStop}%
\bibitem [{Note5()}]{Note5}%
  \BibitemOpen
  \bibinfo {note} {This behavior mirrors that of standard bosonic cat states, whose mean and variance deviate from the expected Poissonian value $\alpha ^2$ for small $\alpha $. The deviation vanishes exponentially with $\alpha ^2$.}\BibitemShut {Stop}%
\bibitem [{\citenamefont {Conway}\ and\ \citenamefont {Maxwell}(1962)}]{conway1961queueing}%
  \BibitemOpen
  \bibfield  {author} {\bibinfo {author} {\bibfnamefont {R.~W.}\ \bibnamefont {Conway}}\ and\ \bibinfo {author} {\bibfnamefont {W.~L.}\ \bibnamefont {Maxwell}},\ }\bibfield  {title} {\bibinfo {title} {A queuing model with state dependent service rates},\ }\href@noop {} {\bibfield  {journal} {\bibinfo  {journal} {J. Ind. Eng.}\ }\textbf {\bibinfo {volume} {12}},\ \bibinfo {pages} {132–136} (\bibinfo {year} {1962})}\BibitemShut {NoStop}%
\bibitem [{\citenamefont {Penrose}(1955)}]{penrose_generalized_1955}%
  \BibitemOpen
  \bibfield  {author} {\bibinfo {author} {\bibfnamefont {R.}~\bibnamefont {Penrose}},\ }\bibfield  {title} {\bibinfo {title} {A generalized inverse for matrices},\ }in\ \href@noop {} {\emph {\bibinfo {booktitle} {Mathematical proceedings of the Cambridge philosophical society}}},\ Vol.~\bibinfo {volume} {51}\ (\bibinfo {organization} {Cambridge University Press},\ \bibinfo {year} {1955})\ pp.\ \bibinfo {pages} {406--413}\BibitemShut {NoStop}%
\bibitem [{\citenamefont {Puri}\ \emph {et~al.}(2017)\citenamefont {Puri}, \citenamefont {Boutin},\ and\ \citenamefont {Blais}}]{puri_engineering_2017}%
  \BibitemOpen
  \bibfield  {author} {\bibinfo {author} {\bibfnamefont {S.}~\bibnamefont {Puri}}, \bibinfo {author} {\bibfnamefont {S.}~\bibnamefont {Boutin}},\ and\ \bibinfo {author} {\bibfnamefont {A.}~\bibnamefont {Blais}},\ }\bibfield  {title} {\bibinfo {title} {Engineering the quantum states of light in a {Kerr}-nonlinear resonator by two-photon driving},\ }\href {https://doi.org/10.1038/s41534-017-0019-1} {\bibfield  {journal} {\bibinfo  {journal} {npj Quantum Inf.}\ }\textbf {\bibinfo {volume} {3}},\ \bibinfo {pages} {1} (\bibinfo {year} {2017})}\BibitemShut {NoStop}%
\bibitem [{\citenamefont {Lieu}\ \emph {et~al.}(2020)\citenamefont {Lieu}, \citenamefont {Belyansky}, \citenamefont {Young}, \citenamefont {Lundgren}, \citenamefont {Albert},\ and\ \citenamefont {Gorshkov}}]{lieu_symmetry_2020}%
  \BibitemOpen
  \bibfield  {author} {\bibinfo {author} {\bibfnamefont {S.}~\bibnamefont {Lieu}}, \bibinfo {author} {\bibfnamefont {R.}~\bibnamefont {Belyansky}}, \bibinfo {author} {\bibfnamefont {J.~T.}\ \bibnamefont {Young}}, \bibinfo {author} {\bibfnamefont {R.}~\bibnamefont {Lundgren}}, \bibinfo {author} {\bibfnamefont {V.~V.}\ \bibnamefont {Albert}},\ and\ \bibinfo {author} {\bibfnamefont {A.~V.}\ \bibnamefont {Gorshkov}},\ }\bibfield  {title} {\bibinfo {title} {Symmetry {Breaking} and {Error} {Correction} in {Open} {Quantum} {Systems}},\ }\href {https://doi.org/10.1103/PhysRevLett.125.240405} {\bibfield  {journal} {\bibinfo  {journal} {Phys. Rev. Lett.}\ }\textbf {\bibinfo {volume} {125}},\ \bibinfo {pages} {240405} (\bibinfo {year} {2020})}\BibitemShut {NoStop}%
\bibitem [{\citenamefont {Lebreuilly}\ \emph {et~al.}(2021)\citenamefont {Lebreuilly}, \citenamefont {Noh}, \citenamefont {Wang}, \citenamefont {Girvin},\ and\ \citenamefont {Jiang}}]{lebreuilly_autonomous_2021}%
  \BibitemOpen
  \bibfield  {author} {\bibinfo {author} {\bibfnamefont {J.}~\bibnamefont {Lebreuilly}}, \bibinfo {author} {\bibfnamefont {K.}~\bibnamefont {Noh}}, \bibinfo {author} {\bibfnamefont {C.-H.}\ \bibnamefont {Wang}}, \bibinfo {author} {\bibfnamefont {S.~M.}\ \bibnamefont {Girvin}},\ and\ \bibinfo {author} {\bibfnamefont {L.}~\bibnamefont {Jiang}},\ }\bibfield  {title} {\bibinfo {title} {Autonomous quantum error correction and quantum computation},\ }\href {http://arxiv.org/abs/2103.05007} {\bibfield  {journal} {\bibinfo  {journal} {arXiv:quant-ph/2103.05007}\ } (\bibinfo {year} {2021})}\BibitemShut {NoStop}%
\bibitem [{\citenamefont {Lieu}\ \emph {et~al.}(2024)\citenamefont {Lieu}, \citenamefont {Liu},\ and\ \citenamefont {Gorshkov}}]{lieu_candidate_2023}%
  \BibitemOpen
  \bibfield  {author} {\bibinfo {author} {\bibfnamefont {S.}~\bibnamefont {Lieu}}, \bibinfo {author} {\bibfnamefont {Y.-J.}\ \bibnamefont {Liu}},\ and\ \bibinfo {author} {\bibfnamefont {A.~V.}\ \bibnamefont {Gorshkov}},\ }\bibfield  {title} {\bibinfo {title} {Candidate for a {Passively} {Protected} {Quantum} {Memory} in {Two} {Dimensions}},\ }\href {https://doi.org/10.1103/PhysRevLett.133.030601} {\bibfield  {journal} {\bibinfo  {journal} {Phys. Rev. Lett.}\ }\textbf {\bibinfo {volume} {133}},\ \bibinfo {pages} {030601} (\bibinfo {year} {2024})}\BibitemShut {NoStop}%
\bibitem [{\citenamefont {Baumgartner}\ and\ \citenamefont {Narnhofer}(2008)}]{baumgartner_analysis_2008}%
  \BibitemOpen
  \bibfield  {author} {\bibinfo {author} {\bibfnamefont {B.}~\bibnamefont {Baumgartner}}\ and\ \bibinfo {author} {\bibfnamefont {H.}~\bibnamefont {Narnhofer}},\ }\bibfield  {title} {\bibinfo {title} {Analysis of quantum semigroups with {GKS}–{Lindblad} generators: {II}. {General}},\ }\href {https://doi.org/10.1088/1751-8113/41/39/395303} {\bibfield  {journal} {\bibinfo  {journal} {J. Phys. A: Math. Theor.}\ }\textbf {\bibinfo {volume} {41}},\ \bibinfo {pages} {395303} (\bibinfo {year} {2008})}\BibitemShut {NoStop}%
\bibitem [{\citenamefont {Buča}\ and\ \citenamefont {Prosen}(2012)}]{buca_note_2012}%
  \BibitemOpen
  \bibfield  {author} {\bibinfo {author} {\bibfnamefont {B.}~\bibnamefont {Buča}}\ and\ \bibinfo {author} {\bibfnamefont {T.}~\bibnamefont {Prosen}},\ }\bibfield  {title} {\bibinfo {title} {A note on symmetry reductions of the {Lindblad} equation: transport in constrained open spin chains},\ }\href {https://doi.org/10.1088/1367-2630/14/7/073007} {\bibfield  {journal} {\bibinfo  {journal} {New J. Phys.}\ }\textbf {\bibinfo {volume} {14}},\ \bibinfo {pages} {073007} (\bibinfo {year} {2012})}\BibitemShut {NoStop}%
\bibitem [{\citenamefont {Pantaleoni}\ \emph {et~al.}(2020)\citenamefont {Pantaleoni}, \citenamefont {Baragiola},\ and\ \citenamefont {Menicucci}}]{pantaleoni_modular_2020}%
  \BibitemOpen
  \bibfield  {author} {\bibinfo {author} {\bibfnamefont {G.}~\bibnamefont {Pantaleoni}}, \bibinfo {author} {\bibfnamefont {B.~Q.}\ \bibnamefont {Baragiola}},\ and\ \bibinfo {author} {\bibfnamefont {N.~C.}\ \bibnamefont {Menicucci}},\ }\bibfield  {title} {\bibinfo {title} {Modular {Bosonic} {Subsystem} {Codes}},\ }\href {https://doi.org/10.1103/PhysRevLett.125.040501} {\bibfield  {journal} {\bibinfo  {journal} {Phys. Rev. Lett.}\ }\textbf {\bibinfo {volume} {125}},\ \bibinfo {pages} {040501} (\bibinfo {year} {2020})}\BibitemShut {NoStop}%
\bibitem [{\citenamefont {Glancy}\ and\ \citenamefont {Knill}(2006)}]{glancy_error_2006}%
  \BibitemOpen
  \bibfield  {author} {\bibinfo {author} {\bibfnamefont {S.}~\bibnamefont {Glancy}}\ and\ \bibinfo {author} {\bibfnamefont {E.}~\bibnamefont {Knill}},\ }\bibfield  {title} {\bibinfo {title} {Error analysis for encoding a qubit in an oscillator},\ }\href {https://doi.org/10.1103/PhysRevA.73.012325} {\bibfield  {journal} {\bibinfo  {journal} {Phys. Rev. A}\ }\textbf {\bibinfo {volume} {73}},\ \bibinfo {pages} {012325} (\bibinfo {year} {2006})}\BibitemShut {NoStop}%
\bibitem [{\citenamefont {Raynal}\ \emph {et~al.}(2010)\citenamefont {Raynal}, \citenamefont {Kalev}, \citenamefont {Suzuki},\ and\ \citenamefont {Englert}}]{raynal_encoding_2010}%
  \BibitemOpen
  \bibfield  {author} {\bibinfo {author} {\bibfnamefont {P.}~\bibnamefont {Raynal}}, \bibinfo {author} {\bibfnamefont {A.}~\bibnamefont {Kalev}}, \bibinfo {author} {\bibfnamefont {J.}~\bibnamefont {Suzuki}},\ and\ \bibinfo {author} {\bibfnamefont {B.-G.}\ \bibnamefont {Englert}},\ }\bibfield  {title} {\bibinfo {title} {Encoding many qubits in a rotor},\ }\href {https://doi.org/10.1103/PhysRevA.81.052327} {\bibfield  {journal} {\bibinfo  {journal} {Phys. Rev. A}\ }\textbf {\bibinfo {volume} {81}},\ \bibinfo {pages} {052327} (\bibinfo {year} {2010})}\BibitemShut {NoStop}%
\bibitem [{\citenamefont {Ketterer}\ \emph {et~al.}(2016)\citenamefont {Ketterer}, \citenamefont {Keller}, \citenamefont {Walborn}, \citenamefont {Coudreau},\ and\ \citenamefont {Milman}}]{ketterer_quantum_2016}%
  \BibitemOpen
  \bibfield  {author} {\bibinfo {author} {\bibfnamefont {A.}~\bibnamefont {Ketterer}}, \bibinfo {author} {\bibfnamefont {A.}~\bibnamefont {Keller}}, \bibinfo {author} {\bibfnamefont {S.~P.}\ \bibnamefont {Walborn}}, \bibinfo {author} {\bibfnamefont {T.}~\bibnamefont {Coudreau}},\ and\ \bibinfo {author} {\bibfnamefont {P.}~\bibnamefont {Milman}},\ }\bibfield  {title} {\bibinfo {title} {Quantum information processing in phase space: {A} modular variables approach},\ }\href {https://doi.org/10.1103/PhysRevA.94.022325} {\bibfield  {journal} {\bibinfo  {journal} {Phys. Rev. A}\ }\textbf {\bibinfo {volume} {94}},\ \bibinfo {pages} {022325} (\bibinfo {year} {2016})}\BibitemShut {NoStop}%
\bibitem [{\citenamefont {Duivenvoorden}\ \emph {et~al.}(2017)\citenamefont {Duivenvoorden}, \citenamefont {Terhal},\ and\ \citenamefont {Weigand}}]{duivenvoorden_single_2017}%
  \BibitemOpen
  \bibfield  {author} {\bibinfo {author} {\bibfnamefont {K.}~\bibnamefont {Duivenvoorden}}, \bibinfo {author} {\bibfnamefont {B.~M.}\ \bibnamefont {Terhal}},\ and\ \bibinfo {author} {\bibfnamefont {D.}~\bibnamefont {Weigand}},\ }\bibfield  {title} {\bibinfo {title} {Single-mode displacement sensor},\ }\href {https://doi.org/10.1103/PhysRevA.95.012305} {\bibfield  {journal} {\bibinfo  {journal} {Phys. Rev. A}\ }\textbf {\bibinfo {volume} {95}},\ \bibinfo {pages} {012305} (\bibinfo {year} {2017})}\BibitemShut {NoStop}%
\bibitem [{\citenamefont {Weigand}\ and\ \citenamefont {Terhal}(2018)}]{weigand_generating_2018}%
  \BibitemOpen
  \bibfield  {author} {\bibinfo {author} {\bibfnamefont {D.~J.}\ \bibnamefont {Weigand}}\ and\ \bibinfo {author} {\bibfnamefont {B.~M.}\ \bibnamefont {Terhal}},\ }\bibfield  {title} {\bibinfo {title} {Generating grid states from {Schr}{\textbackslash}"odinger-cat states without postselection},\ }\href {https://doi.org/10.1103/PhysRevA.97.022341} {\bibfield  {journal} {\bibinfo  {journal} {Phys. Rev. A}\ }\textbf {\bibinfo {volume} {97}},\ \bibinfo {pages} {022341} (\bibinfo {year} {2018})}\BibitemShut {NoStop}%
\bibitem [{\citenamefont {Matsuura}\ \emph {et~al.}(2020)\citenamefont {Matsuura}, \citenamefont {Yamasaki},\ and\ \citenamefont {Koashi}}]{matsuura_equivalence_2020}%
  \BibitemOpen
  \bibfield  {author} {\bibinfo {author} {\bibfnamefont {T.}~\bibnamefont {Matsuura}}, \bibinfo {author} {\bibfnamefont {H.}~\bibnamefont {Yamasaki}},\ and\ \bibinfo {author} {\bibfnamefont {M.}~\bibnamefont {Koashi}},\ }\bibfield  {title} {\bibinfo {title} {Equivalence of approximate {Gottesman}-{Kitaev}-{Preskill} codes},\ }\href {https://doi.org/10.1103/PhysRevA.102.032408} {\bibfield  {journal} {\bibinfo  {journal} {Phys. Rev. A}\ }\textbf {\bibinfo {volume} {102}},\ \bibinfo {pages} {032408} (\bibinfo {year} {2020})}\BibitemShut {NoStop}%
\bibitem [{\citenamefont {Albert}\ \emph {et~al.}(2020)\citenamefont {Albert}, \citenamefont {Covey},\ and\ \citenamefont {Preskill}}]{albert_robust_2020}%
  \BibitemOpen
  \bibfield  {author} {\bibinfo {author} {\bibfnamefont {V.~V.}\ \bibnamefont {Albert}}, \bibinfo {author} {\bibfnamefont {J.~P.}\ \bibnamefont {Covey}},\ and\ \bibinfo {author} {\bibfnamefont {J.}~\bibnamefont {Preskill}},\ }\bibfield  {title} {\bibinfo {title} {Robust {Encoding} of a {Qubit} in a {Molecule}},\ }\href {https://doi.org/10.1103/PhysRevX.10.031050} {\bibfield  {journal} {\bibinfo  {journal} {Phys. Rev. X}\ }\textbf {\bibinfo {volume} {10}},\ \bibinfo {pages} {031050} (\bibinfo {year} {2020})}\BibitemShut {NoStop}%
\bibitem [{Note6()}]{Note6}%
  \BibitemOpen
  \bibinfo {note} {Using the number operator $\protect \hat {n}$ as an input of these probability density functions is strictly speaking an abuse of notation which should be understood as ${\protect \mathbb {P}[K=\protect \hat {n}]=\DOTSB \sum@ \slimits@ _k\protect \mathbb {P}[K=k]\dyad {k}}$. The reshaping operator given in Eq.~\protect \eqref {eq:reshaping_func} is then a well-defined non-singular diagonal operator.}\BibitemShut {Stop}%
\bibitem [{\citenamefont {Azouit}\ \emph {et~al.}(2016)\citenamefont {Azouit}, \citenamefont {Sarlette},\ and\ \citenamefont {Rouchon}}]{azouit_well_posedness_2016}%
  \BibitemOpen
  \bibfield  {author} {\bibinfo {author} {\bibfnamefont {R.}~\bibnamefont {Azouit}}, \bibinfo {author} {\bibfnamefont {A.}~\bibnamefont {Sarlette}},\ and\ \bibinfo {author} {\bibfnamefont {P.}~\bibnamefont {Rouchon}},\ }\bibfield  {title} {\bibinfo {title} {Well-posedness and convergence of the {Lindblad} master equation for a quantum harmonic oscillator with multi-photon drive and damping},\ }\href {https://doi.org/10.1051/cocv/2016050} {\bibfield  {journal} {\bibinfo  {journal} {ESAIM: COCV}\ }\textbf {\bibinfo {volume} {22}},\ \bibinfo {pages} {1353} (\bibinfo {year} {2016})}\BibitemShut {NoStop}%
\bibitem [{\citenamefont {Guillaud}\ and\ \citenamefont {Mirrahimi}(2019)}]{guillaud_repetition_2019}%
  \BibitemOpen
  \bibfield  {author} {\bibinfo {author} {\bibfnamefont {J.}~\bibnamefont {Guillaud}}\ and\ \bibinfo {author} {\bibfnamefont {M.}~\bibnamefont {Mirrahimi}},\ }\bibfield  {title} {\bibinfo {title} {Repetition {Cat} {Qubits} for {Fault}-{Tolerant} {Quantum} {Computation}},\ }\href {https://doi.org/10.1103/PhysRevX.9.041053} {\bibfield  {journal} {\bibinfo  {journal} {Phys. Rev. X}\ }\textbf {\bibinfo {volume} {9}},\ \bibinfo {pages} {041053} (\bibinfo {year} {2019})}\BibitemShut {NoStop}%
\bibitem [{\citenamefont {Ruiz}\ \emph {et~al.}(2025)\citenamefont {Ruiz}, \citenamefont {Guillaud}, \citenamefont {Leverrier}, \citenamefont {Mirrahimi},\ and\ \citenamefont {Vuillot}}]{ruiz_ldpc-cat_2025}%
  \BibitemOpen
  \bibfield  {author} {\bibinfo {author} {\bibfnamefont {D.}~\bibnamefont {Ruiz}}, \bibinfo {author} {\bibfnamefont {J.}~\bibnamefont {Guillaud}}, \bibinfo {author} {\bibfnamefont {A.}~\bibnamefont {Leverrier}}, \bibinfo {author} {\bibfnamefont {M.}~\bibnamefont {Mirrahimi}},\ and\ \bibinfo {author} {\bibfnamefont {C.}~\bibnamefont {Vuillot}},\ }\bibfield  {title} {\bibinfo {title} {{LDPC}-cat codes for low-overhead quantum computing in {2D}},\ }\href {https://doi.org/10.1038/s41467-025-56298-8} {\bibfield  {journal} {\bibinfo  {journal} {Nat. Commun.}\ }\textbf {\bibinfo {volume} {16}},\ \bibinfo {pages} {1040} (\bibinfo {year} {2025})}\BibitemShut {NoStop}%
\bibitem [{\citenamefont {Vogel}\ and\ \citenamefont {de~Matos~Filho}(1995)}]{vogel_nonlinear_1995}%
  \BibitemOpen
  \bibfield  {author} {\bibinfo {author} {\bibfnamefont {W.}~\bibnamefont {Vogel}}\ and\ \bibinfo {author} {\bibfnamefont {R.~L.}\ \bibnamefont {de~Matos~Filho}},\ }\bibfield  {title} {\bibinfo {title} {Nonlinear {Jaynes}-{Cummings} dynamics of a trapped ion},\ }\href {https://doi.org/10.1103/PhysRevA.52.4214} {\bibfield  {journal} {\bibinfo  {journal} {Phys. Rev. A}\ }\textbf {\bibinfo {volume} {52}},\ \bibinfo {pages} {4214} (\bibinfo {year} {1995})}\BibitemShut {NoStop}%
\bibitem [{\citenamefont {Morigi}\ \emph {et~al.}(1997)\citenamefont {Morigi}, \citenamefont {Cirac}, \citenamefont {Lewenstein},\ and\ \citenamefont {Zoller}}]{morigi_ground_state_1997}%
  \BibitemOpen
  \bibfield  {author} {\bibinfo {author} {\bibfnamefont {G.}~\bibnamefont {Morigi}}, \bibinfo {author} {\bibfnamefont {J.~I.}\ \bibnamefont {Cirac}}, \bibinfo {author} {\bibfnamefont {M.}~\bibnamefont {Lewenstein}},\ and\ \bibinfo {author} {\bibfnamefont {P.}~\bibnamefont {Zoller}},\ }\bibfield  {title} {\bibinfo {title} {Ground-state laser cooling beyond the {Lamb}-{Dicke} limit},\ }\href {https://doi.org/10.1209/epl/i1997-00306-3} {\bibfield  {journal} {\bibinfo  {journal} {Europhys. Lett.}\ }\textbf {\bibinfo {volume} {39}},\ \bibinfo {pages} {13} (\bibinfo {year} {1997})}\BibitemShut {NoStop}%
\bibitem [{\citenamefont {Morigi}\ \emph {et~al.}(1999)\citenamefont {Morigi}, \citenamefont {Eschner}, \citenamefont {Cirac},\ and\ \citenamefont {Zoller}}]{morigi_laser_1999}%
  \BibitemOpen
  \bibfield  {author} {\bibinfo {author} {\bibfnamefont {G.}~\bibnamefont {Morigi}}, \bibinfo {author} {\bibfnamefont {J.}~\bibnamefont {Eschner}}, \bibinfo {author} {\bibfnamefont {J.~I.}\ \bibnamefont {Cirac}},\ and\ \bibinfo {author} {\bibfnamefont {P.}~\bibnamefont {Zoller}},\ }\bibfield  {title} {\bibinfo {title} {Laser cooling of two trapped ions: {Sideband} cooling beyond the {Lamb}-{Dicke} limit},\ }\href {https://doi.org/10.1103/PhysRevA.59.3797} {\bibfield  {journal} {\bibinfo  {journal} {Phys. Rev. A}\ }\textbf {\bibinfo {volume} {59}},\ \bibinfo {pages} {3797} (\bibinfo {year} {1999})}\BibitemShut {NoStop}%
\bibitem [{\citenamefont {Wallentowitz}\ \emph {et~al.}(1999)\citenamefont {Wallentowitz}, \citenamefont {Vogel},\ and\ \citenamefont {Knight}}]{wallentowitz_high_order_1999}%
  \BibitemOpen
  \bibfield  {author} {\bibinfo {author} {\bibfnamefont {S.}~\bibnamefont {Wallentowitz}}, \bibinfo {author} {\bibfnamefont {W.}~\bibnamefont {Vogel}},\ and\ \bibinfo {author} {\bibfnamefont {P.~L.}\ \bibnamefont {Knight}},\ }\bibfield  {title} {\bibinfo {title} {High-order nonlinearities in the motion of a trapped atom},\ }\href {https://doi.org/10.1103/PhysRevA.59.531} {\bibfield  {journal} {\bibinfo  {journal} {Phys. Rev. A}\ }\textbf {\bibinfo {volume} {59}},\ \bibinfo {pages} {531} (\bibinfo {year} {1999})}\BibitemShut {NoStop}%
\bibitem [{\citenamefont {Carvalho}\ \emph {et~al.}(2001)\citenamefont {Carvalho}, \citenamefont {Milman}, \citenamefont {de~Matos~Filho},\ and\ \citenamefont {Davidovich}}]{carvalho_decoherence_2001}%
  \BibitemOpen
  \bibfield  {author} {\bibinfo {author} {\bibfnamefont {A.}~\bibnamefont {Carvalho}}, \bibinfo {author} {\bibfnamefont {P.}~\bibnamefont {Milman}}, \bibinfo {author} {\bibfnamefont {R.}~\bibnamefont {de~Matos~Filho}},\ and\ \bibinfo {author} {\bibfnamefont {L.}~\bibnamefont {Davidovich}},\ }\bibfield  {title} {\bibinfo {title} {Decoherence, pointer engineering and quantum state protection},\ }in\ \href@noop {} {\emph {\bibinfo {booktitle} {Modern {Challenges} in {Quantum} {Optics}: {Selected} {Papers} of the {First} {International} {Meeting} in {Quantum} {Optics} {Held} in {Santiago}, {Chile}, 13–16 {August} 2000}}}\ (\bibinfo  {publisher} {Springer},\ \bibinfo {year} {2001})\ pp.\ \bibinfo {pages} {65--79}\BibitemShut {NoStop}%
\bibitem [{\citenamefont {Cheng}\ \emph {et~al.}(2018)\citenamefont {Cheng}, \citenamefont {Arrazola}, \citenamefont {Pedernales}, \citenamefont {Lamata}, \citenamefont {Chen},\ and\ \citenamefont {Solano}}]{cheng_nonlinear_2018}%
  \BibitemOpen
  \bibfield  {author} {\bibinfo {author} {\bibfnamefont {X.-H.}\ \bibnamefont {Cheng}}, \bibinfo {author} {\bibfnamefont {I.}~\bibnamefont {Arrazola}}, \bibinfo {author} {\bibfnamefont {J.~S.}\ \bibnamefont {Pedernales}}, \bibinfo {author} {\bibfnamefont {L.}~\bibnamefont {Lamata}}, \bibinfo {author} {\bibfnamefont {X.}~\bibnamefont {Chen}},\ and\ \bibinfo {author} {\bibfnamefont {E.}~\bibnamefont {Solano}},\ }\bibfield  {title} {\bibinfo {title} {Nonlinear quantum {Rabi} model in trapped ions},\ }\href {https://doi.org/10.1103/PhysRevA.97.023624} {\bibfield  {journal} {\bibinfo  {journal} {Phys. Rev. A}\ }\textbf {\bibinfo {volume} {97}},\ \bibinfo {pages} {023624} (\bibinfo {year} {2018})}\BibitemShut {NoStop}%
\bibitem [{\citenamefont {Joshi}\ \emph {et~al.}(2019)\citenamefont {Joshi}, \citenamefont {Hrmo}, \citenamefont {Jarlaud}, \citenamefont {Oehl},\ and\ \citenamefont {Thompson}}]{joshi_population_2019}%
  \BibitemOpen
  \bibfield  {author} {\bibinfo {author} {\bibfnamefont {M.~K.}\ \bibnamefont {Joshi}}, \bibinfo {author} {\bibfnamefont {P.}~\bibnamefont {Hrmo}}, \bibinfo {author} {\bibfnamefont {V.}~\bibnamefont {Jarlaud}}, \bibinfo {author} {\bibfnamefont {F.}~\bibnamefont {Oehl}},\ and\ \bibinfo {author} {\bibfnamefont {R.~C.}\ \bibnamefont {Thompson}},\ }\bibfield  {title} {\bibinfo {title} {Population dynamics in sideband cooling of trapped ions outside the {Lamb}-{Dicke} regime},\ }\href {https://doi.org/10.1103/PhysRevA.99.013423} {\bibfield  {journal} {\bibinfo  {journal} {Phys. Rev. A}\ }\textbf {\bibinfo {volume} {99}},\ \bibinfo {pages} {013423} (\bibinfo {year} {2019})}\BibitemShut {NoStop}%
\bibitem [{\citenamefont {Puebla}\ \emph {et~al.}(2019)\citenamefont {Puebla}, \citenamefont {Casanova}, \citenamefont {Houhou}, \citenamefont {Solano},\ and\ \citenamefont {Paternostro}}]{puebla_quantum_2019}%
  \BibitemOpen
  \bibfield  {author} {\bibinfo {author} {\bibfnamefont {R.}~\bibnamefont {Puebla}}, \bibinfo {author} {\bibfnamefont {J.}~\bibnamefont {Casanova}}, \bibinfo {author} {\bibfnamefont {O.}~\bibnamefont {Houhou}}, \bibinfo {author} {\bibfnamefont {E.}~\bibnamefont {Solano}},\ and\ \bibinfo {author} {\bibfnamefont {M.}~\bibnamefont {Paternostro}},\ }\bibfield  {title} {\bibinfo {title} {Quantum simulation of multiphoton and nonlinear dissipative spin-boson models},\ }\href {https://doi.org/10.1103/PhysRevA.99.032303} {\bibfield  {journal} {\bibinfo  {journal} {Phys. Rev. A}\ }\textbf {\bibinfo {volume} {99}},\ \bibinfo {pages} {032303} (\bibinfo {year} {2019})}\BibitemShut {NoStop}%
\bibitem [{\citenamefont {Jain}\ \emph {et~al.}(2024)\citenamefont {Jain}, \citenamefont {Sägesser}, \citenamefont {Hrmo}, \citenamefont {Torkzaban}, \citenamefont {Stadler}, \citenamefont {Oswald}, \citenamefont {Axline}, \citenamefont {Bautista-Salvador}, \citenamefont {Ospelkaus}, \citenamefont {Kienzler},\ and\ \citenamefont {Home}}]{jain_penning_2024}%
  \BibitemOpen
  \bibfield  {author} {\bibinfo {author} {\bibfnamefont {S.}~\bibnamefont {Jain}}, \bibinfo {author} {\bibfnamefont {T.}~\bibnamefont {Sägesser}}, \bibinfo {author} {\bibfnamefont {P.}~\bibnamefont {Hrmo}}, \bibinfo {author} {\bibfnamefont {C.}~\bibnamefont {Torkzaban}}, \bibinfo {author} {\bibfnamefont {M.}~\bibnamefont {Stadler}}, \bibinfo {author} {\bibfnamefont {R.}~\bibnamefont {Oswald}}, \bibinfo {author} {\bibfnamefont {C.}~\bibnamefont {Axline}}, \bibinfo {author} {\bibfnamefont {A.}~\bibnamefont {Bautista-Salvador}}, \bibinfo {author} {\bibfnamefont {C.}~\bibnamefont {Ospelkaus}}, \bibinfo {author} {\bibfnamefont {D.}~\bibnamefont {Kienzler}},\ and\ \bibinfo {author} {\bibfnamefont {J.}~\bibnamefont {Home}},\ }\bibfield  {title} {\bibinfo {title} {Penning micro-trap for quantum computing},\ }\href {https://doi.org/10.1038/s41586-024-07111-x} {\bibfield  {journal} {\bibinfo  {journal} {Nature}\ }\textbf {\bibinfo {volume} {627}},\ \bibinfo {pages} {510} (\bibinfo {year} {2024})}\BibitemShut {NoStop}%
\bibitem [{\citenamefont {Stenholm}(1986)}]{stenholm_semiclassical_1986}%
  \BibitemOpen
  \bibfield  {author} {\bibinfo {author} {\bibfnamefont {S.}~\bibnamefont {Stenholm}},\ }\bibfield  {title} {\bibinfo {title} {The semiclassical theory of laser cooling},\ }\href {https://doi.org/10.1103/RevModPhys.58.699} {\bibfield  {journal} {\bibinfo  {journal} {Rev. Mod. Phys.}\ }\textbf {\bibinfo {volume} {58}},\ \bibinfo {pages} {699} (\bibinfo {year} {1986})}\BibitemShut {NoStop}%
\bibitem [{\citenamefont {Cirac}\ \emph {et~al.}(1992)\citenamefont {Cirac}, \citenamefont {Blatt}, \citenamefont {Zoller},\ and\ \citenamefont {Phillips}}]{cirac_laser_1992}%
  \BibitemOpen
  \bibfield  {author} {\bibinfo {author} {\bibfnamefont {J.~I.}\ \bibnamefont {Cirac}}, \bibinfo {author} {\bibfnamefont {R.}~\bibnamefont {Blatt}}, \bibinfo {author} {\bibfnamefont {P.}~\bibnamefont {Zoller}},\ and\ \bibinfo {author} {\bibfnamefont {W.~D.}\ \bibnamefont {Phillips}},\ }\bibfield  {title} {\bibinfo {title} {Laser cooling of trapped ions in a standing wave},\ }\href {https://doi.org/10.1103/PhysRevA.46.2668} {\bibfield  {journal} {\bibinfo  {journal} {Phys. Rev. A}\ }\textbf {\bibinfo {volume} {46}},\ \bibinfo {pages} {2668} (\bibinfo {year} {1992})}\BibitemShut {NoStop}%
\bibitem [{\citenamefont {Kienzler}\ \emph {et~al.}(2016)\citenamefont {Kienzler}, \citenamefont {Fl\"uhmann}, \citenamefont {Negnevitsky}, \citenamefont {Lo}, \citenamefont {Marinelli}, \citenamefont {Nadlinger},\ and\ \citenamefont {Home}}]{kienzler_observation_2016}%
  \BibitemOpen
  \bibfield  {author} {\bibinfo {author} {\bibfnamefont {D.}~\bibnamefont {Kienzler}}, \bibinfo {author} {\bibfnamefont {C.}~\bibnamefont {Fl\"uhmann}}, \bibinfo {author} {\bibfnamefont {V.}~\bibnamefont {Negnevitsky}}, \bibinfo {author} {\bibfnamefont {H.-Y.}\ \bibnamefont {Lo}}, \bibinfo {author} {\bibfnamefont {M.}~\bibnamefont {Marinelli}}, \bibinfo {author} {\bibfnamefont {D.}~\bibnamefont {Nadlinger}},\ and\ \bibinfo {author} {\bibfnamefont {J.~P.}\ \bibnamefont {Home}},\ }\bibfield  {title} {\bibinfo {title} {Observation of {Quantum} {Interference} between {Separated} {Mechanical} {Oscillator} {Wave} {Packets}},\ }\href {https://doi.org/10.1103/PhysRevLett.116.140402} {\bibfield  {journal} {\bibinfo  {journal} {Phys. Rev. Lett.}\ }\textbf {\bibinfo {volume} {116}},\ \bibinfo {pages} {140402} (\bibinfo {year} {2016})}\BibitemShut {NoStop}%
\bibitem [{\citenamefont {Fl\"uhmann}\ \emph {et~al.}(2018)\citenamefont {Fl\"uhmann}, \citenamefont {Negnevitsky}, \citenamefont {Marinelli},\ and\ \citenamefont {Home}}]{fluhmann_sequential_2018}%
  \BibitemOpen
  \bibfield  {author} {\bibinfo {author} {\bibfnamefont {C.}~\bibnamefont {Fl\"uhmann}}, \bibinfo {author} {\bibfnamefont {V.}~\bibnamefont {Negnevitsky}}, \bibinfo {author} {\bibfnamefont {M.}~\bibnamefont {Marinelli}},\ and\ \bibinfo {author} {\bibfnamefont {J.~P.}\ \bibnamefont {Home}},\ }\bibfield  {title} {\bibinfo {title} {Sequential {Modular} {Position} and {Momentum} {Measurements} of a {Trapped} {Ion} {Mechanical} {Oscillator}},\ }\href {https://doi.org/10.1103/PhysRevX.8.021001} {\bibfield  {journal} {\bibinfo  {journal} {Phys. Rev. X}\ }\textbf {\bibinfo {volume} {8}},\ \bibinfo {pages} {021001} (\bibinfo {year} {2018})}\BibitemShut {NoStop}%
\bibitem [{Note7()}]{Note7}%
  \BibitemOpen
  \bibinfo {note} {We consider $\protect \hat {a}$ and $\protect \hat {c}$ to be already the dressed modes of the circuit.}\BibitemShut {Stop}%
\bibitem [{\citenamefont {Armour}\ \emph {et~al.}(2013)\citenamefont {Armour}, \citenamefont {Blencowe}, \citenamefont {Brahimi},\ and\ \citenamefont {Rimberg}}]{armour_universal_2013}%
  \BibitemOpen
  \bibfield  {author} {\bibinfo {author} {\bibfnamefont {A.~D.}\ \bibnamefont {Armour}}, \bibinfo {author} {\bibfnamefont {M.~P.}\ \bibnamefont {Blencowe}}, \bibinfo {author} {\bibfnamefont {E.}~\bibnamefont {Brahimi}},\ and\ \bibinfo {author} {\bibfnamefont {A.~J.}\ \bibnamefont {Rimberg}},\ }\bibfield  {title} {\bibinfo {title} {Universal {Quantum} {Fluctuations} of a {Cavity} {Mode} {Driven} by a {Josephson} {Junction}},\ }\href {https://doi.org/10.1103/PhysRevLett.111.247001} {\bibfield  {journal} {\bibinfo  {journal} {Phys. Rev. Lett.}\ }\textbf {\bibinfo {volume} {111}},\ \bibinfo {pages} {247001} (\bibinfo {year} {2013})}\BibitemShut {NoStop}%
\bibitem [{\citenamefont {Gramich}\ \emph {et~al.}(2013)\citenamefont {Gramich}, \citenamefont {Kubala}, \citenamefont {Rohrer},\ and\ \citenamefont {Ankerhold}}]{gramich_coulomb-blockade_2013}%
  \BibitemOpen
  \bibfield  {author} {\bibinfo {author} {\bibfnamefont {V.}~\bibnamefont {Gramich}}, \bibinfo {author} {\bibfnamefont {B.}~\bibnamefont {Kubala}}, \bibinfo {author} {\bibfnamefont {S.}~\bibnamefont {Rohrer}},\ and\ \bibinfo {author} {\bibfnamefont {J.}~\bibnamefont {Ankerhold}},\ }\bibfield  {title} {\bibinfo {title} {From {Coulomb}-{Blockade} to {Nonlinear} {Quantum} {Dynamics} in a {Superconducting} {Circuit} with a {Resonator}},\ }\href {https://doi.org/10.1103/PhysRevLett.111.247002} {\bibfield  {journal} {\bibinfo  {journal} {Phys. Rev. Lett.}\ }\textbf {\bibinfo {volume} {111}},\ \bibinfo {pages} {247002} (\bibinfo {year} {2013})}\BibitemShut {NoStop}%
\bibitem [{\citenamefont {Trif}\ and\ \citenamefont {Simon}(2015)}]{trif_photon_2015}%
  \BibitemOpen
  \bibfield  {author} {\bibinfo {author} {\bibfnamefont {M.}~\bibnamefont {Trif}}\ and\ \bibinfo {author} {\bibfnamefont {P.}~\bibnamefont {Simon}},\ }\bibfield  {title} {\bibinfo {title} {Photon cross-correlations emitted by a {Josephson} junction in two microwave cavities},\ }\href {https://doi.org/10.1103/PhysRevB.92.014503} {\bibfield  {journal} {\bibinfo  {journal} {Phys. Rev. B}\ }\textbf {\bibinfo {volume} {92}},\ \bibinfo {pages} {014503} (\bibinfo {year} {2015})}\BibitemShut {NoStop}%
\bibitem [{\citenamefont {Hofer}\ \emph {et~al.}(2016)\citenamefont {Hofer}, \citenamefont {Souquet},\ and\ \citenamefont {Clerk}}]{hofer_quantum_2016}%
  \BibitemOpen
  \bibfield  {author} {\bibinfo {author} {\bibfnamefont {P.~P.}\ \bibnamefont {Hofer}}, \bibinfo {author} {\bibfnamefont {J.-R.}\ \bibnamefont {Souquet}},\ and\ \bibinfo {author} {\bibfnamefont {A.~A.}\ \bibnamefont {Clerk}},\ }\bibfield  {title} {\bibinfo {title} {Quantum heat engine based on photon-assisted cooper pair tunneling},\ }\href {https://doi.org/10.1103/PhysRevB.93.041418} {\bibfield  {journal} {\bibinfo  {journal} {Phys. Rev. B}\ }\textbf {\bibinfo {volume} {93}},\ \bibinfo {pages} {041418} (\bibinfo {year} {2016})}\BibitemShut {NoStop}%
\bibitem [{\citenamefont {Souquet}\ and\ \citenamefont {Clerk}(2016)}]{souquet_fock-state_2016}%
  \BibitemOpen
  \bibfield  {author} {\bibinfo {author} {\bibfnamefont {J.-R.}\ \bibnamefont {Souquet}}\ and\ \bibinfo {author} {\bibfnamefont {A.~A.}\ \bibnamefont {Clerk}},\ }\bibfield  {title} {\bibinfo {title} {Fock-state stabilization and emission in superconducting circuits using dc-biased josephson junctions},\ }\href {https://doi.org/10.1103/PhysRevA.93.060301} {\bibfield  {journal} {\bibinfo  {journal} {Phys. Rev. A}\ }\textbf {\bibinfo {volume} {93}},\ \bibinfo {pages} {060301} (\bibinfo {year} {2016})}\BibitemShut {NoStop}%
\bibitem [{\citenamefont {Nathan}\ \emph {et~al.}(2025)\citenamefont {Nathan}, \citenamefont {O’Brien}, \citenamefont {Noh}, \citenamefont {Matheny}, \citenamefont {Grimsmo}, \citenamefont {Jiang},\ and\ \citenamefont {Refael}}]{nathan_self-correcting_2025}%
  \BibitemOpen
  \bibfield  {author} {\bibinfo {author} {\bibfnamefont {F.}~\bibnamefont {Nathan}}, \bibinfo {author} {\bibfnamefont {L.}~\bibnamefont {O’Brien}}, \bibinfo {author} {\bibfnamefont {K.}~\bibnamefont {Noh}}, \bibinfo {author} {\bibfnamefont {M.~H.}\ \bibnamefont {Matheny}}, \bibinfo {author} {\bibfnamefont {A.~L.}\ \bibnamefont {Grimsmo}}, \bibinfo {author} {\bibfnamefont {L.}~\bibnamefont {Jiang}},\ and\ \bibinfo {author} {\bibfnamefont {G.}~\bibnamefont {Refael}},\ }\bibfield  {title} {\bibinfo {title} {Self-{Correcting} {Gottesman}-{Kitaev}-{Preskill} {Qubit} and {Gates} in a {Driven}-{Dissipative} {Circuit}},\ }\href {https://doi.org/10.1103/ykqb-m52z} {\bibfield  {journal} {\bibinfo  {journal} {PRX Quantum}\ }\textbf {\bibinfo {volume} {6}},\ \bibinfo {pages} {030352} (\bibinfo {year} {2025})}\BibitemShut {NoStop}%
\bibitem [{\citenamefont {Weedbrook}\ \emph {et~al.}(2012)\citenamefont {Weedbrook}, \citenamefont {Pirandola}, \citenamefont {Garc\'{\i}a-Patr\'on}, \citenamefont {Cerf}, \citenamefont {Ralph}, \citenamefont {Shapiro},\ and\ \citenamefont {Lloyd}}]{weedbrook_gaussian_2012}%
  \BibitemOpen
  \bibfield  {author} {\bibinfo {author} {\bibfnamefont {C.}~\bibnamefont {Weedbrook}}, \bibinfo {author} {\bibfnamefont {S.}~\bibnamefont {Pirandola}}, \bibinfo {author} {\bibfnamefont {R.}~\bibnamefont {Garc\'{\i}a-Patr\'on}}, \bibinfo {author} {\bibfnamefont {N.~J.}\ \bibnamefont {Cerf}}, \bibinfo {author} {\bibfnamefont {T.~C.}\ \bibnamefont {Ralph}}, \bibinfo {author} {\bibfnamefont {J.~H.}\ \bibnamefont {Shapiro}},\ and\ \bibinfo {author} {\bibfnamefont {S.}~\bibnamefont {Lloyd}},\ }\bibfield  {title} {\bibinfo {title} {Gaussian quantum information},\ }\href {https://doi.org/10.1103/RevModPhys.84.621} {\bibfield  {journal} {\bibinfo  {journal} {Rev. Mod. Phys.}\ }\textbf {\bibinfo {volume} {84}},\ \bibinfo {pages} {621} (\bibinfo {year} {2012})}\BibitemShut {NoStop}%
\bibitem [{\citenamefont {Royer}\ \emph {et~al.}(2020)\citenamefont {Royer}, \citenamefont {Singh},\ and\ \citenamefont {Girvin}}]{royer_stabilization_2020}%
  \BibitemOpen
  \bibfield  {author} {\bibinfo {author} {\bibfnamefont {B.}~\bibnamefont {Royer}}, \bibinfo {author} {\bibfnamefont {S.}~\bibnamefont {Singh}},\ and\ \bibinfo {author} {\bibfnamefont {S.~M.}\ \bibnamefont {Girvin}},\ }\bibfield  {title} {\bibinfo {title} {Stabilization of {Finite}-{Energy} {Gottesman}-{Kitaev}-{Preskill} {States}},\ }\href {https://doi.org/10.1103/PhysRevLett.125.260509} {\bibfield  {journal} {\bibinfo  {journal} {Phys. Rev. Lett.}\ }\textbf {\bibinfo {volume} {125}},\ \bibinfo {pages} {260509} (\bibinfo {year} {2020})}\BibitemShut {NoStop}%
\bibitem [{\citenamefont {Rosenblum}\ \emph {et~al.}(2018)\citenamefont {Rosenblum}, \citenamefont {Reinhold}, \citenamefont {Mirrahimi}, \citenamefont {Jiang}, \citenamefont {Frunzio},\ and\ \citenamefont {Schoelkopf}}]{rosenblum_fault_2018}%
  \BibitemOpen
  \bibfield  {author} {\bibinfo {author} {\bibfnamefont {S.}~\bibnamefont {Rosenblum}}, \bibinfo {author} {\bibfnamefont {P.}~\bibnamefont {Reinhold}}, \bibinfo {author} {\bibfnamefont {M.}~\bibnamefont {Mirrahimi}}, \bibinfo {author} {\bibfnamefont {L.}~\bibnamefont {Jiang}}, \bibinfo {author} {\bibfnamefont {L.}~\bibnamefont {Frunzio}},\ and\ \bibinfo {author} {\bibfnamefont {R.~J.}\ \bibnamefont {Schoelkopf}},\ }\bibfield  {title} {\bibinfo {title} {Fault-tolerant detection of a quantum error},\ }\href {https://doi.org/10.1126/science.aat3996} {\bibfield  {journal} {\bibinfo  {journal} {Science}\ }\textbf {\bibinfo {volume} {361}},\ \bibinfo {pages} {266} (\bibinfo {year} {2018})}\BibitemShut {NoStop}%
\bibitem [{\citenamefont {Bild}\ \emph {et~al.}(2023)\citenamefont {Bild}, \citenamefont {Fadel}, \citenamefont {Yang}, \citenamefont {von L{\"u}pke}, \citenamefont {Martin}, \citenamefont {Bruno},\ and\ \citenamefont {Chu}}]{bild_schrodinger_2023}%
  \BibitemOpen
  \bibfield  {author} {\bibinfo {author} {\bibfnamefont {M.}~\bibnamefont {Bild}}, \bibinfo {author} {\bibfnamefont {M.}~\bibnamefont {Fadel}}, \bibinfo {author} {\bibfnamefont {Y.}~\bibnamefont {Yang}}, \bibinfo {author} {\bibfnamefont {U.}~\bibnamefont {von L{\"u}pke}}, \bibinfo {author} {\bibfnamefont {P.}~\bibnamefont {Martin}}, \bibinfo {author} {\bibfnamefont {A.}~\bibnamefont {Bruno}},\ and\ \bibinfo {author} {\bibfnamefont {Y.}~\bibnamefont {Chu}},\ }\bibfield  {title} {\bibinfo {title} {Schrödinger cat states of a 16-microgram mechanical oscillator},\ }\href {https://doi.org/10.1126/science.adf7553} {\bibfield  {journal} {\bibinfo  {journal} {Science}\ }\textbf {\bibinfo {volume} {380}},\ \bibinfo {pages} {274} (\bibinfo {year} {2023})}\BibitemShut {NoStop}%
\bibitem [{\citenamefont {Marti}\ \emph {et~al.}(2024)\citenamefont {Marti}, \citenamefont {von L{\"u}pke}, \citenamefont {Joshi}, \citenamefont {Yang}, \citenamefont {Bild}, \citenamefont {Omahen}, \citenamefont {Chu},\ and\ \citenamefont {Fadel}}]{marti_quantum_2024}%
  \BibitemOpen
  \bibfield  {author} {\bibinfo {author} {\bibfnamefont {S.}~\bibnamefont {Marti}}, \bibinfo {author} {\bibfnamefont {U.}~\bibnamefont {von L{\"u}pke}}, \bibinfo {author} {\bibfnamefont {O.}~\bibnamefont {Joshi}}, \bibinfo {author} {\bibfnamefont {Y.}~\bibnamefont {Yang}}, \bibinfo {author} {\bibfnamefont {M.}~\bibnamefont {Bild}}, \bibinfo {author} {\bibfnamefont {A.}~\bibnamefont {Omahen}}, \bibinfo {author} {\bibfnamefont {Y.}~\bibnamefont {Chu}},\ and\ \bibinfo {author} {\bibfnamefont {M.}~\bibnamefont {Fadel}},\ }\bibfield  {title} {\bibinfo {title} {Quantum squeezing in a nonlinear mechanical oscillator},\ }\href {https://doi.org/10.1038/s41567-024-02545-6} {\bibfield  {journal} {\bibinfo  {journal} {Nat. Phys.}\ ,\ \bibinfo {pages} {1}} (\bibinfo {year} {2024})}\BibitemShut {NoStop}%
\bibitem [{\citenamefont {Shi}\ \emph {et~al.}(2018)\citenamefont {Shi}, \citenamefont {Demler},\ and\ \citenamefont {Ignacio~Cirac}}]{shi_variational_2018}%
  \BibitemOpen
  \bibfield  {author} {\bibinfo {author} {\bibfnamefont {T.}~\bibnamefont {Shi}}, \bibinfo {author} {\bibfnamefont {E.}~\bibnamefont {Demler}},\ and\ \bibinfo {author} {\bibfnamefont {J.}~\bibnamefont {Ignacio~Cirac}},\ }\bibfield  {title} {\bibinfo {title} {Variational study of fermionic and bosonic systems with non-{Gaussian} states: {Theory} and applications},\ }\href {https://doi.org/10.1016/j.aop.2017.11.014} {\bibfield  {journal} {\bibinfo  {journal} {Ann. Phys.}\ }\textbf {\bibinfo {volume} {390}},\ \bibinfo {pages} {245} (\bibinfo {year} {2018})}\BibitemShut {NoStop}%
\bibitem [{\citenamefont {Walschaers}(2021)}]{walschaers_non-gaussian_2021}%
  \BibitemOpen
  \bibfield  {author} {\bibinfo {author} {\bibfnamefont {M.}~\bibnamefont {Walschaers}},\ }\bibfield  {title} {\bibinfo {title} {Non-{Gaussian} {Quantum} {States} and {Where} to {Find} {Them}},\ }\href {https://doi.org/10.1103/PRXQuantum.2.030204} {\bibfield  {journal} {\bibinfo  {journal} {PRX Quantum}\ }\textbf {\bibinfo {volume} {2}},\ \bibinfo {pages} {030204} (\bibinfo {year} {2021})}\BibitemShut {NoStop}%
\bibitem [{\citenamefont {Wetherbee}\ \emph {et~al.}(2025)\citenamefont {Wetherbee}, \citenamefont {Roy}, \citenamefont {Royer},\ and\ \citenamefont {Fatemi}}]{wetherbee_mathematical_2025}%
  \BibitemOpen
  \bibfield  {author} {\bibinfo {author} {\bibfnamefont {O.~C.}\ \bibnamefont {Wetherbee}}, \bibinfo {author} {\bibfnamefont {S.}~\bibnamefont {Roy}}, \bibinfo {author} {\bibfnamefont {B.}~\bibnamefont {Royer}},\ and\ \bibinfo {author} {\bibfnamefont {V.}~\bibnamefont {Fatemi}},\ }\bibfield  {title} {\bibinfo {title} {A {Mathematical} {Structure} for {Amplitude}-{Mixing} {Error}-{Transparent} {Gates} for {Binomial} {Codes}},\ }\href {https://doi.org/10.22331/q-2025-10-21-1890} {\bibfield  {journal} {\bibinfo  {journal} {Quantum}\ }\textbf {\bibinfo {volume} {9}},\ \bibinfo {pages} {1890} (\bibinfo {year} {2025})}\BibitemShut {NoStop}%
\bibitem [{\citenamefont {Guo}\ \emph {et~al.}(2025)\citenamefont {Guo}, \citenamefont {Huang},\ and\ \citenamefont {Du}}]{guo_engineering_2025}%
  \BibitemOpen
  \bibfield  {author} {\bibinfo {author} {\bibfnamefont {L.}~\bibnamefont {Guo}}, \bibinfo {author} {\bibfnamefont {T.}~\bibnamefont {Huang}},\ and\ \bibinfo {author} {\bibfnamefont {L.}~\bibnamefont {Du}},\ }\bibfield  {title} {\bibinfo {title} {Engineering bosonic codes with quantum lattice gates},\ }\href {https://doi.org/10.1038/s42005-025-02354-0} {\bibfield  {journal} {\bibinfo  {journal} {Communications Physics}\ }\textbf {\bibinfo {volume} {8}},\ \bibinfo {pages} {414} (\bibinfo {year} {2025})}\BibitemShut {NoStop}%
\bibitem [{\citenamefont {Aissaoui}\ \emph {et~al.}(2024)\citenamefont {Aissaoui}, \citenamefont {Murani}, \citenamefont {Lescanne},\ and\ \citenamefont {Sarlette}}]{aissaoui_cat_2024}%
  \BibitemOpen
  \bibfield  {author} {\bibinfo {author} {\bibfnamefont {T.}~\bibnamefont {Aissaoui}}, \bibinfo {author} {\bibfnamefont {A.}~\bibnamefont {Murani}}, \bibinfo {author} {\bibfnamefont {R.}~\bibnamefont {Lescanne}},\ and\ \bibinfo {author} {\bibfnamefont {A.}~\bibnamefont {Sarlette}},\ }\bibfield  {title} {\bibinfo {title} {A cat qubit stabilization scheme using a voltage biased {Josephson} junction},\ }\bibfield  {journal} {\bibinfo  {journal} {arXiv:2411.08132}\ }\href {https://doi.org/10.48550/arXiv.2411.08132} {10.48550/arXiv.2411.08132} (\bibinfo {year} {2024})\BibitemShut {NoStop}%
\bibitem [{\citenamefont {Danner}\ \emph {et~al.}(2025)\citenamefont {Danner}, \citenamefont {Höhe}, \citenamefont {Padurariu}, \citenamefont {Ankerhold},\ and\ \citenamefont {Kubala}}]{danner_quantum_2025}%
  \BibitemOpen
  \bibfield  {author} {\bibinfo {author} {\bibfnamefont {L.}~\bibnamefont {Danner}}, \bibinfo {author} {\bibfnamefont {F.}~\bibnamefont {Höhe}}, \bibinfo {author} {\bibfnamefont {C.}~\bibnamefont {Padurariu}}, \bibinfo {author} {\bibfnamefont {J.}~\bibnamefont {Ankerhold}},\ and\ \bibinfo {author} {\bibfnamefont {B.}~\bibnamefont {Kubala}},\ }\bibfield  {title} {\bibinfo {title} {Quantum microwaves: {Stabilizing} squeezed light by phase locking},\ }\href {https://doi.org/10.1103/PhysRevB.111.184519} {\bibfield  {journal} {\bibinfo  {journal} {Phys. Rev. B}\ }\textbf {\bibinfo {volume} {111}},\ \bibinfo {pages} {184519} (\bibinfo {year} {2025})}\BibitemShut {NoStop}%
\bibitem [{\citenamefont {Rousseau}\ \emph {et~al.}(2025)\citenamefont {Rousseau}, \citenamefont {Ruiz}, \citenamefont {Albertinale}, \citenamefont {d'Avezac}, \citenamefont {Banys}, \citenamefont {Blandin}, \citenamefont {Bourdaud}, \citenamefont {Campanaro}, \citenamefont {Cardoso}, \citenamefont {Cottet}, \citenamefont {Cullip}, \citenamefont {Del{\'e}glise}, \citenamefont {Devanz}, \citenamefont {Devulder}, \citenamefont {Essig}, \citenamefont {F{\'e}vrier}, \citenamefont {Gicquel}, \citenamefont {Gouzien}, \citenamefont {Gras}, \citenamefont {Guillaud}, \citenamefont {G{\"u}m{\"u}{\c s}}, \citenamefont {Hall{\'e}n}, \citenamefont {Jacob}, \citenamefont {Magnard}, \citenamefont {Marquet}, \citenamefont {Miklass}, \citenamefont {Peronnin}, \citenamefont {Polis}, \citenamefont {Rautschke}, \citenamefont {R{\'e}glade}, \citenamefont {Roul}, \citenamefont {Stevens}, \citenamefont {Solard}, \citenamefont {Thomas}, \citenamefont {Ville}, \citenamefont {Wan-Fat}, \citenamefont {Lescanne}, \citenamefont {Leghtas}, \citenamefont {Cohen}, \citenamefont {Jezouin},\ and\ \citenamefont {Murani}}]{rousseau_enhancing_2025}%
  \BibitemOpen
  \bibfield  {author} {\bibinfo {author} {\bibfnamefont {R.}~\bibnamefont {Rousseau}}, \bibinfo {author} {\bibfnamefont {D.}~\bibnamefont {Ruiz}}, \bibinfo {author} {\bibfnamefont {E.}~\bibnamefont {Albertinale}}, \bibinfo {author} {\bibfnamefont {P.}~\bibnamefont {d'Avezac}}, \bibinfo {author} {\bibfnamefont {D.}~\bibnamefont {Banys}}, \bibinfo {author} {\bibfnamefont {U.}~\bibnamefont {Blandin}}, \bibinfo {author} {\bibfnamefont {N.}~\bibnamefont {Bourdaud}}, \bibinfo {author} {\bibfnamefont {G.}~\bibnamefont {Campanaro}}, \bibinfo {author} {\bibfnamefont {G.}~\bibnamefont {Cardoso}}, \bibinfo {author} {\bibfnamefont {N.}~\bibnamefont {Cottet}}, \bibinfo {author} {\bibfnamefont {C.}~\bibnamefont {Cullip}}, \bibinfo {author} {\bibfnamefont {S.}~\bibnamefont {Del{\'e}glise}}, \bibinfo {author} {\bibfnamefont {L.}~\bibnamefont {Devanz}}, \bibinfo {author} {\bibfnamefont {A.}~\bibnamefont {Devulder}}, \bibinfo {author} {\bibfnamefont {A.}~\bibnamefont {Essig}}, \bibinfo {author} {\bibfnamefont {P.}~\bibnamefont {F{\'e}vrier}}, \bibinfo {author} {\bibfnamefont {A.}~\bibnamefont {Gicquel}}, \bibinfo {author} {\bibfnamefont {{\'E}.}~\bibnamefont {Gouzien}}, \bibinfo {author} {\bibfnamefont {A.}~\bibnamefont {Gras}}, \bibinfo {author} {\bibfnamefont {J.}~\bibnamefont {Guillaud}}, \bibinfo {author} {\bibfnamefont {E.}~\bibnamefont {G{\"u}m{\"u}{\c s}}}, \bibinfo {author} {\bibfnamefont {M.}~\bibnamefont {Hall{\'e}n}}, \bibinfo {author} {\bibfnamefont {A.}~\bibnamefont {Jacob}}, \bibinfo {author} {\bibfnamefont {P.}~\bibnamefont {Magnard}}, \bibinfo {author} {\bibfnamefont {A.}~\bibnamefont {Marquet}}, \bibinfo {author} {\bibfnamefont {S.}~\bibnamefont {Miklass}}, \bibinfo {author} {\bibfnamefont {T.}~\bibnamefont {Peronnin}}, \bibinfo {author} {\bibfnamefont {S.}~\bibnamefont {Polis}}, \bibinfo {author} {\bibfnamefont {F.}~\bibnamefont {Rautschke}}, \bibinfo {author} {\bibfnamefont {U.}~\bibnamefont {R{\'e}glade}}, \bibinfo {author} {\bibfnamefont {J.}~\bibnamefont {Roul}}, \bibinfo {author} {\bibfnamefont {J.}~\bibnamefont {Stevens}}, \bibinfo {author} {\bibfnamefont {J.}~\bibnamefont {Solard}}, \bibinfo {author} {\bibfnamefont {A.}~\bibnamefont {Thomas}}, \bibinfo {author} {\bibfnamefont {J.-L.}\ \bibnamefont {Ville}}, \bibinfo {author} {\bibfnamefont {P.}~\bibnamefont {Wan-Fat}}, \bibinfo {author} {\bibfnamefont {R.}~\bibnamefont {Lescanne}}, \bibinfo {author} {\bibfnamefont {Z.}~\bibnamefont {Leghtas}}, \bibinfo {author} {\bibfnamefont {J.}~\bibnamefont {Cohen}}, \bibinfo {author} {\bibfnamefont {S.}~\bibnamefont {Jezouin}},\ and\ \bibinfo {author} {\bibfnamefont {A.}~\bibnamefont {Murani}},\ }\bibfield  {title} {\bibinfo {title} {Enhancing dissipative cat qubit protection by squeezing},\ }\bibfield  {journal} {\bibinfo  {journal} {arXiv:2502.07892}\ }\href {https://doi.org/10.48550/arXiv.2502.07892} {10.48550/arXiv.2502.07892} (\bibinfo {year} {2025})\BibitemShut {NoStop}%
\bibitem [{\citenamefont {Simoni}\ \emph {et~al.}(2025)\citenamefont {Simoni}, \citenamefont {Rojkov}, \citenamefont {Mazzanti}, \citenamefont {Adamczyk}, \citenamefont {Ferk}, \citenamefont {Hrmo}, \citenamefont {Jain}, \citenamefont {Sägesser}, \citenamefont {Kienzler},\ and\ \citenamefont {Home}}]{simoni_non_linear_2025}%
  \BibitemOpen
  \bibfield  {author} {\bibinfo {author} {\bibfnamefont {M.}~\bibnamefont {Simoni}}, \bibinfo {author} {\bibfnamefont {I.}~\bibnamefont {Rojkov}}, \bibinfo {author} {\bibfnamefont {M.}~\bibnamefont {Mazzanti}}, \bibinfo {author} {\bibfnamefont {W.}~\bibnamefont {Adamczyk}}, \bibinfo {author} {\bibfnamefont {A.}~\bibnamefont {Ferk}}, \bibinfo {author} {\bibfnamefont {P.}~\bibnamefont {Hrmo}}, \bibinfo {author} {\bibfnamefont {S.}~\bibnamefont {Jain}}, \bibinfo {author} {\bibfnamefont {T.}~\bibnamefont {Sägesser}}, \bibinfo {author} {\bibfnamefont {D.}~\bibnamefont {Kienzler}},\ and\ \bibinfo {author} {\bibfnamefont {J.}~\bibnamefont {Home}},\ }\bibfield  {title} {\bibinfo {title} {Non-linear cooling and control of a mechanical quantum harmonic oscillator},\ }\bibfield  {journal} {\bibinfo  {journal} {arXiv:2509.05734}\ }\href {https://doi.org/10.48550/arXiv.2509.05734} {10.48550/arXiv.2509.05734} (\bibinfo {year} {2025})\BibitemShut {NoStop}%
\bibitem [{\citenamefont {Ofek}\ \emph {et~al.}(2016)\citenamefont {Ofek}, \citenamefont {Petrenko}, \citenamefont {Heeres}, \citenamefont {Reinhold}, \citenamefont {Leghtas}, \citenamefont {Vlastakis}, \citenamefont {Liu}, \citenamefont {Frunzio}, \citenamefont {Girvin}, \citenamefont {Jiang}, \citenamefont {Mirrahimi}, \citenamefont {Devoret},\ and\ \citenamefont {Schoelkopf}}]{ofek_extending_2016}%
  \BibitemOpen
  \bibfield  {author} {\bibinfo {author} {\bibfnamefont {N.}~\bibnamefont {Ofek}}, \bibinfo {author} {\bibfnamefont {A.}~\bibnamefont {Petrenko}}, \bibinfo {author} {\bibfnamefont {R.}~\bibnamefont {Heeres}}, \bibinfo {author} {\bibfnamefont {P.}~\bibnamefont {Reinhold}}, \bibinfo {author} {\bibfnamefont {Z.}~\bibnamefont {Leghtas}}, \bibinfo {author} {\bibfnamefont {B.}~\bibnamefont {Vlastakis}}, \bibinfo {author} {\bibfnamefont {Y.}~\bibnamefont {Liu}}, \bibinfo {author} {\bibfnamefont {L.}~\bibnamefont {Frunzio}}, \bibinfo {author} {\bibfnamefont {S.~M.}\ \bibnamefont {Girvin}}, \bibinfo {author} {\bibfnamefont {L.}~\bibnamefont {Jiang}}, \bibinfo {author} {\bibfnamefont {M.}~\bibnamefont {Mirrahimi}}, \bibinfo {author} {\bibfnamefont {M.~H.}\ \bibnamefont {Devoret}},\ and\ \bibinfo {author} {\bibfnamefont {R.~J.}\ \bibnamefont {Schoelkopf}},\ }\bibfield  {title} {\bibinfo {title} {Extending the lifetime of a quantum bit with error correction in superconducting circuits},\ }\href {https://doi.org/10.1038/nature18949} {\bibfield  {journal} {\bibinfo  {journal} {Nature}\ }\textbf {\bibinfo {volume} {536}},\ \bibinfo {pages} {441} (\bibinfo {year} {2016})}\BibitemShut {NoStop}%
\bibitem [{\citenamefont {Vanselow}\ \emph {et~al.}(2025)\citenamefont {Vanselow}, \citenamefont {Beauseigneur}, \citenamefont {Lattier}, \citenamefont {Villiers}, \citenamefont {Denis}, \citenamefont {Morfin}, \citenamefont {Leghtas},\ and\ \citenamefont {Campagne-Ibarcq}}]{vanselow_dissipating_2025}%
  \BibitemOpen
  \bibfield  {author} {\bibinfo {author} {\bibfnamefont {A.}~\bibnamefont {Vanselow}}, \bibinfo {author} {\bibfnamefont {B.}~\bibnamefont {Beauseigneur}}, \bibinfo {author} {\bibfnamefont {L.}~\bibnamefont {Lattier}}, \bibinfo {author} {\bibfnamefont {M.}~\bibnamefont {Villiers}}, \bibinfo {author} {\bibfnamefont {A.}~\bibnamefont {Denis}}, \bibinfo {author} {\bibfnamefont {P.}~\bibnamefont {Morfin}}, \bibinfo {author} {\bibfnamefont {Z.}~\bibnamefont {Leghtas}},\ and\ \bibinfo {author} {\bibfnamefont {P.}~\bibnamefont {Campagne-Ibarcq}},\ }\bibfield  {title} {\bibinfo {title} {Dissipating quartets of excitations in a superconducting circuit},\ }\bibfield  {journal} {\bibinfo  {journal} {arXiv:2501.05960}\ }\href {https://doi.org/10.48550/arXiv.2501.05960} {10.48550/arXiv.2501.05960} (\bibinfo {year} {2025})\BibitemShut {NoStop}%
\bibitem [{\citenamefont {Groszkowski}\ \emph {et~al.}(2022)\citenamefont {Groszkowski}, \citenamefont {Koppenhöfer}, \citenamefont {Lau},\ and\ \citenamefont {Clerk}}]{groszkowski_reservoir_2022}%
  \BibitemOpen
  \bibfield  {author} {\bibinfo {author} {\bibfnamefont {P.}~\bibnamefont {Groszkowski}}, \bibinfo {author} {\bibfnamefont {M.}~\bibnamefont {Koppenhöfer}}, \bibinfo {author} {\bibfnamefont {H.-K.}\ \bibnamefont {Lau}},\ and\ \bibinfo {author} {\bibfnamefont {A.~A.}\ \bibnamefont {Clerk}},\ }\bibfield  {title} {\bibinfo {title} {Reservoir-{Engineered} {Spin} {Squeezing}: {Macroscopic} {Even}-{Odd} {Effects} and {Hybrid}-{Systems} {Implementations}},\ }\href {https://doi.org/10.1103/PhysRevX.12.011015} {\bibfield  {journal} {\bibinfo  {journal} {Phys. Rev. X}\ }\textbf {\bibinfo {volume} {12}},\ \bibinfo {pages} {011015} (\bibinfo {year} {2022})}\BibitemShut {NoStop}%
\bibitem [{\citenamefont {Cahill}\ and\ \citenamefont {Glauber}(1969)}]{cahill_ordered_1969}%
  \BibitemOpen
  \bibfield  {author} {\bibinfo {author} {\bibfnamefont {K.~E.}\ \bibnamefont {Cahill}}\ and\ \bibinfo {author} {\bibfnamefont {R.~J.}\ \bibnamefont {Glauber}},\ }\bibfield  {title} {\bibinfo {title} {Ordered {Expansions} in {Boson} {Amplitude} {Operators}},\ }\href {https://doi.org/10.1103/PhysRev.177.1857} {\bibfield  {journal} {\bibinfo  {journal} {Phys. Rev.}\ }\textbf {\bibinfo {volume} {177}},\ \bibinfo {pages} {1857} (\bibinfo {year} {1969})}\BibitemShut {NoStop}%
\bibitem [{\citenamefont {Szegő}(1939)}]{szego_orthogonal_1939}%
  \BibitemOpen
  \bibfield  {author} {\bibinfo {author} {\bibfnamefont {G.}~\bibnamefont {Szegő}},\ }\href {https://doi.org/10.1090/coll/023} {\emph {\bibinfo {title} {Orthogonal {Polynomials}}}},\ \bibinfo {series} {Colloquium {Publications}}, Vol.~\bibinfo {volume} {23}\ (\bibinfo  {publisher} {American Mathematical Society},\ \bibinfo {address} {Providence, Rhode Island},\ \bibinfo {year} {1939})\BibitemShut {NoStop}%
\bibitem [{\citenamefont {Meekhof}\ \emph {et~al.}(1996)\citenamefont {Meekhof}, \citenamefont {Monroe}, \citenamefont {King}, \citenamefont {Itano},\ and\ \citenamefont {Wineland}}]{meekhof_generation_1996}%
  \BibitemOpen
  \bibfield  {author} {\bibinfo {author} {\bibfnamefont {D.~M.}\ \bibnamefont {Meekhof}}, \bibinfo {author} {\bibfnamefont {C.}~\bibnamefont {Monroe}}, \bibinfo {author} {\bibfnamefont {B.~E.}\ \bibnamefont {King}}, \bibinfo {author} {\bibfnamefont {W.~M.}\ \bibnamefont {Itano}},\ and\ \bibinfo {author} {\bibfnamefont {D.~J.}\ \bibnamefont {Wineland}},\ }\bibfield  {title} {\bibinfo {title} {Generation of {Nonclassical} {Motional} {States} of a {Trapped} {Atom}},\ }\href {https://doi.org/10.1103/PhysRevLett.76.1796} {\bibfield  {journal} {\bibinfo  {journal} {Phys. Rev. Lett.}\ }\textbf {\bibinfo {volume} {76}},\ \bibinfo {pages} {1796} (\bibinfo {year} {1996})}\BibitemShut {NoStop}%
\bibitem [{\citenamefont {Varró}(2022)}]{varro_coherent_2022}%
  \BibitemOpen
  \bibfield  {author} {\bibinfo {author} {\bibfnamefont {S.}~\bibnamefont {Varró}},\ }\bibfield  {title} {\bibinfo {title} {Coherent and incoherent superposition of transition matrix elements of the squeezing operator},\ }\href {https://doi.org/10.1088/1367-2630/ac6b4d} {\bibfield  {journal} {\bibinfo  {journal} {New J. Phys.}\ }\textbf {\bibinfo {volume} {24}},\ \bibinfo {pages} {053035} (\bibinfo {year} {2022})}\BibitemShut {NoStop}%
\end{thebibliography}%

\end{document}